\documentclass[dvips,fleqn,openright,twoside]{report}
\usepackage[latin1]{inputenc}
\usepackage{aeguill}
\usepackage[frenchb]{babel}
\usepackage{amsmath,amssymb,graphicx,color,epsfig}

\newcommand{\pubclusters}{P1}
\newcommand{\pubmatrix}{P2}
\newcommand{\pubsphe}{P3}
\newcommand{\pubwsat}{P4}
\newcommand{\pubjsp}{P5}
\newcommand{\pubbethe}{P6}
\newcommand{\pubsat}{C1}
\newcommand{\pubptac}{C2}

\begin{document}
%%\bibliographystyle{unsrt}
%% faire setenv BSTINPUTS /users/theo2/guilhem/These/
\bibliographystyle{myunsrt}

\thispagestyle{empty}
\enlargethispage{7cm}
\voffset=-2cm \hspace{-2cm} \includegraphics{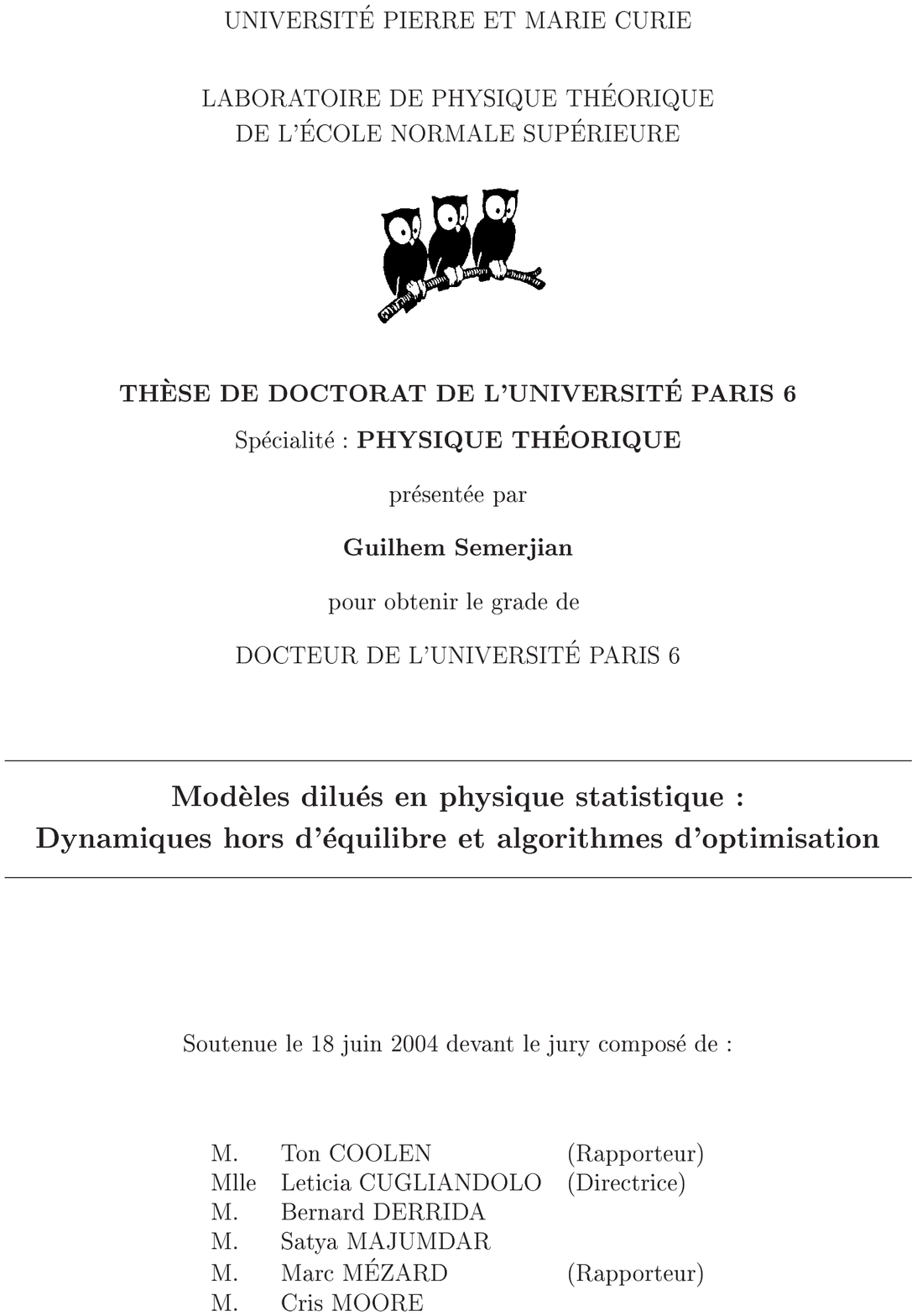}

\newpage
\voffset=0cm

\thispagestyle{empty}
\mbox{}
\newpage
\setcounter{page}{1}

\pagestyle{empty}

\noindent {\Huge \bf Remerciements}

\vspace{1cm}

Cette th\`ese s'est d\'eroul\'ee de septembre 2000 \`a juin 2004 au
laboratoire de physique th\'eorique de l'Ecole Normale Sup\'erieure.
Je remercie ses directeurs successifs Jean Iliopoulos et Eug\`ene Cremmer
d'avoir bien voulu m'y accueillir, ainsi que les
secr\'etaires Michelle Leliepvre, Marcelle Martin et Nicole 
Ribet, dont la gentillesse, la comp\'etence et le d\'evouement sont pour
beaucoup dans la qualit\'e de vie au laboratoire.

\vspace{3mm}

Je ne saurais trop insister ici sur la gratitude que j'\'eprouve envers
ma directrice de th\`ese, Leticia Cugliandolo. Ses comp\'etences scientifiques
et le temps qu'elle a consacr\'e \`a m'en faire b\'en\'eficier m'ont \'et\'e 
extr\^emement profitables. La confiance et le soutien qu'elle m'a prodigu\'es
m'ont aussi particuli\`erement aid\'e tout au long de ces quatre ann\'ees.

\vspace{3mm}

J'ai eu aussi le plaisir de collaborer directement avec Simona Cocco, R\'emi 
Monasson, Andrea Montanari et Martin Weigt. De tous j'ai beaucoup appris et je 
les remercie tr\`es chaleureusement de m'avoir fait partager leur exp\'erience.

\vspace{3mm}

Mon initiation \`a la physique statistique des syst\`emes d\'esordonn\'es
a eu lieu au cours d'un stage \`a l'universit\'e d'Oxford, je
suis tr\`es reconnaissant \`a David Sherrington d'avoir accompagn\'e
ces premiers pas.

\vspace{3mm}

Je tiens aussi \`a exprimer ma gratitude \`a l'\'egard de Bernard Derrida, 
Satya Majumdar et Cris Moore pour avoir accept\'e de participer au jury
de cette th\`ese, ainsi que mes plus vifs remerciements \`a Ton Coolen et
Marc M\'ezard pour avoir assur\'e l'exigeante t\^ache de rapporteur.

\vspace{3mm}

Parmi les chercheurs avec qui j'ai eu la chance de discuter, je tiens \`a
remercier en particulier Alain Barrat, Michel Bauer, Giulio Biroli, 
Jean-Philippe Bouchaud, Werner Krauth et Jorge Kurchan.

\vspace{3mm}

J'ai assur\'e au cours des deux derni\`eres ann\'ees des travaux dirig\'es au
sein du magist\`ere de l'Ecole Normale. Mes tr\`es sinc\`eres remerciements
vont \`a ses responsables Jean-Marc Berroir et Christophe Dupraz de m'avoir
accueilli au sein de leur \'equipe, ainsi qu'aux enseignants et charg\'es de
TD avec qui j'ai travaill\'e~: Henk Hilhorst, Jean Iliopoulos, Chi-Tuong Pham, 
Nicolas R\'egnault, Vincent Rivasseau et Jean Zinn-Justin.

\vspace{3mm}

Un grand merci aussi aux th\'esards avec qui la cohabitation dans le m\^eme 
bureau fut un plaisir, \`a savoir Christophe Deroulers, Serge Florens, Steffen 
Metzger, S\'ebastien Ray, et tout particuli\`erement Gr\'egory Schehr.

\newpage

\tableofcontents

\pagestyle{myheadings}

\chapter{Introduction}
\markboth{\hspace{3mm} \hrulefill \hspace{3mm} Ch. 1~: Introduction}{Ch. 1~: Introduction\hspace{3mm} \hrulefill \hspace{3mm}}

\section{Les verres de spin}

Le domaine de la physique dans lequel cette th\`ese prend place est n\'e 
de l'effort th\'eorique fourni pour expliquer les
r\'esultats d'exp\'eriences sur les verres de spin, compos\'es produits
en laboratoire \`a partir des ann\'ees 70~\cite{Mydosh}.
Exp\'erimentalement, un verre de spin est un alliage entre un \'el\'ement 
noble, non magn\'etique, et une faible fraction d'atomes dot\'es de 
propri\'et\'es magn\'etiques. Ce m\'elange est
effectu\'e \`a haute temp\'erature, les deux esp\`eces formant alors une phase 
liquide.
Lorsque l'alliage est refroidi, le compos\'e se solidifie; le
m\'etal noble cristallise dans un r\'eseau r\'egulier, tandis que les
\'el\'ements magn\'etiques, largement minoritaires,
jouent le r\^ole d'impuret\'es plac\'es al\'eatoirement au sein
du r\'eseau r\'egulier. 

Dans une telle situation, les interactions entre moments magn\'etiques des 
impuret\'es d\'epend de la distance les s\'eparant d'une mani\`ere tr\`es 
particuli\`ere~\footnote{Ce ph\'enom\`ene, dit interaction RKKY d'apr\`es les
noms de Ruderman, Kittel, Kasuya et Yosida, est expliqu\'e dans la plupart
des livres de physique du solide, par exemple~\cite{Ashcroft}.}. 
Comme on peut s'y attendre, l'intensit\'e de l'interaction d\'ecro\^it 
pour des atomes de plus en plus \'eloign\'es. Ce qui est notable, c'est que le 
\emph{signe} de l'interaction est une fonction oscillante de la distance les
s\'eparant. Autrement dit, certaines des interactions entre moments
magn\'etiques sont ferromagn\'etiques, tendant \`a les aligner dans
le m\^eme sens, alors que d'autres sont antiferromagn\'etiques et
favorisent donc les configurations antiparall\`eles.

La pr\'esence simultan\'ee de ces deux types d'interaction va provoquer 
une d\'ependance originale des propri\'et\'es du syst\`eme par rapport \`a 
la temp\'erature~\footnote{On reste toujours en
dessous de la temp\'erature de fusion du verre de spin, les positions des
impuret\'es magn\'etiques, et donc la valeur des interactions entre elles,
restent gel\'ees \`a partir de la pr\'eparation de l'\'echantillon.}. Rappelons
en effet la situation dans le cas o\`u toutes les interactions sont 
ferromagn\'etiques. Les propri\'et\'es d'\'equilibre d'un tel syst\`eme
r\'esultent d'une comp\'etition entre un effet \'energ\'etique, les
interactions tendant \`a aligner tous les spins dans la m\^eme direction,
et un effet entropique, l'agitation thermique favorisant plut\^ot un
\'etat d\'esordonn\'e. A haute temp\'erature c'est la contribution entropique
\`a l'\'energie libre qui est dominante, on a donc un \'etat paramagn\'etique
dans lequel les spins 
fluctuent autour d'une valeur moyenne nulle. Quand on abaisse la temp\'erature
on assiste \`a une transition de phase~: en dessous d'une temp\'erature
critique, les effets \'energ\'etiques l'emportent, et le syst\`eme acquiert une
aimantation macroscopique, tous les spins \'etant align\'es dans une direction
commune. L'aimantation macroscopique est ici un param\`etre d'ordre, qui
cro\^it continument de z\'ero quand on diminue la temp\'erature en dessous
de sa valeur critique.

Que devient cette image dans le cadre d'un verre de spin~? A haute
temp\'erature la phase paramagn\'etique n'est pas modifi\'ee, mais la phase
de basse temp\'erature est diff\'erente. Le syst\`eme cherche toujours \`a se
bloquer autour d'une configuration des spins qui minimise son \'energie, mais
cette configuration ne peut pas \^etre celle avec tous les spins align\'es,
puisqu'une partie des interactions sont antiferromagn\'etiques. On a donc
une transition de phase o\`u certains degr\'es de libert\'e se figent (ce
qui se traduit par une singularit\'e sur la chaleur sp\'ecifique de 
l'\'echantillon), mais sans apparition d'une aimantation macroscopique 
(ni m\^eme d'aucun ordre magn\'etique r\'egulier). 

Une premi\`ere mod\'elisation des verres de spin a \'et\'e propos\'ee par Edwards
et Anderson en 1975~\cite{EdAn}. Elle consiste \`a placer $N$ spins d'Ising 
sur les sommets d'un r\'eseau en dimension finie $d$, avec des couplages entre
proches voisins de signe quelconque. L'hamiltonien du mod\`ele est donc~:
\begin{equation}
H = - \sum_{\langle i ,j \rangle} J_{ij} \sigma_i \sigma_j \ , \qquad 
\sigma_i = \pm 1 \ .
\end{equation}
Les indices $i$ d\'esignent les sites du r\'eseau, et la somme porte uniquement
sur les couples de sites voisins.
Insistons sur le caract\`ere diff\'erent des variables $\sigma_i$ et $J_{ij}$. 
Ces derni\`eres sont suppos\'ees fix\'ees, ou gel\'ees {\em (quenched)}
de la m\^eme fa\c con que les impuret\'es magn\'etiques ne se
d\'eplacent pas dans un \'echantillon de verre de spin tant qu'on ne 
le fait pas fondre. Les variables $\sigma_i$ sont par contre soumises 
\`a des fluctuations thermiques, selon la loi de probabilit\'e de 
Gibbs-Boltzmann. Le param\`etre d'ordre qui d\'ecrit la transition vers
la phase de basse temp\'erature est d\'efini \`a partir des magn\'etisations 
locales $m_i = \langle \sigma_i \rangle$ (les crochets d\'esignent la
moyenne avec le poids de Gibbs-Boltzmann) comme
\begin{equation}
q_{EA} = \frac{1}{N} \sum_{i} m_i^2 \ .
\end{equation}
A cause du signe fluctuant des interactions, l'aimantation totale
$\sum_i m_i$ reste nulle dans la phase de basse temp\'erature. Le
param\`etre d'Edwards-Anderson que l'on vient de d\'efinir sera par
contre positif, chaque spin acqui\`erant une magn\'etisation non nulle, 
dont la direction fluctue d'un site \`a l'autre.

S'il \'etait n\'ecessaire de conna\^itre les positions et les
couplages entre impuret\'es magn\'etiques pour pr\'edire le comportement
d'un \'echantillon macroscopique de verre de spin, tout travail de
mod\'elisation serait vou\'e \`a l'echec, tout comme si l'on devait 
conna\^itre les positions de toutes les mol\'ecules d'un gaz pour \'etablir 
son \'equation d'\'etat. Heureusement ce n'est pas le cas~: deux 
\'echantillons de verres de spins pr\'epar\'es selon le m\^eme protocole 
exp\'erimental seront certes tr\`es diff\'erents au niveau microscopique, 
mais l'on s'attend \`a ce que leurs propri\'et\'es macroscopiques 
(chaleur sp\'ecifique, temp\'erature de transition,\dots) soient identiques. 
D'un point de vue th\'eorique, cela sugg\`ere de d\'efinir des
ensembles d'\'echantillons microscopiques qui correspondent au m\^eme
proc\'ed\'e exp\'erimental de fabrication. Au sein de cet ensemble
les variables gel\'ees microscopiques varient d'un \'echantillon \`a
l'autre, mais les observables macroscopiques sont toutes quasiment
identiques. On peut donc identifier les valeurs moyennes de ces observables 
avec les valeurs typiquement observ\'ees pour un \'echantillon donn\'e.
Dans le cadre du mod\`ele d'Edwards-Anderson par exemple, les $J_{ij}$ seront 
des variables al\'eatoires ind\'ependantes. Leur loi de probabilit\'e 
autorise des couplages positifs et n\'egatifs de mani\`ere \`a reproduire la
frustration pr\'esente dans les verres de spin. Il reste ensuite
\`a calculer la valeur moyenne de l'\'energie libre par rapport \`a cette
distribution pour pr\'edire les propri\'et\'es thermodynamiques du
syst\`eme.

\section{Mod\`eles compl\`etement connect\'es}

La r\'esolution exacte du mod\`ele d'Edwards-Anderson \`a trois dimensions
semble une t\^ache impossible~: pour le cas purement ferromagn\'etique
le calcul de la fonction de partition n'a \'et\'e effectu\'e 
qu'\`a deux dimensions, et la pr\'esence
de d\'esordre dans les interactions rend le probl\`eme encore plus difficile.
Une simplification du mod\`ele, de type champ moyen, a \'et\'e introduite
par Sherrington et Kirkpatrick (SK)~\cite{ShKi-1,ShKi-2}. Leur mod\`ele est
un analogue de celui de Curie-Weiss du ferromagn\'etisme~: chacun
des $N$ spins d'Ising du mod\`ele interagit avec tous les autres, d'o\`u le nom
de \og compl\`etement connect\'e\fg\ que l'on attribue \`a ce type de mod\`ele.
L'hamiltonien consid\'er\'e s'\'ecrit alors
\begin{equation}
H=- \sum_{i<j} J_{ij} \sigma_i \sigma_j \ ,
\label{eq:in-H_SK}
\end{equation}
la somme portant sur toutes les paires de spins. Les couplages gel\'es
$J_{ij}$ sont des variables al\'eatoires positives ou n\'egatives, 
leur variance \'etant d'ordre $N^{-1/2}$ pour que l'hamiltonien soit 
extensif~\footnote{Le \og volume\fg\ du syst\`eme est ici le nombre $N$ de 
spins.}.
Alors que la r\'esolution du mod\`ele ferromagn\'etique compl\`etement
connect\'e est triviale (cf. la section \ref{ch:curieweiss}), 
celle du mod\`ele de Sherrington-Kirkpatrick s'est av\'er\'ee tr\`es subtile 
et a conduit \`a l'introduction de concepts nouveaux en physique statistique. 
L'\'etude des mod\`eles d\'esordonn\'es de champ moyen repose souvent sur
la m\'ethode des r\'epliques. Cette m\'ethode n'est pas, dans sa formulation
originelle, compl\`etement rigoureuse d'un point de vue math\'ematique~:
une fonction calcul\'ee pour un nombre $n$ de r\'epliques doit
\^etre prolong\'ee dans la limite $n \to 0$. Comme $n$ est a priori entier,
ce prolongement n'est pas unique et n\'ecessite l'intoduction d'hypoth\`eses
suppl\'ementaires. La plus naturelle, utilis\'ee par Sherrington et Kirpatrick,
est dite \og sym\'etrique dans les r\'epliques\fg\ (RS). Cet ansatz n'est
pas correct \`a basse temp\'erature car il pr\'edit une entropie de 
temp\'erature nulle n\'egative, ce qui est impossible pour un mod\`ele dont
les degr\'es de libert\'e sont discrets.
Un calcul de stabilit\'e effectu\'e par de Almeida et Thouless~\cite{AlTh}
a montr\'e que c'\'etait l'hypoth\`ese RS qui \'etait fautive~: elle 
entra\^inait l'apparition de directions de fluctuations instables dans une
int\'egrale calcul\'ee par la m\'ethode du col.

On doit \`a Parisi \cite{Pa-RSB1,Pa-RSB2,Pa-RSB3} la formulation de l'ansatz 
correct pour le mod\`ele SK. Celui-ci brise la sym\'etrie entre les 
r\'epliques (RSB), dans un sch\'ema it\'eratif~: les $n$ r\'epliques sont
divis\'es en groupes de $m$ r\'epliques, eux-m\^emes sous-divis\'es, et ainsi
de suite. Pour une reproduction des articles importants
de cette \'epoque ainsi que pour une discussion de la signification du
ph\'enom\`ene de RSB, on pourra se reporter \`a \cite{Beyond}. Mentionnons
simplement que cette brisure de sym\'etrie est reli\'ee \`a la nature 
particuli\`ere de la phase de basse temp\'erature des verres de spin. 
Un syst\`eme d'Ising ferromagn\'etique poss\`ede deux \og \'etats purs\fg\ 
en dessous de la temp\'erature critique, correspondant aux deux signes 
possibles de la magn\'etisation~: tous les moments magn\'etiques s'alignent
dans une direction donn\'ee de fa\c con \`a minimiser l'\'energie du syst\`eme.
Dans un verre de spins tel que le mod\`ele SK, la situation est beaucoup plus 
compliqu\'ee~: la frustration induite par les signes al\'eatoires des 
couplages entra\^ine une d\'eg\'enerescence des configurations de 
basse \'energie. On a donc un grand nombre d'\'etats purs, l'ansatz de Parisi
traduisant leur organisation dans l'espace des configurations du syst\`eme. 

Une preuve rigoureuse de l'exactitude de l'\'energie libre pr\'edite par 
l'ansatz de Parisi pour le mod\`ele SK est apparue tr\`es r\'ecemment.
Cette preuve, finalis\'ee par Talagrand~\cite{Ta-preuveSK}, s'appuie sur une 
m\'ethode d'interpolation due \`a Guerra~\cite{Gu-interpolation}.

On peut reformuler le probl\`eme des verres de spin d'une mani\`ere
l\'eg\`erement diff\'erente. A tr\`es basse temp\'erature, le comportement
d'un syst\`eme physique est, de mani\`ere g\'en\'erale, d\'etermin\'e
par ses configurations de plus basse \'energie. Pour un mod\`ele avec
$N$ spins d'Ising, il y a $2^N$ configurations $\vec{\sigma}$, 
chacune avec une \'energie $E_{\vec{\sigma}}$. Ces \'energies sont, dans un 
mod\`ele d\'esordonn\'e, des variables al\'eatoires~: elles d\'ependent en 
effet des interactions gel\'ees $J_{ij}$. Il conviendrait donc, pour 
comprendre le syst\`eme \`a basse temp\'erature, d'\'etudier les propri\'et\'es
statistiques du minimum de ces $2^N$ variables al\'eatoires, ou plus 
g\'en\'eralement des $k$ plus petites. Il est assez facile de d\'eterminer
les propri\'et\'es des extr\`emes de variables al\'eatoires ind\'ependantes. 
La difficult\'e dans un verre de spin est 
que les \'energies des configurations sont des variables al\'eatoires 
\emph{corr\'el\'ees}~: elles d\'ependent toutes d'un nombre beaucoup plus 
petit de variables ind\'ependantes, les $N^2$ couplages $J_{ij}$. 
Derrida~\cite{De-REM} a introduit le mod\`ele \`a \'energies al\'eatoires 
(REM) comme une simplification d'un probl\`eme de verres de spin, dans
lequel les \'energies des $2^N$ configurations sont ind\'ependantes. 
Ce mod\`ele pr\'esente une transition vers une phase de basse temp\'erature 
dans laquelle le syst\`eme se g\`ele sur un petit nombre de configurations
de basse \'energie. Dans ce m\^eme article il est aussi montr\'e que le REM
peut s'obtenir \`a partir de la g\'en\'eralisation suivante du mod\`ele SK.
Dans ce dernier, les variables $\sigma_i$ interagissent par paires.
Dans les mod\`eles dits \og $p$-spin\fg , les variables interagissent
$p$ par $p$. L'hamiltonien de champ moyen correspondant est donc~:
\begin{equation}
H=- \sum_{i_1<\dots<i_p} J_{i_1 \dots i_p} \sigma_{i_1} \dots \sigma_{i_p} \ .
\end{equation}
La somme porte sur tous les $p$-uplets de variables possibles, et les
intensit\'es des interactions $J_{i_1 \dots i_p}$ sont des variables
al\'eatoires ind\'ependantes. Dans la limite $p\to\infty$ (en r\'e\'echellant
judicieusement l'intensit\'e des couplages), les \'energies de deux 
configurations perdent toute corr\'elation, et on retrouve donc la d\'efinition
du REM.
Cette limite a aussi \'et\'e \'etudi\'ee par Gross et M\'ezard~\cite{GrMe},
qui ont trouv\'e que la phase de basse temp\'erature \'etait alors d\'ecrite
par une version simplifi\'ee de l'ansatz de Parisi~: un seul pas dans
le sch\'ema it\'eratif d\'ecrit plus haut est n\'ecessaire, on parle alors
de 1RSB~\footnote{Gardner~\cite{Ga} a montr\'e que pour $p$ fini, il existe une
deuxi\`eme temp\'erature critique plus basse, en dessous de laquelle le 
syst\`eme doit \^etre d\'ecrit par une brisure compl\`ete de la sym\'etrie des 
r\'epliques.}, alors que dans le mod\`ele SK il faut faire un nombre infini
de pas de brisure des sous-groupes de r\'epliques.

Une interpr\'etation de la brisure \`a un pas de la sym\'etrie des r\'epliques
en termes de propri\'et\'es extr\'emales de variables al\'eatoires 
corr\'el\'ees a \'et\'e donn\'e par Bouchaud et M\'ezard~\cite{BoMe}.

\section{Analogies avec des probl\`emes d'optimisation}

Il existe une analogie, dont l'exploitation s'est av\'er\'ee tr\`es 
fructueuse, entre
les syst\`emes physiques du type verres de spin et les probl\`emes
d'optimisation \'etudi\'es en math\'ematiques et en informatique. 

Un des plus c\'el\`ebres exemples de probl\`eme d'optimisation est
s\^urement celui du voyageur de commerce~: \'etant donn\'ees les positions
de $N$ villes, on voudrait conna\^itre le chemin ferm\'e qui passe une et
une seule fois par toutes les villes, et qui minimise le kilom\'etrage
parcouru par le repr\'esentant. Autrement dit, on veut minimiser une
fonction de co\^ut (la distance totale \`a parcourir) par rapport \`a certains
degr\'es de libert\'e (l'ordre dans lequel on visite les villes), en maintenant
certains param\`etres (la position des villes) fixes. Si l'on remplace 
\og fonction de co\^ut\fg\ par \'energie, \og position des villes\fg\ par
interactions gel\'ees, et \og trajet du voyageur\fg\ par configuration des
variables thermiques, on a traduit le probl\`eme d'optimisation en la recherche
du fondamental d'un syst\`eme de physique statistique, c'est-\`a-dire en
l'\'etude de ses propri\'et\'es de basse temp\'erature. 

Un probl\`eme d'optimisation est d\'efini par certaines r\`egles (ici, chercher
le chemin de longueur minimale) et par des donn\'ees qui caract\'erisent une
instance particuli\`ere du probl\`eme (la position sp\'ecifique de chacune des
villes). \og Instance\fg\ est donc la traduction d'\'echantillon dans ce
nouveau langage. De plus, il est souvent int\'eressant de d\'efinir un 
ensemble d'instances muni d'une loi de probabilit\'e (par exemple en 
r\'epartissant uniform\'ement $N$ villes dans un carr\'e de c\^ot\'e $L$), et
de s'int\'eresser aux propri\'et\'es statistiques du probl\`eme d'optimisation
correspondant (notamment la distribution des longueurs optimales des tourn\'ees
du voyageur de commerce). Ceci correspond aux distributions du d\'esordre 
gel\'e utilis\'ees dans les mod\'elisations des verres de spin.

Cette analogie a \'et\'e exploit\'ee d\`es les ann\'ees 80 dans deux directions
compl\'ementaires~: les m\'ethodes analytiques d\'evelopp\'ees pour l'\'etude
des verres de spin ont \'et\'e r\'eemploy\'ees dans ce contexte, voir par
exemple~\cite{MePa-TSP} pour une \'etude du probl\`eme du voyageur de 
commerce. Il a d'autre part \'et\'e sugg\'er\'e d'utiliser 
les proc\'edures num\'eriques du type Monte Carlo pour trouver des solutions
aux probl\`emes d'optimisation, une d\'emarche intitul\'ee \og simulated 
annealing\fg ~\cite{KiGeVe} . L'id\'ee consiste \`a introduire une 
temp\'erature fictive et 
\`a \'echantillonner l'ensemble des configurations avec le poids de Boltzmann 
correspondant. La limite de temp\'erature nulle de cet \'echantillonnage doit 
conduire au fondamental du syst\`eme, c'est-\`a-dire \`a la solution du 
probl\`eme d'optimisation. En fait l'espace des configurations des
probl\`emes d'optimisation est souvent tr\`es irr\'egulier, et cette
proc\'edure peut rester bloqu\'ee dans des \'etats m\'etastables d'\'energie
plus \'elev\'ee que celle de la configuration optimale.

\section{Mod\`eles dilu\'es}

Dans le mod\`ele SK, chaque degr\'e de libert\'e $\sigma_i$ interagit avec
tous les autres. Cette connectivit\'e infinie est tr\`es \'eloign\'ee 
de celle d'un syst\`eme r\'eel, ou d'un mod\`ele sur un r\'eseau 
g\'eom\'etrique de dimension $d$, chaque variable n'ayant alors qu'un nombre 
fini de voisins ($2d$ pour un r\'eseau hypercubique). Viana et 
Bray~\cite{ViBr} ont introduit un mod\`ele dans lequel la connectivit\'e des 
variables reste finie dans la limite thermodynamique. L'hamiltonien est 
toujours de la forme (\ref{eq:in-H_SK}), mais la loi de distribution des 
couplages est maintenant
\begin{equation}
\mbox{Prob}(J_{ij}) = \left( 1 - \frac{c}{N} \right) \delta(J_{ij}) + 
\frac{c}{N} \pi(J_{ij}) \ ,
\label{eq:in-VB}
\end{equation}
o\`u $\delta$ est la distribution de Dirac et $\pi$ une distribution de
probabilit\'e r\'eguli\`ere. Pour un site $i$ donn\'e, il y a en moyenne 
$c$ interactions $J_{ij}$ non nulles, autrement dit le site interagit avec 
$c$ voisins. Bien que la connectivit\'e soit finie, ce mod\`ele 
est de type champ moyen~: les voisins sont choisis al\'eatoirement parmi les 
$N$ sites du syst\`eme, il n'y a donc pas de notion de distance euclidienne 
entre sites. L'essentiel du travail de th\`ese pr\'esent\'e ici concerne de
tels \og mod\`eles dilu\'es\fg\ ayant une connectivit\'e locale finie sans
pour autant respecter une g\'eom\'etrie euclidienne. On rencontrera d'autres 
exemples de ces mod\`eles dans la suite. L'int\'er\^et port\'e \`a ces
probl\`emes peut \^etre motiv\'e par quelques remarques~:

\vspace{3mm}

\begin{itemize}

\item Une question ouverte concerne la pertinence en dimension finie de 
l'image obtenue pour les mod\`eles compl\`etement connect\'es par la m\'ethode 
des r\'epliques~\footnote{Une approche assez radicalement diff\'erente des 
verres de spin en dimension finie est donn\'e par l'image des 
gouttelettes~\cite{FiHu1,FiHu2}, dans laquelle la phase de basse temp\'erature 
ne comporte qu'un nombre fini d'\'etats purs.}. Les mod\`eles dilu\'es sont 
certes toujours de type champ moyen, mais corrigent la peu vraisemblable 
connectivit\'e infinie de leurs pr\'ed\'ecesseurs. On peut donc esp\'erer 
que leurs propri\'et\'es seront plus proches de celles des syst\`emes de 
dimension finie.

\item Ils comportent en particulier de nouveaux ingr\'edients physiques par
rapport aux mod\`eles compl\`etement connect\'es. En effet,
dans le mod\`ele de Viana-Bray la connectivit\'e locale d'un site est une
variable fluctuante. Ceci va entra\^iner l'apparition de ph\'enom\`enes de
type phase de Griffiths~\cite{Griffiths,BrHu}, \`a cause d'\'ev\`enements 
rares 
concernant des zones du syst\`eme qui interagissent plus fortement que la 
moyenne. En cons\'equence, la relaxation vers l'\'equilibre dans une telle 
phase est anormalement lente, \`a cause d'une distribution large des temps 
d'\'equilibration des sous-syst\`emes~\cite{Br-Griffiths,RaSePa}.

\item Une derni\`ere motivation r\'eside dans l'analogie d\'ej\`a \'evoqu\'ee 
avec les probl\`emes d'optimisation. Il se trouve que les probl\`emes
centraux en th\'eorie de l'optimisation combinatoire conduisent, une fois
traduits en termes physiques, \`a des mod\`eles de spin qui ont une 
connectivit\'e finie. L'exemple le plus frappant est celui de la
\og satisfiabilit\'e\fg , auquel on trouvera une tr\`es bonne introduction
dans~\cite{Hayes}. Une instance de ce probl\`eme, appel\'ee formule, est 
d\'efinie par un jeu de contraintes logiques, dites clauses, sur des variables 
bool\'eenes. Le probl\`eme consiste \`a d\'eterminer l'existence ou pas d'une 
configuration des variables satisfaisant toutes les contraintes, autrement dit
une solution de la formule. 
La satisfiabilit\'e joue un r\^ole central dans la th\'eorie de la 
complexit\'e computationnelle, c'est en effet le premier probl\`eme dont
la NP-compl\'etude ait \'et\'e d\'emontr\'ee~\cite{Cook}. 
Par ailleurs un ensemble al\'eatoire de formules aux propri\'et\'es 
remarquables a \'et\'e d\'ecouvert~\cite{MiSeLe-rsat}.
Celui-ci est d\'efini par deux param\`etres, le nombre de variables $N$ et
un ratio de contraintes par variables $\alpha$. Quand $\alpha$ est tr\`es petit
les formules sont peu contraintes et poss\`edent beaucoup de
solutions. Si au contraire $\alpha$ est tr\`es grand des contradictions
logiques apparaissent et les formules ne peuvent plus \^etre satisfaites.
Le point remarquable est que dans la \og limite thermodynamique\fg\
$N \to \infty$, le passage d'un r\'egime \`a l'autre se fait de mani\`ere 
abrupte~: il existe une valeur seuil $\alpha_c$ s\'eparant les formules
presque toujours~\footnote{C'est \`a dire avec une probabilit\'e tendant vers 
un dans la limite thermodynamique.} satisfiables de celles presque toujours 
insatisfiables. En termes physiques il existe une transition de phase pour 
cette valeur du param\`etre de contr\^ole. On trouvera des d\'efinitions 
pr\'ecises du probl\`eme de la satisfiabilit\'e
et de l'ensemble al\'eatoire de formules dans la partie \ref{sec:walksat},
accompagn\'ees de r\'ef\'erences plus compl\`etes.

\end{itemize}

\vspace{3mm}

La contrepartie de cette richesse physique des mod\`eles dilu\'es est une
plus grande difficult\'e technique par rapport aux mod\`eles compl\`etement
connect\'es~\footnote{D'aucuns classeront cette 
difficult\'e suppl\'ementaire au rang des motivations.}. 
Comparons en effet grossi\`erement le mod\`ele de 
Sherrington-Kirpatrick \`a celui de Viana-Bray. Dans le premier,
chaque spin subit une faible influence de la part d'un grand nombre de voisins,
alors que dans le deuxi\`eme il a un nombre fini de voisins avec qui  
il interagit fortement. La premi\`ere situation est typique
du th\'eor\`eme central limite sur les sommes de variables al\'eatoires, 
l'\og influence\fg\ (plus pr\'ecis\'ement le champ magn\'etique effectif) 
ressentie par un spin est donc une variable al\'eatoire gaussienne. 
Dans le cas du mod\`ele de Viana-Bray, comme dans celui des autres mod\`eles
dilu\'es, cette simplification dispara\^it. Ceci explique que la mise au point 
de l'ansatz brisant la sym\'etrie des r\'epliques a \'et\'e bien plus tardif 
que pour le mod\`ele SK. En effet, m\^eme dans l'approximation de sym\'etrie 
des r\'epliques, le param\`etre d'ordre, qui \'etait un simple nombre 
$q_{\rm EA}$ pour le mod\`ele SK, devient une fonction dans les mod\`eles 
dilu\'es. Ce param\`etre d'ordre RS, solution d'une \'equation fonctionnelle, 
a \'et\'e d'abord calcul\'e pour le mod\`ele de Viana-Bray
au voisinage de la ligne de transition~\cite{ViBr} 
et \`a temp\'erature nulle~\cite{KaSo}. 

Le probl\`eme de la satisfiabilit\'e al\'eatoire \'evoqu\'e ci-dessus a \'et\'e
introduit dans la communaut\'e de physique statistique par Monasson et 
Zecchina~\cite{MoZe-KSAT}. Suivant l'analogie habituelle, les variables 
bool\'eennes sont repr\'esent\'ees par des spins d'Ising et l'on
introduit une fonction \'energie qui compte le nombre de contraintes non
satisfaites. Une formule est donc satisfiable (poss\`ede des solutions)
si le fondamental a une \'energie nulle, et le ph\'enom\`ene de seuil \`a 
$\alpha_c$ se traduit par une transition de phase vers un r\'egime 
o\`u l'\'energie du fondamental est non-nulle. Le mod\`ele de
spin ainsi obtenu a une connectivit\'e finie, \`a l'instar de celui de 
Viana-Bray.

Le traitement de ce probl\`eme dans le cadre de l'ansatz sym\'etrique des
r\'epliques reproduit l'existence d'une transition de satisfiabilit\'e, 
mais la valeur du seuil pr\'edit n'est pas en accord avec les simulations 
num\'eriques. Il faut donc briser la sym\'etrie des r\'epliques, une t\^ache 
particuli\`erement difficile pour ces syst\`emes dilu\'es. 
Le param\`etre d'ordre, qui est d\'ej\`a une fonction au niveau RS, devient 
une fonctionnelle au niveau du premier pas de brisure de la sym\'etrie des 
r\'epliques~\cite{deDoMo,Mo-c-sigma}. Une
r\'esolution approch\'ee de ces \'equations 1RSB a \'et\'e obtenue par
Biroli, Monasson et Weigt~\cite{BiMoWe} \`a l'aide d'une approche 
variationnelle. Celle-ci a notamment conduit \`a une image plus raffin\'ee
des propri\'et\'es des formules al\'eatoires. En effet, en plus de la
transition de satisfiabilit\'e \`a $\alpha_c$, une deuxi\`eme valeur du
param\`etre $\alpha$ s\'epare un r\'egime satisfiable o\`u les solutions de la
formule sont r\'eparties uniform\'ement dans l'espace des configurations d'un
autre o\`u elles se regroupent en groupes de solutions nettement s\'epar\'es 
les uns des autres.

Un nouveau cap dans la compr\'ehension des syst\`emes d\'esordonn\'es
dilu\'es a \'et\'e franchi par M\'ezard et Parisi~\cite{MePa-bethe}.
Ces auteurs ont reconsid\'er\'e le probl\`eme des verres de spin \`a 
connectivit\'e finie par la m\'ethode de la cavit\'e, une m\'ethode
\'equivalente \`a celle des r\'epliques mais qui, dans le cas des
syst\`emes dilu\'es, conduit \`a des \'equations ayant une forme plus
facile \`a traiter. En particulier, celles obtenues au niveau du premier
pas de RSB peuvent \^etre r\'esolues num\'eriquement par une m\'ethode
de dynamique de populations. Cette approche a ensuite \'et\'e
utilis\'ee dans le cas de la satisfiabilit\'e par M\'ezard et 
Zecchina~\cite{MeZe-SP}, qui ont calcul\'e le seuil $\alpha_c$ et mis \`a
profit l'image de l'espace des configurations sugg\'er\'e par les calculs
1RSB pour proposer un nouvel algorithme de r\'esolution des formules, 
intitul\'e \og survey propagation\fg . Plus r\'ecemment des calculs de
stabilit\'e de la solution 1RSB par Montanari et 
Ricci-Tersenghi~\cite{MoRi-stab,MoPaRi-stab} ont montr\'e que la solution 
1RSB n'\'etait pas stable pour toutes les valeurs de $\alpha$, et que dans
certaines r\'egions une brisure compl\`ete de la sym\'etrie des r\'epliques 
\'etait n\'ecessaire. 

Cette m\'ethode de la cavit\'e a \'et\'e tr\`es f\'econde, un grand nombre de
mod\`eles ont \'et\'e \'etudi\'es le long de ces lignes. Je citerai notamment
le probl\`eme du coloriage de graphes~\cite{MuPaWeZe-col} et le probl\`eme
de la XORSAT~\cite{CrDa-XORSAT}. Ce dernier est une variante du probl\`eme
de la satisfiabilit\'e, qui a \'et\'e ind\'ependamment introduit en physique
et en informatique. Dans le contexte de la physique, il correspond \`a la
version dilu\'ee du mod\`ele $p$-spin. Un r\'esultat tr\`es int\'eressant
est la preuve de l'exactitude du sch\'ema 1RSB pour ce mod\`ele \`a
temp\'erature nulle~\cite{CoDuMaMo-XOR,MeRiZe-XOR}.

\section{Aspects dynamiques}

\subsection{Dynamiques physiques}

Tournons-nous maintenant vers les propri\'et\'es dynamiques des
verres de spin et de leurs mod\'elisations de champ moyen, qui
seront l'objet principal de ce manuscrit. Rappelons d'abord deux propri\'et\'es
auxquelles on s'attend pour un syst\`eme \og normal\fg~:

\vspace{3mm}

\begin{itemize}

\item Lorsqu'on le met en contact avec un thermostat, le syst\`eme se met
rapidement \`a l'\'equilibre avec le bain thermique ext\'erieur. Il perd alors
la m\'emoire de son instant de pr\'eparation et sa dynamique devient
stationnaire~: les r\'esultats d'une exp\'erience sont ind\'ependants de 
l'instant auquel on l'effectue.

\item Les fonctions d'auto-corr\'elation et de r\'eponse \`a
une perturbation ext\'erieure sont reli\'ees par le th\'eor\`eme de 
fluctuation-dissipation (FDT). Ce th\'eor\`eme, dont une des versions les plus
c\'el\`ebres est la relation de Stokes-Einstein entre coefficients de 
diffusion et
de tra\^in\'ee d'une particule brownienne, est tr\`es g\'en\'eral. Il ne
fait pas intervenir les d\'etails microscopiques du syst\`eme, mais seulement
la temp\'erature du thermostat ext\'erieur.
\end{itemize}

\vspace{3mm}

On dit qu'une dynamique est \og d'\'equilibre\fg\ quand elle
v\'erifie ces deux propri\'et\'es. Pour un syst\`eme \og normal\fg\ ce
r\'egime dynamique est atteint rapidement, sur l'\'echelle des temps
exp\'erimentaux, apr\`es sa mise en contact avec un thermostat.

Une \'etude de la dynamique du mod\`ele SK \'etait pr\'esente dans l'article
originel~\cite{ShKi-2}, et fut compl\'et\'ee par Sompolinsky et 
Zippelius~\cite{SoZi}. Cependant ces approches n'\'etaient pas coh\'erentes 
dans la phase de basse temp\'erature pour la raison suivante~: les mod\`eles
de verres de spin ne sont pas \og normaux\fg , ils restent hors d'\'equilibre
pendant des temps tr\`es longs~\footnote{Plus pr\'ecis\'ement, le temps
d'\'equilibration du syst\`eme diverge avec la taille du syst\`eme. Si la
limite thermodynamique est prise avant la limite des temps longs, on reste 
toujours hors d'\'equilibre.}. Il faut donc abandonner l'hypoth\`ese
de stationnarit\'e du syst\`eme et le th\'eor\`eme de fluctuation-dissipation.

La premi\`ere \'etude de la dynamique de basse temp\'erature des mod\`eles
de verres de spin en champ moyen qui tienne compte de cette situation
est due \`a Cugliandolo et Kurchan~\cite{CuKu-sphe}.
Elle portait sur la dynamique de Langevin du mod\`ele $p$-spin 
sph\'erique\footnote{Dans un mod\`ele sph\'erique les variables d'Ising 
$\sigma_i$ sont remplac\'ees par des variables continues, on reviendra en 
d\'etail sur ce point dans le corps du manuscrit.}, et mettait en lumi\`ere
les modifications suivantes des propri\'et\'es d'\'equilibre~:

\vspace{3mm}

\begin{itemize}
\item ce mod\`ele n'est pas stationnaire \`a basse temp\'erature, on dit qu'il
vieillit~\cite{Struik}~: 
sa dynamique d\'epend toujours de son \og \^age\fg , c'est-\`a-dire
du temps \'ecoul\'e depuis sa mise en contact avec le thermostat ext\'erieur.
En termes plus techniques, les fonctions de corr\'elation et de r\'eponse
\`a deux temps d\'ependent r\'eellement des deux temps, et non pas seulement
de la diff\'erence entre les deux comme dans une dynamique stationnaire.

De plus, la d\'ependance de la dynamique vis-\`a-vis de l'\^age du syst\`eme
n'est pas compl\`etement arbitraire~: le syst\`eme reste certes hors 
d'\'equilibre \`a tous les temps, mais son \'evolution est de plus en plus
lente. Cette particularit\'e permet des simplifications dans le traitement
analytique de ce type de syst\`eme.

\item le th\'eor\`eme de fluctuation-dissipation n'est pas respect\'e par
les fonctions de corr\'elation et de r\'eponse \`a deux temps. Cependant,
ces derni\`eres sont reli\'ees par une modification relativement simple du
FDT, le caract\`ere hors d'\'equilibre se traduisant par l'apparition d'une
\emph{temp\'erature effective} diff\'erente de celle du thermostat ext\'erieur.

\end{itemize}

\vspace{3mm}

Les mod\`eles de verres de spin de champ moyen constituent donc une classe
particuli\`ere de syst\`emes hors d'\'equilibre, pour lesquels la violation
des propri\'et\'es d'\'equilibre ob\'eit \`a un sc\'enario assez 
pr\'ecis~\cite{BoCuKuMe-random,Cu-Houches}.
J'en donnerai plus de d\'etails dans la partie \ref{sec:ft-outofeq}.

Les ph\'enom\`enes de vieillissement et de violation du FDT ne sont
pas des artefacts de la mod\'elisation th\'eorique~: ils ont \'et\'e
au contraire observ\'es dans un grand nombre d'exp\'eriences.
Celles-ci sont notamment conduites sur des verres de spin~\cite{Sitges},
pour lesquels une mesure directe de la violation du th\'eor\`eme
de fluctuation-dissipation a \'et\'e obtenue r\'ecemment par H\'erisson et
Ocio~\cite{HeOc}. Les verres structuraux pr\'esentent aussi un comportement
similaire~\cite{Kob-Houches}, avec une phase vitreuse vieillissante \`a basse
temp\'erature. Signalons dans ce dernier cas que la th\'eorie de couplage
de modes~\cite{Go-MCT} utilis\'ee pour la description des liquides surfondus
au dessus de la temp\'erature de transition vitreuse est intimement
reli\'ee aux mod\`eles $p$-spin introduits ci-dessus. Ce lien a \'et\'e
d\'evoil\'e par Kirkpatrick, Thirumalai et Wolynes~\cite{KiTh1,KiTh2,KiWo} et 
approfondi dans~\cite{BoCuKuMe-MCA}.

Les travaux sur la dynamique des mod\`eles dilu\'es sont relativement rares
compar\'es \`a ceux sur les mod\`eles compl\`etement connect\'es. On peut
citer en particulier les investigations num\'eriques de Barrat et 
Zecchina~\cite{BaZe}, et de Montanari et 
Ricci-Tersenghi~\cite{MoRi-aging,MoRi-cooling}. Ces \'etudes ont mis en
\'evidence la richesse du comportement dynamique de ces mod\`eles, notamment
\`a cause des fluctuations locales de connectivit\'e qui les rendent
tr\`es h\'et\'erog\`enes.

\subsection{Algorithmes d'optimisation}

La section pr\'ec\'edente \'etait intitul\'ee \og dynamiques physiques\fg\
car elles concernaient des mod\'elisations cens\'ees repr\'esenter 
l'influence d'un bain thermique ext\'erieur au syst\`eme qui lui impose
sa temp\'erature. L'\'evolution des degr\'es de libert\'e du syst\`eme
v\'erifient alors certaines r\`egles \og physiques\fg\ (condition de balance 
d\'etaill\'ee pour des spins discrets, \'equations de Langevin pour des
variables continues) telles que l'\'equilibre de Gibbs-Boltzmann est un
point fixe de l'\'evolution. 

Au cours de cette th\`ese je me suis int\'eress\'e aussi \`a une autre famille
de dynamiques, reli\'ee aux probl\`emes d'optimisation. Dans ce contexte, il
n'y a aucune raison a priori d'imposer les m\^emes r\`egles aux lois
microscopiques d'\'evolution~: le but est de r\'epondre le plus rapidement
possible \`a une question, par exemple l'existence d'une solution \`a un
probl\`eme de satisfiabilit\'e, et non d'\'echantillonner l'espace des
configurations avec le poids de Gibbs. De plus, certains types d'algorithme
n'ont rien \`a voir avec les dynamiques stochastiques locales dans l'espace 
des configurations qui sont habituellement utilis\'ees en physique. 
Les mouvements d'une configuration \`a l'autre peuvent \^etre arbitrairement 
grands, ou bien s'effectuer dans un espace peu naturel du point de vue
physique. Certains algorithmes de r\'esolution de la satisfiabilit\'e 
proc\'edent par construction d'un arbre de recherche, dans lequel chaque
n\oe ud est associ\'e \`a un ensemble de configurations des variables 
bool\'eennes.

On pourrait alors se demander quelle est la pertinence des outils de la
physique statistique pour \'etudier de tels probl\`emes. Les travaux initi\'es
par Cocco et Monasson~\cite{CoMo} ont cependant montr\'e qu'une telle approche
\'etait possible et fructueuse, compl\'etant les \'etudes rigoureuses des
math\'ematiciens et des informaticiens. On trouvera dans la publication 
\pubptac\ une revue des travaux de la communaut\'e de physique statistique
sur ces probl\`emes d'algorithmes d'optimisation.

\section{R\'esum\'e du travail de th\`ese}

On va pr\'esenter dans la suite du manuscrit les r\'esultats de travaux
plus ou moins directement reli\'es au m\^eme objectif~: une meilleure 
compr\'ehension analytique des dynamiques hors d'\'equilibre dans les 
mod\`eles dilu\'es. Au del\`a du d\'efi technique que cet objectif
repr\'esente, il serait appr\'eciable de capturer par une approche
analytique certains des traits nouveaux de ces mod\`eles qui \'etaient
absents dans le cas compl\`etement connect\'e.
Une deuxi\`eme direction de travail a consist\'e \`a appliquer des
m\'ethodes de physique statistique pour d\'ecrire le comportement d'un
algorithme de recherche locale de solutions du probl\`eme de la 
satisfiabilit\'e.

Le manuscrit est organis\'e de la mani\`ere suivante. Un premier chapitre
pr\'ecise les d\'efinitions et les propri\'et\'es g\'eom\'etriques des
mod\`eles dilu\'es. A cette occasion on d\'ecrira une m\'ethode g\'en\'erale
de d\'eveloppement \`a faible connectivit\'e, qui a fait l'objet de la 
publication \pubclusters\ et qui a \'et\'e mise \`a profit dans d'autres
parties de la th\`ese.

Les deux chapitres suivants ont \'et\'e divis\'es selon la nature (continue
ou discr\`ete) des variables des mod\`eles. Cette division est un peu 
arbitraire, mais correspond \`a des formalismes et des m\'ethodes 
d'approximation diff\'erents. 
Le chapitre \ref{sec:continu} expose les r\'esultats des publications
\pubmatrix , \pubsphe\ et une partie de la publication \pubjsp . 
Ces diff\'erentes \'etudes sont reli\'ees ainsi~: un des mod\`ele dilu\'es les
plus simples que l'on peut imaginer correspond \`a la version sph\'erique
du mod\`ele de Viana-Bray. On montre d'abord que ce probl\`eme se ram\`ene
\`a la d\'etermination du spectre d'un certain type de matrices al\'eatoires.
On adapte alors la m\'ethode d'approximation \og \`a un seul d\'efaut\fg\ de
Biroli et Monasson~\cite{BiMo-matrix} \`a l'ensemble de matrices qui nous
int\'eresse ici (publication \pubmatrix). On tire ensuite
les cons\'equences des propri\'et\'es de ces matrices al\'eatoires sur la 
dynamique du mod\`ele (\pubsphe). 
Finalement un formalisme g\'en\'eral apte \`a traiter la
dynamique de tous les mod\`eles de champ moyen (compl\`etement connect\'es ou
dilu\'es) pour des variables continues est d\'evelopp\'e, en s'appuyant sur
une analogie avec le probl\`eme de matrices al\'eatoires. Dans le cas dilu\'e
le r\'esultat obtenu est trop formel pour pouvoir en tirer directement 
des pr\'edictions physiques, de possibles approximations sont sugg\'er\'ees.
Une diff\'erence appara\^it entre mod\`eles compl\`etement 
connect\'es et mod\`eles dilu\'es~: ces derniers ne sont pas caract\'eris\'es
uniquement par leurs fonctions de corr\'elation et de r\'eponse \`a deux temps.

Le chapitre \ref{sec:discret} regroupe deux travaux de natures a 
priori diff\'erentes~:
le premier (publication \pubbethe ) concerne la dynamique d'un mod\`ele 
ferromagn\'etique dilu\'e, le deuxi\`eme (\pubwsat\ et \pubsat ) traite 
d'un algorithme de r\'esolution de formules de satisfiabilit\'e. 
Une approche commune
est d'abord pr\'esent\'ee en termes g\'en\'eriques, avant d'\^etre appliqu\'ee
\`a ces deux probl\`emes. On fera notamment
le lien avec la m\'ethode dynamique des r\'epliques de Coolen et 
Sherrington~\cite{DRT1,DRT2}. Trois appendices \`a ce chapitre exposent des
r\'esultats num\'eriques et analytiques non publi\'es sur des variantes du
probl\`eme d'optimisation.

Le chapitre \ref{sec:ch-ft} verra la r\'econciliation des variables continues 
et discr\`etes~: les propri\'et\'es d'\'equilibre des fonctions de 
corr\'elation avec un nombre quelconque de temps y seront explor\'ees. 
La publication \pubjsp\
traitait le cas continu, on fera dans le manuscrit les d\'emonstrations dans
le cas discret pour insister sur la g\'en\'eralit\'e de ces r\'esultats.
On pr\'esente \'egalement une version du th\'eor\`eme de fluctuation qui
r\'esume ces propri\'et\'es. Finalement des conjectures sur leurs 
g\'en\'eralisations dans les situations hors d'\'equilibre du type verres
de spin dilu\'es sont avanc\'ees.

\vspace{3mm}

J'ai essay\'e, dans la mesure du possible, de faire de ce manuscrit un
ensemble coh\'erent en ne pr\'esentant pas les publications dans un ordre 
chronologique. L'organisation retenue permettra, je l'esp\`ere, d'insister
sur les liens entre ces diff\'erents travaux. Je me suis donc efforc\'e de
pr\'esenter les m\'ethodes dans une certaine g\'en\'eralit\'e avant de les
appliquer aux cas particuliers.
Pour cette raison je me suis permis de changer certaines
notations par rapport \`a celles utilis\'ees dans les articles, et
de pr\'esenter dans certaines parties les d\'emonstrations avec peut-\^etre
trop de d\'etails. L'utilisation des m\^emes lettres pour des quantit\'es
diff\'erentes d'un chapitre sur l'autre n'a pu \^etre compl\`etement 
\'evit\'ee, j'esp\`ere que la lisibilit\'e du manuscrit n'en sera pas trop 
affect\'ee. 

\chapter{Propri\'et\'es g\'eom\'etriques des mod\`eles dilu\'es}
\label{sec:graphes}
\markboth{\hspace{3mm} \hrulefill \hspace{3mm} Ch. 2~: Propri\'et\'es g\'eom\'etriques}{Ch. 2~: Propri\'et\'es g\'eom\'etriques\hspace{3mm} \hrulefill \hspace{3mm} }

La plupart des \'etudes pr\'esent\'ees dans ce manuscrit partagent
une structure sous-jacente commune, qui porte en physique le nom g\'en\'erique 
de mod\`ele dilu\'e. Ce chapitre est consacr\'e \`a quelques propri\'et\'es
\og g\'eom\'etriques\fg\ de ces syst\`emes, \'etudi\'es en math\'ematiques
sous le nom de graphes et d'hypergraphes al\'eatoires. L'adjectif 
g\'eom\'etrique n'est pas \`a prendre au sens strict ici. En effet ces 
structures ne sont pas d\'efinies \`a partir d'un espace euclidien de 
dimension finie, et par nombre de leurs caract\'eristiques elles appartiennent
\`a la famille des probl\`emes de champ moyen. 

Dans les premi\`eres parties de ce chapitre on trouvera une introduction
sommaire \`a quelques mod\`eles de graphes et d'hypergraphes al\'eatoires.
Suit la pr\'esentation d'une m\'ethode syst\'ematique de d\'eveloppement
dans un r\'egime de faible concentration, qui a fait l'objet de la publication
\pubclusters\ et que l'on retrouvera \`a plusieurs reprises dans la suite 
de la th\`ese.

\section{Le graphe al\'eatoire d'Erd\"os et R\'enyi}

Ce mod\`ele, introduit en math\'ematiques dans les ann\'ees 60~\cite{ErRe}, est
l'arch\'etype des syst\`emes dilu\'es. Il a \'et\'e tr\`es largement
\'etudi\'e par les math\'ematiciens, un grand nombre de ses propri\'et\'es
sont connues rigoureusement et avec une tr\`es grande finesse (on pourra 
se reporter par exemple \`a \cite{Bo-bookrandom,Ja-book,Ja-birth,AlSp-proba}).
On se contente ici d'une approche non rigoureuse et de quelques r\'esultats 
utiles pour la suite.

\subsection{D\'efinition d'un graphe}
Du point de vue math\'ematique, un graphe de taille $N$ est constitu\'e de~:
\begin{itemize}
\item un ensemble $V$ de sommets ({\em vertices}), 
de cardinalit\'e (nombre d'\'el\'ements de l'ensemble) $|V|=N$. On peut
donc prendre $V=\{1,\dots,N\}$ sans perdre en g\'en\'eralit\'e.
\item un ensemble $E$ de liens ({\em edges}), c'est-\`a-dire de paires non
orient\'ees de sommets, $\{i,j\}\in V^2$. 
\end{itemize}

Selon les cas on peut autoriser ou non les liens $\{i,i\}$ d'un sommet \`a 
lui-m\^eme, ainsi que les r\'ep\'etitions du m\^eme lien dans l'ensemble $E$.
Cette distinction entre \og graphes simples\fg\ et \og multi-graphes\fg\ ne 
sera pas utilis\'ee dans la suite, pas plus que la notion de graphe 
orient\'e, pour lequel les liens portent une direction.

Il est clair qu'une repr\'esentation naturelle d'un graphe ainsi d\'efini 
consiste \`a dessiner les sommets comme des points d'un plan, et les liens
comme des courbes reliant ces sommets. En g\'en\'eral on est
oblig\'e de faire se croiser certains liens, si l'on peut dessiner le graphe
sans qu'aucun lien ne se croise il est dit planaire.

D\'efinissons quelques notions g\'en\'erales sur les graphes, avant de parler
d'ensembles al\'eatoires. La plupart de ces d\'efinitions sont intuitives,
mais il sera utile pour la suite de les formaliser un peu.

\begin{itemize}
\item Deux sommets $x$ et $y$ sont dits adjacents dans un graphe si le lien
$\{x,y\}$ appartient \`a l'ensemble $E$.

\item Deux sommets sont dits connect\'es s'il existe une suite de sommets
successivement adjacents (i.e. un chemin) qui les relient. Un graphe est 
dit connexe si toute paire de ses sommets est connect\'ee. Une composante
connexe d'un graphe est un sous-graphe maximal (au sens de l'inclusion)
connexe. Il sera loisible dans la suite de consid\'erer un sommet isol\'e
comme une composante connexe du graphe.

\item Une boucle est un chemin ferm\'e de sites adjacents. Un arbre est un
graphe connexe sans boucles. On peut en donner une caract\'erisation plus
simple~: si un graphe connexe a $n$ sommets et $m$ liens, c'est un arbre 
pour $m=n-1$.

\item Le th\'eor\`eme de Cayley affirme qu'il y a $n^{n-2}$ arbres distincts
avec $n$ sommets. \og Distincts\fg\ est \`a prendre ici au sens des graphes 
\'etiquet\'es, c'est \`a dire que chacun des $n$ sommets porte un indice de 
$[1,\dots,n]$, et deux \'etiquetages sont diff\'erents si et seulement si 
l'ensemble des liens correspondants est diff\'erent. Par exemple pour un arbre
\`a trois sommets, les trois \'etiquetages distincts 
sont $1-2-3$, $1-3-2$ et $2-1-3$.

\item Deux graphes \'etiquet\'es $G=(V,E)$ et $G'=(V',E')$ sont isomorphes 
s'il existe une bijection $\phi$ entre $V$ et $V'$ telle que 
$\{x,y\} \in E$ si et seulement si $\{\phi(x),\phi(y)\} \in E'$. Cette
d\'efinition correspond \`a la notion intuitive de forme d'un graphe.
Par exemple, les trois arbres \`a trois sommets cit\'es pr\'ec\'edemment 
sont isomorphes. Pour $n=4$ on a par contre deux types diff\'erents d'arbres 
non isomorphes. La figure \ref{fig:gr-extypes} illustre ces d\'efinitions 
avec les diff\'erents types d'arbres pour $n$ entre 1 et 4. On a not\'e
$V_t$ le nombre d'\'etiquetages distincts pour chacun des types. Pour $n=4$,
l'arbre avec un site central et trois voisins a quatre \'etiquetages 
distincts, selon le choix du site central. L'arbre lin\'eaire a lui douze 
\'etiquetages distincts, et donc conform\'ement au th\'eor\`eme de Cayley
on a bien $16$ arbres \'etiquet\'es \`a quatre sommets.
\begin{figure}
\begin{center}
\includegraphics{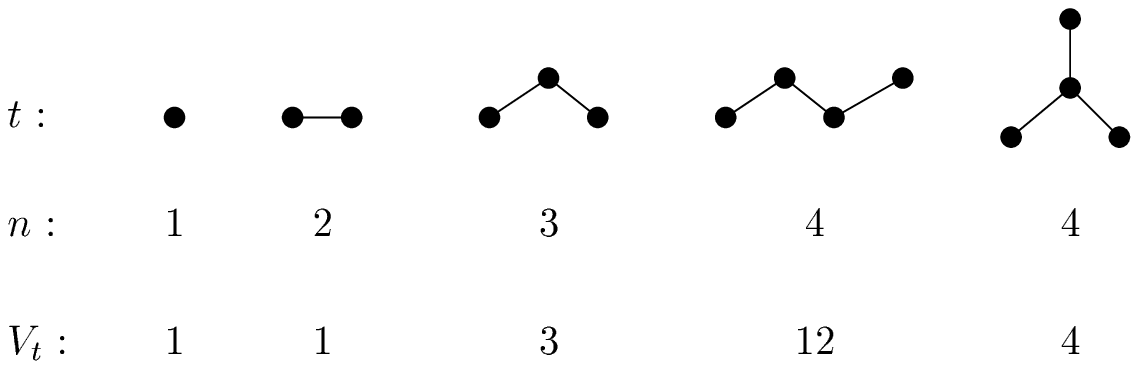}
\end{center}
\label{fig:gr-extypes}
\caption{Les diff\'erents types d'arbres \`a $n$ sommets, pour $n$ entre 1 et 4.
$V_t$ est le nombre d'\'etiquetages distincts.}
\end{figure}

\item Une clique \`a $n$ sommets est un graphe compl\`etement connect\'e,
c'est-\`a-dire dont les ${n \choose 2}$ liens sont pr\'esents. Le nombre 
chromatique d'un graphe est le nombre minimal de couleurs
n\'ecessaires pour colorier les sommets de fa\c con telle qu'aucun lien ne
relie deux sommets de la m\^eme couleur. Le nombre de clique d'un graphe 
est la taille de la plus grande clique contenue comme sous-graphe. Ces deux
exemples de propri\'et\'es ne seront pas utilis\'ees dans la suite, mais
illustrent le type de questions d'int\'er\^et en math\'ematiques.
\end{itemize}

\subsection{Graphe al\'eatoire}
Muni de ces d\'efinitions formelles, on constate que le nombre
de graphes de taille $N$ est fini (il vaut $2^{N(N-1)/2}$) , on peut donc 
d\'efinir sans difficult\'e une loi de probabilit\'e $\mbox{Prob}(G)$
sur l'ensemble des graphes $G$ de taille $N$, et se poser
des questions probabilistes sur cet ensemble. Si on a une
propri\'et\'e qu'on peut d\'efinir pour tout graphe $G$, par exemple
\og$G$ contient une clique de cinq sommets\fg\ , on peut se demander quelle 
est la probabilit\'e que cette propri\'et\'e soit v\'erifi\'ee quand on tire au
hasard un graphe selon la loi $\mbox{Prob}(G)$.

La loi de probabilit\'e la plus \'etudi\'ee, qu'on nomme graphe al\'eatoire
de Erd\"os et R\'enyi, consiste \`a consid\'erer ind\'ependamment les
${N \choose 2}$ liens possibles parmi les $N$ sommets,
et \`a prendre chacun de ces liens pr\'esent (resp. absent) avec 
probabilit\'e $\frac{c}{N}$ (resp. $1-\frac{c}{N}$). Autrement dit, si l'on
note $M(G)=|E(G)|$ le nombre de liens dans un graphe donn\'e $G$, 
on peut \'ecrire la loi de probabilit\'e des graphes comme
\begin{equation}
\mbox{Prob}(G)= 
\left( \frac{c}{N} \right)^{M(G)} 
\left( 1-\frac{c}{N} \right)^{\frac{N(N-1)}{2} - M(G)} \ .
\label{eq:gr-probG}
\end{equation}
Comme il y a ${\frac{N(N-1)}{2} \choose M(G)}$ graphes avec $M(G)$ liens,
cette loi est bien normalis\'ee.

Le choix d'une probabilit\'e de pr\'esence de lien qui d\'epend de la taille
du syst\`eme comme $1/N$ n'est \'evidemment pas le fruit du hasard. Comme on
va le voir, ce r\'egime permet d'obtenir une limite thermodynamique 
($N\to \infty$) int\'eressante.

Dans la suite du chapitre on notera les moyennes sur les ensembles de graphe
comme
\begin{equation}
[\bullet] = \sum_G \bullet \ \mbox{Prob}(G) \ .
\end{equation}
Toutes les limites et les \'equivalents sont \`a comprendre dans le sens de 
la limite thermodynamique $N\to \infty$ sauf mention explicite du contraire.

\subsection{Ses propri\'et\'es}
On va \'etudier ici quelques propri\'et\'es simples du graphe al\'eatoire
d\'efini par la loi (\ref{eq:gr-probG}).
\begin{itemize}
\item Consid\'erons tout d'abord $M(G)$, le nombre de liens pr\'esents
dans un graphe. C'est ici une variable al\'eatoire binomiale, dont on calcule
ais\'ement la moyenne et l'\'ecart quadratique moyen,

\begin{eqnarray}
&&[M]= \frac{c(N-1)}{2} \sim N \frac{c}{2} \quad , \\
&&\sqrt{[M^2]-[M]^2} = \sqrt{(N-1)\frac{c}{2} \left( 1 - \frac{c}{N} \right)}
 \sim N^{1/2} \sqrt{\frac{c}{2}} \ .
\end{eqnarray}

\vspace{2mm}

$M(G)/N$ se concentre donc dans la limite thermodynamique
autour de sa valeur moyenne $c/2$, \`a des fluctuations d'ordre $N^{-1/2}$
pr\`es. On reviendra sur ce point au cours de la discussion
des autres mod\`eles de
graphes al\'eatoires. Notons que le choix de la d\'ependance en $1/N$ de
la probabilit\'e de pr\'esence d'un lien permet d'obtenir un nombre extensif
(proportionnel au \og volume\fg\ $N$ du syst\`eme) de liens en moyenne.

\item On peut aussi s'int\'eresser aux propri\'et\'es locales d'un graphe
al\'eatoire. Cherchons par exemple la probabilit\'e $p_k$ (par rapport \`a la 
distribution (\ref{eq:gr-probG})) qu'un sommet donn\'e ait exactement 
$k$ voisins. On parle de connectivit\'e (ou de degr\'e) du sommet \'egal 
\`a $k$. On a
\begin{equation}
p_k= {N-1 \choose k} \left(\frac{c}{N} \right)^k 
\left(1-\frac{c}{N} \right)^{N-1-k} \to \frac{e^{-c} c^k}{k!} \ ,
\end{equation}
la limite thermodynamique \'etant prise avec $k$ fix\'e.
En effet, on est libre de choisir les $k$ sites voisins parmi les $N-1$ autres
sites, chacun de ces $k$ liens doit \^etre pr\'esent, et le site central ne 
doit pas \^etre reli\'e \`a d'autres sites. On constate que la connectivit\'e
d'un site devient une loi de Poisson avec param\`etre $c$ dans la limite 
thermodynamique. On appelle souvent le graphe al\'eatoire d'Erd\"os-R\'enyi un
graphe poissonien \`a cause de cette propri\'et\'e. La loi de Poisson est 
typique d'un nombre d'\'ev\`enements
se r\'ealisant avec une probabilit\'e individuelle faible (${\cal O}(1/N)$ 
ici), mais sur un grand nombre de tentatives (${\cal O}(N)$).

\item On peut aussi calculer la probabilit\'e $\tilde{p}_k$ d'observer un site 
avec $k+1$
voisins, conditionn\'ee \`a ce qu'il en ait au moins un, ce que l'on peut
se repr\'esenter plus facilement avec la figure \ref{fig:gr-kcond}.
\begin{equation}
\tilde{p}_k= {N-2 \choose k} \left(\frac{c}{N} \right)^k 
\left(1-\frac{c}{N} \right)^{N-2-k} \to 
\frac{e^{-c} c^k}{k!} \ .
\label{eq:gr-kcond}
\end{equation}
Dans la limite thermodynamique $\tilde{p}_k$ est aussi une loi poissonnienne
de param\`etre $c$. Un petit argument permet de pressentir \`a partir de
ce r\'esultat que la valeur $c=1$ va \^etre particuli\`ere. Supposons en
effet que l'on choisisse dans le graphe al\'eatoire un site racine au hasard, 
et que l'on explore le graphe en suivant les liens (ou branches) qui en
\'emergent. Si l'on appelle g\'en\'eration le nombre de pas que l'on a fait
au cours de l'exploration depuis la racine, on a $k$ descendants de
premi\`ere g\'en\'eration avec probabilit\'e $p_k$. Chacun de ces descendants
va avoir $l$ descendants, qui seront donc de deuxi\`eme g\'en\'eration, avec
probabilit\'e $\tilde{p}_l$, car les sites de premi\`ere g\'en\'eration ont 
\'et\'e, par d\'efinition, atteints par un lien pr\'esent entre eux et la
racine. En continuant l'exploration on construit ainsi un arbre o\`u le nombre
de descendants est \`a chaque g\'en\'eration d\'etermin\'ee par la loi 
conditionnelle $\tilde{p}$. Si $c<1$, ce processus de branchement meurt 
rapidement, alors que pour $c>1$ il continue \'eternellement. On va voir plus
bas que cette diff\'erence de comportement s'interpr\`ete ici comme une 
transition de percolation.

\begin{figure}
\begin{center}
\includegraphics{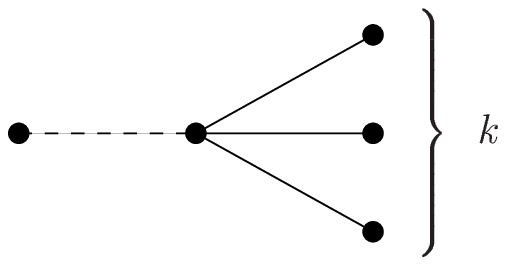}
\end{center}
\caption{Dans le calcul de la probabilit\'e conditionnelle (\ref{eq:gr-kcond}),
on suppose la pr\'esence d'un lien, en tirets ici, et l'on cherche la 
probabilit\'e d'avoir $k$ voisins suppl\'ementaires.}
\label{fig:gr-kcond}
\end{figure}

\item En \'elargissant le champ des questions pos\'ees, on peut maintenant
se demander quelle va \^etre la probabilit\'e que la composante connexe d'un 
site donn\'e soit un certain type de graphe $t$, avec $n_t$ sites et 
$m_t$ liens. Un peu de d\'enombrement conduit \`a

\begin{eqnarray}
P_t &=& {N-1 \choose n_t-1} V_t \left(\frac{c}{N} \right)^{m_t}
\left(1- \frac{c}{N} \right)^{n_t(N-n_t) +\frac{n_t(n_t-1)}{2} - m_t} \\
&\sim& 
\frac{1}{N^{m_t - n_t +1}} c^{m_t} e^{-n_t c} \frac{V_t}{(n_t-1)!} \ .
\label{eq:proba_tree_K2}
\end{eqnarray}

\vspace{2mm}

Expliquons ces diff\'erents facteurs. On doit d'abord choisir les $n_t-1$ 
autres sites de la composante connexe, parmi les $N-1$ sites du graphe, puis
une des $V_t$ diff\'erentes fa\c cons d'\'etiqueter la composante connexe.
Les $m_t$ liens doivent \^etre pr\'esents, avec donc la probabilit\'e 
$(c/N)^{m_t}$. Il faut finalement exclure les autres liens pouvant reliant 
les $n_t$ sommets entre eux, ainsi que ceux relieraient les $n_t$ sommets 
au reste du graphe. Notons que l'\'equivalent a \'et\'e pris en supposant
que $n_t$ et $m_t$ restaient finis dans la limite thermodynamique, cette
expression n'est donc valable que pour des composantes connexes de taille
finie.

On constate que si $m_t>n_t-1$, cette expression tend vers 0 dans la 
limite thermodynamique. Or pour un graphe connexe, $m \ge n-1$, avec
\'egalit\'e si et seulement si le graphe est un arbre. Plus pr\'ecis\'ement, 
la probabilit\'e qu'un site appartienne \`a une composante connexe de taille 
finie et qui contient des boucles est d'ordre $N^{-l}$, o\`u $l$ est le 
nombre de boucles 
ind\'ependantes. On peut de la m\^eme fa\c con montrer que la probabilit\'e 
qu'un site appartienne \`a une boucle de taille finie (sans imposer que sa 
composante connexe soit de taille finie) est d'ordre $N^{-1}$. Remarquons que 
cela ne signifie pas qu'il n'y a aucune boucle de taille finie dans la limite 
thermodynamique~: la probabilit\'e \'etant d'ordre $1/N$, mais le nombre de 
sites \'etant $N$, il y en a en moyenne un nombre fini.

\item Dans le cas particulier o\`u l'on cherche la probabilit\'e d'appartenance
\`a un arbre quelconque \`a $n$ sommets, on peut simplifier la formule 
(\ref{eq:proba_tree_K2}) en utilisant le th\'eor\`eme de Cayley
\begin{equation}
\sum_{t|n_t=n} V_t = n^{n-2} \ ,
\end{equation}
pour obtenir
\begin{equation}
P_n= \frac{e^{-cn} (cn)^{n-1}}{n!} \ .
\label{eq:gr_Pn}
\end{equation}

Comme on a vu que les seules composantes connexes de taille finie qui ont
une probabilit\'e (par site) finie dans la limite thermodynamique sont des
arbres, la somme 
$\sum_{n=0}^\infty P_n$ compte la fraction de sites qui sont dans des 
composantes de taille finie. On peut montrer que cette somme converge
vers 1 pour $c \le 1$, dans ce cas presque tous les sites sont dans des
composantes de taille finie. Par contre, quand $c>1$, la somme vaut
$1-P_\infty(c)$, o\`u $P_\infty(c)$ est la solution non nulle de 
l'\'equation $1-P_\infty=e^{-c P_\infty}$, repr\'esent\'ee sur la figure
\ref{fig:gr-Lambert}. $P_\infty$ est donc la fraction des sites qui ne
sont pas dans des composantes de taille finie, autrement dit c'est la 
fraction de sites dans l'\og amas infini\fg\ de percolation qui envahit un
nombre extensif de sites \`a partir de $c=1$. On peut en fait justifier 
l'\'equation sur $P_\infty$ de la mani\`ere suivante~: un site appartient 
\`a l'amas infini d\`es qu'un de ses voisins y appartient. 
R\'eciproquement, pour qu'un site n'y appartienne pas, il faut
qu'aucun de ses voisins n'y appartienne~:
\begin{equation}
1-P_\infty = \sum_{k=0}^\infty p_k (1- P_\infty)^k =
 \sum_{k=0}^\infty \frac{e^{-c} c^k}{k!} (1- P_\infty)^k = e^{-c P_\infty} \ .
\label{eq:gr-perco}
\end{equation}
Un tel raisonnement, o\`u l'on n\'eglige les corr\'elations entre les
probabilit\'es d'appartenance \`a l'amas infini des voisins d'un site donn\'e,
serait faux en dimension finie. Il est correct ici gr\^ace
au caract\`ere champ moyen du mod\`ele, qui ne repose pas sur un r\'eseau
g\'eom\'etrique r\'egulier.
\end{itemize}

\begin{figure}
\begin{center}
\includegraphics{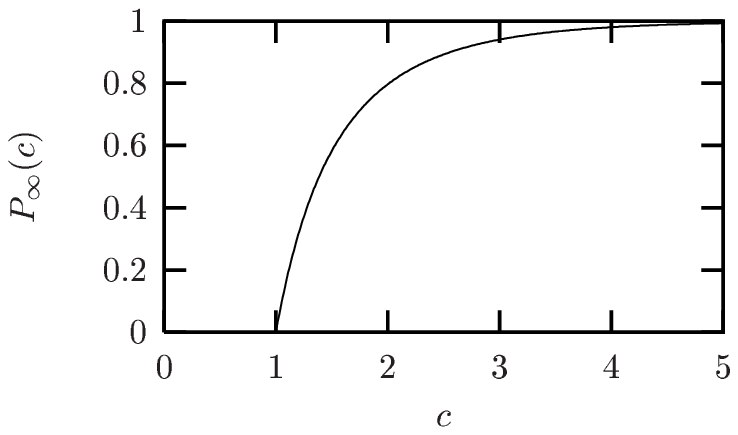}
\end{center}
\caption{La fraction de sites dans l'amas g\'eant en fonction de
la connectivit\'e moyenne pour le graphe d'Erd\"os-R\'enyi, solution de 
l'\'equation (\ref{eq:gr-perco}).}
\label{fig:gr-Lambert}
\end{figure}

Pour r\'esumer cette \'etude sommaire, on a vu que la connectivit\'e d'un site
est une loi de Poisson avec param\`etre $c$, que si on regarde un graphe 
al\'eatoire sur une \'echelle finie dans la limite thermodynamique on voit
toujours une structure en arbre avec grande probabilit\'e, et qu'il y a une
transition de percolation \`a $c=1$. Pour des connectivit\'es plus faibles,
une fraction des sites qui tend vers 1 dans la limite thermodynamique sont
contenues dans des composantes connexes de taille finie, par contre quand 
$c>1$ une composante connexe de taille extensive appara\^it. 

Les \'etudes
math\'ematiques de ce probl\`eme ont conduit \`a de nombreux autres r\'esultats
tr\`es pr\'ecis. Citons par exemple qu'\`a la transition ($c=1$), la taille
de la plus grande composante connexe diverge dans la limite thermodynamique
comme $N^{2/3}$ et que pour $c<1$ la plus grande composante est de taille
${\cal O}(\ln N)$. De plus les graphes al\'eatoires ne sont en arbre que sur
des \'echelles finies, on trouve en fait un grand nombre de boucles de 
longueur $\ln N$. Cette taille peut se comprendre \`a partir de l'argument
sur la descendance d'un processus de branchement poissonien expos\'e
pr\'ec\'edemment. Pour $c>1$ le nombre de sites a la g\'en\'eration $g$ cro\^it
typiquement comme $c^g$. Quand $g$ est d'ordre $\ln N$ ce nombre de sites 
devient d'ordre $N$, on est donc oblig\'e alors de retrouver des sites d\'ej\`a
pr\'esents dans le processus de branchement, ce qui implique la pr\'esence de
boucles.

On a consid\'er\'e ici seulement les propri\'et\'es typiques des graphes
al\'eatoires. R\'ecemment des m\'ethodes de physique 
statistique ont \'et\'e utilis\'ees~\cite{EnMoHa,Ri} pour \'etudier 
des propri\'et\'es atypiques, dans un r\'egime de grande d\'eviation, 
de ces objets.

\section{Hypergraphes}
\label{sec:gr-hyper}
Une g\'en\'eralisation naturelle du point de vue de la physique consiste
\`a remplacer les liens par des \og hyperliens\fg\ qui joignent un nombre 
$K \ge 2$ arbitraire de sommets, les graphes habituels correspondants \`a 
$K=2$. Un hypergraphe est alors la donn\'ee d'un ensemble de sommets et d'un
ensemble d'hyperliens. C'est ce
type de g\'en\'eralisation qui conduit du mod\`ele de Sherrington-Kirkpatrick
aux mod\`eles dits $p$-spin. Pour une valeur de $K$ donn\'ee, il y a 
${N \choose K}$ hyperliens possibles. On peut par exemple d\'efinir une loi
de probabilit\'e sur les $K$-hypergraphes en prenant chacun des hyperliens
ind\'ependamment
pr\'esent avec probabilit\'e $\frac{c}{N^{K-1}}$, absent avec probabilit\'e
$1-\frac{c}{N^{K-1}}$. Notant toujours $G$ un hypergraphe, et $M(G)$ le
nombre d'hyperliens pr\'esents, la loi de probabilit\'e est
\begin{equation}
\mbox{Prob}(G) = \left( \frac{c}{N^{K-1}} \right)^{M(G)}
\left(1- \frac{c}{N^{K-1}} \right)^{{N \choose K} - M(G)} \ .
\label{eq:loi-hyperg}
\end{equation}
A nouveau la d\'ependance en $N$ a \'et\'e choisie de
mani\`ere \`a avoir un nombre moyen de liens pr\'esents qui soit extensif,
$[M] \sim (c/K!) N$ dans la limite thermodynamique. On notera dans la suite 
de ce paragraphe $\alpha=c/K!$ pour simplifier certaines \'ecritures. Dans le 
m\^eme but on omettra le pr\'efixe \og hyper\fg\ quand il n'y a pas de 
confusion possible.

Des raisonnements combinatoires similaires
\`a ceux pr\'esent\'es dans le cas du graphe al\'eatoire conduisent \`a~:
\begin{itemize}
\item La probabilit\'e d'avoir $k$ liens autour d'un site donn\'e est,
dans la limite thermodynamique, une loi de Poisson de param\`etre $\alpha K$.
On appellera aussi cet ensemble \og hypergraphe poissonien\fg\ .

\item La probabilit\'e d'avoir $k+1$ liens autour d'un site atteint
par un hyperlien d\'ej\`a pr\'esent est aussi une loi de Poisson de 
param\`etre $\alpha K$. G\'en\'eralisant l'argument qualitatif pr\'esent\'e 
pour $K=2$, on rencontre
$\alpha K (K-1)$ sites \`a chaque nouvelle g\'en\'eration explor\'ee, 
on peut donc penser que le seuil de la transition de percolation sera ici 
$\alpha_p=1/(K(K-1))$.

\item En effet, pour qu'un site n'appartienne pas \`a l'amas infini il faut
qu'aucun des sites voisins n'y appartienne. Un site de degr\'e $k$ ayant
$k(K-1)$ voisins, on obtient en notant $P_\infty$ la probabilit\'e
d'appartenance \`a l'amas infini~:

\begin{eqnarray}
1-P_\infty&=& \sum_{k=0}^\infty \frac{e^{-\alpha K}(\alpha K)^k}{k!} 
(1-P_\infty)^{k(K-1)} \nonumber \\
&=& \exp \left[ -\alpha K + \alpha K(1-P_\infty)^{K-1} 
\right] \ ,
\end{eqnarray}

\vspace{2mm}

\'equation qui a une solution non triviale pour $\alpha > \alpha_p=1/(K(K-1))$.

\item On g\'en\'eralise sans difficult\'es la notion de composante connexe
et de boucle \`a un hypergraphe. Un graphe connexe 
avec $n_t$ sommets et $m_t$ liens 
est en arbre si $m_t (K-1) = n_t - 1$. On trouve comme pour $K=2$ que les 
composantes connexes finies avec des boucles ont une probabilit\'e 
n\'egligeable dans la limite thermodynamique.
On a alors la g\'en\'eralisation de (\ref{eq:proba_tree_K2}) 
pour la probabilit\'e qu'un site donn\'ee appartienne \`a une composante 
connexe en arbre $t$ ,
\begin{equation}
P_t = (\alpha K!)^{m_t} e^{-n_t \alpha K} \frac{V_t}{(n_t-1)!} \ .
\label{eq:proba_tree_Kg}
\end{equation}
$V_t$ est \`a nouveau le nombre d'\'etiquetage distincts de l'arbre $t$.
On en donnera des exemples dans la partie \ref{sec:devclusters}.
\end{itemize}

\section{Autres types d'ensemble}
\label{sec:gr-autres}
L'ensemble des graphes (resp. hypergraphes) peut \^etre muni d'une structure 
probabiliste avec des loi diff\'erentes de (\ref{eq:gr-probG}) 
(resp. (\ref{eq:loi-hyperg})). On mentionne ici quelques
possibilit\'es. 

\begin{itemize}
\item Une variante relativement inoffensive consiste \`a fixer le nombre $M$
de liens pr\'esents dans le syst\`eme, \`a une valeur not\'ee
traditionnellement $\alpha N$. Les graphes al\'eatoires sont alors 
g\'en\'er\'es
en choisissant $M$ fois un $K$-uplet de sommets de mani\`ere ind\'ependante et
non biais\'ee. On s'attend \`a ce que dans la limite thermodynamique, les
propri\'et\'es typiques de cet ensemble al\'eatoire soient les m\^emes 
que celles d\'ecrites dans la partie \ref{sec:gr-hyper}, du moment que
$\alpha$ et $c$ sont tels que le nombre moyen de liens dans le premier 
ensemble soit \'egal \`a celui (fix\'e strictement) dans le deuxi\`eme. C'est
une variation du type ensemble canonique {\em vs} microcanonique en
m\'ecanique statistique. Il faut tout de m\^eme garder \`a l'esprit que 
certaines propri\'et\'es ne vont pas \^etre \'equivalentes dans les deux 
ensembles, les
fluctuations et les corrections de taille finie notamment. L'exemple trivial
du nombre de liens, strictement fix\'e dans un cas, avec des fluctuations 
relatives 
d'ordre $N^{-1/2}$ dans l'autre, suffit \`a illustrer le probl\`eme.
Certains calculs ou simulations num\'eriques pouvant s'av\'erer plus simple
dans un ensemble que dans l'autre, on pourra \^etre amen\'e \`a utiliser les
deux.

\item 
Comme on l'a vu, la connectivit\'e locale des graphes d\'efinis ci-dessus 
ont des lois de probabilit\'e de Poisson, qui d\'ecroissent donc vite pour les
grandes connectivit\'es.
Un certain nombre d'\'etudes exp\'erimentales dans des domaines aussi divers
que la structure de la toile Internet, les collaborations scientifiques ou
d'acteurs de cin\'ema, j'en passe et des meilleures, \'etablissent des
r\'eseaux, ou graphes, \`a partir de ces donn\'ees (voir
\cite{AlBa-review,Ne-review} pour des revues).
Les sommets correspondent
par exemple aux diff\'erents acteurs, un lien entre deux acteurs \'etant 
pr\'esent s'ils ont particip\'e \`a un tournage en commun. Il se trouve que 
dans un grand nombre de ces situations, la loi de probabilit\'e des 
connectivit\'es des sommets est tr\`es \'eloign\'e d'une poisonnienne. En 
particulier, le comportement pour les tr\`es grandes connectivit\'es est du
type loi de puissance, d'o\`u le nom de \og scale-free\fg\ associ\'e \`a ces
r\'eseaux. A la suite de ces \'etudes statistiques, un certain nombre de 
mod\`eles ont \'et\'e introduits qui permettaient de reproduire ce
type de comportement. D'une part, certains mod\`eles sont dits dynamiques, 
les sommets sont introduits un par un avec des lois d'attachement 
pr\'ef\'erentiel \`a certains sites, qui privil\'egient les sites ayant 
d\'ej\`a une grande connectivit\'e. On trouvera une construction rigoureuse
d'un tel mod\`ele dans \cite{BoRiTuSp}. D'autre part, des mod\`eles
statiques consistent \`a consid\'erer l'ensemble des graphes pr\'esentant 
une distribution de connectivit\'e donn\'ee comme \'equiprobables~\cite{MoRe}. 
On peut alors par exemple \'etudier la transition de percolation de ces 
graphes.
Signalons aussi que le mod\`ele d'Ising ferromagn\'etique d\'efini sur ces
graphes avec des distributions de connectivit\'e arbitraire a \'et\'e 
\'etudi\'e dans \cite{Ising-networks1,Ising-networks2}. 

\item Comme cas tr\`es particulier de graphes dont on fixe la distribution
empirique de connectivit\'es, on va rencontrer dans la suite les mod\`eles
dilu\'es \`a connectivit\'e fixe. C'est donc un ensemble al\'eatoire o\`u
l'on garde les graphes (resp. hypergraphes) tels que chaque sommet appartient 
\`a un nombre fix\'e de liens (resp. hyperliens). Localement, c'est \`a dire
sur une \'echelle petite devant $\ln N$, ces graphes sont des arbres 
r\'eguliers, mais sur des \'echelles plus grandes on s'aper\c coit qu'ils 
contiennent des boucles, qui traduisent le caract\`ere al\'eatoire
de leur d\'efinition. On peut se poser la question de l'int\'er\^et d'une telle
construction, alors qu'il semblerait plus simple de consid\'erer des arbres
parfaitement r\'eguliers, sans boucle. Le probl\`eme de ce deuxi\`eme point
de vue est que le nombre de sites \`a la \og surface\fg\ d'un arbre r\'egulier
est du m\^eme ordre que le volume int\'erieur dans la limite thermodynamique,
ce qui conduit \`a des effets de bords tr\`es importants. Si l'on peut traiter
ces effets de bords de mani\`ere relativement simple pour un mod\`ele
ferromagn\'etique (en n'\'etudiant que la magn\'etisation du site central 
par exemple), la situation est assez inextricable pour un mod\`ele de
verre de spin~: la frustration sur un arbre ne peut venir que des conditions
aux bords \`a la surface, qu'il faut donc traiter avec beaucoup de soin. 
La d\'efinition du graphe \`a 
connectivit\'e fixe permet de s'affranchir de ce probl\`eme, le graphe n'a 
plus de surface puisque tous les sites sont statistiquement \'equivalents.
La frustration vient dans ce cas des boucles al\'eatoires. On trouvera une
discussion plus d\'etaill\'ee de ce sujet dans \cite{MePa-bethe}.
\end{itemize}

\section{D\'eveloppements en clusters pour les graphes poissonniens}
\label{sec:devclusters}
Une m\'ethode tr\`es \'el\'ementaire, qui a fait l'objet de la 
publication \pubclusters , s'est r\'ev\'el\'ee utile pour l'\'etude
de diff\'erents probl\`emes pr\'esent\'es dans cette th\`ese. On va
l'exposer ici sous une forme g\'en\'erique. 
Signalons que l'id\'ee de cette m\'ethode est pr\'esente, bien que peu 
explicit\'ee, dans un travail ant\'erieur de Hartmann et Weigt sur le vertex 
cover~\cite{HaWe-vc}. Dans une perspective plus large on pourrait la rattacher
aux d\'eveloppements de basse densit\'e dans les syst\`emes de particules,
la connectivit\'e rempla\c cant ici la densit\'e.

\subsection{Formulation g\'en\'erale}
Notons $G$ un \'el\'ement de l'ensemble al\'eatoire d'hypergraphes, muni
de la loi de probabilit\'e (\ref{eq:loi-hyperg}).
Chaque hypergraphe $G$ peut se d\'ecomposer comme l'union disjointe de ses 
$r(G)$ composantes connexes (appel\'es ici \og clusters\fg), 
$G=\bigcup_{i=1}^{r(G)} G_i$. 
Consid\'erons une fonction $F(G)$ qui \`a un graphe associe un nombre r\'eel,
avec les propri\'et\'es suivantes~:
\begin{itemize}
\item additivit\'e vis-\`a-vis de la d\'ecomposition en clusters, 
$F(G)= \sum_{i=1}^{r(G)} F(G_i)$.
\item ind\'ependance par rapport \`a l'\'etiquetage du graphe, autrement dit
$F$ renvoie la m\^eme valeur pour deux graphes isomorphes.
\end{itemize}

\begin{figure}
\includegraphics{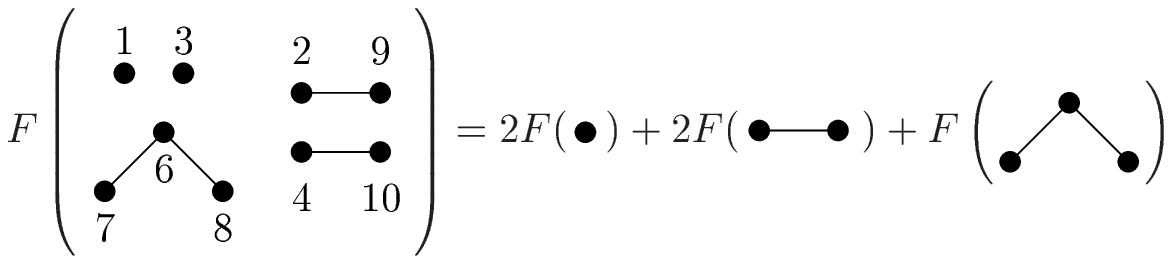}
\caption{Un exemple de d\'ecomposition pour une fonction additive.}
\label{fig:gr-exF}
\end{figure}

L'exemple pr\'esent\'e sur la figure \ref{fig:gr-exF} (pour $K=2$) devrait
clarifier ces d\'efinitions.

Ces deux propri\'et\'es permettent d'\'ecrire
\begin{equation}
F(G)=\sum_t {\cal N}_t(G) F_t \ ,
\end{equation}
o\`u la somme porte sur les diff\'erents types $t$ de composantes connexes,
$F_t$ est la valeur que prend la fonction sur un graphe isomorphe \`a $t$,
et ${\cal N}_t(G)$ est le nombre de clusters de type $t$ dans le graphe $G$.

On s'int\'eresse \`a la valeur moyenne d'une telle fonction
sur l'ensemble al\'eatoire. On d\'efinit donc la densit\'e associ\'ee $f$~:
\begin{equation}
f(\alpha) = \frac{1}{N} [F(G)] = \sum_t \frac{{\cal N}_t}{N} F_t \ ,
\label{eq:gr-def_falpha}
\end{equation}
avec ${\cal N}_t = [{\cal N}_t(G)]$ le nombre moyen de clusters de type $t$.
On peut facilement se convaincre que ${\cal N}_t/N = P_t / n_t$, o\`u $n_t$ 
est le nombre de sommets d'un cluster de type $t$, et $P_t$ la probabilit\'e 
qu'un site donn\'e soit dans un cluster de type $t$. Le calcul exact de cette
somme est a priori impossible pour une fonction $F$ compliqu\'ee. On peut
cependant se simplifier consid\'erablement la t\^ache si l'on se contente
d'un d\'eveloppement de la fonction $f$ en puissances de $\alpha$, autour de
$\alpha=0$. En effet, dans un voisinage de $\alpha=0$, la somme 
(\ref{eq:gr-def_falpha}) est domin\'ee dans la limite thermodynamique par
les contributions des arbres de taille finie~: on a vu qu'en dessous du seuil
de percolation la fraction des sites dans de tels clusters tendait vers 1 
dans la limite thermodynamique. On peut alors utiliser (\ref{eq:proba_tree_Kg})
pour \'ecrire
\begin{equation}
\hat{f}(\alpha) = \sum_t \alpha^{m_t} e^{-n_t \alpha K} V'_t F_t \ , \qquad
V'_t = \frac{(K!)^{m_t} V_t}{n_t!} \ ,
\label{dev_cluster_generique}
\end{equation}
et la somme est prise seulement sur les arbres~\footnote{Je discuterai la
diff\'erence de notation entre $f$ et $\hat{f}$ dans la partie 
\ref{sec:validite}.}.
Rappelons que $m_t$ est le nombre de clauses dans le cluster de type $t$,
avec $m_t (K-1) = n_t - 1$ car $t$ est un arbre. Le facteur de sym\'etrie
$V'_t$ s'av\`ere plus utile que le nombre d'\'etiquetages $V_t$ dont il
d\'ecoule.

On s'aper\c coit finalement que les clusters comportant
un nombre $m$ de liens ne contribuent qu'aux ordres sup\'erieurs ou \'egaux
\`a $m$ dans le d\'eveloppement en puissances de $\alpha$. Pour d\'evelopper
\`a un ordre donn\'e en $\alpha$ il suffit donc de calculer les facteurs
de sym\'etries $V'_t$ pour les premiers arbres, ce qui est un simple exercice 
d'\'enum\'eration, et les valeurs de $F_t$ correspondantes. Selon la fonction
\'etudi\'ee cette deuxi\`eme t\^ache peut se r\'ev\'eler plus ou moins 
fastidieuse, comme on le
verra dans la partie \ref{sec:walksat} o\`u l'on appliquera cette m\'ethode
au calcul du temps mis par un algorithme de recherche locale pour 
r\'esoudre un probl\`eme d'optimisation combinatoire.

\begin{figure}
\begin{center}
\includegraphics[width=11cm]{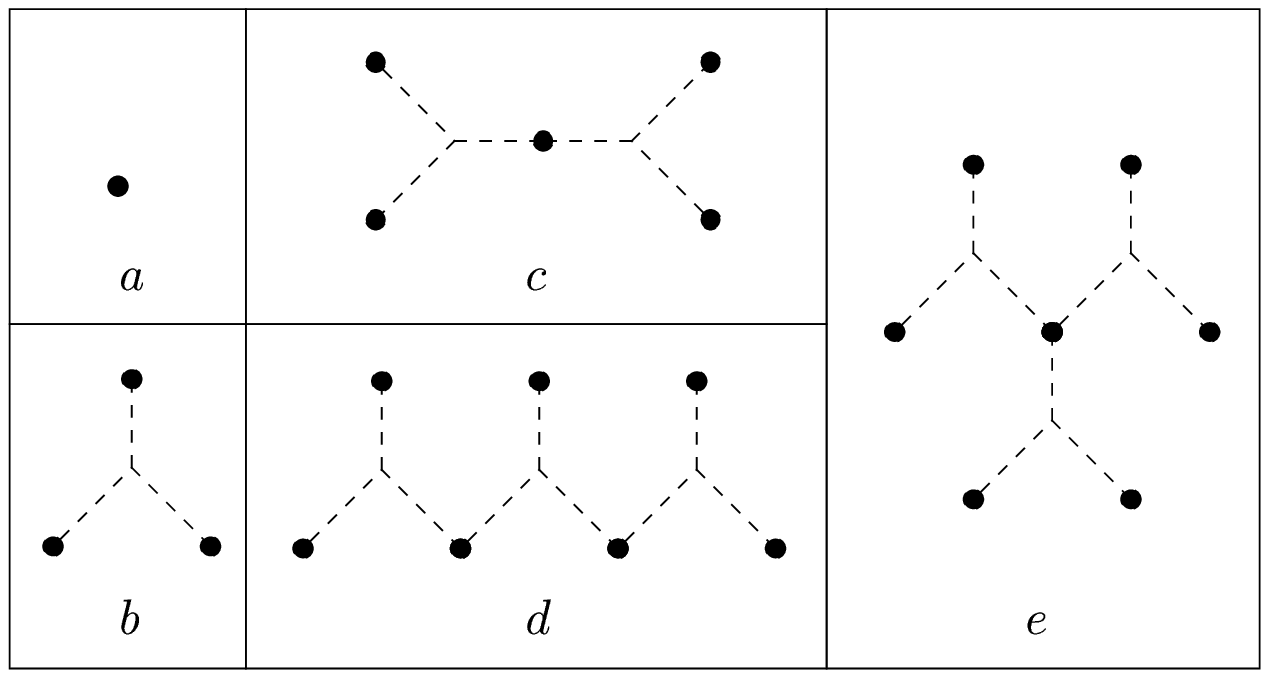}
\end{center}
\caption{Les clusters en arbre ayant entre 0 et 3 hyperliens utilis\'es dans
l'\'equation (\ref{eq:gr-devalpha3}). Les hyperliens sont ici repr\'esent\'es
pour $K=3$ avec des \'etoiles en pointill\'e.}
\label{fig:gr_clusters}
\end{figure}

Donnons le d\'eveloppement \`a l'ordre $\alpha^3$ pour une fonction $F$ 
quelconque~:
\begin{eqnarray}
f(\alpha) = F_a &+& \alpha (F_b - K F_a) + \frac{\alpha^2}{2} K^2 
(F_c - 2 F_b + F_a) \label{eq:gr-devalpha3} \\ 
&+& \frac{\alpha^3}{6} K^3 ( F_e + 3(K-1) F_d - 3(2K-1) F_c
+3 K F_b - F_a) + {\cal O}(\alpha^4) \nonumber \ .
\end{eqnarray}
Les clusters $t=a,b,c,d,e$ sont repr\'esent\'es sur la figure 
\ref{fig:gr_clusters}, et on trouvera dans la table 
\ref{table:gr_clusters} les facteurs de sym\'etrie qui ont
\'et\'e utilis\'es pour obtenir (\ref{eq:gr-devalpha3}).

\begin{table}
\renewcommand{\arraystretch}{1.5}
$$\begin{array}{| c | c c c |}
\hline
\phantom{a} {\mbox{Type}} \phantom{a} & \phantom{a} m_t \phantom{a} & 
\phantom{a} n_t \phantom{a} &  \phantom{a} V'_t \phantom{a}  \\
\hline 
a & 0 & 1 & 1 \\ 
b & 1 & K & 1 \\ 
c & 2 & 2K-1 & \frac{K^2}{2} \\ 
d & 3 & 3K-2 & \frac{K^3 (K-1)}{2} \\ 
e & 3 & 3K-2 &  \frac{K^3}{6} \\
\hline
\end{array}$$
\renewcommand{\arraystretch}{.66}
\caption{Facteurs de sym\'etrie du d\'eveloppement en clusters, se reporter
\`a la figure \ref{fig:gr_clusters} pour la nomenclature des types.}
\label{table:gr_clusters}
\end{table}

\subsection{L'\'energie libre du mod\`ele de Viana-Bray \`a faible 
connectivit\'e}

Cette m\'ethode a \'et\'e appliqu\'ee dans la publication \pubclusters\ 
au calcul de 
l'entropie du fondamental d'un probl\`eme de satisfiabilit\'e, et \`a celui
de l'\'energie libre du mod\`ele de Viana-Bray~\cite{ViBr} dans 
la phase de basse connectivit\'e. Revenons ici sur la deuxi\`eme de ces 
applications.

On consid\`ere donc un graphe al\'eatoire d'Erd\"os et R\'enyi de
connectivit\'e moyenne $c$, et pour chacun des liens pr\'esents entre les 
sites $i$ et $j$
on tire al\'eatoirement une interaction $J_{ij}$ avec la m\^eme
loi de probabilit\'e $\pi$. On notera $\overline{\bullet}$ les moyennes
sur la loi $\pi$. On place sur les sommets du graphe des spins d'Ising 
$\sigma_i$ et on d\'efinit l'hamiltonien
\begin{equation}
H=- \sum_{i<j} J_{ij} \ \sigma_i \sigma_j \ ,
\end{equation}
avec $J_{ij}=0$ si le lien entre les sommets $i$ et $j$ est absent.
Viana et Bray ont introduit ce mod\`ele afin d'expliquer les propri\'et\'es
du compos\'e ${\rm Eu}_x {\rm Sr}_{1-x}{\rm S}$ qui, selon la concentration
$x$, peut pr\'esenter diff\'erents types de transition. La connectivit\'e $c$
dans cette mod\'elisation permet de reproduire ce ph\'enom\`ene de dilution.
La temp\'erature de transition va notamment d\'ependre de $c$. Dans l'article
originel~\cite{ViBr}, le probl\`eme \'etait trait\'e au niveau sym\'etrique des
r\'epliques, pr\`es de la ligne de transition. Kanter et 
Sompolinsky~\cite{KaSo} ont \'etudi\'e la limite de temp\'erature nulle, 
toujours avec l'hypoth\`ese RS.

On peut appliquer la m\'ethode pr\'esent\'ee dans cette partie au calcul
de l'\'energie libre pour de petites concentrations. En effet, la fonction de 
partition du syst\`eme peut s'\'ecrire
comme un produit de fonctions de partition pour chacune des composantes
du graphe. De plus, si l'on fait la moyenne $\overline{\bullet}$ sur les 
intensit\'es des interactions pour un graphe donn\'e, on obtient une 
\'energie libre qui v\'erifie les propri\'et\'es suffisantes pour
\'etablir le d\'eveloppement en clusters. On a une simplification
suppl\'ementaire ici~: il est facile de montrer par r\'ecurrence
que la fonction de partition d'un syst\`eme d'Ising sur un arbre se
factorise comme un produit de termes de liens et de termes de sites. Une
fois que la moyenne sur la loi $\pi$ est prise, tous les arbres de m\^eme
taille contribuent de la m\^eme fa\c con, quelque soit leur forme. Ceci
permet de resommer le d\'eveloppement en clusters dans toute la phase non
percolante $c<1$,
\begin{equation}
-\beta f(\beta) = \ln 2 + \frac{c}{2} \, \overline{\ln \cosh(\beta J)} \ ,
\end{equation}
o\`u $f=-\left[ \, \overline{\ln Z} \, \right]/(N\beta)$ est la 
densit\'e 
d'\'energie libre moyenn\'ee sur la distribution des graphes et sur 
l'intensit\'e des couplages.
Ce r\'esultat est bien s\^ur trivial pour une phase paramagn\'etique, il ne
peut pas y avoir de transition \`a temp\'erature finie car presque tous les
sites sont dans des composantes connexes de taille finie.
Cependant la m\'ethode pourrait s\^urement \^etre rendue rigoureuse et 
fournir des bornes de concentration sur l'\'energie libre dans ce cas-l\`a. 
En effet la m\'ethode dite du deuxi\`eme moment en math\'ematiques 
permet de montrer que le nombre de clusters d'un type donn\'e devient 
tr\`es piqu\'e autour de sa valeur moyenne dans la limite thermodynamique. 
Il faut pour cela calculer l'\'ecart quadratique moyen de ${\cal N}_t$ et
montrer qu'il est n\'egligeable devant sa valeur moyenne. Une autre approche 
bien plus g\'en\'erale et puissante
repose sur l'id\'ee de l'interpolation de Guerra~\cite{Gu-interpolation}, 
appliqu\'ee au mod\`ele de
Viana-Bray par Guerra et Toninelli~\cite{GuTo-VB}.

On peut facilement \'etendre le d\'eveloppement en clusters pour calculer
les corrections de taille finie \`a une grandeur extensive, toujours dans la
limite de faible connectivit\'e. Il faut pour cela consid\'erer d'une part les
corrections d'ordre $1/N$ \`a la probabilit\'e d'apparition d'un cluster
en arbre, et d'autre part tenir compte aussi
des contributions des clusters contenant des boucles.
Ce calcul \'etait pr\'esent\'e dans la publication \pubclusters\
 pour l'\'energie
libre du mod\`ele de Viana-Bray,
malheureusement le r\'esultat \'etait entach\'e d'une erreur~\cite{Fabio},
j'en donne donc ici une version moins fausse~:
\begin{eqnarray}
&&-\beta f(\beta) = \ln 2 + \frac{c}{2} \,
\overline{\ln \cosh(\beta J)} \\ &+& \frac{1}{N} 
\left(-\frac{c}{2} \, \overline{\ln \cosh \beta J} + \frac{c^3}{6} \,
\overline{\ln (1 + \tanh \beta J_1 \tanh \beta J_2 \tanh \beta J_3)}
+ {\cal O}(c^4) \right) + {\cal O} \left(\frac{1}{N^2} \right) \nonumber
\end{eqnarray}

\subsection{Domaine de validit\'e de la m\'ethode}
\label{sec:validite}
Je voudrais revenir maintenant sur le probl\`eme de la validit\'e de
cette m\'ethode de d\'eveloppement en clusters. Mon point de vue sur la 
question a sensiblement \'evolu\'e depuis la r\'edaction de la publication 
\pubclusters\ dans laquelle on argumentait en faveur d'une singularit\'e de
ces d\'eveloppements au seuil $\alpha_p$ de percolation du graphe. Il
convient d'\^etre un peu plus pr\'ecis dans cette discussion.

Consid\'erons d'abord l'expression (\ref{eq:gr-def_falpha}) qui d\'efinit
la fonction $f(\alpha)$. Cette s\'erie doit prendre en compte toutes les
composantes connexes, il faudrait donc la calculer pour une taille $N$ finie,
puis prendre la limite thermodynamique \emph{apr\`es} que la somme sur les 
clusters ait \'et\'e effectu\'ee. La fonction $\hat{f}(\alpha)$ de l'\'equation
(\ref{dev_cluster_generique}) est au contraire obtenue en intervertissant
ces deux op\'erations~: on a simplifi\'e l'expression de la probabilit\'e
d'un cluster de taille finie dans la limite thermodynamique \emph{avant} de 
faire la somme sur les diff\'erents types d'arbre. On doit donc avoir
$f(\alpha)=\hat{f}(\alpha)$ pour $\alpha < \alpha_p$ puisque dans ce r\'egime
presque tous les sites sont dans des composantes de taille finie~; par contre
ces deux fonctions seront diff\'erentes pour des connectivit\'es plus grandes,
$\hat{f}$ n\'egligeant la contribution de l'amas infini.

On peut donner un exemple tr\`es \'el\'ementaire de cette distinction~: pour 
$F_t=n_t$, la fonction $F$ compte le nombre total de sites dans un graphe, 
qui est bien s\^ur $N$. On a donc $f=1$ quelque soit la valeur de $\alpha$, 
alors que $\hat{f}$ ne compte que la fraction de sites dans des composantes 
de taille finie. Cette derni\`ere fonction est donc \'egale \`a 1 en dessous
du seuil de percolation et vaut $1-P_\infty(\alpha)$ au del\`a de $\alpha_p$, 
$P_\infty(\alpha)$ d\'esignant la fraction de sites dans le cluster infini.

Revenons maintenant au cas d'une fonction $F$ g\'en\'erique. Si
l'on savait calculer $\hat{f}$ en sommant la s\'erie 
(\ref{dev_cluster_generique}), cette fonction aurait une singularit\'e
\`a $\alpha_p$ et sa valeur ne nous saurait d'aucune utilit\'e pour
pr\'edire la valeur de $f(\alpha)$ dans le r\'egime de percolation de l'amas 
infini. La plupart du temps (sauf dans des cas simples comme le mod\`ele de
Viana-Bray dans la phase de basse connectivit\'e) cette resommation est
impossible. On se contente donc de couper la s\'erie apr\`es quelques termes 
{\em et} de d\'evelopper les exponentielles de $\alpha$  pour r\'eordonner le 
d\'eveloppement en puissances de $\alpha$. C'est cette op\'eration qui a 
conduit au r\'esultat final (\ref{eq:gr-devalpha3}). Autrement dit, on
a calcul\'e le d\'ebut du d\'eveloppement de Taylor de $\hat{f}(\alpha)$ au 
voisinage de $\alpha=0$. Comme $f$ et $\hat{f}$ co\"incident sur
un intervalle fini $[0,\alpha_p[$, c'est aussi le d\'eveloppement de Taylor de
la fonction $f$, objet de notre \'etude. Si cette fonction, a priori inconnue,
est bien d\'efinie et r\'eguli\`ere sur $[0,\alpha_r]$ , avec 
$\alpha_r > \alpha_p$ (m\^eme si l'on n'a pas de justification pour la
r\'egularit\'e de $f$ \`a $\alpha_p$), le d\'eveloppement en cluster peut 
tr\`es bien \^etre convergent jusqu'\`a $\alpha_r$.

Un exemple tr\`es clair de ce ph\'enom\`ene est l'entropie de temp\'erature 
nulle du mod\`ele $p$-spin dilu\'e (alias XORSAT), qui
a \'et\'e calcul\'ee rigoureusement~\cite{CoDuMaMo-XOR,MeRiZe-XOR}. 
Pour $\alpha<0.818$ (avec $K=3$), 
l'entropie est une fonction lin\'eaire de $\alpha$, alors que 
$\alpha_p(K=3)=1/6$. Le calcul des premiers ordres du d\'eveloppement en 
clusters est en accord avec ce r\'esultat rigoureux, dont le domaine de 
validit\'e est bien plus \'etendu que la phase non percol\'ee.

On verra aussi dans la partie \ref{sec:walksat} un autre exemple d'application
de la m\'ethode au calcul du temps de
r\'esolution d'un algorithme de recherche locale pour le probl\`eme de la
satisfiabilit\'e. Dans ce cas il n'y a pas de r\'esultats exacts, mais on
peut faire des simulations num\'eriques avec des syst\`emes de tr\`es grande 
taille (dans le r\'egime int\'eressant pour cette question, la complexit\'e
cro\^it seulement lin\'eairement avec la taille du syst\`eme, au contraire
du probl\`eme exponentiel de l'entropie consid\'er\'e dans la publication 
\pubclusters ).
Ces simulations sont tr\`es convaincantes en faveur de l'absence de 
singularit\'e \`a $\alpha_p$, la fonction $f(\alpha)$ \'etant r\'eguli\`ere 
jusqu'\`a une valeur de $\alpha$ de l'ordre de $2.7$ (pour $K=3$) o\`u 
elle diverge.

A titre de remarque, soulignons finalement la similitude entre le 
d\'eveloppement (\ref{eq:gr-devalpha3}) et un principe 
d'\og inclusion-exclusion\fg~: le coefficient du terme $\alpha^m$ est donn\'e,
\`a des facteurs de sym\'etrie pr\`es, par la contribution des graphes de $m$
clauses, auquel on soustrait celle des sous-composantes de taille inf\'erieure
pour \'eviter un double comptage. Ceci provient du d\'eveloppement des
termes $e^{-n_t \alpha K}$ de (\ref{dev_cluster_generique}) qui imposait aux
arbres de $n_t$ sites d'\^etre d\'econnect\'es du reste du graphe. Dans le
d\'eveloppement final cette condition n'est plus impos\'ee, ce qui permet
d'\'etendre la validit\'e du r\'esultat au del\`a de $\alpha_p$.

\newpage
\pagestyle{empty}
\mbox{}
\newpage
\chapter{Dynamiques de spins continus}
\label{sec:continu}
\markboth{\hspace{3mm} \hrulefill \hspace{3mm} Ch. 3~: Dynamiques de spins continus}{Ch. 3~: Dynamiques de spins continus\hspace{3mm} \hrulefill \hspace{3mm} }

Ce chapitre s'articule autour de l'\'etude du mod\`ele de Viana-Bray dans
sa version sph\'erique. On commence par rappeler quelques g\'en\'eralit\'es 
sur les mod\`eles sph\'eriques, dont les propri\'et\'es sont facilement
d\'eduites du spectre de leur matrice d'interaction. Avec cette motivation
en t\^ete on fera ensuite un d\'etour du c\^ot\'e des matrices al\'eatoires. 
Les cons\'equences de cette investigation sur la dynamique seront 
alors pr\'esent\'ees, en insistant sur les nouvelles propri\'et\'es du mod\`ele
dilu\'e par rapport au cas compl\`etement connect\'e.
Le cas particulier
\'etudi\'e ici souffrant de certaines pathologies, on introduit finalement 
un formalisme plus g\'en\'eral qui constitue un premier pas vers le 
traitement syst\'ematique de la dynamique des mod\`eles dilu\'es \`a
variables continues.

Ces travaux ont fait l'objet des publications \pubmatrix\ pour la partie 
concernant les matrices al\'eatoires, \pubsphe\ pour le mod\`ele sph\'erique, 
et d'une partie de \pubjsp\ pour la g\'en\'eralisation.

\section{G\'en\'eralit\'es sur le mod\`ele sph\'erique}

\subsection{Statique}

Consid\'erons un syst\`eme de
$N$ spins d'Ising $\sigma_i = \pm 1$, interagissant par paires avec 
l'hamiltonien
\begin{equation}
H= -\frac{1}{2} \sum_{i,j} J_{ij} \sigma_i \sigma_j \ .
\end{equation}
La matrice $J_{ij}$ d\'efinit les couplages entre les spins. Pour un 
ferromagn\'etique en dimension finie par exemple, l'indice $i$ repr\'esente 
les 
coordonn\'ees du site sur un r\'eseau \`a $d$ dimensions, $J_{ij}$ \'etant
positif si $i$ et $j$ sont des sites voisins du r\'eseau, nul sinon.
Pour un verre de spins le signe des interactions est al\'eatoire.

Hormis quelques cas particuliers (probl\`emes unidimensionnels, 
ferromagn\'etique bidimensionnel, graphes compl\`etement connect\'es), on
ne sait pas calculer exactement la fonction de partition d'un tel mod\`ele.
Une simplification possible consiste \`a modifier la nature des variables
$\sigma_i$. Berlin et Kac~\cite{BeKa} ont introduit en 1952 le 
mod\`ele sph\'erique, dans lequel les $\sigma_i$ deviennent des variables 
continues, soumises \`a la
contrainte globale $\sum_i \sigma_i^2=N$. 
L'espace des configurations qui \'etait constitu\'e des sommets de 
l'hypercube \`a 
$N$ dimensions pour le mod\`ele d'Ising est
ainsi \'etendu \`a l'hypersph\`ere passant par ces sommets. 
Cette modification est a priori arbitraire et introduit une 
interaction entre tous les spins par l'interm\'ediaire de la contrainte. 
Stanley~\cite{St} a cependant montr\'e que pour des interactions 
ferromagn\'etiques en dimension finie, le mod\`ele sph\'erique 
est la limite du mod\`ele
dit $O(m)$, o\`u chaque spin appartient \`a une sph\`ere 
$m$-dimensionnelle, quand $m \to \infty$. Le mod\`ele d'Ising correspond
\`a $m=1$ dans cette classification, le mod\`ele XY \`a $m=2$ et celui
d'Heisenberg \`a $m=3$. Le fait qu'on puisse obtenir le mod\`ele de
Berlin et Kac comme la limite d'une famille de mod\`eles plus
r\'ealistes constituait un argument en sa faveur. On trouvera une discussion
plus d\'etaill\'ee de cette \'equivalence dans~\cite{Pastur}. 

\pagestyle{myheadings}

Les propri\'et\'es statiques du mod\`ele sph\'erique sont faciles \`a 
calculer car elles font
intervenir des int\'egrales gaussiennes \`a la place des sommes sur les 
spins d'Ising. La fonction de partition s'\'ecrit
\begin{eqnarray}
Z(\beta) &=& \int d\sigma_i \ \delta \left(N - \sum_i \sigma_i^2 \right) 
\ \exp\left[ \frac{1}{2}\beta \sum_{ij}J_{ij} \sigma_i \sigma_j\right] \\
&=& \int \frac{dz}{2\pi} d\sigma_i \ \exp\left[N z - \frac{1}{2} \sum_{ij} 
(2 z \delta_{ij} - \beta J_{ij}) \sigma_i \sigma_j\right] \ .
\label{eq:sphe-Z}
\end{eqnarray}
L'int\'egrale sur $z$ se fait dans le plan complexe, parall\`element
\`a l'axe imaginaire et dans le domaine tel que
les valeurs propres de $(2 z \delta_{ij} - \beta J_{ij})$ aient toutes
une partie r\'eelle positive. Dans ce cas l'int\'egrale gaussienne \`a $N$
dimensions converge, et l'on a 
\begin{equation}
Z(\beta) = \int \frac{dz}{2 \pi} (2 \pi)^{\frac{N}{2}} 
\exp\left[ N z -\frac{1}{2} \sum_k
\ln ( 2 z - \beta \lambda_k) \right] \ , \label{eq:Zsphe}
\end{equation}
o\`u les $\lambda_k$ sont les valeurs propres de la matrice $J_{ij}$. 
Dans la limite thermodynamique on peut finalement calculer 
cette int\'egrale par la m\'ethode du col.

On voit
ici la simplification par rapport au mod\`ele d'Ising~: quelque soit le type
d'interaction, la seule information sur la matrice $J_{ij}$ dont on a besoin 
est la distribution de ses valeurs propres. 
Pour un syst\`eme de spins d'Ising cela n'est pas suffisant, il faut
aussi des quantit\'es impliquant les vecteurs propres de la matrice qui
sont plus difficiles \`a obtenir.

\subsection{Dynamique}
\label{sec:sphe-genedyn}
On mod\'elise g\'en\'eralement l'\'evolution dynamique d'un syst\`eme de
spins continus en contact avec un thermostat de temp\'erature $T$ par
l'\'equation de Langevin,
\begin{equation}
\frac{d}{dt} \sigma_i (t) = -\frac{\partial H}{\partial \sigma_i} + \xi_i(t) 
\ \ \ \forall i \ ,
\label{eq:sphe-lan}
\end{equation}
o\`u $\xi_i$ est un bruit blanc gaussien avec
\begin{equation}
\langle \xi_i (t) \rangle=0 \quad \mbox{et} \qquad 
\langle \xi_i(t) \xi_j(t') \rangle=2 T \delta_{ij} \delta(t-t') \ .
\label{eq:sphe-defxi}
\end{equation} 
Dans cette partie les moyennes sur les histoires
du bruit thermique sont not\'ees $\langle \bullet \rangle$.
Ici et dans tout le manuscrit la constante de Boltzmann $k_B$ est prise
\'egale \`a 1. La mod\'elisation de l'influence du thermostat par des
\'equations de Langevin trouve sa justification dans le fait qu'elles 
conduisent aux temps longs, pour un syst\`eme de taille finie, 
\`a l'\'equilibre de Gibbs-Boltzmann.

En toute rigueur l'\'equation (\ref{eq:sphe-lan}) 
n'est pas bien d\'efinie sous cette forme~: les
bruits blancs $\xi_i(t)$ sont si irr\'eguliers que $\sigma_i (t)$ n'est
d\'erivable nulle part, ce qui rend la signification du membre de gauche 
douteuse. En fait il faut se donner une convention de lecture de ces
\'equations (les plus connues \'etant celles d'Ito et de Stratanovitch), en
discr\'etisant l'axe des temps. Dans la suite on \'eludera ce probl\`eme,
objet d'\'etudes math\'ematiques sous le nom d'\og \'equations 
diff\'erentielles stochastiques\fg .

Dans le cas du mod\`ele sph\'erique consid\'er\'e ici l'\'equation de Langevin
devient
\begin{equation}
\frac{d}{dt} \sigma_i (t) = \sum_j J_{ij} \sigma_j (t) - z(t) \sigma_i(t) +
\xi_i(t) \ ,
\end{equation}
o\`u $z(t)$ est un multiplicateur de Lagrange dynamique destin\'e \`a
imposer la contrainte sph\'erique $\sum_i \sigma_i^2(t) = N$. 

La matrice $J_{ij}$ \'etant sym\'etrique r\'eelle, on peut la diagonaliser
par un changement de base orthogonal. Notons $\tilde{\sigma}_k$ la
coordonn\'ee de $\vec{\sigma}=\{\sigma_1,\dots,\sigma_N\}$ dans la direction 
du vecteur propre de $J$
associ\'e \`a la valeur propre $\lambda_k$. Le jeu d'\'equations de Langevin
devient dans cette base~:
\begin{equation}
\frac{d}{dt} \tilde{\sigma}_k (t) = \lambda_k \tilde{\sigma}_k (t) 
- z(t) \tilde{\sigma}_k(t) + \tilde{\xi}_k(t) \ .
\label{eq:sphe-Langevintourne}
\end{equation}
Comme le changement de base est orthogonal, $\tilde{\xi}_k(t)$ est encore un 
bruit blanc gaussien avec les m\^emes cumulants que $\xi_i(t)$, 
cf. (\ref{eq:sphe-defxi}).
Chaque mode $\tilde{\sigma}_k$ v\'erifie donc ind\'ependamment l'\'equation
correspondant au mouvement d'une particule dans le potentiel harmonique 
$(-\lambda_k+z(t)) \sigma^2$. 
Les modes sont coupl\'ees implicitement par le multiplicateur de Lagrange 
$z(t)$.

Ces \'equations s'int\`egrent sans difficult\'es en
\begin{equation}
\tilde{\sigma}_k(t)=\tilde{\sigma}_k(0) e^{\lambda_k t - \int_0^t dt' z(t')} 
+ \int_0^t dt'' \ e^{\lambda_k(t-t'') - \int_{t''}^{t} dt' z(t') } 
\tilde{\xi}_k(t'') \ ,
\end{equation}
l'instant initial ayant \'et\'e fix\'e \`a $t=0$.
Introduisant la notation
$\Gamma(t)=\exp[2 \int_0^t dt' z(t')]$, on peut mettre ce r\'esultat sous
la forme
\begin{equation}
\tilde{\sigma}_k(t) = \frac{1}{\sqrt{\Gamma(t)}} \left[ \tilde{\sigma}_k(0)
e^{\lambda_k t} + \int_0^t dt' e^{\lambda_k (t-t')} \sqrt{\Gamma(t')} 
\tilde{\xi}_k(t') \right] \ .
\end{equation}
Il ne reste plus qu'\`a d\'eterminer le multiplicateur de Lagrange $z(t)$,
ou de mani\`ere \'equivalente sa version int\'egr\'ee $\Gamma(t)$ pour
avoir une solution explicite de la dynamique du syst\`eme. Exprimons donc
la fonction de corr\'elation
\begin{equation}
C(t_1,t_2)= \frac{1}{N} \sum_i \langle \sigma_i(t_1) \sigma_i (t_2) \rangle
= \frac{1}{N} \sum_k \langle \tilde{\sigma}_k(t_1) \tilde{\sigma}_k (t_2) 
\rangle \ ,
\end{equation}
o\`u l'on a utilis\'e l'orthogonalit\'e de la matrice de passage pour \'etablir
la deuxi\`eme \'egalit\'e. En supposant que la condition initiale est 
al\'eatoire avec $\tilde{\sigma}_k(0)=\pm 1$\footnote{On mod\'elise donc
une trempe instantan\'e d'une tr\`es haute temp\'erature vers 
la temp\'erature $T$ du bain ext\'erieur.}
, et en notant 
$f(t)=(1/N)\sum_k \exp[2 \lambda_k t]$, on obtient
\begin{equation}
C(t_1,t_2) = \frac{1}{\sqrt{\Gamma(t_1) \Gamma(t_2)}} \left[ 
f\left(\frac{t_1+t_2}{2}\right) + 2 T \int_0^{ {\rm min}(t_1,t_2)} dt' \
f\left(\frac{t_1+t_2}{2} - t'\right) \Gamma(t') \right] \ .
\label{eq:genedyn-C}
\end{equation}
La condition de sph\'ericit\'e s'\'ecrit alors $C(t,t)=1$, soit
\begin{equation}
\Gamma(t) = f(t) + 2 T \int_0^t dt' \ f(t-t') \Gamma(t') \ ,
\label{eq:Volterra}
\end{equation}
ce qui est une \'equation int\'egrale de Volterra.

A partir de la solution explicite des \'equations du mouvement, on peut 
exprimer toutes les quantit\'es int\'eressantes en fonction de $\Gamma$ et $f$.
Par exemple la fonction de r\'eponse \`a un champ ext\'erieur et l'\'energie 
s'\'ecrivent
\begin{eqnarray}
R(t_1;t_2) &=& \sqrt{\frac{\Gamma(t_2)}{\Gamma(t_1)}} 
f\left(\frac{t_1-t_2}{2} \right) \ , \label{eq:genedyn-G} \\
e(t) &=& \frac{T - z(t)}{2} =
\frac{T}{2} - \frac{1}{4} \frac{d}{dt} \ln \Gamma(t) \ . \label{eq:genedyn-ene}
\end{eqnarray}

Notons que la condition de sph\'ericit\'e n'est impos\'ee ici qu'en moyenne 
par rapport aux histoires du bruit thermique et non pour chacune de ses
r\'ealisations. C'est la version dite \og mean spherical\fg\ du
mod\`ele.

On peut conclure de ces g\'en\'eralit\'es que tant la statique que
la dynamique de ces mod\`eles sph\'eriques sont d\'etermin\'ees par
la distribution de valeurs propres de la matrice d'interaction.
Dans le cas compl\`etement connect\'e, c'est-\`a-dire la
version sph\'erique du mod\`ele de Sherrington-Kirkpatrick,
la matrice d'interaction appartient \`a l'ensemble gaussien orthogonal, pour
lequel les valeurs propres sont distribu\'es selon la loi du demi-cercle de 
Wigner. Il est naturel de s'int\'eresser aussi \`a la version sph\'erique du
mod\`ele de Viana-Bray, qui est potentiellement un des plus simples mod\`eles 
dilu\'es. D'apr\`es ce que l'on vient de dire, il convient donc
de d\'eterminer la distribution des valeurs propres de la matrice
d'interaction d\'efinie sur le graphe al\'eatoire poissonnien.
Cet objectif est \`a l'origine de la publication \pubmatrix\ que l'on va
exposer dans la section suivante.

\newpage

\section{Un probl\`eme de matrices al\'eatoires}

\subsection{Introduction}

Les matrices al\'eatoires ont fait leur apparition en physique
dans les ann\'ees 50 avec les travaux de Wigner et Dyson sur les niveaux
d'excitation des noyaux complexes. Elles ont depuis envahies un tel nombre
de domaines de la physique qu'il serait difficile de seulement les mentionner
tous. Citons simplement parmi les sujets connexes \`a celui que l'on va 
d\'evelopper ici le probl\`eme de la localisation d'Anderson dans les 
syst\`emes d\'esordonn\'es~\cite{Th-loc,AbAnTh,DeRo-loc}.
Je renvoie le lecteur int\'eress\'e au livre classique de 
Mehta~\cite{Mehta} et \`a une collection de revues~\cite{JPA-specialmatrix} 
pour une discussion des d\'eveloppements r\'ecents du sujet. 

Comme son nom l'indique, la th\'eorie des matrices al\'eatoires consiste
\`a munir un ensemble de matrices d'une loi de probabilit\'e.
On cherche alors \`a d\'eterminer les propri\'et\'es
statistiques de certaines grandeurs, par exemple la densit\'e moyenne 
de valeurs propres. Des quantit\'es plus fines sont aussi \'etudi\'ees, comme 
la distribution de la plus grande
valeur propre, ou encore la distribution des intervalles entre valeurs propres 
successives. 

On va s'int\'eresser ici \`a un cas particulier, o\`u les matrices
$J$ (de taille $N \times N$) que l'on \'etudie sont r\'eelles sym\'etriques 
(elles sont donc diagonalisables, avec $N$ valeurs propres r\'eelles).
Les \'el\'ements de matrice sont tir\'ees ind\'ependamment avec la m\^eme loi
de probabilit\'e (on distingue seulement les \'el\'ements diagonaux des
autres),
\begin{equation}
\mbox{Prob}(J) = \prod_{i < j} P_1(J_{ij}) \prod_{i} P_2(J_{ii})    \ . 
\label{eq:matrix-loi-factor}
\end{equation}
On notera dans cette partie $[\bullet]$ les moyennes sur l'ensemble de 
matrices. L'exemple le plus connu dans cette famille est l'ensemble 
Gaussien Orthogonal, pour lequel $P_1$ et $P_2$ sont des lois gaussiennes 
de moyenne nulle et de variance respectivement $J_0^2/N$ et $2 J_0^2/N$.
$J_0$ est une grandeur finie, la d\'ependance en $N$ de ces variances est 
choisie de mani\`ere \`a ce que le
spectre des valeurs propres soit born\'e dans la limite thermodynamique.
Une forme \'equivalente pour la loi de probabilit\'e de la matrice est
alors
\begin{equation}
\mbox{Prob}(J) = \exp\left(-\frac{N}{4 J_0^2} \mbox{Tr}(J^2)\right) \ ,
\end{equation}
\`a une constante de normalisation pr\`es. 

Les consid\'erations g\'en\'erales sur le mod\`ele sph\'erique incitent
\`a s'int\'eresser aux spectres de ces matrices.
Si l'on note $\lambda_k$ les valeurs propres pour une r\'ealisation donn\'ee
de la matrice $J$, on d\'efinit la densit\'e de valeurs propres comme
\begin{equation}
\rho_J(\lambda) = \frac{1}{N} \sum_{k=1}^N \delta (\lambda - \lambda_k) \ .
\end{equation}
Sa valeur moyenne sur l'ensemble de matrices sera not\'ee 
$\rho(\lambda)=[\rho_J(\lambda)]$. Dans
le cas de l'ensemble Gaussien Orthogonal, il est bien connu que 
$\rho(\lambda)$ tend dans la limite thermodynamique vers la loi du demi-cercle 
de Wigner. Une d\'emonstration heuristique par la m\'ethode des 
r\'epliques est donn\'ee dans~\cite{EdJo}, et l'on retrouvera ce r\'esultat
comme cas particulier dans la suite de ce chapitre. 
Pour une preuve rigoureuse et des
r\'esultats plus forts sur le type de convergence on pourra 
consulter~\cite{Pastur}.

Notons que d'autres mod\`eles de matrice, notamment dans le cadre
de la gravitation bidimensionnelle~\cite{2dgrav}, utilisent des lois de
probabilit\'e de la forme~:
\begin{equation}
\mbox{Prob}(J) = {\cal N} \exp\left(-\frac{1}{N} \mbox{Tr} \ V(J)\right) \ ,
\end{equation}
avec $V$ un polyn\^ome quelconque. Quand $V$ a
des termes d'ordre sup\'erieur \`a 2, les \'el\'ements de matrice ne sont pas 
ind\'ependants~; ces mod\`eles sortent donc du cadre de l'\'etude pr\'esent\'ee
ici.

\subsection{Matrices dilu\'ees}
\label{sec:mat-dilue}
Le graphe al\'eatoire poissonien d'Erd\"os-R\'enyi conduit naturellement
\`a la d\'efinition d'un ensemble de matrices al\'eatoires dilu\'ees. Pour
cela, il suffit de prendre $J_{ij}=0$ si le lien entre les sommets $i$ et
$j$ est absent du graphe,  et de tirer la valeur de $J_{ij}$
avec une loi de probabilit\'e $\pi$ si $i$ et $j$ sont des sommets adjacents.
On peut aussi poser $J_{ii}=0$ car on consid\`ere qu'il n'y a pas
de liens entre un sommet et lui-m\^eme.
En appelant $p$ la connectivit\'e moyenne du graphe, on a avec les notations 
de la section pr\'ec\'edente
\begin{equation}
P_1(J_{ij}) = \left(1-\frac{p}{N}\right) \delta (J_{ij}) 
+ \frac{p}{N} \pi(J_{ij}) \ , \qquad P_2(J_{ij}) = \delta (J_{ij}) \ ,
\label{eq:mat-P1-dilue}
\end{equation}
o\`u $\pi$ ne contient pas de delta de Dirac en $0$ (cela revient sinon 
\`a modifier la d\'efinition de $p$). Si $\pi(J_{ij})=\delta(J_{ij}-1)$,
autrement dit si les \'el\'ements non nuls de la matrice sont \'egaux \`a
$1$, on a construit la matrice d'adjacence du graphe. Dans la 
suite on va supposer plus g\'en\'eralement que 
\begin{equation}
\pi(J_{ij})=a \, \delta(J_{ij}-J_0) + (1-a) \delta(J_{ij}+J_0) \ .
\end{equation}

Commen\c cons par quelques remarques simples \`a la lumi\`ere de la discussion
sur la g\'eom\'etrie du graphe al\'eatoire pr\'esent\'ee au 
chapitre~\ref{sec:graphes}. Si l'on renomme les sommets de fa\c con \`a les
regrouper selon leur appartenance aux diff\'erentes composantes connexes du
graphe, il est clair que la matrice va se d\'ecomposer sous une forme 
bloc-diagonale, avec un bloc pour chacune des composantes connexes. La 
d\'etermination des valeurs propres de la matrice peut donc se faire
ind\'ependamment pour chacune des composantes du graphe, et la densit\'e
de valeurs propres est une fonction additive par rapport \`a la d\'ecomposition
en clusters.

On a vu que pour $p<1$, c'est-\`a-dire en dessous du seuil de percolation,
une fraction qui tend vers 1 dans la limite thermodynamique de sites sont dans 
des composantes connexes de taille finie, sans boucles. Consid\'erons deux 
types d'arbre pour lesquels on peut facilement d\'eterminer les valeurs 
propres de la matrice qui leur est associ\'ee.

\begin{figure}
\centerline{
\epsfig{file=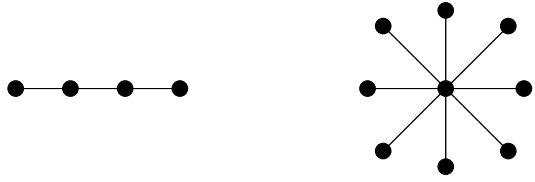,width=7cm}}
\caption{Deux exemples de composantes connexes~: lin\'eaire ($n=4$) 
et en \'etoile ($k=8$).}
\label{fig:matrix-examples}
\end{figure}

Prenons d'abord un graphe lin\'eaire de $n$ sites 
(partie de gauche de la figure~\ref{fig:matrix-examples}). La matrice 
correspondante est tridiagonale et ses $n$
valeurs propres se calculent facilement,
\begin{equation}
\lambda_m = 2 J_0 \cos\left( \frac{m \pi}{n+1} \right) \ , \qquad m 
\in [1,n] \ ,
\end{equation}
qui sont born\'ees sur $]-2 J_0,2J_0[$ quelque soit la valeur de $n$. 

Si l'on consid\`ere au contraire un graphe \og en \'etoile\fg\ o\`u un site 
central est reli\'e \`a $k$ voisins (partie de droite de la 
figure~\ref{fig:matrix-examples}),
la matrice associ\'ee poss\`ede $k-1$ valeurs propres nulles, et deux valeurs 
propres en $\pm J_0 \sqrt{k}$.

En choisissant un site racine au sein d'un arbre, que l'on d\'ecompose 
en plusieurs branches partant de cette racine, on peut assez facilement 
obtenir des relations de r\'ecurrence entre les
polyn\^omes caract\'eristiques des matrices correspondantes.
Cette m\'ethode, expliqu\'ee dans la publication \pubmatrix , permet
notamment de montrer que le spectre de tout arbre est invariant si l'on 
change les signes des \'el\'ements non nuls de la matrice correspondante.
L'ind\'ependance vis-\`a-vis du param\`etre $a$ de la densit\'e moyenne de
valeurs propres est donc prouv\'e, dans la limite thermodynamique, pour
$p<1$. Golinelli~\cite{Go-arbres} a mis \`a profit ces r\'ecurrences pour
calculer le spectre des arbres form\'es d'un squelette lin\'eaire de $n$ sites,
sur chacun desquels on greffe un bouquet de $k$ feuilles. En consid\'erant
toutes les valeurs possibles de $n$ et $k$, il a ainsi montr\'e que l'ensemble
des valeurs propres associ\'ees aux arbres de taille finie \'etait dense dans
l'ensemble des r\'eels.

Tirons les cons\'equences de ces remarques~:
\begin{itemize}
\item Pour $p<1$, tous les vecteurs propres sont localis\'es sur des
composantes connexes de taille finie.
\item Un graphe poissonien comporte un nombre extensif de clusters en 
\'etoiles, pour toutes les connectivit\'es du site central $k$. 
La densit\'e de valeurs propres comporte donc des pics \`a toutes les 
valeurs $\pm J_0 \sqrt{k}$, elle est donc non born\'ee. Ceci reste
d'ailleurs vrai pour toute valeur de $p$~: il y a toujours un nombre
extensifs de ces clusters, que $p$ soit plus grand ou plus petit que le
seuil de percolation.
\item Le r\'esultat de Golinelli implique de plus que la densit\'e de
valeurs propres est form\'ee d'une somme dense de pics de Dirac.
\end{itemize}

Quand la connectivit\'e moyenne $p$ diverge, si l'on r\'e\'echelle 
correctement l'amplitude $J_0$,
on doit retrouver l'ensemble gaussien orthogonal (on rendra cette
remarque plus pr\'ecise dans la section suivante). Pour ce dernier, le
comportement du spectre est tr\`es diff\'erent, et en particulier tous
les vecteurs propres sont \'etendus. On s'attend donc \`a voir une 
transition de d\'elocalisation d'une partie du spectre pour une valeur $p_q>1$ 
(cette transition de d\'elocalisation a \'et\'e
estim\'ee num\'eriquement \`a $p_q \approx 1.4$ dans \cite{EvEc}). Le fait
que les vecteurs propres soient localis\'es ou \'etendus ne se traduit pas
directement dans le caract\`ere continu ou discret de la densit\'e de
valeurs propres, mais dans des quantit\'es plus fines comme les corr\'elations
entre valeurs propres successives, ou les produits de fonctions de Green.

Dans la section suivante on pr\'esente les r\'esultats d'une investigation
de la limite $p \gg 1$ (mais fini par rapport \`a $N$).
Ce probl\`eme a \'et\'e \'etudi\'e \`a plusieurs 
reprises~\cite{BrRo-adj,RodeDo,MiFo-universal,BaGo-moments}, 
on l'a reconsid\'er\'e dans la publication \pubmatrix\
\`a l'aide d'une m\'ethode d\'evelopp\'ee par Biroli et 
Monasson~\cite{BiMo-matrix}. 

\subsection{M\'ethode des r\'epliques}
\label{sec:mat-repliques}
En utilisant l'identit\'e
\begin{equation}
\delta(x)=-\frac{1}{\pi} {\rm Im} \frac{1}{x+i\epsilon} \ ,
\end{equation}
o\`u $\epsilon$ est positif et infinit\'esimal, on peut mettre la
densit\'e de valeurs propres d'une matrice $J$ sous la forme
\begin{equation}
\rho_J(\mu) = \frac{1}{N \pi} {\rm Im} {\rm Tr} \left((J- \mu \mathbb{I})^{-1}
\right) \ ,
\end{equation}
avec $\mathbb{I}$ la matrice identit\'e $N \times N$. Il est sous-entendu
\`a partir de maintenant que $\mu$ a une partie imaginaire infinit\'esimale 
positive. Les propri\'et\'es des int\'egrales gaussiennes permettent
de reformuler cette expression comme un probl\`eme de m\'ecanique statistique,
\begin{eqnarray}
Z_J(\mu) &=& \int \prod_{i=1}^N d\phi_i 
\exp\left(\frac{i \mu}{2} \sum_i \phi_i^2
- \frac{i}{2} \sum_{i,j} J_{ij} \phi_i \phi_j \right) \ , \\
\rho_{J}(\mu) &=& \frac{2}{N \pi} \mbox{Im} 
\frac{\partial}{\partial \mu} \ln Z_J(\mu) \ . \label{eq:mat-rhoJ}
\end{eqnarray}
La convergence de l'int\'egrale gaussienne est assur\'ee par
le choix des exposants complexes gr\^ace \`a la partie imaginaire de $\mu$.
Afin d'obtenir la densit\'e moyenne de valeurs propres, il faut donc 
calculer $[\ln Z_J(\mu)]$,
en moyennant sur les \og variables gel\'ees\fg\ $J_{ij}$. Le calcul
direct de la moyenne d'un logarithme \'etant difficile, on utilise la m\'ethode
des r\'epliques qui repose sur l'identit\'e
\begin{equation}
[ \ln Z ] = \lim_{n \to 0}\frac{1}{n} \ln [Z^n] \ . \label{eq:mat-repliques}
\end{equation}
Le passage \`a la limite des valeurs de $n$ enti\`eres vers 0 peut
dans certains cas n\'ecessiter la prise en compte d'effets subtils
de brisure de sym\'etrie des r\'epliques. L'exemple le plus fameux
est donn\'e par la phase de basse temp\'erature du mod\`ele de
Sherrington-Kirkpatrick. Dans le cas pr\'esent du calcul
d'une densit\'e de valeurs propres de tels effets ne sont pas 
attendus~\cite{BrRo-adj,MiFo-universal} (il faudrait par contre briser
la sym\'etrie des r\'epliques pour calculer les corr\'elations entre valeurs
propres, cf. \cite{KaMe-matrix}).

$Z^n$ va s'exprimer comme l'int\'egrale sur des champ \og r\'epliqu\'es\fg\
$n$ fois, que l'on notera $\vec{\phi}_i$.
La moyenne sur l'ensemble des matrices fait appara\^itre un couplage entre
les diff\'erentes r\'epliques. Dans le cadre des syst\`emes dilu\'es, le
param\`etre d'ordre global qui s'est av\'er\'e utile est
la fraction des sites portant un champ donn\'e~\cite{Mo-c-sigma} ,
\begin{equation}
c(\vec{\phi})=\frac{1}{N} \sum_i \delta(\vec{\phi}_i - \vec{\phi}) \ .
\label{eq:mat-defc_phi}
\end{equation}
On va faire le calcul avec une loi de probabilit\'e des \'el\'ements de
matrice $P_1$ quelconque. Cela permettra de retrouver
la loi du demi-cercle dans le cas de l'ensemble gaussien orthogonal et de
d\'emontrer son \og universalit\'e\fg\ avant de traiter le cas dilu\'e. On
doit calculer
\begin{equation}
[Z(\mu)^n] = \int d\vec{\phi}_i \, Dc(\vec{\phi}) \, 
\delta \left( Nc(\vec{\phi}) - \sum_i \delta(\vec{\phi}-\vec{\phi}_i) \right) 
e^{\frac{i\mu}{2} \sum_i \vec{\phi}_i^2 } 
\left[e^{-\frac{i}{2} \sum_{ij} J_{ij} 
\vec{\phi}_i \cdot \vec{\phi}_j } \right] \ .
\end{equation}
L'int\'egration sur $c$ et le $\delta$ 
doivent \^etre compris ici dans un sens fonctionnel~: l'\'egalit\'e
(\ref{eq:mat-defc_phi}) est impos\'ee pour toutes les valeurs du
champ $\vec{\phi}$. Comme les \'el\'ements de matrice pour $i<j$
sont ind\'ependants, la moyenne sur l'ensemble des matrices se factorise en
un produit sur toutes les paires de sites. En d\'efinissant
\begin{equation}
g(x) = N \ln \left( \int dJ \ P_1(J) e^{-iJx} \right) 
\label{eq:mat-defg}
\end{equation}
il vient
\begin{eqnarray}
[Z(\mu)^n]&=& \int d\vec{\phi}_i \, Dc(\vec{\phi}) \, 
\delta \left( Nc(\vec{\phi}) - \sum_i \delta(\vec{\phi}-\vec{\phi}_i) \right) 
\label{eq:mat-Zn}\\ 
&& \hspace{1cm} \exp\left[ N \left(   
\frac{i \mu}{2} \int d\vec{\phi} \  c(\vec{\phi}) \vec{\phi}^2 +
\frac{1}{2} \int d\vec{\phi} \, d\vec{\psi} \ c(\vec{\phi}) c(\vec{\psi})
g(\vec{\phi} \cdot \vec{\psi}) \right) \right] \nonumber \ .
\end{eqnarray}
Il reste \`a effectuer l'int\'egrale sur les champs initiaux $\vec{\phi}_i$,
ce qui va faire appara\^itre un terme entropique due \`a la multiplicit\'e
des configurations des variables $\vec{\phi}_i$ qui conduisent au m\^eme
$c(\vec{\phi})$. Une fa\c con de faire ce calcul consiste \`a introduire
une fonctionnelle conjugu\'ee $\hat{c}(\vec{\phi})$ pour exponentier 
la contrainte, puis effectuer l'int\'egrale sur
les champs $\vec{\phi}_i$, et finalement celle sur $\hat{c}$ avec la m\'ethode
du col~:
\begin{eqnarray}
& & \hspace{-1cm}
\int d\vec{\phi}_i \, D\hat{c}(\vec{\phi}) \, 
\exp \left[\int d\vec{\phi} \, \hat{c}(\vec{\phi})
\left(N c(\vec{\phi}) - \sum_i \delta(\phi - \phi_i)\right) \right]
= \nonumber \\ & & \hspace{1cm} \int D\hat{c}(\vec{\phi}) 
\exp \left[N\left(\int d\vec{\phi} \, \hat{c}(\vec{\phi}) c(\vec{\phi}) 
+ \ln \left( \int d\vec{\phi} \, e^{-\hat{c}(\vec{\phi}) } \right) \right) 
\right] \ .
\end{eqnarray}
L'\'equation de col pour cette derni\`ere int\'egrale s'\'ecrit
\begin{equation}
c(\vec{\phi}) = e^{-\hat{c}(\vec{\phi}) } \left( \int d\vec{\psi} \,
e^{-\hat{c}(\vec{\psi}) } \right)^{-1} \ .
\end{equation}
En ins\'erant ce r\'esultat dans (\ref{eq:mat-Zn}), on obtient finalement
\begin{eqnarray}
& &[Z(\mu)^n] = \int Dc(\vec{\phi}) \exp[N S(c)] \ , \label{eq:mat-Sdec} \\
& & 
S(c) = - \int d\vec{\phi} \ c(\vec{\phi}) \ln c(\vec{\phi}) + \frac{i \mu}{2}
\int d\vec{\phi} \ \vec{\phi}^2 c(\vec{\phi}) + \frac{1}{2}
\int d\vec{\phi} \, d\vec{\psi} \ c(\vec{\phi}) c(\vec{\psi}) 
g( \vec{\phi} \cdot \vec{\psi}) \ , \nonumber
\end{eqnarray}
o\`u l'on reconna\^it le terme entropique en $c \ln c$. Cette int\'egrale
fonctionnelle, dont le domaine d'int\'egration doit \^etre restreint aux
$c(\vec{\phi})$ normalis\'ees, peut se calculer par la m\'ethode du col
dans la limite thermodynamique. Le col
$c_*$ est solution de
\begin{equation}
c_*(\vec{\phi}) = {\cal N} \exp \left[ \frac{i \mu}{2} \vec{\phi}^2 + \int
d \vec{\psi} \ c_*(\vec{\psi}) g(\vec{\phi} \cdot \vec{\psi}) \right] \ .
\label{eq:mat-spc}
\end{equation}
Une fois cette \'equation r\'esolue, la densit\'e moyenne de valeurs propres
d\'ecoule de (\ref{eq:mat-rhoJ}) et (\ref{eq:mat-repliques})~:
\begin{equation}
\rho (\mu) =\lim_{n \to 0} \frac{1}{n \pi} {\rm Re} \int d\vec{\phi} \ 
c_*(\vec{\phi}) \vec{\phi}^2 \ .
\end{equation}

\vspace{8mm}

\noindent 
\underline{\emph{L'ensemble gaussien orthogonal et son universalit\'e}}

\vspace{4mm}
La d\'emarche pr\'esent\'ee jusqu'ici est valable quelque soit l'ensemble de 
matrices utilis\'e, les diff\'erents ensembles conduisant \`a diff\'erentes
formes de la fonction $g(x)$. Commen\c cons par traiter le cas de 
l'ensemble gaussien orthogonal, pour lequel les \'el\'ements
de matrice sont tir\'ees avec une loi gaussienne de moyenne nulle et de
variance $[J_{ij}^2]=J_0^2/N$. On trouve alors que $g$ est quadratique,
$g(x)=-J_0^2 x^2/2$. La solution de l'\'equation de col (\ref{eq:mat-spc})
est obtenue avec un $c_*$ gaussien,
\begin{equation}
c_*(\vec{\phi}) = (2i\pi \sigma(\mu))^{-n/2} e^{-\frac{1}{2} 
\frac{\vec{\phi}^2}{i \sigma(\mu)}} \ ,
\label{eq:mat-cgaussien}
\end{equation}
dont la variance $\sigma(\mu)$ est solution de l'\'equation du deuxi\`eme 
ordre~:
\begin{equation}
J_0^2 \sigma^2 - \mu \sigma + 1 =0 \ .
\end{equation}
La densit\'e de valeurs propres est alors donn\'ee par
$\rho(\mu)= -(1/\pi) {\rm Im}\, \sigma(\mu)$. En r\'esolvant
l'\'equation sur $\sigma$, on trouve le r\'esultat attendu~: pour $\mu$ \`a
l'ext\'erieur de $[-2 J_0,2 J_0]$ $\sigma$ est r\'eel et donc $\rho$ est nul.
Sur cet intervalle les valeurs propres sont par contre distribu\'ees selon 
la loi du demi-cercle,
\begin{equation}
\rho(\mu) = \frac{1}{2 \pi J_0^2} \sqrt{4 J_0^2 - \mu^2} \ .
\end{equation}

Discutons la g\'en\'eralit\'e de ce r\'esultat (l'argumentation est
adapt\'ee de \cite{MiFo-universal}). Il est clair que d\`es que $g(x)$ est
quadratique, on obtient une telle densit\'e d'\'etats. De plus, si l'on veut
que le spectre soit ind\'ependant de $N$ dans la limite $N \to \infty$, il
faut que $g(x)$ soit d'ordre $1$ dans cette limite. D'apr\`es la d\'efinition
(\ref{eq:mat-defg}), si $P_1(J)$ ne contient pas de delta en 0, il faut
que la distribution soit support\'e par les $J$ d'ordre $N^{-1/2}$ pour
que $g$ soit d'ordre 1, et alors elle est forc\'ement quadratique. Dans ce
cas on avait tous les \'el\'ements de la matrice non-nuls, et d'ordre 
$N^{-1/2}$. On peut essayer de \og diluer\fg\ la matrice, c'est-\`a-dire
de ne prendre en moyenne que ${\cal O}(N^{1-\alpha})$ termes non nuls par 
ligne, avec $\alpha \in [0,1]$. Plus pr\'ecis\'ement, posons
\begin{equation}
P_1(J) = \left( 1 - \frac{p}{N^\alpha} \right) \delta(J) + \frac{p}{N^\alpha}
\pi(J) \ ,
\end{equation}
avec $\pi$ une loi de probabilit\'e paire, sans Dirac en 0. Si $\alpha<1$,
il faut que $\pi$ soit significative pour des $J$ d'ordre $N^{(\alpha-1)/2}$,
toujours pour avoir $g$ d'ordre 1, et dans ce cas-l\`a \`a nouveau seul le
terme quadratique de $g$ survit dans la limite thermodynamique.

Il ne reste en fait que le cas $\alpha=1$ pour \'echapper \`a la loi du
demi-cercle. En effet, $\pi$ est alors support\'e par les $J$ d'ordre 1, et
donc $g$ peut \^etre quelconque. Dans la situation $\alpha=1$, le graphe
associ\'e aux \'el\'ements de matrice non nuls est pr\'ecis\'ement un graphe
al\'eatoire d'Erd\"os et R\'enyi, auquel on va s'int\'eresser dans la suite de
cette partie. On suivra la m\'ethode dite d'approximation \`a un seul d\'efaut,
introduite par Biroli et Monasson~\cite{BiMo-matrix}.

Notons avant cela que l'on a suppos\'e que la loi de probabilit\'e des $J_{ij}$
\'etait paire. Si ce n'est pas le cas, $g(x)$ poss\`ede un terme lin\'eaire,
ce qui peut entra\^iner l'apparition d'une valeur propre 
isol\'ee~\cite{EdJo}. Enfin, le raisonnement ci-dessus est pris en d\'efaut
quand les \'el\'ements de matrice sont distribu\'ees avec une loi qui ne
d\'ecro\^it qu'alg\'ebriquement \`a l'infini~: c'est le cas des matrices de
L\'evy qui ont \'et\'e \'etudi\'ees par Cizeau et Bouchaud~\cite{CiBo}.
L'argument fait ici supposait que la variance des $J_{ij}$ \'etait bien
d\'efinie, hypoth\`ese viol\'ee par les lois larges de L\'evy.

\vspace{8mm}

\noindent 
\underline{\emph{L'approximation du milieu effectif (EMA)}}

\vspace{4mm} 

On reprend \`a partir de maintenant la forme 
(\ref{eq:mat-P1-dilue}), avec $\pi(J_{ij})=a \delta(J_{ij}-J_0) + (1-a) 
\delta(J_{ij}+J_0)$, ce qui conduit \`a
\begin{equation}
g(x)= - p + p \left( a e^{-i J_0 x} + (1-a) e^{i J_0 x} \right) \ .
\end{equation}
Dans ce cas on ne peut pas avoir une r\'esolution analytique exacte de 
l'\'equation (\ref{eq:mat-spc}). On peut cependant, inspir\'e par la
r\'esolution du cas compl\`etement connect\'e, faire un ansatz gaussien
pour $c(\vec{\phi})$, en utilisant la forme (\ref{eq:mat-cgaussien}).
On cherche donc un point col dans le sous-ensemble des param\`etres d'ordre
de cette forme. Si l'on ins\`ere cet ansatz gaussien dans l'expression de 
l'action (\ref{eq:mat-Sdec}), on obtient une expression qui n'est plus fonction
que du param\`etre variationnel $\sigma$
\begin{equation}
S(\sigma) \sim_{n \to 0} \frac{n}{2} \left[ 1+ \ln (2 i \pi \sigma) 
- \mu \sigma - p \ln (1- J_0^2 \sigma^2) \right] \ .
\end{equation}
Remarquons que cette expression est ind\'ependante du param\`etre de
biais $a$.

L'extremum de l'action dans le sous-espace correspondant \`a cet
ansatz est atteint quand $\sigma$ v\'erifie l'\'equation cubique suivante~:
\begin{equation}
\sigma^3 + \frac{p-1}{\mu} \sigma^2 - \frac{1}{J_0^2} \sigma 
+ \frac{1}{\mu J_0^2} = 0 \ .
\label{eq:mat-sigmaEMA}
\end{equation}
On peut v\'erifier que dans la limite $p\to \infty$ avec $J_0\sim p^{-1/2}$ on
retrouve l'\'equation quadratique de l'ensemble gaussien orthogonal. Pour $p$
fini on peut r\'esoudre cette \'equation cubique, et trouver une valeur
$\lambda_c(p)$ qui s\'epare deux r\'egimes~:
\begin{itemize}
\item \`a l'ext\'erieur de l'intervalle $[-\lambda_c,\lambda_c]$,
 $\sigma$ est r\'eelle, et donc la densit\'e de valeurs propres s'annule.
\item \`a l'int\'erieur de cet intervalle $\sigma$ a une partie imaginaire
non nulle, on a donc une densit\'e de valeurs propres $\rho$ positive,
qui s'annule \`a $\lambda_c$ comme une racine carr\'ee.
\end{itemize}

\begin{figure}
\includegraphics{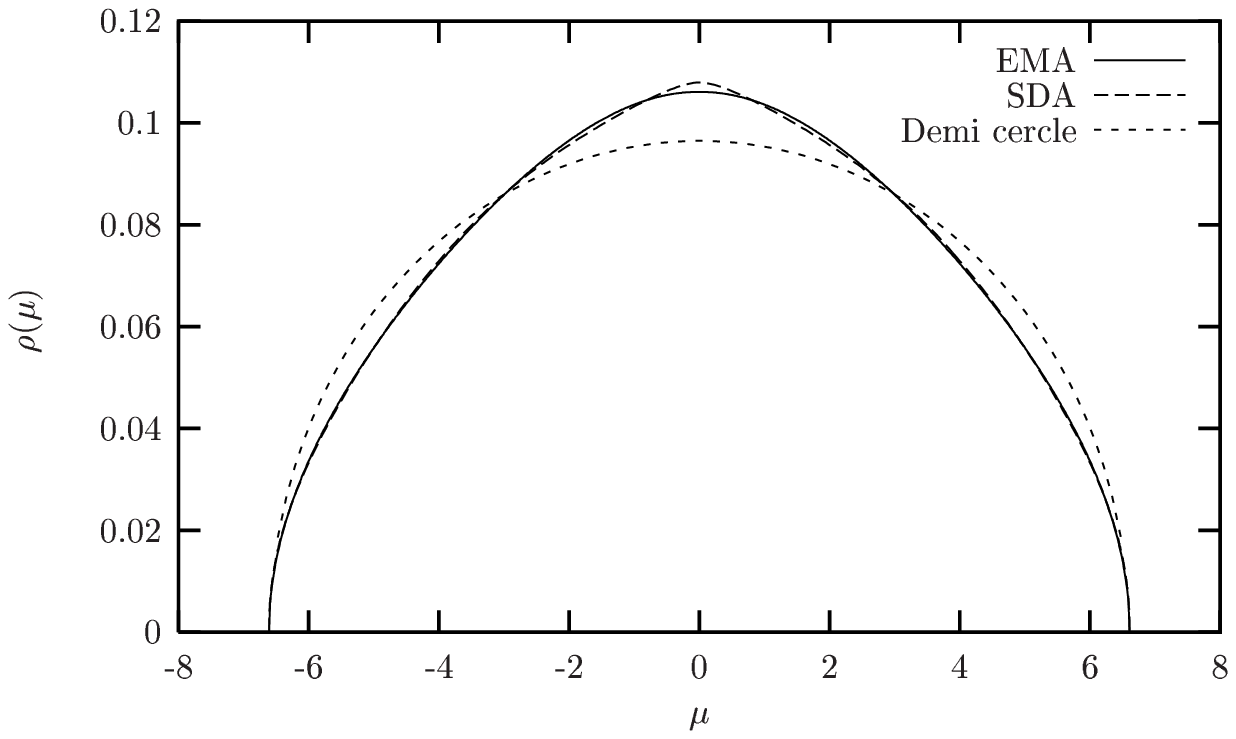}
\caption{Les pr\'edictions de l'approximation du milieu effectif, et de
l'approximation \`a un seul d\'efaut pour $p=10$.
La loi du demi-cercle de m\^eme largeur est l\`a \`a titre de comparaison.}
\label{fig:mat-rhoana}
\end{figure}

L'expression de $\lambda_c$ et de $\rho$ n'\'etant pas particuli\`erement
\'eclairantes, je ne les reproduit pas ici. L'allure de la densit\'e
d'\'etats ainsi pr\'edites est repr\'esent\'ee sur la 
figure~\ref{fig:mat-rhoana}.

L'ansatz gaussien pour $c$ n'est pas justifi\'e par
un argument variationnel au sens strict~: on n'a pas une borne sur 
l'action $S(c)$ qui justifierait de chercher un extremum sur un 
sous-ensemble de l'espace des param\`etres d'ordre. La justification tient
plut\^ot dans la co\"incidence avec le r\'esultat correct (loi du demi-cercle)
dans la limite o\`u $p$ diverge.

Notons finalement que ce r\'esultat est clairement en d\'esaccord avec les 
remarques
qualitatives de la section pr\'ec\'edente~: le spectre est born\'e
sur $[-\lambda_c,\lambda_c]$, alors que l'on avait montr\'e qu'il devait
s'\'etendre sur tout l'axe r\'eel. Cette diff\'erence s'explique simplement~:
les tr\`es grandes valeurs propres sont dues aux sites avec une grande
connectivit\'e. Or l'approximation du milieu effectif consiste justement
\`a traiter tous les sites sur le m\^eme pied, et donc \`a n\'egliger les
fluctuations de la connectivit\'e.

\vspace{8mm}

\underline{\emph{L'approximation \`a un seul d\'efaut (SDA)}}

\vspace{4mm}

On peut aller au-del\`a de l'approximation du milieu effectif, avec
l'objectif de gommer la contradiction que l'on vient de mentionner.

L'id\'ee consiste \`a se servir de l'expression gaussienne pour $c(\vec{\phi})$
obtenue avec l'EMA comme le point de 
d\'epart d'une s\'erie d'it\'erations qui devraient conduire \`a des 
am\'eliorations successives de la pr\'ediction pour $\rho(\mu)$. 
Reformulons l'\'equation du point col (\ref{eq:mat-spc}) sous la forme
\begin{equation}
c(\vec{\phi}) = {\cal N} e^{\frac{i \mu}{2} \vec{\phi}^2} \sum_{k=0}^\infty
\frac{e^{-p} p^k}{k!} 
\left[\int d\vec{\psi} \, c(\vec{\phi}) \left( 
a \, e^{-i J_0 \vec{\phi} \cdot \vec{\psi}} 
+ (1-a) e^{i J_0 \vec{\phi} \cdot \vec{\psi}}
\right)
\right]^k \ ,
\end{equation} 
o\`u ${\cal N}$ est un facteur de normalisation. On injecte alors la
forme gaussienne approch\'ee de $c$ dans le membre de droite, et le membre de
gauche en fournit une nouvelle forme, que l'on esp\`ere meilleure. On 
peut penser qu'en r\'ep\'etant ces it\'erations un certain nombre de fois,
on va se rapprocher du vrai point col.

Consid\'erons le r\'esultat de la premi\`ere it\'eration. On obtient 
pour $c(\vec{\phi})$ une somme de gaussiennes, qui conduisent \`a 
l'expression de la densit\'e de valeurs propres~:
\begin{equation}
\rho(\mu) = - \frac{1}{\pi}\sum_{k=0}^\infty \frac{e^{-p} p^k}{k!}  {\rm Im}
\frac{1}{\mu - k J_0^2 \sigma(\mu)} \ ,
\label{eq:mat-rhosda}
\end{equation}
o\`u $\sigma(\mu)$ est la solution de l'\'equation cubique obtenue dans
l'approximation du milieu effectif.

L'interpr\'etation de cette \'equation est la suivante~: un site donn\'e a 
connectivit\'e $k$ avec une loi poissonnienne de param\`etre $p$, et ses 
$k$ voisins sont d\'ecrits
de mani\`ere approch\'ee par l'interm\'ediaire du $\sigma$ calcul\'e 
pr\'ec\'edemment dans l'approximation gaussienne. Ceci explique le
nom d'approximation \`a un seul d\'efaut, on traite exactement un site
(d\'efaut) au milieu d'un r\'eseau homog\`ene effectif. A nouveau le
param\`etre de biais $a$ a disparu de l'expression des grandeurs physiques.

Dans la zone $[-\lambda_c,\lambda_c]$ o\`u l'approximation gaussienne 
pr\'edisait une densit\'e d'\'etats non nulle, $\sigma$ \'etait d\'ej\`a 
imaginaire, la nouvelle expression (\ref{eq:mat-rhosda}) modifie un peu la 
forme de $\rho$ (cf. figure \ref{fig:mat-rhoana}).

Une grande diff\'erence entre les deux niveaux d'approximations appara\^it
dans la zone $|\mu|>\lambda_c$~: alors que l'EMA pr\'edisait une densit\'e
d'\'etats nulle, ici on a une s\'erie de pics de Dirac quand le d\'enominateur
de (\ref{eq:mat-rhosda}) s'annule (rappelons que $\mu$ a une partie imaginaire
infinit\'esimale). Leur position est donc $\pm \mu_k$, et
leur poids $w_k$, avec
\begin{equation}
\sigma(\mu_k) = \frac{\mu_k}{k J_0^2} \ , \qquad
w_k = \frac{e^{-p} p^k}{k!} \frac{1}{1-k J_0^2 \sigma'(\mu_k)} \ .
\end{equation}
On peut en particulier s'int\'eresser au r\'egime asymptotique 
$|\mu| \to \infty$. Il est facile de montrer \`a partir de 
(\ref{eq:mat-sigmaEMA}) que dans cette limite $\sigma(\mu) \sim \mu^{-1}$.
On trouve donc que les pics sont situ\'es asymptotiquement en 
$\pm J_0 \sqrt{k}$, avec un poids donn\'e par la moiti\'e 
(\`a cause des deux signes possibles)
de la loi de Poisson de param\`etre $p$. En remarquant que 
$\sqrt{k}-\sqrt{k-1} \to 0$ quand $k \to \infty$, on peut formuler une
approximation continue pour la densit\'e d'\'etats dans ce r\'egime,
\begin{equation}
\rho(J_0 \sqrt{k}) \times J_0 (\sqrt{k}-\sqrt{k-1}) 
\sim \frac{1}{2} \frac{e^{-p} p^k}{k!} \ ,
\end{equation}
soit en utilisant la formule de Stirling et en changeant de variables
\begin{equation}
\rho(\mu) \sim \frac{1}{J_0 \sqrt{2\pi}} \exp \left[ - p - \frac{\mu^2}{J_0^2} 
\ln \left( \frac{\mu^2}{J_0^2 e p} \right) \right] \ .
\label{eq:mat-rhoas}
\end{equation}
Cette expression avait \'et\'e obtenue par Rodgers et Bray~\cite{BrRo-adj} 
apr\`es un traitement assez subtil d'une \'equation int\'egrale,
inspir\'e par un travail de Kim et Harris~\cite{KiHa}.
Ce r\'esultat prend ici un sens g\'eom\'etrique tr\`es simple~:
les sites dont la connectivit\'e $k$ est
tr\`es sup\'erieure \`a la connectivit\'e moyenne $p$ portent des
vecteurs propres fortement localis\'es sur eux, qu'ils soient strictement 
isol\'es du reste du graphe comme dans un cluster en \'etoile, ou que leur
environnement soit remplac\'e par un milieu effectif comme l'on vient de le 
faire.

\begin{figure}
\includegraphics{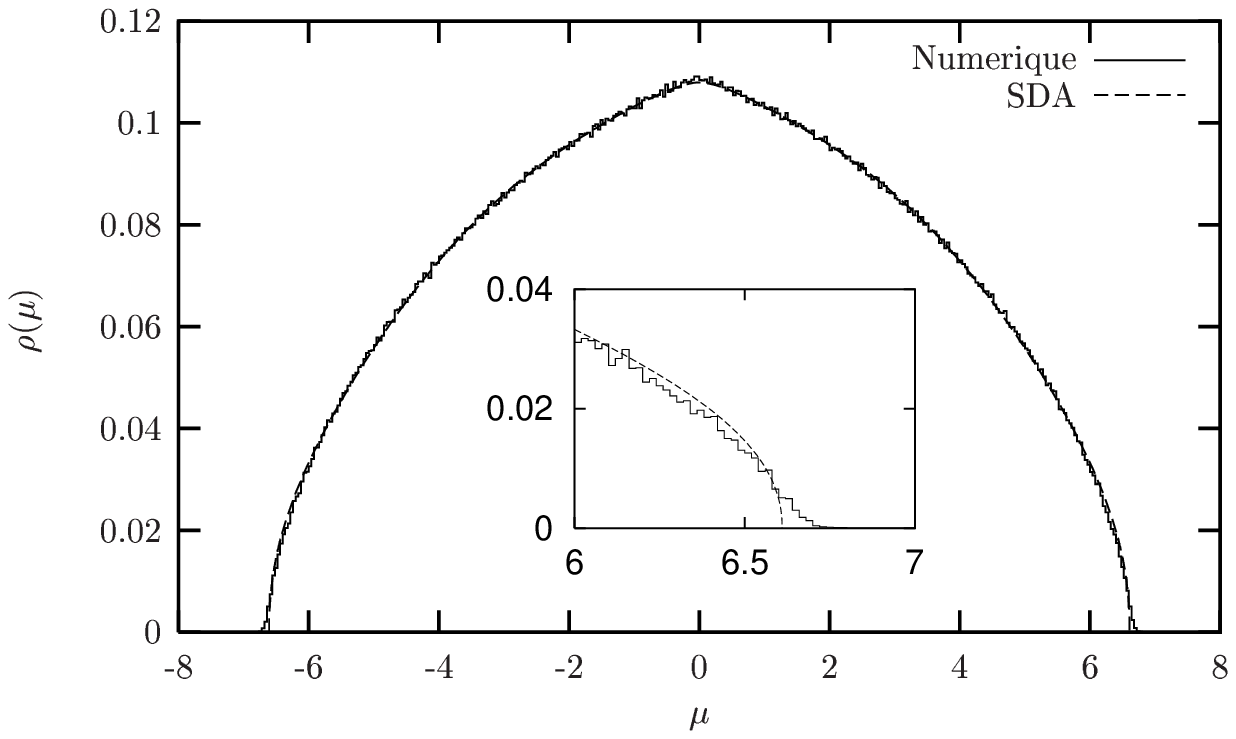}
\caption{L'approximation \`a un seul d\'efaut est compar\'ee aux
r\'esultats de diagonalisation num\'erique pour des matrices de taille
$N=2000$. Les courbes sont quasiment superpos\'ees dans toute la partie
centrale. L'inset montre les d\'eviations au voisinage de $\lambda_c$.}
\label{fig:mat-rhonum}
\end{figure}

La figure \ref{fig:mat-rhonum} pr\'esente les r\'esultats d'une \'etude
num\'erique, o\`u l'on a diagonalis\'e des matrices tir\'ees al\'eatoirement
avec la loi de probabilit\'e \'etudi\'ee ici. L'accord avec l'approximation
\`a un seul d\'efaut est tr\`es bon dans la partie centrale du spectre.
On constate aussi qu'il y a une queue s'\'etendant au del\`a de $\lambda_c$.
Cependant le calcul SDA n'est pas capable de pr\'edire quantitativement 
la densit\'e de valeurs propres au voisinage de $\lambda_c$, il faudrait
pour cela \^etre capable d'aller aux niveaux sup\'erieurs d'it\'erations
dans ce sch\'ema d'approximation. L'expression (\ref{eq:mat-rhoas}) n'est
en effet valable que dans la limite $|\mu| \to \infty$. Or le poids dans
ces queues, dues \`a des \'ev\`enements rares (grandes fluctuations dans la
connectivit\'e) est tr\`es faible, et donc quasiment impossible \`a observer
dans ces simulations num\'eriques o\`u l'on g\'en\`ere seulement des graphes
typiques. Une possibilit\'e pour explorer num\'eriquement ce r\'egime de
grande d\'eviation~\cite{Werner-pc} consisterait \`a biaiser la 
g\'en\'eration des graphes en faveur de ceux qui pr\'esentent des grandes 
valeurs propres, une m\'ethode d\'ej\`a utilis\'ee dans un cadre un peu 
diff\'erent~\cite{Werner}.

\subsection{Perspectives}

Plusieurs questions restent ouvertes sur ce probl\`eme du spectre des 
matrices d'adjacence de graphes al\'eatoires d'Erd\"os-R\'enyi. La premi\`ere
concerne la validit\'e des r\'esultats obtenus ici par une m\'ethode 
it\'erative dont on ne contr\^ole pas explicitement la convergence vers
la vraie solution. Comme on l'a vu, dans la limite de grande connectivit\'e
moyenne $p$, le premier niveau d'approximation est correct~: l'approximation
du milieu effectif redonne en effet la loi du demi-cercle dans cette limite.
A $p$ fini, le d\'eveloppement asymptotique pour les grandes valeurs propres
(\ref{eq:mat-rhoas}) est aussi s\^urement correct. La co\"incidence avec
le r\'esultat obtenu par Rodgers et Bray~\cite{BrRo-adj} \`a l'aide d'une 
m\'ethode un peu 
diff\'erente est rassurante de ce point de vue, ainsi que la simplicit\'e
de l'argument g\'eom\'etrique dont il provient. Les vecteurs propres 
correspondants \'etant tr\`es fortement localis\'es autour de sites 
particuli\`erement connect\'es, ils sont asymptotiquement insensibles \`a
leur environnement. Bauer et Golinelli~\cite{BaGo-moments} ont \'etablis des 
relations de r\'ecurrence sur les moments de la densit\'e d'\'etats de ces 
matrices, il serait peut-\^etre possible d'en tirer une autre preuve du 
d\'eveloppement asymptotique (\ref{eq:mat-rhoas}). Signalons au passage
que leurs r\'esultats justifient l'ind\'ependance par rapport au biais 
$a$ que l'on a constat\'e ici ordre par ordre dans la r\'esolution it\'erative.
En effet, le calcul d'un moment d'ordre fini ne d\'epend que de l'environnement
\`a distance finie d'un site. Dans la limite thermodynamique les graphes
al\'eatoires \'etant localement des arbres, le signe des interactions n'est pas
pertinent.

Dans une perspective plus rigoureuse, il serait aussi int\'eressant de 
conna\^itre la nature du spectre \`a une \'echelle plus fine. Comme on l'a
discut\'e dans la partie \ref{sec:mat-dilue}, la densit\'e de valeurs
propres comporte une infinit\'e de pics de Dirac \`a toutes les positions
correspondant aux valeurs propres d'arbres de taille finie, c'est-\`a-dire
un ensemble qui est dense dans les r\'eels~\cite{Go-arbres}. 
Les vecteurs propres correspondants sont fortement localis\'es sur un nombre
fini de sites. Apparaissent aussi, pour une valeur de $p$ suffisamment grande, 
des vecteurs propres \'etendus sur l'amas infini de percolation. 
Le seuil $\lambda_c$ calcul\'e ici est
une estimation approch\'ee d'un seuil de mobilit\'e s\'eparant
une r\'egion $|\mu|<\lambda_c$ o\`u coexistent des vecteurs propres localis\'es
et \'etendus d'une r\'egion ext\'erieure o\`u tous les vecteurs propres
sont localis\'es. A ma connaissance le seul r\'esultat analytique sur ce 
probl\`eme~\cite{BaGo-conductor} concerne le comportement de la valeur propre 
nulle de la matrice d'adjacence, qui pr\'esente un ph\'enom\`ene de
d\'elocalisation et de relocalisation \`a deux valeurs de $p$.

La m\'ethode it\'erative utilis\'ee ici a \'et\'e introduite par
Biroli et Monasson~\cite{BiMo-matrix} pour l'\'etude des matrices dites
Laplaciennes~: les \'el\'ements diagonaux de ces matrices sont ajust\'es en
fonctions des \'el\'ements hors-diagonale, de mani\`ere telle que la somme
des \'el\'ements sur une ligne s'annule. Ce probl\`eme a \'et\'e aussi
\'etudi\'e dans \cite{BrRo-diffusion,Dean-matrix}. Des matrices
similaires apparaissent aussi dans l'\'etude des matrices al\'eatoires
euclidiennes~\cite{euclideanRMT}, en rapport avec l'\'etude des modes
instantan\'es de vibration dans les liquides surfondus~\cite{CaGiPa-INM}.
Dans ce dernier article notamment une r\'esolution num\'erique d'une
\'equation de col proche de celle rencontr\'ee ici \'etait propos\'ee.
Signalons finalement que le spectre de matrice d'adjacence
des graphes \og scale-free\fg\ a \'et\'e l'objet de travaux num\'eriques
et analytiques~\cite{sf-spectra1,sf-spectra2,Papa-eigen}.

\newpage

\section{Cons\'equences sur le mod\`ele sph\'erique}

\subsection{Rappels sur le cas compl\`etement connect\'e}
\label{sec:stadynp2}

Afin de faciliter l'exposition des r\'esultats dans le cas dilu\'e, 
je vais commencer par rappeler bri\`evement le comportement du mod\`ele 
sph\'erique dans le cas o\`u la matrice d'interaction appartient \`a l'ensemble
gaussien orthogonal. La partie statique a \'et\'e trait\'ee dans~\cite{KoThJo}
et celle dynamique dans~\cite{CidePa,CuDe,ZiKuHo}, pour une approche 
math\'ematiquement rigoureuse on pourra se reporter \`a~\cite{p2-matheux}. 

En red\'efinissant l'\'echelle de temp\'erature, on se ram\`ene \`a
une densit\'e de valeurs propres distribu\'ees selon la loi du demi-cercle 
sur $[-2,2]$,
\begin{equation}
\rho(\mu)=\frac{1}{2\pi} \sqrt{4-\mu^2} \ .
\end{equation}
La fonction de partition du mod\`ele est obtenue dans la limite thermodynamique
en \'evaluant l'int\'egrale (\ref{eq:Zsphe}) par la m\'ethode du col.
Les sommes sur les valeurs propres sont alors remplac\'ees par des 
int\'egrales,
\begin{equation}
\frac{1}{N} \sum_k F(\lambda_k) \to \int d\mu \, \rho(\mu) F(\mu) \ .
\end{equation}
Le col $z_*$ de l'int\'egrale (\ref{eq:Zsphe}) v\'erifie l'\'equation
suivante~:
\begin{equation}
1=\int d\mu \, \rho(\mu) \, \frac{1}{2 z_* - \beta \mu} \ .
\end{equation}
On doit par ailleurs imposer $z_* > \beta$ pour que l'int\'egrale 
gaussienne initiale (cf. (\ref{eq:sphe-Z})) soit convergente.

A haute temp\'erature l'\'equation de col a une solution qui v\'erifie cette
condition~: $z_*=(1+\beta^2)/2$. Quand on r\'eduit la temp\'erature le point
col se rapproche du point de branchement de l'int\'egrale en $z=\beta$,
qui est atteint pour $\beta_c=1$. A des temp\'eratures plus basses que cette
temp\'erature critique, l'int\'egrale sur $z$ est domin\'e par le voisinage
du point de branchement, le chemin d'int\'egration reste \og coll\'e\fg\
\`a la coupure. De plus, cette transition de phase se traduit
par une \og condensation\fg\ un peu similaire \`a la transition de 
Bose-Einstein pour un syst\`eme de bosons. En effet, la projection de la
configuration des spins $\sigma_i$ sur le vecteur propre de plus grande
valeur propre devient d'ordre $\sqrt{N}$ \`a basse temp\'erature. Le
pr\'efacteur, qui mesure le taux de condensation sur ce vecteur propre,
cro\^it contin\^ument et lin\'eairement de 0 \`a la temp\'erature critique
jusqu'\`a atteindre 1 \`a temp\'erature nulle. L'\'equilibre \`a basse
temp\'erature correspond donc \`a une condensation macroscopique sur
le mode de plus grande valeur propre.

La dynamique de ce mod\`ele pr\'esente une transition de phase \`a la
m\^eme temp\'erature\footnote{La co\"incidence des temp\'eratures de
transition statique et dynamique est une particularit\'e de ce mod\`ele
o\`u les interactions se font entre paires de spins~: les mod\`eles
$p$-spin avec $p \ge 3$ ont deux temp\'eratures critiques diff\'erentes.}.
Comme on l'a vu dans la partie g\'en\'erale \ref{sec:sphe-genedyn}, 
la premi\`ere quantit\'e 
\`a calculer pour d\'eterminer les propri\'et\'es dynamiques est la fonction 
$f(t)=\int d\mu \ \rho(\mu) e^{2 \mu t}$. Dans le cas d'un densit\'e de valeurs
propres en demi-cercle cette int\'egrale est une repr\'esentation d'une 
fonction de Bessel~; le point le plus important pour la suite est son 
comportement asymptotique,
\begin{equation}
f(t) \sim \frac{1}{4\sqrt{2 \pi}} \frac{e^{4t}}{t^{3/2}} \ .
\end{equation}
On peut le d\'eterminer sans utiliser les propri\'et\'es des fonctions de 
Bessel~: l'int\'egrale d\'efinissant $f(t)$ est domin\'e par le voisinage
de $\mu=2$. Le comportement exponentiel $e^{4t}$ est
d\^u \`a l'annulation de $\rho$ pour $\mu > 2$, et l'exposant de la correction
alg\'ebrique vient de son annulation en racine carr\'ee. Une fois $f(t)$
d\'etermin\'ee, il convient de r\'esoudre l'\'equation int\'egrale 
(\ref{eq:Volterra}) sur $\Gamma(t)$. On peut le faire ici en introduisant les 
transform\'ees de Laplace $\tilde{f}$ et $\tilde{\Gamma}$, 
\begin{equation}
\tilde{f}(s) = \int_0^\infty dt \, f(t) e^{-s t} \ , \qquad
\tilde{\Gamma}(s) = \int_0^\infty dt \, \Gamma(t) e^{-s t} \ .
\end{equation} 
L'\'equation de Volterra prend une forme assez simple en terme de ces
transform\'ees,
\begin{equation}
\tilde{\Gamma}(s) = \tilde{f}(s) + 2 T \tilde{f}(s) \tilde{\Gamma}(s) \ .
\end{equation}

\vspace{8mm}

\noindent 
\underline{\emph{Dynamique \`a haute temp\'erature}}

\vspace{4mm}

Etudions d'abord la situation \`a haute temp\'erature ($T>1$). On trouve
alors que $\tilde{\Gamma}(s)$ a un p\^ole en $s_*=2(T + T^{-1})>4$ 
et une coupure sur $[-4,4]$. Le comportement de $\Gamma(t)$ aux temps longs 
est contr\^ol\'e par la singularit\'e de sa transform\'ee de Laplace qui
a la plus grande partie r\'eelle. C'est donc le p\^ole qui est pertinent ici,
et on a $\Gamma(t) \sim \exp[s_* t]$ \`a un pr\'efacteur constant pr\`es.
On v\'erifie alors ais\'ement \`a partir des \'equations 
(\ref{eq:genedyn-C}), (\ref{eq:genedyn-G}) et (\ref{eq:genedyn-ene}) que~:
\begin{itemize}
\item L'\'energie relaxe exponentiellement vite vers sa valeur d'\'equilibre.
\item Les fonctions de corr\'elation et de r\'eponse sont stationnaires 
(apr\`es un bref r\'egime transitoire)~: $C(t_1+\tau,t_1)=C_{\rm eq}(\tau)$
et $R(t_1+\tau;t_1)=R_{\rm eq}(\tau)$.
\item Elles sont reli\'ees par le th\'eor\`eme de fluctuation-dissipation,
\begin{equation}
R_{\rm eq}(\tau) = -\frac{1}{T} C'_{\rm eq}(\tau) \ .
\end{equation}
\end{itemize}
On a donc \`a haute temp\'erature toutes les caract\'eristiques d'une
dynamique d'\'equilibre.

\vspace{8mm}

\noindent 
\underline{\emph{Dynamique \`a basse temp\'erature}}

\vspace{4mm}

A la temp\'erature de transition le p\^ole de $\tilde{\Gamma}(s)$ rejoint
le bord de la coupure, et cette derni\`ere contr\^ole le comportement
asymptotique de $\Gamma(t)$. Celui-ci prend donc la forme d'une exponentielle
modifi\'ee par un pr\'efacteur alg\'ebrique, $\Gamma(t) \sim 
\exp[4 t]/t^{3/2}$, \`a une constante multiplicative pr\`es.

Ce nouveau comportement pour $\Gamma$ va se traduire par une dynamique
hors d'\'equilibre\footnote{Soulignons ici que la limite thermodynamique 
$N\to \infty$ est prise {\em avant} la limite des temps longs. Dans le cas
d'un syst\`eme fini \'evoluant selon des \'equations de Langevin on tend
asymptotiquement vers l'\'equilibre thermodynamique.}, que l'on peut mettre 
en \'evidence par diff\'erentes observations~:
\begin{itemize}

\item La d\'ecroissance de l'\'energie vers sa valeur d'\'equilibre se fait
avec une loi de puissance, et non plus exponentiellement comme \`a haute
temp\'erature. On ne peut donc plus d\'efinir de temps caract\'eristique de 
relaxation.

\item Les fonctions de corr\'elation et de r\'eponse \`a deux temps 
pr\'esentent le ph\'enom\`ene de vieillissement~: m\^eme dans la limite
des temps longs, elles d\'ependent explicitement des deux temps, et non
de la diff\'erence entre les deux comme pour une dynamique d'\'equilibre.
Plus pr\'ecis\'ement, si l'on consid\`ere $t_1 \gg 1$ et 
$t_2=t_1 + \tau$, on a deux r\'egimes diff\'erents selon la valeur de $\tau$.

Si $\tau \ll t_1$, les fonctions sont quasi-stationnaires, 
$C(t_1+\tau,t_1) \approx C_{\rm st}(\tau)$, $R(t_1+\tau;t_1) \approx 
R_{\rm st}(\tau)$, avec $C_{\rm st}$ et $R_{\rm st}$ reli\'ees par le 
th\'eor\`eme de fluctuation-dissipation.

Par contre quand la s\'eparation des temps $\tau$ est du m\^eme ordre que
le temps d'attente $t_1$ depuis la pr\'eparation du syst\`eme, les 
corr\'elations et r\'eponses d\'ependent des deux temps par l'interm\'ediaire
du ratio $t_1/t_2$~: $C(t_2,t_1) \approx C_{\rm slow}(t_2/t_1)$ et 
$R(t_2;t_1) \approx t_1^{-3/2} R_{\rm slow}(t_2/t_1)$. 
L'\og \^age\fg\ $t_1$ du syst\`eme fixe donc, dans ce r\'egime, l'\'echelle de 
temps sur laquelle le syst\`eme
relaxe. Les fonctions $C_{\rm slow}$ et $G_{\rm slow}$ sont reli\'ees par
une modification du th\'eor\`eme de fluctuation-dissipation o\`u appara\^it
une temp\'erature effective. On reviendra dans de plus grands d\'etails sur ce 
sc\'enario de dynamique hors d'\'equilibre dans la partie \ref{sec:ft-outofeq}.
Signalons simplement que le cas $p=2$ trait\'e ici n'est pas repr\'esentatif 
du comportement g\'en\'erique de la famille des mod\`eles $p$-spin. En 
particulier la temp\'erature effective est ici infinie, ce qui n'est pas
vrai pour $p \ge 3$.

\item Finalement, on peut noter que le multiplicateur de Lagrange $z(t)$
tendant vers 2, l'\'equation de Langevin (\ref{eq:sphe-Langevintourne}) 
r\'egissant l'\'evolution du mode
correspondant \`a la plus grande valeur propre voit son potentiel de 
confinement dispara\^itre aux temps longs. Rappelons que l'\'equilibre \`a
basse temp\'erature correspondant \`a une condensation macroscopique sur
ce mode (c'est-\`a-dire que la projection de la configuration dans cette
direction est d'ordre $\sqrt{N}$). On con\c coit donc que cette situation ne
peut \^etre atteinte que sur des \'echelles de temps divergeant avec la
taille du syst\`eme. La limite thermodynamique ayant ici \'et\'e prise en
premier lieu, le syst\`eme n'atteint jamais cet \'equilibre.

\end{itemize}
  
\subsection{Le cas dilu\'e}

Etudions maintenant le cas o\`u la matrice d'interaction est d\'efini
\`a partir d'un graphe poissonien de connectivit\'e moyenne $p$. On va
se concentrer sur le cas $p \gg 1$ mais fini, la limite $p \to \infty$
correspondant au graphe compl\`etement connect\'e de la section 
pr\'ec\'edente. On prend pour valeur des \'el\'ements de matrice non nuls
$J_0=1/\sqrt{p}$, de mani\`ere \`a obtenir pour $\rho$ la loi du demi-cercle
sur $[-2,2]$ dans la limite $p \to \infty$.
 
\begin{figure}
\includegraphics{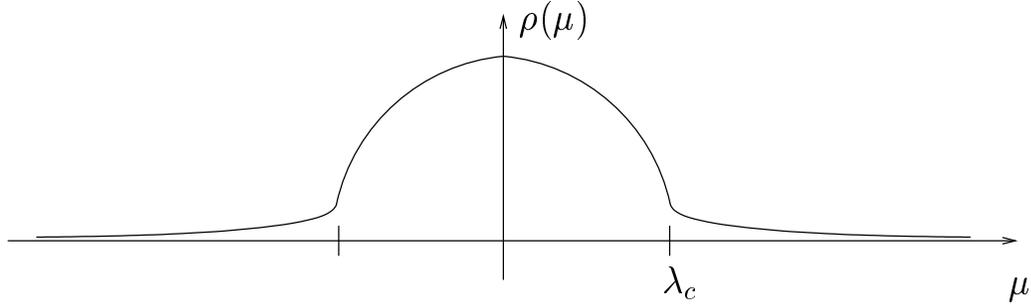}
\caption{Allure de la densit\'e de valeurs propres d'une matrice dilu\'ee.}
\label{fig:dilue-sketchrho}
\end{figure}

R\'esumons les conclusions de l'\'etude de ces matrices (le sch\'ema de la 
figure \ref{fig:dilue-sketchrho} illustre ces diff\'erents points)~:

\begin{itemize}
\item A cause des fluctuations non born\'ees de la connectivit\'e locale,
le spectre des valeurs propres n'est pas born\'e.

\item Quand $p \gg 1$, la densit\'e de valeurs propres comporte une partie 
centrale qui ressemble \`a un demi-cercle, et des queues dues aux 
\'ev\`enements rares de sites tr\`es connect\'es.
Ces queues disparaissent dans la limite $p \to \infty$, elles ont un poids 
non perturbatif par rapport \`a $1/p$, et l'on a trouv\'e l'expression 
asymptotique suivante pour la densit\'e de valeurs propres~:
\begin{equation}
\rho(\mu) \underset{|\mu|\to \infty}{\sim} \ e^{-2p \mu^2 \ln \mu} ,
\label{eq:dilue-rhoas}
\end{equation}
\`a un facteur multiplicatif pr\`es.

\item Le passage d'un r\'egime \`a l'autre se fait autour d'une valeur
$\lambda_c(p) \sim 2 \left(1 + \frac{1}{2p}\right) $. Ce crossover est 
la trace de l'annulation stricte de $\rho$ dans la limite compl\`etement 
connect\'ee.
\end{itemize}

Il faut maintenant examiner les cons\'equences de ces propri\'et\'es sur 
le mod\`ele sph\'erique d\'efini avec de telles interactions. 

La premi\`ere remarque \`a faire est que la statique du mod\`ele n'est pas 
bien d\'efinie dans la limite thermodynamique. En effet, le spectre des 
valeurs propres n'\'etant pas born\'e, la valeur de la plus grande
valeur propre diverge avec $N$. Autrement dit la valeur du multiplicateur 
de Lagrange telle que l'int\'egrale (\ref{eq:sphe-Z}) existe
diverge avec la taille du syst\`eme, et l'\'energie libre n'est pas
extensive dans la limite $N \to \infty$.

On peut cependant \'etudier la dynamique de Langevin d'un tel
syst\`eme. La pathologie des propri\'et\'es statiques se traduira par une 
divergence de l'\'energie aux temps longs. 

\vspace{8mm}

\noindent 
\underline{\emph{Les diff\'erents r\'egimes asymptotiques}}

\vspace{4mm}

Suivant la m\'ethode rappel\'ee dans le cas compl\`etement connect\'ee, 
il nous faut d'abord \'evaluer la fonction $f(t) = \int d\mu \ \rho(\mu) 
e^{2 \mu t}$. La s\'eparation de la densit\'e de valeurs propres en une zone
centrale et des queues entra\^ine l'existence de deux r\'egimes asymptotiques
pour $f(t)$. Le premier est tr\`es similaire \`a celui \'etudi\'e dans le cas
compl\`etement connect\'e~: pour des valeurs de $t$ telles que la contribution 
dominante de l'int\'egrale provient du voisinage de $\lambda_c$, on aura
\begin{equation}
f(t) \approx \frac{e^{2 \lambda_c t}}{t^\alpha} \ ,
\end{equation}
\`a une constante multiplicative pr\`es. L'exposant $\alpha$ de la correction
alg\'ebrique tend vers $3/2$ dans la limite $p \to \infty$. 

Pour des temps encore plus longs, la contribution dominante \`a l'int\'egrale
va venir du domaine des tr\`es grandes valeurs propres. Utilisant la forme
asymptotique (\ref{eq:dilue-rhoas}) pour \'evaluer l'int\'egrale donnant 
$f$ par la m\'ethode du col, on trouve que l'\'equation du col
est asymptotiquement $\mu \ln \mu \sim t/(2p)$. En prenant la r\'eciproque de
ce d\'eveloppement asymptotique, il vient $\mu \sim t/(2 p \ln t)$. On a
donc dans ce r\'egime des tr\`es longs temps l'expression suivante pour $f$,
\begin{equation}
f(t) \approx e^{\frac{t^2}{2 p \ln t}} \ .
\label{eq:dilue-fas}
\end{equation}
La s\'eparation entre les deux r\'egimes asymptotiques n'est \'evidemment pas
franche, ce n'est qu'un crossover quand $p$ est fini. Dans la limite 
$p \to \infty$ le deuxi\`eme r\'egime dispara\^it. On peut donner une 
estimation de $t_{\rm co}$,
le temps o\`u le comportement de $f(t)$ passe d'exponentiel (\`a une
correction alg\'ebrique pr\`es) \`a ce deuxi\`eme r\'egime (\ref{eq:dilue-fas})
plus rapide qu'exponentiel, comme la valeur de $t$ o\`u les arguments
des exponentielles sont \'egaux. Dans la limite $p \to \infty$ on trouve
que $t_{\rm co} \sim 8 p \ln p$~: comme attendu ce temps diverge avec $p$, le 
deuxi\`eme r\'egime asymptotique disparaissant dans cette limite.

Il faut ensuite d\'eterminer $\Gamma (t)$ comme solution de l'\'equation de
Volterra (\ref{eq:Volterra}) (rappelons qu'\`a temp\'erature nulle $\Gamma (t)
=f(t)$). Ici on ne peut pas utiliser les transform\'ees de Laplace, qui ne
sont pas d\'efinies pour des fonctions divergeant \`a l'infini plus vite que
des exponentielles. Une analyse qualitative, confirm\'ee par l'int\'egration
num\'erique de cette \'equation, va suffire pour d\'egager le comportement
du syst\`eme. Remarquons tout d'abord que pour des temps inf\'erieurs \`a
$t_{\rm co}$, les queues dans la densit\'e de valeurs propres jouent un
r\^ole n\'egligeable, et $f$ a en premi\`ere approximation le m\^eme 
comportement que dans le cas compl\`etement connect\'e. La d\'etermination
de $\Gamma(t)$ par l'\'equation de Volterra est causale, autrement dit
$\Gamma(t)$ ne d\'epend que du comportement de $f$ sur $[0,t]$. Il s'ensuit
donc que $\Gamma(t)$ se comporte comme dans le cas compl\`etement connect\'e 
jusqu'au temps de crossover. On a donc une temp\'erature $T_0$ (proche de 1
pour $p$ suffisamment grand) telle que pour $T>T_0$, 
$\Gamma(t) \sim \exp[b(T)t]$ avec $b(T)>2\lambda_c$, au moins jusqu'au temps de
crossover. Pour $T<T_0$, on a par contre $\Gamma(t) \sim \exp[ 2 \lambda_c t]$,
\`a une correction alg\'ebrique pr\`es. Dans ce cas la d\'ependance en 
temp\'erature n'est que dans le pr\'efacteur, pas dans le comportement
exponentiel. Reste \`a d\'eterminer le comportement de $\Gamma$ pour des
temps sup\'erieurs \`a $t_{\rm co}$. Remarquons que l'\'equation de Volterra
implique $\Gamma(t)\ge f(t)$. Dans le cas $T<T_0$, on a donc n\'ecessairement
un changement de comportement de $\Gamma$ \`a $t_{\rm co}$, puisque $f$ se
met alors \`a cro\^itre plus vite qu'un exponentielle. A haute temp\'erature
par contre, la forme $\Gamma(t) \sim \exp[b(T)t]$ reste valable jusqu'\`a
ce que $b(T) t \approx t^2/(2 p \ln t)$, suite \`a quoi $\Gamma(t) \sim f(t)$.
Ce temps de crossover $t_{\rm co}(T)$ peut s'exprimer dans la limite de 
haute temp\'erature o\`u $b(T) \sim 4 T$, $t_{\rm co}(T) \sim 8 p T \ln T$.
Comme on l'a vu dans la section pr\'ec\'edente, on a une dynamique 
d'\'equilibre quand $\Gamma$ a une d\'ependance exponentielle en temps.

\begin{figure}
\includegraphics[width=8cm]{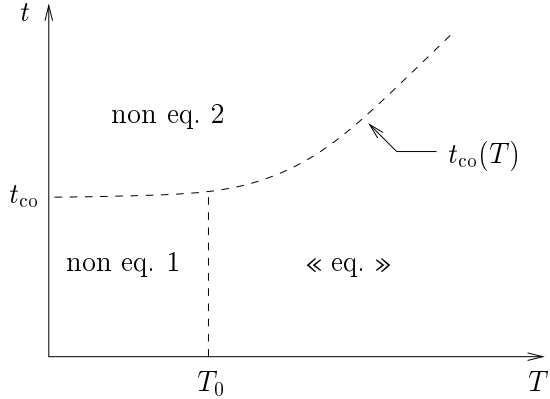}
\caption{Les diff\'erents r\'egimes asymptotiques. Dans la zone d\'enot\'ee
\og non eq. 1\fg\ le comportement est similaire \`a celui du mod\`ele 
compl\`etement connect\'e, \og non eq. 2\fg\ d\'esigne le nouveau r\'egime
d\^u aux queues de la densit\'e de valeurs propres.}
\label{fig:dilue-Ttime}
\end{figure}

La figure \ref{fig:dilue-Ttime} r\'esume l'\'etude que l'on vient de faire~:
\`a haute temp\'erature,
pour des temps interm\'ediaires on a un r\'egime de pseudo-\'equilibre, 
puis au bout
d'un temps d'autant plus grand que la temp\'erature est \'elev\'ee, un
crossover vers un r\'egime hors-\'equilibre contr\^ol\'e par les queues de la
densit\'e de valeurs propres. A basse temp\'erature, on passe d'un r\'egime
hors-\'equilibre ressemblant \`a celui pr\'esent dans le cas compl\`etement
connect\'e (contr\^ol\'e par la partie centrale du spectre) \`a celui d\^u
aux queues. Le temps de crossover est approximativement constant dans la 
phase de basse temp\'erature.

\clearpage

\noindent 
\underline{\emph{Comportement des observables}}

\vspace{4mm}

\begin{figure}
\includegraphics[width=8cm]{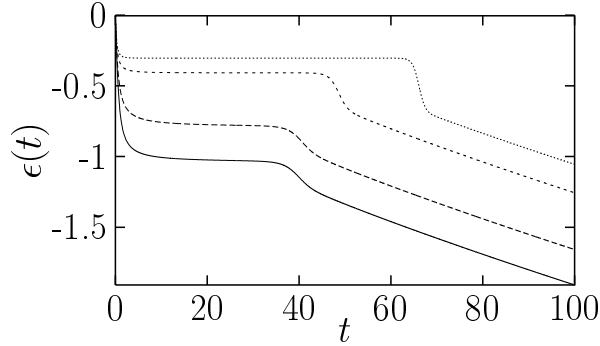}
\caption{L'\'energie en fonction du temps. De haut en bas les temp\'eratures sont
\'egales \`a 0, 0.5, 1.3 et 1.7. Le d\'ecrochement pour les deux premi\`eres
temp\'eratures se fait au bout du m\^eme temps, alors que ce temps de crossover
cro\^it avec la temp\'erature au del\`a de $T_0$ (cf. figure 
\ref{fig:dilue-Ttime}). La courbe a \'et\'e obtenue en calculant
num\'eriquement les fonctions $f$ et $\Gamma$ \`a partir de la forme du
spectre obtenue par l'approximation \`a un seul d\'efaut expliqu\'ee
pr\'ec\'edemment, et par une r\'esolution num\'erique de l'\'equation de
Volterra.}
\label{fig:dilue-ene}
\end{figure}

La figure \ref{fig:dilue-ene} pr\'esente l'\'evolution temporelle de 
l'\'energie du syst\`eme pour diff\'erentes temp\'eratures. On a dans
un premier temps un plateau qui est atteint soit exponentiellement (pour
$T>T_0$) soit alg\'ebriquement (pour $T<T_0$) vite, puis un d\'ecrochement
\`a $t_{\rm co}(T)$ quand le syst\`eme explore les queues de la densit\'e
d'\'etats. Dans la limite des tr\`es longs temps, l'\'energie diverge
comme $-t/(4 p \ln t)$.

On peut \'egalement \'etudier le comportement des fonctions de corr\'elation
et de r\'eponse dans le r\'egime des temps contr\^ol\'es par les queues du
spectre. Leurs formes sont d\'etaill\'ees dans la publication \pubsphe . On 
trouve en particulier qu'elles sont non-stationnaires, et qu'on peut les 
mettre sous la forme
\begin{equation}
C(t_1,t_2) = C_{\rm slow} \left( \frac{l(t_1)}{l(t_2)} \right) \ , \qquad
l(t) = \exp \left( \frac{t}{\sqrt{p \ln t}} \right) \ .
\end{equation}
La fonction $l(t)$ qui d\'efinit un \og \^age\fg\ effectif du syst\`eme est
diff\'erente de celle rencontr\'ee dans l'\'etude du mod\`ele compl\`etement
connect\'e (on a vu en effet qu'alors $l(t)=t$). Dans le cas d'une fonction de 
corr\'elation stationnaire pour laquelle $C(t_1,t_2) = C_{\rm st}(t_1-t_2)$, 
on peut aussi \'ecrire $C_{\rm st}(\tau)={\cal C}(l(t_1 + \tau)/l(t_1)) $, en
utilisant $l(t)=e^t$. 

On est donc ici dans une situation interm\'ediaire, o\`u $l(t)$ diverge plus
vite que dans le cas compl\`etement connect\'e, mais moins vite qu'\`a 
l'\'equilibre. Ce comportement est parfois qualifi\'ee de 
\og sub-aging\fg\ \cite{Sitges} (on en trouvera un autre exemple dans 
\cite{Be-subaging}). 

La temp\'erature effective dans ce deuxi\`eme r\'egime asymptotique est
\'egalement infinie.

\subsection{Perspectives}

Le mod\`ele que l'on vient d'\'etudier pr\'esente certains d\'efauts~: son
comportement statique est pathologique, et la dynamique \`a haute
temp\'erature n'a des propri\'et\'es d'\'equilibre que pendant un temps
fini.

Cependant quelques unes de ses caract\'eristiques sont int\'eressantes, 
et devraient subsister dans des mod\`eles d\'epourvus des d\'efauts 
sus-cit\'ees. Insistons en particulier sur 
l'existence de deux r\'egimes hors d'\'equilibre distincts, l'un provenant
des propri\'et\'es \og moyennes\fg\ du syst\`eme, l'autre des \og 
\'ev\`enements rares\fg . Si ces deux r\'egimes apparaissent ici de mani\`ere
caricaturale, leur pr\'esence constitue s\^urement un ingr\'edient universel
de la dynamique de tout syst\`eme vitreux localement h\'et\'erog\`ene,
que ce soit un mod\`ele dilu\'e ou des cas plus r\'ealistes en dimension
finie. L'exemple des verres structuraux qui pr\'esentent de fortes
fluctuations spatiales de densit\'e et de \og vitesse de r\'earrangement\fg\ 
serait un des plus int\'eressants dans cette perspective.

D'un point de vue plus technique, mentionnons que dans le cas de syst\`emes
inhomog\`enes comme celui consid\'er\'e ici les mod\`eles sph\'eriques et
la limite $m\to \infty$ du mod\`ele $O(m)$ ne sont pas tout \`a fait
\'equivalents (plus pr\'ecis\'ement, les multiplicateurs de Lagrange introduits
pour imposer la contrainte $O(m)$ varient de site \`a site, alors que le 
mod\`ele sph\'erique les suppose homog\`ene).
Ce dernier mod\`ele a alors une statique bien d\'efinie~\cite{BrMo-eigen}, et
sa relaxation \`a haute temp\'erature est anormalement lente \`a cause
des fluctuations de connectivit\'e (cf. \cite{Br-Griffiths} pour des arguments 
similaires en dimension finie). Il serait donc int\'eressant d'\'etudier
la phase de basse temp\'erature de ce mod\`ele, en esp\'erant y trouver une
trace de ces deux r\'egimes hors d'\'equilibre.

\newpage

\section{Vers une m\'ethode g\'en\'erale pour les mod\`eles de spins continus dilu\'es}
\label{sec:msrgene}

La m\'ethode utilis\'ee ci-dessus est limit\'ee \`a ce cas particulier 
o\`u les spins sont continus avec une contrainte sph\'erique globale, 
et o\`u l'hamiltonien ne fait interagir que des paires de spins. C'est 
en effet \`a ces deux conditions que l'on peut diagonaliser le syst\`eme 
d'\'equations de Langevin et r\'esoudre chacune des \'equations 
ind\'ependamment.

On va pr\'esenter dans cette partie une m\'ethode qui permet,
au moins formellement, de traiter les mod\`eles o\`u le nombre de spins
dans chaque interaction est arbitraire,
et comportant des termes du type \og soft-spin\fg\ .
On d\'esigne sous ce vocable des potentiels qui s'appliquent localement
\`a chacun des spins (toujours des variables continues), correspondant
dans l'hamiltonien \`a un terme $\sum_i V(\sigma_i)$, o\`u le potentiel
$V$ est souvent choisi comme 
\begin{equation}
V(\sigma) = \kappa (\sigma^2 - 1 )^2 \ .
\end{equation}
Une telle contribution \`a l'\'energie favorise les configurations o\`u les
$\sigma_i$ sont tous autour de $\pm 1$, on s'attend
donc \`a ce que ces mod\`eles ait des comportements similaires \`a ceux
des mod\`eles d'Ising correspondant. Cela se justifie dans les syst\`emes 
ferromagn\'etiques par des arguments d'universalit\'e au voisinage de la
transition, il est moins \'evident que le rapprochement entre les
mod\`eles soft-spins et Ising soit aussi pertinente pour les verres de 
spins~\cite{Szamel1,Szamel2}. 

On peut aussi faire une remarque sur la valeur du coefficient $\kappa$.
On dit souvent que l'on doit prendre la limite $\kappa \to \infty$ afin
de retrouver le mod\`ele d'Ising. C'est certainement le cas si l'on 
s'int\'eresse \`a la statique et si l'on calcule une fonction de partition.
En ce qui concerne les propri\'et\'es dynamiques, cette affirmation 
m\'erite d'\^etre nuanc\'ee~: le potentiel introduit une barri\`ere
\'energ\'etique $\kappa$ pour passer d'un puits \`a l'autre. Dans la limite
o\`u la hauteur de la barri\`ere diverge, les transitions d'un minimum \`a
l'autre du potentiel soft-spin n'auront lieu
que sur des \'echelles de temps exponentiellement grandes en $\kappa$,
et non sur des temps finis comme on le souhaiterait. Il faudrait donc soit
red\'efinir l'\'echelle de temps, soit, et c'est la voie toujours suivie
\`a ma connaissance, garder une hauteur de barri\`ere $\kappa$ finie (elle
est m\^eme la plupart du temps trait\'ee comme une perturbation).

Ces r\'eserves sur l'\'equivalence avec les mod\`eles du type Ising \'etant 
pos\'ees, on peut consid\'erer les versions soft-spins comme
dignes d'int\'er\^et pour elles-m\^emes. On les prend donc comme point de 
d\'epart dans la suite de cette partie, o\`u l'on pr\'esentera le formalisme
bien connu de Martin-Siggia-Rose, puis sa reformulation en terme de champs 
supersym\'etriques. L'analogie formelle entre ces notations supersym\'etriques 
et les calculs statiques par la m\'ethode des r\'epliques est ensuite 
exploit\'ee pour formuler les \'equations dynamiques des mod\`eles de champ 
moyen tant compl\`etement connect\'es que dilu\'es.
Finalement, une m\'ethode it\'erative de r\'esolution, inspir\'ee de celle 
utilis\'ee dans la partie \ref{sec:mat-repliques}, est sugg\'er\'ee pour le 
cas dilu\'e.

\subsection{Formalisme fonctionnel}

Consid\'erons un syst\`eme d\'ecrit par les variables continues 
$\sigma = \{\sigma_1,\dots,\sigma_N \}$, avec un hamiltonien g\'en\'erique
$H[\sigma]$. On introduit une notation pour les d\'eriv\'ees partielles de $H$,
\begin{equation}
H_i = \frac{\partial H}{\partial \sigma_i} \qquad , \qquad
H_{ij} = \frac{\partial^2 H}{\partial \sigma_i \partial \sigma_j} \ .
\end{equation} 
On cherche \`a \'etudier le comportement de ce syst\`eme lorsque son
\'evolution est r\'egie par des \'equations de Langevin~:
\begin{equation}
\frac{d}{dt} \sigma_i(t) = - H_i[\sigma(t)] + h_i(t) + \xi_i(t) 
\ , 
\end{equation}
avec $\xi$ un bruit blanc gaussien d\'etermin\'e par ses deux premiers
moments, $\langle \xi_i(t) \rangle = 0$, $\langle \xi_i(t) \xi_j(t) \rangle =
2T \delta_{ij} \delta (t-t')$. Les champs $h_i$ sont des champs ext\'erieurs
qui vont permettre de calculer les fonctions de r\'eponse du syst\`eme
(ils correspondent \`a l'ajout de termes $-h_i \sigma_i$ \`a l'hamiltonien,
ils sont donc coupl\'es naturellement aux variables $\sigma_i$).
On prend comme instant initial de l'\'evolution $t=0$, instant auquel la
configuration $\sigma(0)$ a une certaine loi de probabilit\'e. La solution
$\sigma_i(t)$ de l'\'equation de Langevin d\'epend de la condition initiale
$\sigma(0)$, et de toute l'histoire des champs ext\'erieurs $h$ et du bruit 
thermique $\xi$ entre l'instant initial et l'instant $t$. Pour simplifier
les notations dans la suite, on appellera $s_i(t)$ la solution 
des \'equations de Langevin, la d\'ependance en $\sigma(0)$, $h$ et $\xi$
\'etant sous-entendue. 

Le syst\`eme est caract\'eris\'e par ses fonctions de corr\'elation et
de r\'eponse, d\'efinies par
\begin{eqnarray}
C(i_1,t_1,\dots,i_n,t_n) &=& 
\langle s_{i_1}(t_1) \dots 
s_{i_n}(t_n) \rangle |_{h=0}  \\
R(i_1,t_1,\dots,i_k,t_k;i_{k+1},t_{k+1},\dots,i_n,t_n) &=&
\left. 
\frac{\delta^{n-k}}{\delta h_{i_{k+1}}(t_{k+1}) \dots \delta h_{i_n}(t_n)}
\langle s_{i_1}(t_1) \dots 
s_{i_k}(t_k) \rangle \right|_{h=0} \ , \nonumber
\end{eqnarray}
o\`u $\langle \bullet \rangle$ d\'esigne une moyenne sur la condition initiale
et sur les histoires du bruit thermique. 

Une m\'ethode de calcul de ces fonctions, initi\'ee par
Martin, Siggia et Rose~\cite{MaSiRo}, Jannsen~\cite{Ja-MSR} et 
de Dominicis~\cite{deDo-MSR}, consiste \`a \'ecrire une fonctionnelle 
g\'en\'eratrice dynamique pour l'\'evolution des spins,
\begin{equation}
{\cal Z}[\eta,h] = \left< \exp \left[ \int_0^\infty dt \ \eta_i(t) 
s_i(t) \right] \right> \ .
\end{equation}
Les fonctions de corr\'elation et de r\'eponse peuvent alors s'exprimer 
comme des
d\'eriv\'ees fonctionnelles de $\cal Z$ par rapport \`a la \og source\fg\ 
$\eta$ et au champ ext\'erieur $h$,
\begin{eqnarray}
C(i_1,t_1,\dots,i_n,t_n) &=& \left. \frac{\delta^n}{\delta \eta_{i_1}(t_1) 
\dots \delta \eta_{i_n}(t_n)} {\cal Z}[\eta,h] \right|_{\eta=h=0} 
\ , \label{eq:def-CG}\\
R(i_1,t_1,\dots,i_k,t_k;i_{k+1},t_{k+1},\dots,i_n,t_n) &=&
\left. \frac{\delta^n}{\delta \eta_{i_1}(t_1) 
\dots \delta \eta_{i_k}(t_k) \delta h_{i_{k+1}}(t_{k+1}) 
\dots \delta h_{i_n}(t_n)} {\cal Z}[\eta,h] \right|_{\eta=h=0} \ .
\nonumber
\end{eqnarray}

Pour \'eviter d'expliciter la solution $s_i(t)$ des \'equations de Langevin,
on ins\`ere formellement une repr\'esentation de l'identit\'e comme une 
int\'egrale de chemin sur les trajectoires  $\sigma_i(t)$, 
\begin{equation}
1=\int D\sigma \prod_{i,t} \delta (\sigma_i(t)-s_i(t)) \ .
\label{eq:msr-id1}
\end{equation}
La contrainte s\'electionne la seule trajectoire qui nous int\'eresse,
$\sigma_i(t)=s_i(t)$. Le produit sur les instants $t$ doit \^etre compris 
comme une discr\'etisation infinit\'esimale de l'axe temporel. 
On a aussi int\'egr\'e
la loi de probabilit\'e sur la condition initiale $\sigma(0)$ \`a la mesure
fonctionnelle $D\sigma$. 

Introduisons la fonctionnelle de $\sigma$ suivante~:
\begin{equation}
F_i(t) = \dot{\sigma}_i(t) + H_i[\sigma(t)] - h_i(t) \ . 
\end{equation}
Les \'equations de Langevin peuvent alors se r\'e\'ecrire $\xi_i(t)=F_i(t)$, et
l'identit\'e (\ref{eq:msr-id1}) devient apr\`es le changement de variables 
$\sigma \to F$~:
\begin{equation}
1=\int D\sigma \prod_{i,t} \delta (\xi_i(t)-F_i(t)) {\cal J}[\sigma] \ .
\end{equation}
$\cal J$ est le jacobien de ce changement de variables, que l'on explicitera 
dans quelques lignes. On a seulement besoin pour l'instant d'admettre
qu'il est ind\'ependant des champs $h_i$. 
On peut maintenant exponentier la contrainte
$\xi_i(t)=F_i(t)$ \`a l'aide de champs $\hat{\sigma}_i(t)$ (il est sous-entendu
qu'ils sont \`a int\'egrer le long de l'axe imaginaire). Le
bruit $\xi$ appara\^it alors comme 
$\exp[\sum_i \int dt \, \xi_i(t) \hat{\sigma}_i(t)]$. Ce bruit ayant une
distribution gaussienne, l'int\'egrale sur $\xi$ se fait 
imm\'ediatement, pour conduire \`a~:
\begin{eqnarray}
{\cal Z}[\eta,h] &=& \int D\sigma D\hat{\sigma} \, e^{-S[\sigma,\hat{\sigma}] +
\sum_i \int_0^\infty dt (\eta_i(t) \sigma_i(t) + 
h_i(t) \hat{\sigma}_i(t)) } \ , \\ 
-S[\sigma,\hat{\sigma}] &=& \ln ({\cal J}[\sigma]) + \sum_i \int_0^\infty dt \,
\hat{\sigma}_i(t)(T \hat{\sigma}_i(t) - \dot{\sigma}_i(t) - H_i[\sigma(t)])
\ .
\end{eqnarray}
On voit alors \`a partir des expressions (\ref{eq:def-CG}) que les 
fonctions de corr\'elation et de r\'eponse s'expriment comme des moyennes
des champs $\sigma$ et $\hat{\sigma}$ avec le poids 
$\exp[-S[\sigma,\hat{\sigma}]]$, on peut donc supprimer 
\`a partir de maintenant les sources $\eta$ et $h$. 

Explicitons le jacobien ${\cal J}[\sigma]$ \footnote{En toute rigueur il
d\'epend de la r\`egle de lecture utilis\'ee dans la discr\'etisation des
\'equations de Langevin. On passera cette subtilit\'e sous silence.}. 
Il est de la forme
\begin{equation}
\left| \det \left(\frac{\delta F_i}{\delta \sigma_j}\right) \right| \ ,
\end{equation}
o\`u le d\'eterminant est \`a la fois matriciel (\`a cause des indices de 
sites) et op\'eratoriel (\`a cause de la d\'ependance temporelle)~:
\begin{equation}
\frac{\delta F_i(t)}{\delta \sigma_j(t')} = \delta(t-t') \left( \delta_{ij}
\frac{\partial}{\partial t'} + H_{ij}[\sigma(t)] \right) \ .
\end{equation}
Cette expression \'etant ind\'ependante des champs $h_i$, leur abandon dans la
suite est justifi\'e.
En supprimant la valeur absolue dans le jacobien (les \'equations de Langevin
\'etant causales, il y a univocit\'e du changement de variables, le 
d\'eterminant a donc toujours le m\^eme signe), on peut l'exponentier avec
des champs de variables de Grassmann $\overline{\psi}$ et $\psi$. Ceci
conduit \`a un poids $\exp[-S[\sigma,\hat{\sigma},\overline{\psi},\psi]]$ avec
l'action
\begin{eqnarray}
-S[\sigma,\hat{\sigma},\overline{\psi},\psi] &=&
\sum_i \int_0^\infty dt \left[ T \hat{\sigma}_i(t)^2 - 
\hat{\sigma}_i(t) \dot{\sigma}_i(t) + \overline{\psi}_i(t) \dot{\psi}_i(t) 
\right]
\nonumber \\
& & + \int_0^\infty dt\left[ - \sum_i  \hat{\sigma}_i(t) H_i[\sigma(t)] +
\sum_{ij} \overline{\psi}_i(t) H_{ij}[\sigma(t)] \psi_j(t) \right] \ .
\label{eq:S-susy1}
\end{eqnarray}
Les termes de la premi\`ere ligne sont ind\'ependants du mod\`ele, alors
que ceux de la deuxi\`eme varient selon l'hamiltonien $H$ consid\'er\'e.

Les \'equations de Langevin purement 
conservatives, c'est \`a dire dont les forces d\'ecoulent d'un potentiel
comme on l'a suppos\'e ici, conduisent \`a des actions fonctionnelles qui
pr\'esentent un certain nombre de 
sym\'etries~\cite{PaSo-SUSY,Zinn,Jorge-SUSY1}.
Certaines m\'elangeant les champs \og bosoniques\fg\ 
$\{\sigma ,\hat{\sigma}\}$ et ceux \og fermioniques\fg\ 
$\{\psi, \overline{\psi} \}$, on parle souvent de supersym\'etrie (SUSY) \`a
leur \'egard. On va ici se servir de la supersym\'etrie plut\^ot comme une 
notation compacte, et pour exploiter l'analogie qu'elle permet de tracer
avec les calculs par la m\'ethode des r\'epliques.

Etendons donc la coordonn\'ee temporelle $t$ en une \og super-coordonn\'ee\fg\
$a=(t_a,\theta_a,\overline{\theta}_a)$, o\`u $\theta_a$ et 
$\overline{\theta}_a$ sont deux variables de Grassmann. Il s'av\`ere alors 
utile de r\'eunir les
diff\'erents champs $\{\sigma, \hat{\sigma} , \psi, \overline{\psi} \} $
dans un \og super-champ\fg\ $\Phi$ fonction de $a$,
\begin{equation}
\Phi_i(a) = 
\sigma_i(t_a) +  \overline\theta_a \psi_i(t_a) + \overline\psi_i(t_a) \theta_a
+ \hat{\sigma}_i(t_a) \overline{\theta}_a \theta_a \ .
\end{equation}
Notons que chacun de ces termes faisant intervenir un nombre pair de
variables de Grassmann, le superchamp a les propri\'et\'es d'un nombre
commutant habituel. L'action (\ref{eq:S-susy1}) se r\'e\'ecrit en 
termes du superchamp comme 
\begin{equation}
-S[\Phi] = -\frac{1}{2} \sum_i \int da \ \Phi_i(a) {\cal K} \Phi_i(a) 
- \int da \ H[\Phi(a)] \ ,
\label{eq:S-susy2}
\end{equation}
o\`u 
$da= dt_a \, d\theta_a \, d\overline\theta_a$ est l'\'el\'ement d'int\'egration
sur la coordonn\'ee \'etendue, 
et $\cal K$ d\'esigne un op\'erateur diff\'erentiel, 
\begin{equation}
{\cal K}= \frac{\partial}{\partial t_a}
- 2 \theta \frac{\partial^2}{\partial \theta_a\partial t_a}
- 2 T \frac{\partial^2}{\partial \theta_a\partial\overline\theta_a} \ .
\end{equation}
La partie de l'action d\'ependant de l'hamiltonien a pris une forme 
extr\^emement simple.
On peut se convaincre de sa v\'eracit\'e en faisant un d\'eveloppement limit\'e
de $H[\Phi(a)]$ en puissance des variables de Grassmann $\theta_a$ et 
$\overline\theta_a$~: par d\'efinition le d\'eveloppement s'arr\^ete apr\`es
un nombre fini de termes puisque les variables de Grassmann sont nilpotentes.
On constate ensuite que les seuls termes qui survivent \`a l'int\'egrale
sur $\theta_a$ et $\overline\theta_a$ sont ceux pr\'esents dans 
(\ref{eq:S-susy1}). 

On notera \`a partir de maintenant $\langle \bullet \rangle$ les moyennes
sur les super-champs avec le poids donn\'e par l'\'equation (\ref{eq:S-susy2}).
Les super-fonctions de corr\'elation contiennent les fonctions de corr\'elation
et de r\'eponse initiales, puisque $\Phi$ a \'et\'e d\'efini \`a partir
des champs $\sigma$ et $\hat{\sigma}$. En particulier la fonction \`a deux 
points s'\'ecrit
\begin{equation}
Q_{ij}(a,b)=\langle \Phi_i(a) \Phi_j(b) \rangle = C_{ij}(t_a,t_b) +
(\overline{\theta}_b-\overline{\theta}_a)(\theta_b R_{ij}(t_a,t_b) - \theta_a
R_{ji}(t_b,t_a)) \ .
\label{eq:def-QSUSY}
\end{equation}
On a utilis\'e ici certaines sym\'etries de l'action pour transformer les
corr\'elateurs fermioniques en fonctions de r\'eponse~\cite{Jorge-SUSY1}.
Par contre on n'a pas impos\'e l'invariance par translation dans le
temps et le th\'eor\`eme de fluctuation-dissipation (on reviendra sur ces
propri\'et\'es dans la suite).

La forme supersym\'etrique (\ref{eq:S-susy2}) de l'action a
l'avantage d'\^etre d'une part tr\`es compacte, et d'autre part de
pr\'esenter une analogie avec les calculs statiques par la
m\'ethode des r\'epliques (cf.~\cite{Jorge-SUSY2} pour une pr\'esentation
approfondie de cette analogie). Au terme cin\'etique $\int da \; \Phi_i(a) 
{\cal K} \Phi_i(a) $ pr\`es (qui est quadratique et local dans
l'espace des sites), l'action est donn\'ee par l'hamiltonien du syst\`eme
dans lequel on a remplac\'e la variable $\sigma_i$ par le
super-champ $\Phi_i(a)$, et cette expression est somm\'ee sur la 
super-coordonn\'ee $a$. Or la puissance $n$-\`eme de la fonction de partition,
que l'on calcule dans la m\'ethode des r\'epliques, s'\'ecrit~:
\begin{equation}
Z= \mbox{Tr}_{\sigma} e^{-\beta H[\sigma]} \Rightarrow 
Z^n= \mbox{Tr}_{\sigma_1,\dots\sigma_n} e^{-\beta \sum_{a=1}^n H[\sigma_a]} \ .
\end{equation}
On a formellement la m\^eme expression, l'indice des r\'epliques correspondant
\`a la super-coordonn\'ee de l'approche dynamique, le champ r\'epliqu\'e au
super-champ.

Le paragraphe suivant s'appuiera de mani\`ere cruciale sur cette analogie~: 
les calculs seront essentiellement les m\^emes que ceux men\'es pour le
probl\`eme de matrices al\'eatoires, on ne r\'ep\'etera donc pas tous les
d\'etails.

D\'efinissons avant cela deux types de produit pour des fonctions 
supersym\'etriques \`a deux points du type de (\ref{eq:def-QSUSY})~:
\begin{equation}
(F_1 \otimes F_2)(a,b) = \int dc \, F_1(a,c) F_2(c,b) \ , \qquad
(F_1 \bullet F_2)(a,b) = F_1(a,b) F_2(a,b) \ .
\end{equation}
Le premier, dit produit de convolution, s'interpr\`ete comme un produit
de matrices si l'on regarde les coordonn\'ees supersym\'etriques comme
des indices discrets, ce qui est le cas dans l'analogie avec une th\'eorie
r\'epliqu\'ee. Dans ce contexte le produit direct de la deuxi\`eme
d\'efinition correspond \`a un produit de Hadamard.

\subsection{Mod\`eles de champ moyen}

La d\'emonstration de la forme (\ref{eq:S-susy2}) de 
l'action dynamique pr\'esent\'ee ci-dessus est valable quelque soit 
l'hamiltonien du syst\`eme.
Int\'eressons-nous maintenant au cas o\`u l'hamiltonien contient des
variables gel\'ees, et notons $[ \bullet ]$ les moyennes sur ce d\'esordre
gel\'e. On s'attend \`a ce que les fonctions de corr\'elation et de r\'eponse
globales soient auto-moyennantes (concentr\'es autour de leur moyennes
par rapport \`a la distribution du d\'esordre), 
on veut donc calculer leurs valeurs moyennes. Prenons
par exemple un hamiltonien qui contient des termes soft-spins et une partie
d'interaction entre $p$ spins,
\begin{equation}
H(\vec{\sigma}) = H^{\rm (i)}(\vec{\sigma}) + \sum_{i=1}^N V(\sigma_i)
\qquad , \qquad
H^{\rm (i)}(\vec{\sigma}) = -\sum_{i_1 < \dots < i_p} J_{i_1 \dots i_p}
\sigma_{i_1} \dots \sigma_{i_p} \ .
\label{eq:continu-defHi}
\end{equation}
On supposera que les couplages $J_{i_1 \dots i_p}$ sont des variables 
al\'eatoires gel\'ees ind\'ependantes, et identiquement distribu\'es avec 
une loi $P(J)$.

Comme la fonctionnelle g\'en\'eratrice dynamique \'evalu\'ee sans sources
vaut 1 ind\'ependamment de la r\'ealisation du d\'esordre~\cite{deDo-MSR},
il suffit de calculer la moyenne de $\cal Z$ et non celle de son logarithme, 
ce qui \'evite l'introduction de r\'epliques (sauf si la 
condition initiale est correl\'ee avec le d\'esordre gel\'ee, une situation  
consid\'er\'ee par exemple dans \cite{HoJaYo,BaBuMe}). 

Autrement dit, les fonctions de corr\'elation moyenn\'ees sur le bruit
thermique et sur le d\'esordre gel\'e sont donn\'ees par
\begin{equation}
[\langle \bullet \rangle ] = \int D \Phi \bullet \left[ e^{-S[\Phi]}\right] \ .
\end{equation}
La contribution de $e^{-S}$ qui d\'epend du d\'esordre est 
$\exp[-\int da \ H^{\rm (i)}[\Phi(a)]]$. Pour calculer sa moyenne, il est
naturel d'introduire alors l'analogue supersym\'etrique du param\`etre
d'ordre $c(\vec{\phi})$~\cite{Mo-c-sigma} utilis\'e pour le 
calcul de la densit\'e d'\'etat des matrices~:
\begin{equation}
c(\Phi) = \frac{1}{N} \sum_{i=1}^N \delta [ \Phi - \Phi_i ] \ ,
\label{eq:susy-defcphi}
\end{equation}
o\`u $\delta[\dots ]$ impose l'\'egalit\'e $\Phi(a)= \Phi_i(a)$ pour
toutes les super-coordonn\'ees $a$, \`a l'instar de l'\'equation 
(\ref{eq:mat-defc_phi}) o\`u
l'\'egalit\'e \'etait impos\'ee pour tous les indices de r\'eplique.
D\'efinissons alors
\begin{equation}
g(x) = \frac{1}{p!} N^{p-1} \ln \left( \int dJ \ P(J) e^{Jx} \right) \ ,
\label{eq:susy-defg}
\end{equation}
de mani\`ere analogue \`a (\ref{eq:mat-defg}). La moyenne du terme d\'ependant
du d\'esordre dans le poids des trajectoires des super-champs s'exprime donc, 
dans la limite thermodynamique, comme
\begin{equation}
\left[ e^{-\int da \ H^{\rm (i)}[\Phi(a)]}\right]
= \exp \left[ N 
\int D\Psi_1 \dots D\Psi_p  \, c(\Psi_1) \dots c(\Psi_p) \,
g\left(\int da \, \Psi_1(a) \dots \Psi_p(a)  \right)  \right] \nonumber \ .
\end{equation}

On peut alors suivre la m\^eme d\'emarche que dans le calcul du spectre de 
matrices pour int\'egrer sur les champs originels $\Phi$, ce qui conduit \`a~:
\begin{eqnarray}
[\langle \bullet \rangle ] &=& \int Dc \ \bullet \ \exp \left[ N \left(
-\int D\Psi \, c(\Psi) \ln c(\Psi) \right. \right.  \\ 
&& \hspace{2cm} + \int D\Psi \, c(\Psi) \int da \left( -\frac{1}{2}
\Psi(a) {\cal K} \Psi(a) + V(\Psi(a)) \right) \nonumber \\
&&\hspace{2cm} +  \left. \left. \int D\Psi_1 \dots D\Psi_p  \, 
c(\Psi_1) \dots c(\Psi_p) \
g\left(\int da \, \Psi_1(a) \dots \Psi_p(a)  \right) \right) \right] \ . 
\nonumber
\end{eqnarray}
Dans la limite thermodynamique cette int\'egrale est domin\'ee par la 
contribution du col $c_*$, qui v\'erifie
\begin{eqnarray}
c_*(\Phi)&=& {\cal N} \exp \left[ \int da \left( -\frac{1}{2}
\Phi(a) {\cal K} \Phi(a) + V(\Phi(a)) \right) \right.
\label{eq:susy-speq} \\
&& \hspace{1.2cm} \left. + p \int D\Psi_2 \dots D\Psi_p  \, 
c_*(\Psi_2) \dots c_*(\Psi_p) \
g\left(\int da \, \Phi(a) \Psi_2(a) \dots \Psi_p(a) \right) \right] \ ,
\nonumber
\end{eqnarray}
o\`u $\cal N$ est une constante de normalisation. Une fois cette \'equation
r\'esolue, on peut obtenir les fonctions de corr\'elation et de r\'eponse
moyenne globale,
\begin{equation}
\frac{1}{N}\sum_{i=1}^N [\langle \Phi_i(a_1) \dots \Phi_i(a_n) \rangle ]
=\int D \Psi \ c_*(\Psi) \  \Psi(a_1) \dots \Psi(a_n) \ . 
\end{equation}
Il ne reste plus qu'\`a expliciter les composantes bosoniques de ces 
super-corr\'elateurs pour exprimer les fonctions de corr\'elation et
de r\'eponse qui \'etaient l'objectif initial de l'\'etude.

\subsection{Le cas compl\`etement connect\'e}

Afin de rendre l'\'etude pr\'ec\'edente, pour le moins abstraite, un peu
plus explicite, consid\'erons le cas du mod\`ele $p$-spin sph\'erique
compl\`etement connect\'e. Celui-ci a \'et\'e largement \'etudi\'e dans le 
pass\'e, on pourra trouver une revue de ses propri\'et\'es 
dans~\cite{Ba-pspin}. 

Dans l'hamiltonien d'interaction 
(\ref{eq:continu-defHi}), on suppose donc que les couplages $J_{i_1 \dots i_p}$
sont distribu\'es selon une loi gaussienne de moyenne
nulle, avec $[J_{i_1 \dots i_p}^2]=(p! J_0^2)/(2 N^{p-1})$. On a donc selon la
d\'efinition (\ref{eq:susy-defg}) $g(x)=J_0^2 x^2/4$.
Comme le mod\`ele est sph\'erique, la contrainte 
$\sum_i \sigma_i^2(t)=N$ doit \^etre impos\'ee par un multiplicateur de 
Lagrange d\'ependant du temps $\mu(t)$, qui appara\^it dans 
(\ref{eq:continu-defHi}) comme $V(\sigma)=\mu(t) \sigma^2$. Ce terme \'etant
quadratique en $\sigma$, on peut de mani\`ere \'equivalente l'incorporer
dans $\cal K$, et noter ${\cal K}^s = {\cal K} - \mu(t)$ ce nouvel op\'erateur.
$g(x)$ \'etant une fonction quadratique, l'\'equation (\ref{eq:susy-speq})
va admettre une solution $c_*$ gaussienne, en analogie avec le calcul de la
densit\'e de valeurs propres de l'ensemble gaussien orthogonal.

Posons en effet
\begin{equation}
c_*(\Phi) = {\cal N} \exp \left[ -\frac{1}{2} \int da db \ \Phi(a) Q^{-1}(a,b)
\Phi(b) \right] \ ,
\end{equation}
o\`u l'inverse $Q^{-1}$ est par rapport au produit de
convolution $\otimes$. Le membre
de droite de (\ref{eq:susy-speq}) peut alors se calculer, en particulier
\begin{eqnarray}
&&\int D\Psi_2 \dots D\Psi_p  \ 
c_*(\Psi_2) \dots c_*(\Psi_p) \
g\left(\int da \ \Phi(a) \Psi_2(a) \dots \Psi_p(a) \right) \nonumber \\
&& \hspace{3.5cm}  
= \frac{J_0^2}{4} \int da db \ \Phi(a) Q^{\bullet(p-1)}(a,b) \Phi(b) \ . 
\end{eqnarray}
L'\'equation de point col (\ref{eq:susy-speq}) est donc v\'erifi\'ee si 
les noyaux des gaussiennes des deux membres sont \'egaux, autrement dit si
\begin{equation}
Q^{-1}(a,b)= \delta(a-b) {\cal K}^{\rm s} - p \frac{J_0^2}{2} Q^{\bullet(p-1)}
(a,b) 
\ .
\end{equation}
Multiplions (de convolution) les deux membres de cette \'equation par $Q$
pour obtenir finalement~:
\begin{equation}
\delta(a-b)=({\cal K}^{\rm s} \otimes Q)(a,b) - p \frac{J_0^2}{2} 
(Q^{\bullet(p-1)} \otimes Q)(a,b) \ .
\end{equation}
Ceci est la forme supersym\'etrique des \'equations dynamiques pour ce 
mod\`ele~\cite{BoCuKuMe-MCA}, que l'on peut expliciter en termes
des fonctions de corr\'elation et de r\'eponse \`a deux points (on prend
$t_a>t_b$ pour simplifier les notations)~:
\begin{eqnarray}
\frac{\partial}{\partial t_a} R(t_a;t_b) &=& -\mu(t_a) R(t_a;t_b) + 
\int_{t_b}^{t_a} dt \, \Sigma(t_a;t) R(t;t_b) \ ,
\\
\frac{\partial}{\partial t_a} C(t_a,t_b) &=& -\mu(t_a) C(t_a,t_b) +
\int_{0}^{t_b} dt \, \Sigma(t_b;t) C(t,t_a)
+  \int_{0}^{t_a} dt \, D(t_b,t) R(t_a;t) \ , \nonumber 
\end{eqnarray}
o\`u les noyaux $\Sigma$ et $D$ sont donn\'es par
\begin{eqnarray}
D(t,t') &=& p \frac{J_0^2}{2} C(t,t')^{p-1} \ , \nonumber \\
\Sigma(t;t') &=& p (p-1) \frac{J_0^2}{2} C(t,t')^{p-2} R(t;t') \ .
\label{eq:noy-pspin}
\end{eqnarray}
On pourra consulter~\cite{Cu-Houches} pour une discussion d\'etaill\'ee 
des propri\'et\'es de ces \'equations, en particulier le comportement 
hors d'\'equilibre de leur solution \`a basse temp\'erature,
ainsi que pour leur relation avec les versions sch\'ematiques de la th\'eories 
du couplage de modes des liquides surfondus~\cite{Go-MCT}.
On reviendra sur ce sujet dans la partie~\ref{sec:ft-outofeq}.

Une interpr\'etation alternative de ce r\'esultat consiste \`a imaginer 
l'\'evolution d'un seul degr\'e de libert\'e (\og single spin
equation\fg ) qui conduirait aux m\^emes \'equations pour la corr\'elation
et la r\'eponse. Cela correspond \`a une \'equation de Langevin 
g\'en\'eralis\'ee (une d\'emonstration par la m\'ethode de la cavit\'e dans 
le cas du mod\`ele SK se trouve dans~\cite{Beyond}),
\begin{equation}
\frac{d}{dt} \sigma(t) = -\mu(t) \sigma(t) + \int_0^t dt' \, \Sigma(t;t') 
\sigma(t') + \xi(t) + \nu(t) \ .
\label{eq:singlespin}
\end{equation}
$\xi(t)$ est toujours le bruit blanc gaussien mod\'elisant l'interaction
avec le bain thermique \`a la temp\'erature $T$, mais l'influence du reste
du syst\`eme se traduit par l'apparition~:
\begin{itemize}
\item d'un bruit color\'e $\nu(t)$ gaussien, de variance 
$\langle \nu(t) \nu(t') \rangle = D(t,t')$.
\item d'une interaction retard\'ee par l'interm\'ediaire du noyau
$\Sigma(t;t')$.
\end{itemize}
Autrement dit l'\'elimination des $N$ degr\'es de libert\'e du syst\`eme
de d\'epart a fait perdre le caract\`ere markovien des \'equations de
Langevin dont on est parti, et se traduit par l'apparition d'un bain color\'e
dont les propri\'et\'es doivent \^etre d\'etermin\'ees de mani\`ere 
auto-coh\'erente par l'interm\'ediaire des \'equations (\ref{eq:noy-pspin}).

On trouvera dans le chapitre suivant d'autres exemples de ce ph\'enom\`ene~:
assez g\'en\'eralement, l'\'elimination d'une partie des degr\'es de libert\'e 
d'un syst\`eme conduit \`a des \'equations effectives plus compliqu\'ees que 
celles qui d\'ecrivaient la dynamique microscopique originelle.

\subsection{Le cas dilu\'e}

La relative simplicit\'e du cas compl\`etement connect\'e se traduit
par le caract\`ere quadratique de $g(x)$, ce qui permet de trouver une solution
de l'\'equation du col (\ref{eq:susy-speq}) avec une forme 
gaussienne~\footnote{Ceci n'est en toute rigueur vrai que pour les mod\`eles
sph\'eriques~: les interactions soft-spin, m\^eme pour un mod\`ele 
compl\`etement connect\'e, sont non quadratiques. On peut cependant les 
traiter en perturbation en esp\'erant que la physique ne sera pas trop 
modifi\'ee.}. 
On peut s'interroger sur la g\'en\'eralit\'e de cette situation.
En fait le m\^eme ph\'enom\`ene d'\og universalit\'e\fg\
discut\'e dans la partie sur le spectre des matrices al\'eatoires se produit
ici. Rappelons que dans le cadre des matrices al\'eatoires, on avait conclu
que pour des \'el\'ements de matrice distribu\'es ind\'ependamment et de
mani\`ere identique, la seule distribution qui ne conduise pas
\`a la loi du demi-cercle correspondait au graphe al\'eatoire poissonien 
(\`a l'exception des lois larges de L\'evy pour lesquelles l'argument n'est pas
valable). On peut faire ici le m\^eme type de raisonnement pour montrer que 
dans la limite thermodynamique, si l'on veut que la fonction $g(x)$ d\'efinie
par (\ref{eq:susy-defg}) reste finie sans \^etre quadratique, on doit prendre
une loi de probabilit\'e des couplages $J_{i_1 \dots i_p}$ de la forme
\begin{equation}
P(J) = \left(1 - \frac{\alpha p!}{N^{p-1}}\right) \delta(J) + 
\frac{\alpha p!}{N^{p-1}} \pi(J) \ .
\end{equation}
Autrement dit les interactions forment un hypergraphe poissonien avec
en moyenne $\alpha N$ plaquettes pr\'esentes. Rappelons que dans ce cas 
la connectivit\'e d'une variable, {\em i.e.} le nombre d'interactions 
auxquelles elle appartient, est une variable al\'eatoire poissonnienne de 
moyenne $\alpha p$. On trouve alors pour la fonction $g$~:
\begin{equation}
g(x) = -\alpha + \alpha \int dJ \ \pi(J) e^{J x} \ .
\end{equation}
Prenons par exemple une distribution $\pi$ bimodale sym\'etrique en 
$\pm J_0$, pour laquelle $g(x)=-\alpha + \alpha \cosh(J_0 x)$.

Avant de poursuivre cette \'etude, il convient de v\'erifier si les mod\`eles
dilu\'es avec des variables continues vont pr\'esenter le m\^eme genre de
pathologie que celle du mod\`ele de Viana-Bray sph\'erique. Un petit
raisonnement montre que la situation pour des mod\`eles sph\'eriques est
encore plus grave~: les couplages $J_0$ \'etant d'ordre 1 dans le cas
dilu\'e, on peut imaginer que les variables vont se localiser tr\`es
fortement sur un nombre fini de sites. Leurs composantes sur ces sites seront
d'ordre $N^{1/2}$ pour v\'erifier la contrainte sph\'erique, l'\'energie de
ces plaquettes sera donc d'ordre $N^{p/2}$, donc largement favoris\'ee par
rapport aux configurations o\`u toutes les variables sont d'ordre 1, qui
ont une \'energie extensive. Heureusement, ce ph\'enom\`ene de forte
localisation peut \^etre contrecarr\'e par l'introduction de termes soft-spins,
par exemple
\begin{equation}
V(\sigma) = \kappa (\sigma^2-1)^n \ .
\end{equation}
Dans la situation de forte localisation sur un nombre fini de sites, la
contribution de $\sum_i V(\sigma_i)$ \`a l'hamiltonien sera d'ordre $N^n$.
Il suffit donc de prendre $n>p/2$ pour emp\^echer la localisation d'\^etre
\'energ\'etiquement favoris\'e.

Revenons \`a l'\'equation du point col (\ref{eq:susy-speq}). Elle peut se 
r\'e\'ecrire  comme
\begin{eqnarray}
&&c_*(\Phi) = {\cal N} \exp \left[ \int da \left( -\frac{1}{2}
\Phi(a) {\cal K} \Phi(a) + V(\Phi(a)) \right) \right] 
\label{eq:susy-speq-dilue} \\
&& \sum_{k=0}^\infty e^{-\alpha p}\frac{(\alpha p)^k}{k!}
\left[ 
\int D\Psi_2 \dots D\Psi_p \, c_*(\Psi_2) \dots c_*(\Psi_p) \cosh 
\left(J_0 \int da \, \Phi(a) \Psi_2(a) \dots \Psi_p(a) \right) 
\right]^k \ . \nonumber
\end{eqnarray}
On peut interpr\'eter cette \'equation comme ceci~: $c_*(\Phi)$ est la
probabilit\'e que la trajectoire du superchamp $\Phi_i$ d'un site $i$ choisi 
au hasard soit \'egale \`a $\Phi$.
Le premier facteur du membre de droite correspond \`a l'\'evolution 
\og libre\fg\ , c'est-\`a-dire l'influence du 
bain thermique et du potentiel local $V$. Dans la deuxi\`eme ligne, $k$
correspond au nombre d'interactions auxquelles la variable appartient,
pour chacune de ces plaquettes les trajectoires des superchamps des $p-1$
autres variables sont combin\'ees pour exprimer leur influence sur la variable
choisie.

\subsection{Perspectives}

On pourrait, par pure provocation, dire que le probl\`eme de la 
dynamique des mod\`eles
dilu\'es est ici r\'esolu~: il \og suffit\fg\ simplement de r\'esoudre
cette derni\`ere \'equation pour calculer toutes les fonctions de 
corr\'elation et de r\'eponse de ces mod\`eles. Cela n'est bien s\^ur qu'une
boutade, car r\'esoudre exactement cette \'equation semble sans espoir. On peut
cependant sugg\'erer une m\'ethode it\'erative approch\'ee, inspir\'ee de
celle de Biroli et Monasson pour les matrices al\'eatoires. Elle consisterait
\`a chercher une solution approch\'ee avec un $c_*(\Phi)$ gaussien, comme
dans la limite compl\`etement connect\'ee, ins\'erer cette solution approch\'ee
dans le membre de droite de (\ref{eq:susy-speq-dilue}), et prendre le membre
de gauche comme nouvelle forme approch\'ee. On aurait ainsi une prise en compte
successive des fluctuations de connectivit\'e locale du graphe. Pour l'instant
cette id\'ee n'a pas \'et\'e compl\`etement mise en \oe uvre, mais cette piste 
m\'eriterait s\^urement d'\^etre explor\'ee.

Remarquons aussi que l'approche par une \'equation effective single-spin
du type de (\ref{eq:singlespin})
pr\'esente elle aussi des difficult\'es. Si l'on peut \'ecrire formellement
une telle \'equation, elle fait appara\^itre des termes de friction 
retard\'ee avec un nombre arbitraire de temps pr\'ec\'edents, et la
distribution du bruit effectif $\nu$ est quelconque, au lieu d'\^etre 
gaussienne dans le cas compl\`etement connect\'e.

Notons avant de conclure cette partie que la solution $c_*$ de l'\'equation
de col dans le cas dilu\'e n'\'etant pas gaussienne, les fonctions de
corr\'elation et de r\'eponse \`a $n$ points ne s'expriment plus en
fonction uniquement des fonctions \`a deux points, comme c'est le cas
pour des mod\`eles compl\`etement connect\'es. Cette remarque a motiv\'e 
l'\'etude pr\'esent\'ee dans le chapitre \ref{sec:ch-ft}, o\`u l'on 
s'int\'eressera aux propri\'et\'es des fonctions de corr\'elation et de 
r\'eponse \`a plus que deux temps. Ceci devrait faciliter, \`a terme, 
la d\'eduction de pr\'edictions physiques \`a partir de l'\'equation 
(\ref{eq:susy-speq-dilue}).

\chapter{Dynamiques de spins discrets}
\label{sec:discret}

\markboth{\hspace{3mm} \hrulefill \hspace{3mm} Ch. 4~: Dynamiques de spins discrets}{Ch. 4~: Dynamiques de spins discrets\hspace{3mm} \hrulefill \hspace{3mm} }

On va pr\'esenter dans ce chapitre les r\'esultats des publications \pubwsat\ 
sur la dynamique d'un algorithme d'optimisation, et \pubbethe\ qui concerne 
celle d'un ferromagn\'etique dilu\'e.
Ces mod\`eles sont de natures assez diff\'erentes, la dynamique de l'algorithme
ne v\'erifiant pas de condition de balance d\'etaill\'ee. On peut pourtant
\'etudier les deux probl\`emes en suivant des d\'emarches similaires. 

Dans une premi\`ere partie je commencerai par pr\'esenter la m\'ethode de
r\'esolution en termes g\'en\'eriques. Elle permet de se concentrer sur
l'\'evolution d'observables macroscopiques en \og projetant\fg\ les degr\'es
de libert\'e microscopiques. A titre d'illustration j'applique ensuite cette
m\'ethode au cas trivial du mod\`ele de Curie-Weiss, avant de passer \`a
l'\'etude des deux travaux en question.

\section{G\'en\'eralit\'es}
\subsection{Op\'erateurs de projection}
Consid\'erons un syst\`eme d\'ecrit par des configurations microscopiques
$\vec{\sigma}$ que l'on suppose discr\`etes, et qui a une \'evolution
stochastique contr\^ol\'ee par l'\'equation ma\^itresse
\begin{equation}
\mbox{Prob}(\vec{\sigma}',T+1) = \sum_{\vec{\sigma}} 
W(\vec{\sigma}',\vec{\sigma}) \mbox{Prob}(\vec{\sigma},T) \ .
\label{eq:proj-eqM}
\end{equation}
Le temps est discret ici, les $W$ sont donc des probabilit\'es de transition
entre deux configurations. L'exemple canonique est un 
syst\`eme de $N$ spins d'Ising $\sigma_i = \pm 1$, on a alors $2^N$ 
configurations microscopiques. Une mod\'elisation habituelle de la
dynamique d'un syst\`eme physique en contact avec un thermostat consiste
\`a imposer les conditions de balance d\'etaill\'ee aux probabilit\'es de
transition $W$, de mani\`ere \`a atteindre l'\'equilibre thermodynamique
aux temps longs. On reviendra plus en d\'etail sur ce point dans la suite
de ce chapitre ainsi que dans le suivant.

Il est a priori impossible et pas directement int\'eressant de
r\'esoudre le grand nombre d'\'equations coupl\'ees (\ref{eq:proj-eqM}) pour
suivre individuellement les probabilit\'es de chacune des
configurations microscopiques. L'information int\'eressante sur le syst\`eme
est contenue dans un plus petit nombre de grandeurs macroscopiques (\'energie, 
magn\'etisation, densit\'e de particules, ou d'autres selon les situations). 
Petit ne veut pas n\'ecessairement dire fini~: pour un syst\`eme de 
particules sur un 
r\'eseau tridimensionnel par exemple, on peut s'int\'eresser aux fluctuations 
de densit\'e de tout vecteur d'onde. 

On peut se \og d\'ebarrasser\fg\ de l'information microscopique superflue 
et obtenir directement des \'equations d'\'evolution pour les quantit\'es 
macroscopiques, mais le prix \`a payer pour cette perte d'information sera
l'abandon du caract\`ere markovien de l'\'evolution. Cette id\'ee peut 
se formaliser en utilisant des op\'erateurs de projection, une m\'ethode 
attribu\'ee \`a Mori~\cite{Mori} et Zwanzig~\cite{Zwanzig}, et sur laquelle
repose notamment la th\'eorie du couplage de modes pour les liquides
surfondus~\cite{Go-MCT}.

Introduisons \`a cet effet une matrice $\hat{W}$ et un vecteur colonne $p(T)$,
indic\'es par les configurations microscopiques,
\begin{equation}
(\hat{W})_{\vec{\sigma}'\vec{\sigma}}= W(\vec{\sigma}',\vec{\sigma}) \ ,
\qquad (p(T))_{\vec{\sigma}} = \mbox{Prob}(\vec{\sigma},T) \ .
\end{equation}
L'\'equation ma\^itresse se r\'e\'ecrit alors comme un produit matriciel,
\begin{equation}
p(T+1) = \hat{W} p(T) \ .
\end{equation}

Notons $X(\vec{\sigma})$ l'observable macroscopique \`a laquelle on 
s'int\'eresse.  Pour all\'eger les \'ecritures on laisse sous-entendu le fait 
qu'elle pourrait \^etre elle-m\^eme vectorielle. On peut partitionner 
l'ensemble des configurations $\vec{\sigma}$ selon les valeurs de $X$ qui
leur sont associ\'ees. La matrice $\hat{\cal P}$ d\'efinie par
\begin{equation}
(\hat{\cal P})_{\vec{\sigma}'\vec{\sigma}} = \frac{\delta(X(\vec{\sigma}') - 
X(\vec{\sigma}))}{\sum_{\vec{\sigma}''}\delta(X(\vec{\sigma}'') - 
X(\vec{\sigma}))}
\end{equation}
est un projecteur ($\hat{\cal P}^2=\hat{\cal P}$) dont l'action sur un vecteur
colonne consiste \`a le \og lisser\fg\ en faisant une moyenne sur chacune
des partitions engendr\'ees par l'observable $X$. On va poser $\overline{p}(T)=
\hat{\cal P}p(T)$ la loi de probabilit\'e liss\'ee, et 
$q(T)=p(T)-\overline{p}(T)$ son compl\'ement. En projetant l'\'equation
ma\^itresse il vient
\begin{equation}
\begin{cases} 
\overline{p}(T+1)=\hat{\cal P}\hat{W} \overline{p}(T) 
+ \hat{\cal P}\hat{W} q(T)
\\
q(T+1)=(\hat{1}-\hat{\cal P})\hat{W} \overline{p}(T) 
+ (\hat{1}-\hat{\cal P})\hat{W} q(T)
\end{cases} \ ,
\end{equation}
o\`u $\hat{1}$ d\'esigne la matrice identit\'e.
En it\'erant la deuxi\`eme \'equation il est possible d'\'eliminer $q$~:
\begin{equation}
\overline{p}(T+1) = \hat{\cal P}\hat{W} \overline{p}(T) 
+\sum_{T'=1}^T \hat{\cal P}\hat{W} ((\hat{1}-\hat{\cal P})\hat{W})^{T'} 
\overline{p}(T-T') \ ,
\label{eq:proj-nonmarkov}
\end{equation}
en supposant que la probabilit\'e au temps initial $T=0$ est telle que
$q(0)=0$.

Par d\'efinition $\overline{p}_{\vec{\sigma}}$ ne d\'epend de $\vec{\sigma}$
que par l'interm\'ediaire de la valeur de l'observable $X(\vec{\sigma})$. 
L'\'equation projet\'ee (\ref{eq:proj-nonmarkov}) d\'etermine donc
l'\'evolution des probabilit\'es d'observation des diff\'erentes valeurs de
$X$. Comme on pouvait s'y attendre, la perte d'information due \`a la
projection est compens\'ee par l'apparition d'une m\'emoire de
l'\'evolution pass\'ee de l'observable.

On nommera approximation markovienne dans la suite de ce chapitre 
l'approximation qui consiste \`a remplacer (\ref{eq:proj-nonmarkov}) par
\begin{equation}
\overline{p}(T+1) = \hat{\cal P}\hat{W} \overline{p}(T) \ ,
\label{eq:proj-markov}
\end{equation}
c'est-\`a-dire \`a n\'egliger tous les termes de m\'emoire. Cette approximation
est a priori incontr\^ol\'ee, on verra au cas par cas la qualit\'e des
pr\'edictions \`a laquelle elle conduit. Faisons deux remarques~:
\begin{itemize}
\item Les r\'esultats d'une telle approximation d\'ependent bien s\^ur de
l'observable $X$ sur laquelle on projette. Plus celle-ci contient une
description fine de la configuration microscopique, moins on perd 
d'information en passant au processus projet\'e, et meilleurs devraient
\^etre les r\'esultats de l'approximation.
\item En termes plus parlants, cette approximation consiste \`a supposer
qu'\`a chaque instant toutes les configurations microscopiques correspondant
\`a une m\^eme valeur de l'observable sont \'equiprobables. En
effet $q(T)=0$ dans ce cas, et (\ref{eq:proj-markov}) est alors correcte.
Cette interpr\'etation sera utile dans la suite.
\end{itemize}

\subsection{Processus \og markoviens locaux\fg }
\label{sec:pml}

L'\'evolution des observables macroscopiques d'un syst\`eme physique 
pr\'esente souvent des caract\'eristiques particuli\`eres. Pour un
syst\`eme de $N$ spins d'Ising par exemple, la magn\'etisation totale
$M=\sum_i \sigma_i$ est extensive, proportionnelle \`a la taille $N$ du
syst\`eme. De plus on consid\`ere souvent des probabilit\'es de transition
qui relient les configurations ne diff\'erant que par le renversement d'un
spin~: la magn\'etisation ne varie donc que d'une quantit\'e finie sur un
pas de temps \'el\'ementaire. On va \'etudier dans cette partie les
propri\'et\'es de tels processus stochastiques, dans un cadre un peu plus
g\'en\'eral. 

Consid\'erons une variable al\'eatoire $\vec X$ \`a $d$ dimensions, qui prend
des valeurs discr\`etes, et qui \'evolue \`a chaque pas de temps $T \to T+1$
selon l'\'equation ma\^itresse markovienne 
\begin{equation}
\mbox{Prob}(\vec{X}',T+1) = \sum_{\vec{X}} 
W(\vec{X}',\vec{X}) \mbox{Prob}(\vec{X},T) \ .
\end{equation}
Cela pourrait notamment \^etre le r\'esultat de l'approximation markovienne
\`a partir d'une description microscopique~; on a explicit\'e ici la
possibilit\'e pour l'observable $\vec X$ d'\^etre multidimensionnelle, et
on r\'eutilise la notation $W$ en esp\'erant qu'il n'y aura pas de 
confusion avec les probabilit\'es de transition microscopiques.

Supposons que l'on ait dans notre probl\`eme un param\`etre $N \gg 1$,
et que les valeurs typiques de $\vec X$ soient d'ordre $N$. Le processus
est dit ici \og local\fg\ si la variation typique de $\vec X$ sur un pas de 
temps est d'ordre $1$, et si les probabilit\'es de transition $W$ ne varient 
sensiblement que lorsque ses arguments varient sur des quantit\'es d'ordre 
$N$. Autrement dit,
\begin{equation}
W(\vec{X}',\vec{X})=w(\vec{X}'-\vec{X},\vec{X}/N) \ ,
\end{equation}
avec $w(\vec{\Delta},\vec{x})$
non nulle pour $\vec{\Delta}$ d'ordre $1$, et suffisamment r\'eguli\`ere dans 
la deuxi\`eme variable.

Introduisons deux notations~:
\begin{equation}
\langle \bullet \rangle_T = \sum_{\vec{X}} \bullet \ \mbox{Prob}(\vec{X},T)
\quad , \quad
[ \bullet ]_{\vec{x}} = \sum_{\vec{\Delta}} \bullet \ w(\vec{\Delta},\vec{x}) 
\ .
\end{equation}
La premi\`ere correspond \`a une moyenne sur les trajectoires possibles de
la dynamique, la deuxi\`eme sur les transitions possibles de cette marche 
al\'eatoire au voisinage d'un point donn\'e. Cette derni\`ere moyenne est
bien normalis\'ee car l'\'equation ma\^itresse d\'efinie par les $W$ conserve
la probabilit\'e totale.

On va d'abord s'int\'eresser \`a l'\'evolution temporelle de la valeur 
moyenne de $\vec X$. Pour deux instants successifs, il vient tr\`es simplement
\begin{equation}
\langle \vec{X} \rangle_{T+1} = \langle \vec{X} \rangle_{T}
+\langle [\vec{\Delta}]_{\vec{X}/N} \rangle_T \ .
\label{eq:pml-moydis}
\end{equation}
Comme on va le voir dans la suite, $\vec X$ est fortement piqu\'e autour
de sa valeur moyenne dans la limite thermodynamique.
On peut donc intervertir les deux op\'erations $\langle \bullet \rangle$ et 
$[\bullet]$ dans le dernier terme de l'\'equation (\ref{eq:pml-moydis}). 
Il est alors naturel de d\'efinir
un temps (quasi-) continu $t = T/N$, ainsi que la densit\'e de la valeur 
moyenne de $\vec X$,
\begin{equation}
\vec{m}(t) = \frac{1}{N} \langle \vec{X} \rangle_{T=Nt} \ .
\end{equation}
Posant $\vec{v}(\vec{x}) = [\vec{\Delta}]_{\vec{x}}$ la d\'erive moyenne
au voisinage du point $\vec{x}$, on obtient en d\'eveloppant 
(\ref{eq:pml-moydis}) l'\'equation diff\'erentielle ordinaire
\begin{equation}
\frac{d}{dt} \vec{m}(t) = \vec{v}(\vec{m}(t)) \ .
\label{eq:pml-eqsurm}
\end{equation}
En r\'esolvant cette \'equation avec la condition initiale appropri\'ee, on
a d\'etermin\'e le comportement moyen du processus stochastique. C'est en
fait aussi son comportement typique, les fluctuations de $\vec{X}$ autour
de $N\vec{m}$ sont d'ordre $\sqrt{N}$.

Int\'eressons nous maintenant \`a ces d\'eviations autour de l'\'evolution
moyenne. Afin de rendre plus claire la suite de l'expos\'e il est peut-\^etre 
utile de rappeler quelques propri\'et\'es des sommes de variables 
al\'eatoires ind\'ependantes. Soit donc $\vec{X}=\vec{\Delta}_1 + \dots +
\vec{\Delta}_N$ la somme de $N \gg 1$ variables, distribu\'ees selon
la loi $w(\vec{\Delta})$. On utilisera \`a nouveau la 
notation $[ \bullet ]$ pour d\'enoter les moyennes sur cette loi $w$, et on 
pose $\vec{v}=[\vec{\Delta}]$. D'apr\`es le 
th\'eor\`eme central limite, $\vec{X}$ est une variable al\'eatoire gaussienne
centr\'ee sur $N \vec{v}$, avec des fluctuations d'ordre $\sqrt{N}$. Cependant
cette forme gaussienne n'est valable qu'autour de $N \vec{v}$, les queues de
la loi pour les valeurs improbables de $\vec{X}$ ne sont pas d\'ecrites par
le th\'eor\`eme central limite. Ce r\'egime de grande d\'eviation est l'objet
du th\'eor\`eme de Cramer, que l'on va retrouver ici de mani\`ere heuristique.
Notons 
\begin{equation}
e^{\ell(\vec{\lambda})} = \left[e^{\vec{\lambda} \cdot \vec{\Delta}} \right]
\end{equation}
la fonction g\'en\'eratrice de $\vec{\Delta}$. 
Comme $\vec{X}$ est la somme de $N$ variables $\vec{\Delta}_i$ ind\'ependantes,
il vient
\begin{equation}
\sum_{\vec{X}} \mbox{Prob}(\vec{X}) e^{\vec{\lambda} \cdot \vec{X}} = 
e^{N\ell(\vec{\lambda})} \ .
\end{equation}
Dans la limite $N \to \infty$ la somme dans le membre de gauche s'\'evalue
par la m\'ethode du col. En posant $\mbox{Prob}(\vec{X}) \sim 
\exp[- N \pi(\vec{X}/N)]$, on s'aper\c coit que la fonction de grande 
d\'eviation $\pi(\vec{x})$ et la fonction g\'en\'eratrice $\ell(\vec{\lambda})$
sont des transform\'ees de Legendre l'une de l'autre,
\begin{equation}
\pi(\vec{x}) = \min_{\vec{\lambda}}\left[\vec{\lambda} \cdot \vec{x} - 
\ell(\vec{\lambda})\right] \ , \qquad
\ell(\vec{\lambda}) = \max_{\vec{\lambda}}\left[\vec{\lambda} \cdot \vec{x} - 
\pi(\vec{x})\right] \ .
\end{equation}
Le th\'eor\`eme central limite s'obtient \`a partir de ce th\'eor\`eme
plus puissant en d\'eveloppant $\pi$ autour de $\vec{x}=\vec{v}$, ce qui
revient \`a d\'evelopper $\ell$ autour de $\vec{\lambda}=\vec{0}$.

Revenons \`a notre probl\`eme de processus stochastique. La valeur de
$\vec{X}$ \`a l'instant $t$ est aussi la somme d'un grand nombre, $N t$,
de variables al\'eatoires $\vec{\Delta}_i$, mais qui ne sont ni 
ind\'ependantes ni tir\'ees avec 
la m\^eme loi de probabilit\'e puisque $w(\vec{\Delta},\vec{x})$ d\'epend de
la position $\vec{x}$. L'id\'ee dans la suite de cette partie consiste
\`a exploiter les propri\'et\'es de \og localit\'e\fg\ du processus~: 
sur un intervalle de temps $[t,t+\epsilon]$ avec $\epsilon \ll 1$
le d\'eplacement du processus est la somme de $\epsilon N$ variables 
$\vec{\Delta}_i$ que l'on supposera distribu\'ees avec la m\^eme loi 
$w(\vec{\Delta},\vec{x}(t))$. On utilisera donc le th\'eor\`eme de Cramer 
sur des petits intervalles de temps pour lesquels la marche n'a pas trop 
boug\'e.

Il y a deux mani\`eres d'exploiter cette id\'ee. La premi\`ere
consiste \`a \'ecrire une \'equation aux d\'eriv\'ees partielles sur
la fonction g\'en\'eratrice.

Inspir\'e par le r\'esultat du th\'eor\`eme de Cramer, on pose
\begin{equation}
\mbox{Prob}(\vec{X},T) \sim \exp \left[- N \pi\left(\frac{\vec{X}}{N},
\frac{T}{N}\right)\right] \ .
\label{eq:scaling-P}
\end{equation}
Introduisons aussi la fonction g\'en\'eratrice
\begin{equation}
G(\vec{\lambda},T) = \langle e^{\vec{\lambda} \cdot \vec{X}} \rangle_T 
\sim \sum_{\vec X} e^{N(\vec{\lambda} \cdot \vec{x} - \pi(\vec{x},t))}
\sim e^{N g(\vec{\lambda},t)} \ .
\end{equation}
En \'evaluant la somme sur $\vec{X}$ par la m\'ethode du col, on constate
que $g$ est la transform\'ee de Legendre de $\pi$,
$g(\vec{\lambda},t)=\underset{\vec{x}}{\max} 
[\vec{\lambda} \cdot \vec{x} - \pi(\vec{x},t)]$. 
Cherchons maintenant \`a \'ecrire l'\'equation
qui r\'egit l'\'evolution de $g$. On obtient facilement
\begin{equation}
G(\vec{\lambda},T+1) = \langle 
\left[ e^{\vec{\lambda} \cdot \vec{\Delta}} \right]_{\vec{X}/N} 
e^{\vec{\lambda} \cdot \vec{X}} \rangle_T \ .
\label{eq:larged-discret}
\end{equation}
Le membre de gauche devient, en d\'eveloppant l'argument temporel de $g$,
\begin{equation}
G(\vec{\lambda},T) \exp\left(\frac{\partial}{\partial t} g(\vec{\lambda},t)
\right) \ .
\end{equation}
D\'efinissons aussi la fonction g\'en\'eratrice
des d\'eplacements microscopiques autour d'une position donn\'ee,
\begin{equation}
\left[ e^{\vec{\lambda} \cdot \vec{\Delta}} \right]_{\vec{x}} =
e^{\ell (\vec{\lambda},\vec{x})} \ .
\end{equation}
Le membre de droite de (\ref{eq:larged-discret}) peut alors \^etre \'evalu\'e 
par la m\'ethode du col,
\begin{equation}
\int d\vec{x} \ e^{N(\vec{\lambda} \cdot \vec{x} - \pi(\vec{x},t))} 
e^{\ell (\vec{\lambda},\vec{x})} = G(\vec{\lambda},T) 
e^{\ell (\vec{\lambda},\vec{x}(\vec{\lambda},t))} \ ,
\end{equation}
o\`u $\vec{x}(\vec{\lambda},t)$ est le point col de l'int\'egrale. D'apr\`es
les propri\'et\'es des transform\'ees de Legendre, on a en fait 
$\vec{x}(\vec{\lambda},t) = \vec{\nabla}g(\vec{\lambda},t)$. D'o\`u finalement
l'\'equation aux d\'eriv\'ees partielles qui gouverne l'\'evolution de la
fonction g\'en\'eratrice~:
\begin{equation}
\frac{\partial}{\partial t} g(\vec{\lambda},t) = \ell (\vec{\lambda}, 
\vec{\nabla}g(\vec{\lambda},t)) \ .
\label{eq:pml-eqsurg}
\end{equation}
Une fois que cette \'equation, compl\'et\'ee par une condition initiale 
$g(\vec{\lambda},t=0)$, est r\'esolue, il reste \`a effectuer une 
transform\'ee de Legendre inverse pour obtenir la fonction de grande 
d\'eviation $\pi(\vec{x},t)$. On peut facilement v\'erifier deux propri\'et\'es
attendues de (\ref{eq:pml-eqsurg})~:

\vspace{2mm}

\begin{itemize}
\item elle conserve $g(\vec{0},t)=0$, qui traduit la normalisation des
probabilit\'es.
\item elle permet de r\'eobtenir l'\'equation (\ref{eq:pml-eqsurm}) pour 
l'\'evolution de la position moyenne avec
$\vec{m}(t) = \vec{\nabla}g(\vec{\lambda},t)|_{\vec{0}}$.
\end{itemize}

\vspace{2mm}

\noindent
De plus on retrouve naturellement la forme habituelle du th\'eor\`eme de
Cramer quand $\ell (\vec{\lambda},\vec{x})$ est ind\'ependant de $\vec{x}$.

L'\'equation (\ref{eq:pml-eqsurg}) se traduit en une \'equation \'equivalente
sur la fonction de grande d\'eviation~:
\begin{equation}
\frac{\partial}{\partial t} \pi(\vec{x},t) = - \ell 
(\vec{\nabla}\pi(\vec{x},t) ,\vec{x}) \ .
\end{equation}
Cette deuxi\`eme forme est d'une moindre utilit\'e pratique, $\pi$ \'etant 
souvent plus irr\'eguli\`ere que $g$. Par exemple pour une condition initiale
o\`u $\vec{x}$ est fix\'e \`a une certaine valeur $\vec{x}_0$, on a 
$g(\vec{\lambda})=\vec{\lambda} \cdot \vec{x}_0$, alors que $\pi(\vec{x})$ 
vaut $+\infty$ partout sauf en $\vec{x}_0$ o\`u il s'annule.

Une deuxi\`eme approche \`a ce probl\`eme de grande d\'eviation consiste
\`a \'ecrire, sous forme d'int\'egrale de chemin,
la probabilit\'e de toute une trajectoire $\{\vec{x}(t)\}$ pour
$t \in [0,t_f]$.
On l'obtient en d\'ecoupant l'axe des temps en
intervalle de longueur $\epsilon$ et en utilisant le th\'eor\`eme de
Cramer sur chacun des intervalles. Prenant finalement la limite $\epsilon 
\to 0$ il vient
\begin{equation}
\mbox{Prob}[\{\vec{x}(t)\}] \sim \int D\vec{\lambda} \ 
\exp \left[ N \int_0^{t_f}
dt \ \left( -  \vec{\lambda}(t) \cdot \dot{\vec{x}}(t) + \ell(\vec{\lambda}(t),
\vec{x}(t))
\right) \right] \ .
\end{equation}
De cette repr\'esentation en int\'egrale de chemin on peut obtenir
la fonction de grande d\'eviation \`a l'instant $t_f$ comme
\begin{equation}
e^{-N \pi(\vec{x}_f,t_f)} \sim \int D\vec{x} \ D\vec{\lambda} \ 
\exp \left[ N \int_0^{t_f}
dt \ \left( -  \vec{\lambda}(t) \cdot \dot{\vec{x}}(t) + \ell(\vec{\lambda}(t),
\vec{x}(t))
\right) \right] \ ,
\label{eq:pml-pathint}
\end{equation}
o\`u l'int\'egrale fonctionnelle sur les trajectoires $\{\vec{x}(t)\}$ doit
\^etre restreinte \`a celles qui v\'erifient $\vec{x}(t_f)=\vec{x}_f$, et doit
\^etre pond\'er\'ee selon la distribution de probabilit\'e de $\vec{x}$ \`a
l'instant initial.

Le lien entre l'approche par la fonction g\'en\'eratrice (\ref{eq:pml-eqsurg})
et celle de l'int\'egrale de chemin fait appara\^itre, de mani\`ere assez 
frappante, une analogie avec le formalisme de la m\'ecanique analytique.
En effet, l'\'evaluation de l'int\'egrale de chemin (\ref{eq:pml-pathint})
par la m\'ethode du col conduit aux \'equations suivantes pour les
trajectoires dominantes~:
\begin{equation}
\begin{cases}
\dot{\vec{x}}(t) = \vec{\nabla}_{\vec{\lambda}} 
\ell(\vec{\lambda}(t),\vec{x}(t) )
\\
\dot{\vec{\lambda}}(t) = - \vec{\nabla}_{\vec{x}} 
\ell(\vec{\lambda}(t),\vec{x}(t) )
\end{cases} \ ,
\end{equation}
c'est-\`a-dire des \'equations de mouvement classique o\`u $\vec{x}$ et
$\vec{\lambda}$ sont des moments conjugu\'es l'un de l'autre. La fonction 
$\ell$
s'interpr\`ete comme un hamiltonien, et l'\'equation sur la
fonction g\'en\'eratrice (\ref{eq:pml-eqsurg}) correspond \`a
l'\'equation de Hamilton-Jacobi de ce probl\`eme de m\'ecanique.
On peut donc interpr\'eter $g$ comme la fonction g\'en\'eratrice du
changement de variables canonique, qui n'est autre que la transform\'ee
de Legendre de l'action (au sens m\'ecanique du terme). Cette action appara\^it
aussi comme le poids de l'int\'egrale de chemin, ce qui nous ram\`ene \`a
l'interpr\'etation originelle de $\pi$ et $g$ comme transform\'ees de
Legendre l'une de l'autre.

\vspace{2mm}

Pour conclure cette partie, remarquons que la pr\'esentation utilis\'ee
\'etait loin d'\^etre rigoureuse. En particulier la notation $\sim$ ne 
d\'esignait pas des vrais \'equivalents, puisque seulement
les termes exponentiels en $N$ ont \'et\'e conserv\'es, en n\'egligeant tous 
les pr\'efacteurs alg\'ebriques. Ces r\'esultats peuvent pourtant, au prix
d'\'enonc\'es plus pr\'ecis, prendre un sens math\'ematique. La concentration
de l'\'evolution typique autour de l'\'evolution moyenne solution de
l'\'equation diff\'erentielle ordinaire (\ref{eq:pml-eqsurm}) est par
exemple un cas particulier du th\'eor\`eme de Wormald~\cite{Wormald}.
Ce type de raisonnement est souvent utilis\'e en informatique pour prouver des 
bornes inf\'erieures sur le seuil de satisfiabilit\'e~\cite{Ac-bounds}. Pour
un \'enonc\'e rigoureux des principes de grande d\'eviation, on pourra
se r\'ef\'erer \`a~\cite{DeZe-LDP}.

\newpage

\section{Le mod\`ele de Curie-Weiss}
\label{ch:curieweiss}
Dans cette partie on va illustrer ce formalisme g\'en\'erique dans un
cas tr\`es simple. Consid\'erons \`a cet effet le mod\`ele de
Curie-Weiss, alias le ferromagn\'etique sur le graphe complet. Chacun des $N$
spins d'Ising $\sigma_i$ du mod\`ele interagit avec tous les autres, et
avec un champ ext\'erieur d'intensit\'e $h$, ce qui conduit \`a l'hamiltonien
\begin{equation}
H=-\frac{J}{2N} \sum_{i,j} \sigma_i \sigma_j -h \sum_i \sigma_i \ .
\end{equation}
Le couplage entre spins est d'ordre $1/N$ de mani\`ere \`a avoir un 
hamiltonien extensif. On prendra $J=1$ dans la suite, ce qui revient
\`a red\'efinir l'\'echelle de temp\'erature.

Les propri\'et\'es statiques du mod\`ele sont d\'etermin\'ees tr\`es
simplement. L'hamiltonien ne d\'epend en effet de la configuration 
microscopique que par l'interm\'ediaire de la magn\'etisation 
$M=\sum_i \sigma_i$, la fonction de partition s'\'ecrit donc
\begin{equation}
\renewcommand{\arraystretch}{.66}
Z = \sum_{\begin{array}{c} \scriptstyle M=-N \\ \scriptstyle N-M \ {\rm pair} 
\end{array}}^N
{N \choose \frac{N-M}{2}} 
\exp \left[\beta \frac{M^2}{2N} + \beta h M \right] \ .
\renewcommand{\arraystretch}{1.5}
\end{equation}
Dans la limite thermodynamique on peut \'evaluer le coefficient binomial
avec la formule de Stirling et transformer la somme en une int\'egrale,
que l'on calcule par la m\'ethode du col. On aboutit alors \`a l'expression 
de l'\'energie libre par site
\begin{equation}
f = \underset{m}{\rm min} \left[
-\frac{1}{2} m^2 - h m -\frac{1}{\beta} 
\left( - \frac{1+m}{2} \ln \frac{1+m}{2}
- \frac{1-m}{2} \ln \frac{1-m}{2} \right) \right] \ ,
\end{equation}
o\`u $m=M/N$ est la magn\'etisation par site. On reconna\^it dans cette
expression la d\'ecomposition en partie \'energ\'etique et entropique. 
La magn\'etisation spontan\'ee du syst\`eme minimise l'\'energie libre. 
Elle v\'erifie l'\'equation $m=\tanh(\beta (m+h))$, avec donc
une temp\'erature inverse critique en champ nul de $\beta_c=1$.

Les propri\'et\'es dynamiques de ce mod\`ele ont \'et\'e \'etudi\'ees 
dans~\cite{GrWeLa}. On reconsid\`ere ici cette \'etude, en s'appuyant sur
le formalisme d\'evelopp\'e dans la partie pr\'ec\'edente.

Prenons l'\'evolution dynamique habituelle pour des syst\`emes de spins
d'Ising~: \`a chaque pas de temps,
un des $N$ spins est choisi au hasard, on calcule la variation d'\'energie
$\Delta E$ que son renversement induirait dans le syst\`eme, et on effectue
le renversement avec une probabilit\'e $R(\Delta E,\beta)$.
De mani\`ere \`a atteindre l'\'equilibre thermodynamique \`a la 
temp\'erature inverse $\beta$, on impose la condition de balance d\'etaill\'ee
sur $R$, sous la forme
\begin{equation}
R(\Delta E,\beta)=e^{-\beta \Delta E} R(-\Delta E,\beta) \ .
\end{equation}
Les taux de transition les plus connus, qui respectent tous deux cette
condition, sont ceux de Metropolis et de Glauber,
\begin{equation}
R_{\rm Metropolis}= {\rm min} \left(1, e^{-\beta \Delta E} \right)  \ ,
\end{equation}
\begin{equation}
R_{\rm Glauber}= \frac{1}{2} \left( 1 - \tanh \left( \frac{\beta \Delta E}{2}
\right) \right)\ .
\end{equation}

Ici la projection de la dynamique est triviale. Il est clair en
effet que les configurations microscopiques avec la m\^eme 
magn\'etisation sont \'equiprobables \`a tout instant si l'on part
d'une condition initiale uniforme sur les configurations d'une magn\'etisation
donn\'ee.
Autrement dit on ne perd pas d'information en passant \`a la dynamique sur la 
magn\'etisation, qui reste markovienne. De plus l'\'energie d'une configuration
s'exprime en fonction de sa magn\'etisation. Dans une configuration
de magn\'etisation par site $m$, le spin choisi au hasard est de signe 
$\sigma$ avec
la probabilit\'e $(1+m\sigma)/2$, et la variation d'\'energie du syst\`eme
s'il est flipp\'e vaut $\Delta E=2\sigma (m+h)$ (\`a des termes d'ordre
$1/N$ pr\`es). La magn\'etisation totale $M$ varie alors de $-2 \sigma$.
On est donc dans le cadre g\'en\'eral d'un processus markovien local
\'etudi\'e dans la section pr\'ec\'edente, avec
\begin{eqnarray}
w(-2,m) &=& \frac{1+m}{2} R(2(m+h),\beta)\ , \\
w(2,m) &=& \frac{1-m}{2} R(-2(m+h),\beta)\ , \\
w(0,m) &=& 1 - w(-2,m) - w(2,m) \ .
\end{eqnarray}
Suivant toujours les notations g\'en\'erales, la variation moyenne de la
magn\'etisation lors d'un pas de temps \`a partir d'une configuration
de magn\'etisation $m$ s'\'ecrit
\begin{equation}
v(m,\beta)=(1-m) R(-2(m+h),\beta) - (1+m) R(2(m+h),\beta) \ ,
\end{equation}
La magn\'etisation moyenne \'evolue donc selon l'\'equation diff\'erentielle
$\dot{m}(t)=v(m(t),\beta)$. 
Utilisons la condition de balance d\'etaill\'ee pour transformer cette
\'equation en
\begin{equation}
\dot{m}(t) = -  R(2(m(t)+h),\beta) 
\left( (1+m(t)) - (1-m(t)) e^{2 \beta (m(t) + h(t))} \right) \ .
\label{eq:curieweiss-moy}
\end{equation}
Partant d'une condition initiale $m(t=0)=m_0$ quelconque, on atteint aux temps
longs une magn\'etisation stationnaire $m^*$ telle que $v(m^*,\beta)=0$.
Sous la forme (\ref{eq:curieweiss-moy}) il est facile de voir
que $m^*$ v\'erifie $m^*=\tanh(\beta (m^*+h))$. Comme il se doit, la
condition de balance d\'etaill\'ee sur les probabilit\'es de transition
implique que les points fixes de l'\'evolution dynamique sont les
extrema de l'\'energie libre thermodynamique. L'\'equation 
(\ref{eq:curieweiss-moy}) permet en outre de d\'ecrire la relaxation
vers l'\'equilibre \`a partir d'une magn\'etisation initiale qui en est 
arbitrairement \'eloign\'ee.

On va \'etudier plus en d\'etails la stabilit\'e de ces points fixes,
ainsi que le comportement critique au voisinage
de $\beta_c=1$. Prenons pour simplifier $h=0$. On a par d\'efinition
$v(m^*(\beta),\beta)=0$. La stabilit\'e du point fixe $m^*(\beta)$, ainsi
que le comportement du syst\`eme aux temps longs, est contr\^ol\'ee par le 
d\'eveloppement de $v$ autour de $m^*$. En d\'efinissant le temps de
relaxation $\tau$ par $m(t) \sim m^* + C e^{-t/\tau}$, on obtient
\begin{equation}
\tau = \left( R(2m^*,\beta) \left(1+e^{2\beta m^*}(1-2\beta (1-m^*)) \right)
\right)^{-1} \ .
\label{eq:curieweiss-tau}
\end{equation}
En accord avec la stabilit\'e thermodynamique, le point fixe paramagn\'etique 
$m^*=0$ est dynamiquement stable \`a haute temp\'erature ($\tau>0$ pour 
$\beta<\beta_c$), et instable \`a basse temp\'erature ($\tau<0$).
Dans ce dernier cas ce sont les solutions ferromagn\'etiques $\pm m^* \ne 0$
qui attirent la dynamique aux temps longs.

La divergence du temps de relaxation quand on s'approche de $\beta_c$ dans
la phase paramagn\'etique s'obtient facilement \`a partir de cette expression,
\begin{equation}
\tau_p \underset{\beta \to \beta_c^-}{\sim} (\beta_c-\beta)^{-1} 
\frac{1}{2 R(0,\beta_c)} \ .
\end{equation}
Dans la phase ferromagn\'etique, il faut d\'evelopper (\ref{eq:curieweiss-tau})
autour de $m^*(\beta) \sim \sqrt{3(\beta-\beta_c)}$, on trouve alors
\begin{equation}
\tau_f \underset{\beta \to \beta_c^+}{\sim} (\beta-\beta_c)^{-1} 
\frac{1}{4 R(0,\beta_c)} \ .
\end{equation}
L'amplitude de ces divergences d\'epend de la dynamique microscopique
par l'interm\'ediaire de $R(0,\beta_c)$, mais le rapport des deux
amplitudes est universel, il vaut $1/2$ quelque soit la dynamique 
microscopique.

Exactement \`a $\beta=\beta_c$, le premier ordre du d\'eveloppement de $v$
en puissances de $m$ s'annule, le deuxi\`eme est nul pour des raisons de 
sym\'etrie par renversement du signe des spins, c'est donc le troisi\`eme ordre
qui est pertinent. On a aux temps longs $\dot{m} \sim m^3$, ce qui implique 
que la magn\'etisation (si elle \'etait non nulle dans l'\'etat initial)
s'annule comme $t^{-1/2}$. Comme l'\'energie est le carr\'e de la
magn\'etisation, elle va vers sa valeur d'\'equilibre comme
$t^{-1}$.

\begin{figure}
\includegraphics{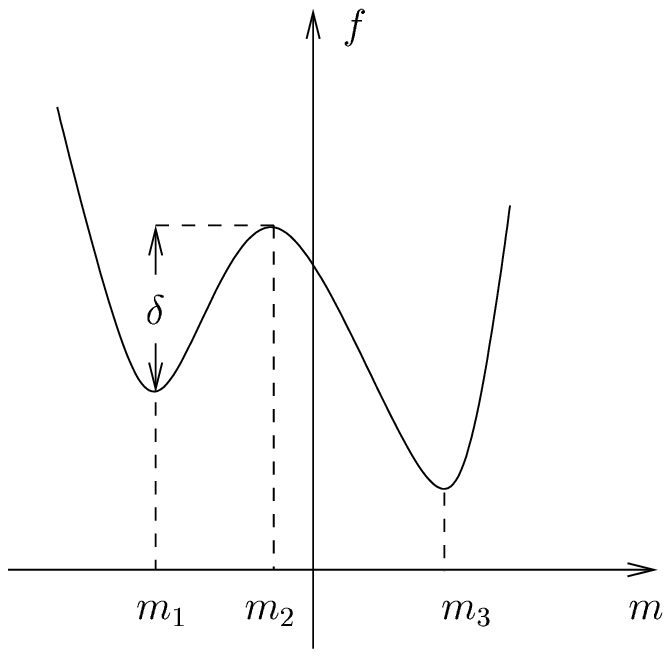}
\caption{L'allure de l'\'energie libre en pr\'esence d'un ph\'enom\`ene de
m\'etastabilit\'e.}
\label{fig:curieweiss-f}
\end{figure}

Il reste un dernier aspect de la dynamique \`a traiter. Supposons que
l'on soit en pr\'esence d'un champ ext\'erieur $h>0$, et que la temp\'erature 
soit suffisamment basse, de telle sorte que l'\'energie libre $f(m)$ ait 
l'allure sch\'ematis\'ee sur la figure \ref{fig:curieweiss-f}. Si la
magn\'etisation initiale $m(t=0)=m_0$ est plus petite que $m_2$, le calcul du
comportement typique pr\'esent\'e ci-dessus conduit \`a la conclusion~:
$m(t) \to m_1$ quand $t \to \infty$. Autrement dit le syst\`eme reste
bloqu\'e dans l'\'etat m\'etastable de magn\'etisation $m_1$. Il est
clair que l'on arrive \`a ce r\'esultat parce que la limite thermodynamique 
$N \to \infty$ a \'et\'e prise avant la limite de temps longs $t \to \infty$.
Pour un syst\`eme de taille $N$ finie, la condition de balance d\'etaill\'ee
implique qu'aux temps suffisamment longs, les configurations sont 
\'echantillonn\'ees par la dynamique selon la distribution d'\'equilibre de
Gibbs-Boltzmann, et donc la magn\'etisation typique vaut $m_3$, quelque soit
la condition initiale. Comme le temps de relaxation du processus de Markov
diverge avec la taille du syst\`eme, quand on prend la limite thermodynamique
avant celle des temps longs l'ergodicit\'e est bris\'ee et le syst\`eme reste
confin\'e dans l'\'etat m\'etastable.

On peut \^etre plus pr\'ecis et calculer la divergence de ce temps
d'ergodicit\'e dans la limite thermodynamique. Il faut pour cela \'etudier
la fonction de grande d\'eviation $\pi(m,t)$. Sa transform\'ee de Legendre
$g(\lambda,t)$ \'evolue selon l'\'equation (\ref{eq:pml-eqsurg}), o\`u
\begin{equation}
\ell(\lambda,m)= \ln \left [1 + \frac{1-m}{2} R(-2(m+h),\beta) (e^{2\lambda}-1)
+ \frac{1+m}{2} R(2(m+h),\beta) (e^{-2\lambda}-1) \right] \ .
\end{equation}
En utilisant la condition de balance d\'etaill\'ee, on peut montrer
que la solution stationnaire atteinte aux temps longs, $g_{\rm as}(\lambda)
= g(\lambda,t \to \infty)$ v\'erifie
\begin{equation}
g_{\rm as}'(\lambda) = \tanh ( \beta (g_{\rm as}'(\lambda) + h) + \lambda ) \ .
\end{equation}
En int\'egrant cette \'equation avec la condition $g_{\rm as}'(0)=m_1$, on
obtient finalement 
\begin{equation}
\pi_{\rm as}(m) = \beta f(m) - \beta f(m_1) \ , \qquad \mbox{pour} \ \ 
m \le m_2 \ .
\end{equation}
Les grandes d\'eviations autour de la magn\'etisation typique m\'etastable sont
contr\^ol\'ees par l'\'energie libre gr\^ace \`a la condition de balance
d\'etaill\'ee. En particulier la probabilit\'e d'une grande d\'eviation
jusqu'au maximum $m_2$ de l'\'energie libre est exponentiellement petite,
en $\exp[-N \beta \delta ]$. Le temps de vie de l'\'etat m\'etastable va
diverger de mani\`ere inversement proportionnelle \`a cette probabilit\'e,
\begin{equation}
t_{\rm erg} \sim \exp [N \beta \delta ] \ .
\end{equation}
A nouveau le signe d'\'equivalence n'est pas utilis\'e dans son sens
math\'ematique, ce r\'esultat est affect\'e d'un pr\'efacteur polynomial en 
$N$, calcul\'e dans~\cite{GrWeLa}.

Ce r\'esultat n'est \'evidemment pas surprenant, puisqu'il correspond au temps
d'activation d'Arrh\'enius pour passer une barri\`ere d'\'energie libre de
hauteur $N \delta$. C'est cependant un des cas, assez rares, o\`u l'on peut
le calculer explicitement.
Un autre exemple o\`u ce calcul est possible est celui du probl\`eme de
Kramers pour une particule brownienne dans un potentiel m\'etastable. Dans
ce dernier cas la divergence des temps prend place \`a la limite de basse
temp\'erature, et non \`a celle thermodynamique comme dans le mod\`ele de
champ moyen consid\'er\'e ici.

\newpage

\section{Le ferromagn\'etique \`a connectivit\'e fixe}
\label{sec:ferro-arbre}

Le mod\`ele de Curie-Weiss que l'on vient d'\'etudier est l'arch\'etype 
des mod\`eles ferromagn\'etiques de type champ moyen.
Le fait qu'il soit compl\`etement connect\'e, c'est \`a dire que chacun
des spins interagisse avec tous les autres, rend sa description tr\`es 
simple, puisque toutes les configurations microscopiques correspondant 
au m\^eme param\`etre d'ordre macroscopique (la magn\'etisation) sont 
\'equivalentes. 

On va consid\'erer dans cette partie un autre mod\`ele, lui aussi 
ferromagn\'etique et de type champ moyen, mais cette fois-ci il sera dilu\'e. 
La connectivit\'e finie
rend le probl\`eme plus difficile, et l'on n'obtiendra pas une solution exacte
de l'\'evolution dynamique. On va donc s'efforcer de d\'evelopper une s\'erie 
d'approximations de plus en plus pr\'ecises. Ce travail a fait l'objet de
la publication \pubbethe .

\subsection{D\'efinition du mod\`ele, propri\'et\'es thermodynamiques}

Le mod\`ele en question est un syst\`eme de $N$ spins d'Ising $\sigma_i$, 
interagissant par des couplages ferromagn\'etiques sur un arbre de Bethe 
de connectivit\'e $L$. Comme il a \'et\'e discut\'e
dans le chapitre \ref{sec:gr-autres}, arbre de Bethe signifie dans ce contexte
graphe al\'eatoire r\'egulier. Autrement dit, le graphe d'interaction est 
localement un arbre o\`u chaque variable a $L$ voisines, 
mais il y a des boucles de longueur ${\cal O}(\log N)$ dans le syst\`eme. 
Ainsi chaque site est statistiquement \'equivalent, et on \'evite le 
probl\`eme des effets de bords puisque le graphe n'a pas de surface.

On prend pour hamiltonien 
\begin{equation}
  H = -\frac 12 \sum_{i<j} J_{ij} (\sigma_i \sigma_j-1) 
= \sum_{i<j} J_{ij} \delta_{\sigma_i,-\sigma_j} \ .
\end{equation}
$H$ est donc un entier positif qui compte le nombre d'interactions
frustr\'ees, c'est \`a dire le nombre de liens dans le graphe dont les 
extr\'emit\'es portent des spins de signe oppos\'es. 
On notera aussi $E=H$ l'\'energie d'une configuration, $M =\sum_i \sigma_i$
la magn\'etisation totale, et $e=E/N$, $m=M/N$ les quantit\'es par spin
correspondantes.

\begin{figure}[h]
\includegraphics[width=10cm]{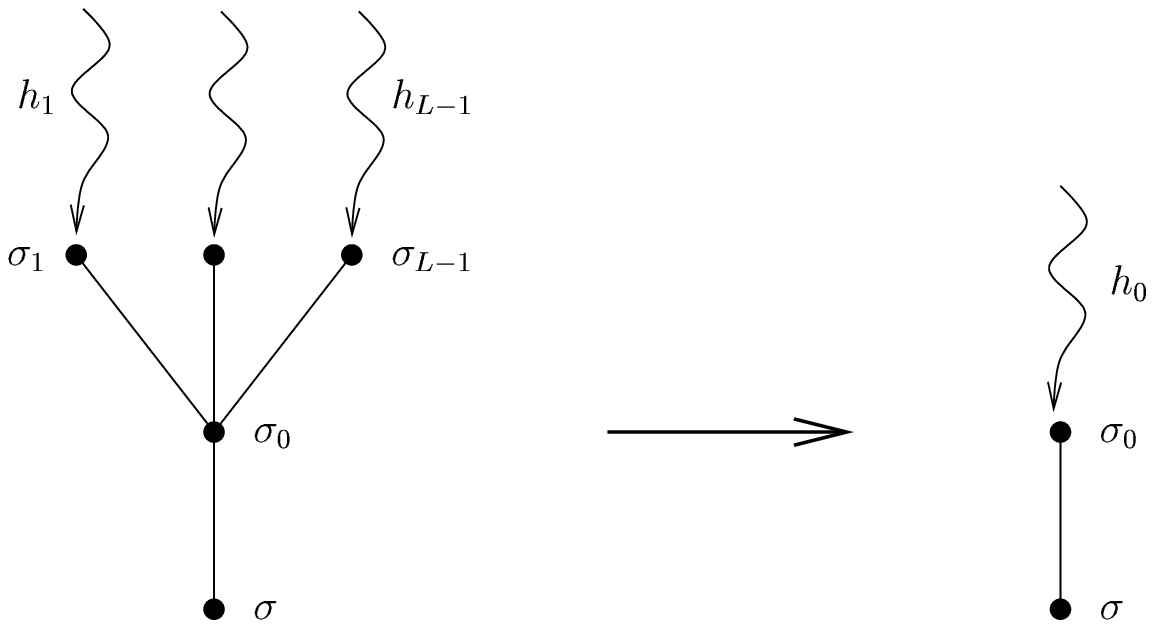}
\caption{Sch\'ematisation de l'\'equation de r\'ecurrence 
(\ref{eq:bethe-recu}).}
\label{fig:bethe-cav}
\end{figure}

D\'eterminons les propri\'et\'es thermodynamiques du mod\`ele par la m\'ethode
de la cavit\'e, utilis\'ee ici dans sa version la plus \'el\'ementaire dite
sym\'etrique dans les r\'epliques. On se reportera \`a la figure 
\ref{fig:bethe-cav} pour une illustration des d\'efinitions. 

\begin{figure}
\includegraphics[width=7cm]{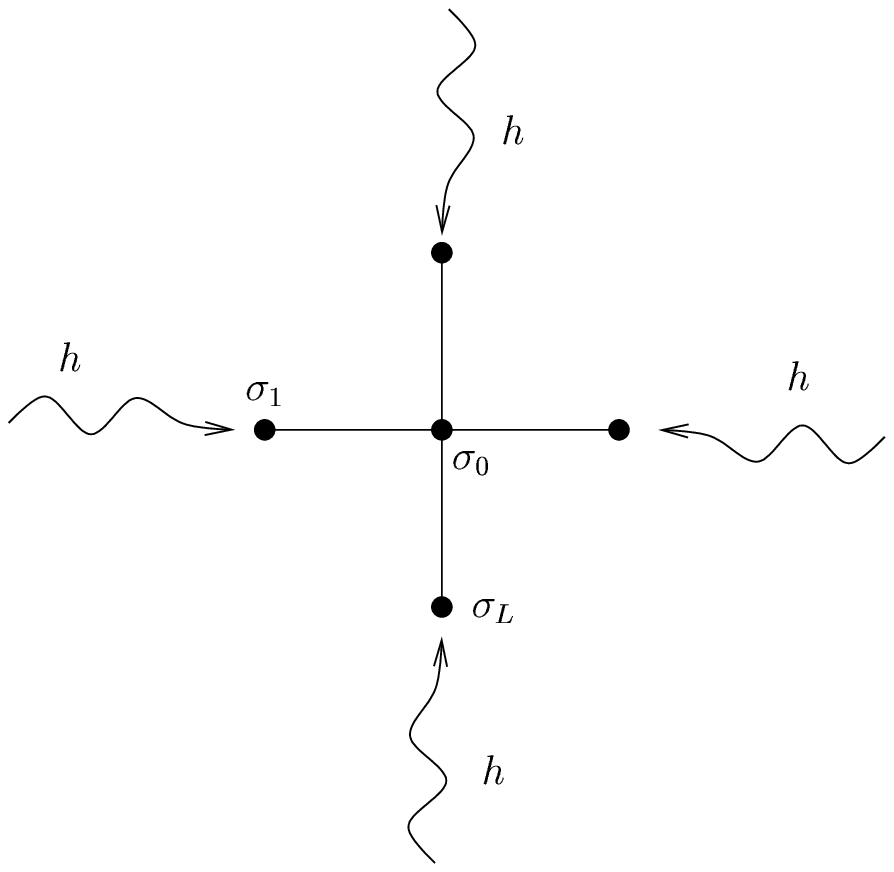}
\caption{Calcul des quantit\'es thermodynamiques par la m\'ethode 
de la cavit\'e.}
\label{fig:bethe-cav2}
\end{figure}

Supposons pour commencer que le graphe d'interactions soit un arbre. On prend
pour site central la variable not\'ee $\sigma_0$, ses $L$ voisins sont 
s\'epar\'ees en un site $\sigma$ qui repr\'esente l'\og aval\fg\ du graphe, 
et $L-1$ sites $\{\sigma_1,\dots,\sigma_{L-1}\}$ en \og amont\fg\ . 
La loi de probabilit\'e d'une configuration microscopique des
$N$ variables $\vec{\sigma}$ est donn\'ee par $P(\vec{\sigma})=
\exp[-\beta H(\vec{\sigma})]/Z$. Le champ de cavit\'e $h_i$ s'exer\c cant 
sur la variable $\sigma_i$ est d\'efini de la mani\`ere suivante. 
Lorsque l'on somme la loi $P(\vec{\sigma})$ sur toutes les variables de la 
branche du graphe en amont de $\sigma_i$, la probabilit\'e marginale 
r\'esultant de cette sommation partielle a une d\'ependance en $\sigma_i$
de la forme
\begin{equation}
{\rm cste} \times e^{-\beta a(\sigma_0,\sigma_i) + \beta h_i \sigma_i} \ , 
\label{eq:bethe-defhcav}
\end{equation}
o\`u $a(\sigma,\sigma') = \delta_{\sigma,-\sigma'}$ est l'\'energie du
lien entre deux spins. L'\'equation (\ref{eq:bethe-defhcav}) constitue
une d\'efinition du champ de cavit\'e $h_i$. On peut maintenant,
en faisant la trace sur les variables $\{\sigma_1,\dots,\sigma_{L-1}\}$,
\'ecrire une \'equation de r\'ecurrence donnant le champ de cavit\'e
$h_0$ en fonction des champs $\{h_1,\dots,h_{L-1}\}$~:
\begin{equation}
e^{2\beta h_0} = \prod_{i=1}^{L-1} \left( \frac{
e^{\beta h_i - \beta a(+,+)} + e^{-\beta h_i - \beta a(+,-)}}{
e^{\beta h_i - \beta a(-,+)} + e^{-\beta h_i - \beta a(-,-)}} \right) \ .
\label{eq:bethe-recu}
\end{equation}
Une fois cette r\'ecurrence \'etablie, on l'applique au graphe al\'eatoire
r\'egulier pour lequel tous les sites sont \'equivalents, et l'on cherche donc
une solution homog\`ene de cette \'equation,
\begin{equation}
  h = \frac{L-1}{2\beta} \ln\left( 
    \frac{e^{\beta h}+e^{-\beta(h+1)}}{e^{\beta(h-1)}+e^{-\beta h}}
  \right)\ .
\label{eq:bethe-hcav}
\end{equation}
Il peut sembler contradictoire d'utiliser ici l'\'equation de r\'ecurrence 
(\ref{eq:bethe-recu}), qui n'est strictement valable que pour un arbre, 
alors que le graphe al\'eatoire poss\`ede des boucles. On peut cependant
esp\'erer que cette approche soit rendue l\'egitime par la divergence de la 
longueur des boucles dans la limite thermodynamique.

On voit facilement que (\ref{eq:bethe-hcav}) ne poss\`ede \`a haute
temp\'erature que la solution triviale $h=0$, alors que
deux solutions $\pm h^*$ se d\'eveloppent continuement en dessous d'une 
temp\'erature critique $T_c$, dont l'inverse vaut $\beta_c= \ln (L/(L-2))$. 

Il reste \`a calculer les grandeurs thermodynamiques du probl\`eme. 
Consid\'erons \`a cet effet la situation sch\'ematis\'ee sur la figure 
\ref{fig:bethe-cav2}, o\`u le spin $\sigma_0$ est entour\'ee de ses
$L$ voisins $\sigma_i$, l'influence du reste du graphe ayant \'et\'e
incorpor\'ee dans les $L$ champs de cavit\'e $h$, solution de
l'\'equation (\ref{eq:bethe-hcav}). La loi de probabilit\'e
de ces $L+1$ variables est, \`a une normalisation pr\`es,
\begin{equation}
\prod_{i=1}^L e^{-\beta a(\sigma_0,\sigma_i) + \beta h \sigma_i } \ .
\label{eq:bethe-cavv}
\end{equation}
De cette forme on peut tirer la magn\'etisation par site
\begin{equation}
m =  \tanh\left( \beta \frac L{L-1} h \right) \ ,
\end{equation}
ainsi que l'\'energie par site,
\begin{equation}
  e = Lh \tanh\left( \beta \frac L{L-1} h \right) - \frac L2\  \frac
  {2h\sinh(2\beta h) -e^{-\beta}}{\cosh(2\beta h) + e^{-\beta}} \ .
\end{equation}
L'apparition d'une solution $h^*$ non triviale correspond donc \`a une
transition ferromagn\'etique. Notons
au passage que la magn\'etisation s'annule \`a la temp\'erature critique
en racine carr\'ee de $\beta - \beta_c$, ce qui redonne bien s\^ur l'exposant
critique de champ moyen.

Il nous sera utile pour la suite de conna\^itre la probabilit\'e
$p_{\sigma}(u)$ qu'un site ait spin $\sigma$ et soit entour\'e de
$u$ liens frustr\'es. A l'\'equilibre, l'expression (\ref{eq:bethe-cavv})
conduit \`a~:
\begin{equation}
p_{\sigma}(u) = {\cal N} {L \choose u} e^{-\beta u} e^{\beta h (L-2u) \sigma}
\ ,
\end{equation}
o\`u $\cal N$ est une constante de normalisation. On peut transformer
cette expression en utilisant les densit\'es d'\'energie et de magn\'etisation
d'\'equilibre, ce qui conduit \`a
\begin{equation}
p_{\sigma}(u)=\frac{1+\sigma m}{2} {L \choose u} \left(\frac{2e}
{L(1+\sigma m)}\right)^u \left(1-\frac{2e}{L(1+\sigma m)}\right)^{L-u}
\ .
\label{eq:bethe-psigeq}
\end{equation}

\subsection{Les descriptions approch\'ees de la dynamique}
\label{sec:bethe-approx}
On va \'etudier la m\^eme dynamique que celle utilis\'ee sur le mod\`ele de
Curie-Weiss~: \`a chaque pas de temps un des spins est s\'electionn\'e
au hasard, puis renvers\'e avec une probabilit\'e d\'ependant de la variation
d'\'energie que ce renversement induit dans le syst\`eme. On s'int\'eressera
uniquement \`a des taux de transition qui v\'erifient la condition de balance 
d\'etaill\'ee, de mani\`ere \`a assurer la convergence aux temps longs
vers l'\'equilibre thermodynamique.

\vspace{8mm}

\noindent \underline{\emph{Premier niveau~: l'approximation binomiale}}

\vspace{4mm}

Comme ce sont les deux param\`etres qui permettent de d\'ecrire compl\`etement
l'\'etat d'\'equilibre, il est naturel d'essayer de b\^atir une description
dynamique en termes de l'\'energie et de la magn\'etisation. Suivant les
id\'ees g\'en\'erales d\'ecrites pr\'ec\'edemment, il faudrait pour cela 
\^etre capable de d\'eterminer les probabilit\'es de variation de ces deux
grandeurs au cours d'un pas de temps \'el\'ementaire. Supposons
connue la probabilit\'e $p_{\sigma}(u;t)$ que la variable choisie
au cours du pas $T=Nt \to T+1$ porte un spin $\sigma=\pm 1$ et soit 
entour\'ee de $u$ liens frustr\'es, c'est-\`a-dire que $u$ parmi les 
$L$ variables voisines portent un spin $-\sigma$. Alors la variation 
d'\'energie si le renversement de la variable est effectu\'e vaut 
$\Delta E= L-2u$ (les liens frustr\'es deviennent satisfaits et vice versa), 
et la variation de la magn\'etisation totale est $\Delta M = - 2 \sigma$. 
Dans cette partie $W(u,\beta)=R(L-2u,\beta)$ d\'esignera la probabilit\'e que 
le flip (renversement) soit accept\'e. La condition de balance d\'etaill\'ee
s'exprime donc comme
\begin{equation}
W(u,\beta) = W(L-u,\beta) e^{-\beta(L-2u)} \ .
\end{equation}
Dans la limite thermodynamique le temps devient continu et l'on obtient les 
\'equations diff\'erentielles qui r\'egissent l'\'evolution 
des moyennes de l'\'energie et de la magn\'etisation~:
\begin{eqnarray}
\frac{d}{dt} e(t) &=& 
\sum_{u=0}^L W(u,\beta) (L-2u) [p_{-}(u;t) + p_{+}(u;t)]\ , \nonumber
\\
\frac{d}{dt} m(t) &=& 
2 \sum_{u=0}^L W(u,\beta) [p_{-}(u;t) - p_{+}(u;t)] \ .
\label{eq:bethe-evbino}
\end{eqnarray}
A nouveau les trajectoires typiques 
seront proches de ces valeurs moyennes dans la limite thermodynamique.

Les \'equations d'\'evolution de $e(t)$ et $m(t)$ font intervenir la fonction
$p_{\sigma}(u,t)$, qui a priori contient plus d'informations. Cela n'est
pas surprenant au vu de la discussion du d\'ebut du chapitre sur les 
propri\'et\'es des dynamiques projet\'ees. Il va donc falloir faire une
approximation pour fermer les \'equations (\ref{eq:bethe-evbino}) en termes 
de $e(t)$ et $m(t)$ seulement. Faisons le raisonnement suivant~: par 
d\'efinition de la magn\'etisation, la variable choisi al\'eatoirement aura 
spin $\sigma$ avec probabilit\'e $(1+ m(t) \sigma)/2$. A ce niveau de 
description, on peut seulement supposer que les $L$ liens autour de la 
variable en question sont frustr\'es avec la m\^eme probabilit\'e 
$\alpha_\sigma(t)$. Ceci conduit donc \`a une loi binomiale (d'o\`u le nom
de l'approximation),
\begin{equation}
p_{\sigma}(u;t) =
\frac{1+\sigma m(t)}{2} {L \choose u} \alpha_\sigma (t)^u 
(1 - \alpha_\sigma (t))^{L-u} \ .
\end{equation}
Reste \`a d\'eterminer $\alpha_\sigma$. Pour cela, remarquons que chaque
lien frustr\'e relie un spin $+1$ et un spin $-1$. L'\'energie doit donc 
s'exprimer, de mani\`ere coh\'erente, comme le nombre de variables $-1$ 
autour de celles $+1$, et vice versa. Autrement dit,
\begin{eqnarray}
e(t) &=& \sum_{u=0}^L u \ p_+ (u;t) = \frac{1+m(t)}{2} L \alpha_+(t) \\
&=& \sum_{u=0}^L u \ p_- (u;t) = \frac{1-m(t)}{2} L \alpha_-(t) \ .
\end{eqnarray}
Cette \'equation d\'etermine $\alpha_\sigma(t)$ en fonction de $e(t)$ et 
$m(t)$, on obtient donc finalement l'expression approch\'ee de 
$p_{\sigma}(u;t)$ comme
\begin{equation}
p_{\sigma}(u;t)=\frac{1+\sigma m(t)}{2} {L \choose u} \left(\frac{2e(t)}
{L(1+\sigma m(t))}\right)^u \left(1-\frac{2e(t)}{L(1+\sigma m(t))}\right)^{L-u}
\ .
\label{eq:bethe-psigbin}
\end{equation}
Cette forme est en fait exacte \`a l'\'equilibre (cf. (\ref{eq:bethe-psigeq})),
l'approximation consiste \`a supposer qu'elle reste vraie en rempla\c cant
$e$ et $m$ par leurs valeurs $e(t)$ et $m(t)$, en g\'en\'eral diff\'erentes
de celles d'\'equilibre.

Une fois admise cette forme approch\'ee de $p_{\sigma}(u;t)$, les \'equations
sur $e$ et $m$ se ferment. On s'est ramen\'e \`a un syst\`eme de deux 
\'equations diff\'erentielles ordinaires, du premier ordre en temps, 
coupl\'ees. On discutera dans la section suivante leur comportement et 
la pr\'ecision d'une telle approximation. Notons au passage que ce niveau
d'approximation devrait devenir exact dans la limite $L\to \infty$, en
prenant des intensit\'es de couplage entre spins d'ordre $1/L$~: on retrouve
dans cette limite le mod\`ele de Curie-Weiss, pour lequel l'\'evolution de
la magn\'etisation est markovienne.

R\'esumons ce que l'on vient de faire~: partant d'une description minimale
de l'\'etat du syst\`eme par deux observables macroscopiques 
($e(t)$ et $m(t)$), on a eu besoin d'une nouvelle quantit\'e macroscopique 
($p_\sigma(u;t)$) pour \'ecrire les \'equations d'\'evolution exactes de nos
observables de d\'epart. Une approximation a alors \'et\'e n\'ecessaire pour
fermer les \'equations sur $e$ et $m$ seulement. Une autre possibilit\'e 
consiste \`a \'ecrire une \'equation d'\'evolution pour $p_\sigma(u;t)$
elle-m\^eme, qui va faire intervenir une nouvelle observable contenant plus
de d\'etails sur l'\'etat microscopique du syst\`eme, et l'on devra \`a nouveau
faire une approximation pour fermer l'\'equation sur $p_\sigma(u;t)$, ou
bien \'ecrire une \'equation pour la nouvelle observable... A chaque 
it\'eration de ce processus apparaissent des observables de plus en plus 
pr\'ecises mais qui ob\'eissent \`a des \'equations de plus en plus 
compliqu\'ees, et que l'on doit de toute fa\c con couper de mani\`ere 
approch\'ee \`a un certain stade, \`a moins de vouloir suivre l'\'evolution
microscopique originelle, ce que l'on cherche \`a \'eviter depuis le d\'ebut.
Cette situation rappelle bien s\^ur un grand nombre de probl\`emes
physiques o\`u apparaissent des hi\'erarchies de fonctions que l'on doit
tronquer \`a un certain ordre, notamment la 
hi\'erarchie dite BBGKY (d'apr\`es Bogoliubov, Born, Green, Kirkwood et Yvon)
pour la dynamique des syst\`emes de particules en interaction.

Dans la publication \pubbethe\ nous avons explor\'e les deux \'etapes 
suivant l'approximation binomiale dans cette hi\'erarchie. Je ne discuterai 
ici, par souci de simplicit\'e, que la premi\`ere des deux.

On utilisera dans la suite la notation 
\begin{equation}
\langle \bullet \rangle_\sigma = \sum_{u=0}^L \bullet \ p_\sigma(u;t) \ .
\end{equation}
Contrairement aux apparences, ce n'est pas une moyenne normalis\'ee pour
une valeur donn\'ee de $\sigma$, mais seulement quand on ajoute les
deux valeurs possibles, $\langle 1 \rangle_+ + \langle 1 \rangle_-=1$.

\vspace{8mm}

\noindent \underline{\emph{Deuxi\`eme niveau~: l'approximation des
voisins ind\'ependants}}

\vspace{4mm}

On veut donc \'ecrire une \'equation d'\'evolution pour $p_\sigma(u;t)$.
Au cours d'un pas de temps $t \to t+(1/N)$, ces quantit\'es \'evoluent 
de deux mani\`eres
diff\'erentes. Tout d'abord, supposons que la variable s\'electionn\'ee pour
un renversement potentiel porte un
spin $\tilde \sigma$ et soit entour\'ee de $\tilde u$ liens frustr\'es. Si
elle est renvers\'ee, elle devient de type $(-\tilde{\sigma},L-\tilde{u})$.
On a donc $p_{\tilde{\sigma}}(\tilde{u})$ qui augmente de $1/N$, et 
$p_{-\tilde{\sigma}}(L-\tilde{u})$ qui diminue de $1/N$. Ce n'est pas la
seule contribution~: les $L$ variables voisines de la variable flipp\'ee
ont leur spin qui reste inchang\'e, mais le nombre de liens frustr\'es autour
d'elles augmente ou diminue de $1$. 

Suivons d'abord un des $\tilde u$ liens
frustr\'es autour de la variable flipp\'ee. La variable ainsi atteinte a
n\'ecessairement un spin $-\tilde \sigma$. Quel est le nombre $u'$ de liens
frustr\'es autour de cette nouvelle variable~? On doit faire ici une 
approximation pour fermer l'\'equation sur $p_\sigma(u;t)$ seulement. Elle
consiste \`a supposer que la variable ainsi atteinte est de type $u'$ avec
une probabilit\'e proportionnelle \`a $u' p_{-\tilde{\sigma}}(u';t)$,
donc par normalisation cette probabilit\'e vaut $u' p_{-\tilde{\sigma}}(u';t) /
\langle u \rangle_{-\tilde{\sigma}}$. Si l'on prenait un lien frustr\'e
au hasard dans le graphe, c'est avec cette probabilit\'e que la variable
de spin $-\tilde{\sigma}$ aurait $u'$ liens frustr\'es autour d'elle.
L'approximation consiste donc \`a oublier les corr\'elations entre $\tilde{u}$
et $u'$. La variable que l'on 
atteint ainsi voit le nombre de liens frustr\'es autour d'elle diminuer de
1 au cours du renversement, ce nombre passe donc de $u'$ \`a $u'-1$.
En faisant le m\^eme raisonnement pour les $L-\tilde{u}$ liens non frustr\'es
autour de la variable flipp\'e, on obtient
\begin{eqnarray}
\frac{d}{dt}p_\sigma(u;t) &=& 
- W(u,\beta) p_\sigma(u;t) 
+ W(L-u,\beta) p_{-\sigma}(L-u;t)  \\
&+& \sum_{\tilde{u}, \tilde{\sigma}} p_{\tilde \sigma}(\tilde{u},t) 
W(\tilde{u},\beta)
\left[ \tilde{u} \left(\sum_{u'} \frac{u' p_{-\tilde \sigma}(u',t)}
{\langle u \rangle_{-\tilde{\sigma}}} ( \delta_{u',u+1} - \delta_{u',u})
\delta_{\sigma,-\tilde{\sigma}}\right)
\right. \nonumber
\\
&& \hspace{2cm} +
\left. (L-\tilde{u}) \left(\sum_{u'} \frac{(L-u') p_{\tilde \sigma}(u',t)}
{\langle L-u \rangle_{\tilde{\sigma}}} (\delta_{u',u-1} - \delta_{u',u})
\delta_{\sigma,\tilde{\sigma}}\right)
\right] 
\nonumber
\end{eqnarray}
On peut effectuer les sommes, ce qui conduit \`a
\begin{eqnarray}
  \frac d{dt} p_\sigma (u;t) 
  &=& - W(u,\beta)\ p_\sigma(u;t) + W(L-u,\beta)\ p_{-\sigma}(L-u;t)
  \nonumber\\
  &&+\frac{\langle(L-\tilde u)\ W(\tilde u,\beta)\rangle_\sigma}
  {\langle L-\tilde{u} \rangle_\sigma}  
  \left[- (L-u)\ p_\sigma(u;t) + (L-u+1) p_\sigma(u-1;t) \right]
  \nonumber\\ 
  &&+\frac{\langle \tilde u \ W(\tilde u,\beta)\rangle_{-\sigma}}
  {\langle \tilde u \rangle_\sigma}  
  \left[- u\ p_\sigma(u;t) + (u+1) \ p_\sigma(u+1;t) \right]  
\label{eq:bethe-approx2}
\end{eqnarray}
On a donc obtenu, de mani\`ere approch\'ee, un jeu d'\'equations 
diff\'erentielles coupl\'ees pour ces fonctions $p_\sigma(u;t)$. Comme
ici on travaille sur un graphe dont la connectivit\'e est fixe, $u$ ne
peut prendre qu'un nombre fini de valeurs.
L'expression (\ref{eq:bethe-approx2}) r\'esume donc un jeu de $2(L+1)$ 
\'equations coupl\'ees.

Une fois ces \'equations r\'esolues, on peut en d\'eduire 
l'\'energie et la magn\'etisation du syst\`eme, avec 
$m(t)=\langle 1 \rangle_+ - \langle 1 \rangle_-$, et 
$e(t)=\langle u \rangle_+ = \langle u \rangle_-$. 
Notons que la derni\`ere \'equation impose une condition de coh\'erence sur
les fonctions $p_\sigma (u;t)$. On peut v\'erifier que si cette
condition est respect\'ee au temps initial, elle sera conserv\'ee
par les \'equations (\ref{eq:bethe-approx2}).

\vspace{8mm}

\noindent \underline{\emph{Interpr\'etation des hypoth\`eses de
fermeture}}

\vspace{4mm}

Les raisonnement pr\'esent\'es ci-dessus pour fermer approximativement
les \'equations dynamiques peuvent sembler assez vagues. On va
pr\'eciser ici les hypoth\`eses faites implicitement. Si l'on raisonne
en termes d'op\'erateurs de projection, les diff\'erents niveaux de
troncature correspondent chacun \`a une approximation markovienne, o\`u
l'observable sur laquelle on projette est $(e,m)$ pour l'approximation
binomiale, $p_{\sigma}(u)$ pour la suivante, et ainsi de suite. En effet,
on a indiqu\'e pr\'ec\'edemment que l'approximation markovienne consiste
\`a supposer que toutes les configurations microscopiques avec la m\^eme 
valeur de l'observable macroscopique sont \'equiprobables. Dans la publication
\pubbethe\ on a montr\'e explicitement que les
fermetures d\'eriv\'ees heuristiquement peuvent \'egalement s'obtenir 
\`a partir de cette hypoth\`ese d'\'equiprobabilit\'e des configurations 
microscopiques. 

Cette v\'erification permet aussi de comprendre la profonde similitude
de cette approche avec la th\'eorie dynamique des r\'epliques (DRT) de
Coolen et Sherrington~\cite{DRT1,DRT2}. La DRT a \'et\'e appliqu\'ee au
calcul des propri\'et\'es dynamiques du mod\`ele de Sherrington-Kirkpatrick,
en faisant aussi des hypoth\`eses d'\'equipartition des probabilit\'es
microscopiques sur les sous-ensembles d\'efinies par des observables
macroscopiques. Dans un premier niveau d'approximation~\cite{DRT1} la 
projection de la dynamique se fait sur la magn\'etisation et l'\'energie, 
puis une version plus raffin\'ee~\cite{DRT2} a \'et\'e d\'evelopp\'ee, dans
laquelle le param\`etre d'ordre est fonctionnel. Dans ces articles la m\'ethode
des r\'epliques \'etait utilis\'ee pour faire des moyennes sur le d\'esordre
avec l'hypoth\`ese d'\'equiprobabilit\'e microscopique, et conduisait \`a des
calculs relativement lourds. De mani\`ere assez paradoxale, l'\'etude des 
mod\`eles dilu\'es par cette m\'ethode s'av\`ere plus simple que pour
des mod\`eles compl\`etement connect\'es~:
comme on l'a vu, on peut d\'eriver les relations de fermeture par des
raisonnements combinatoires, et l'utilisation de la m\'ethode de la cavit\'e
\`a la place de celle des r\'epliques simplifie les v\'erifications explicites.

\subsection{R\'esultats}

A chacun des niveaux successifs de cette hi\'erarchie d'approximation,
on se ram\`ene \`a un jeu d'\'equations diff\'erentielles coupl\'ees 
pour un nombre fini d'observables macroscopiques~: au premier niveau
on suit seulement l'\'energie et la magn\'etisation, au deuxi\`eme niveau
la fonction $p_\sigma(u;t)$, au troisi\`eme (non d\'ecrit ici
mais que l'on trouvera expos\'e dans la publication) un objet un peu plus
complexe $p_{\sigma_1 \sigma_2} (u_1,u_2)$. Ces \'equations diff\'erentielles
peuvent \^etre sans difficult\'e int\'egr\'ees num\'eriquement, et leurs
pr\'edictions compar\'ees aux r\'esultats de simulations num\'eriques 
Monte-Carlo de la dynamique microscopique originelle. On s'est int\'eress\'e
en particulier \`a la relaxation vers l'\'equilibre \`a partir de conditions
initiales qui en sont arbitrairement \'eloign\'ees. Une mani\`ere
simple de les g\'en\'erer consiste \`a tirer la valeur des spins
al\'eatoirement et de mani\`ere ind\'ependante, avec un biais 
pour donner au syst\`eme une certaine magn\'etisation.

\begin{figure}
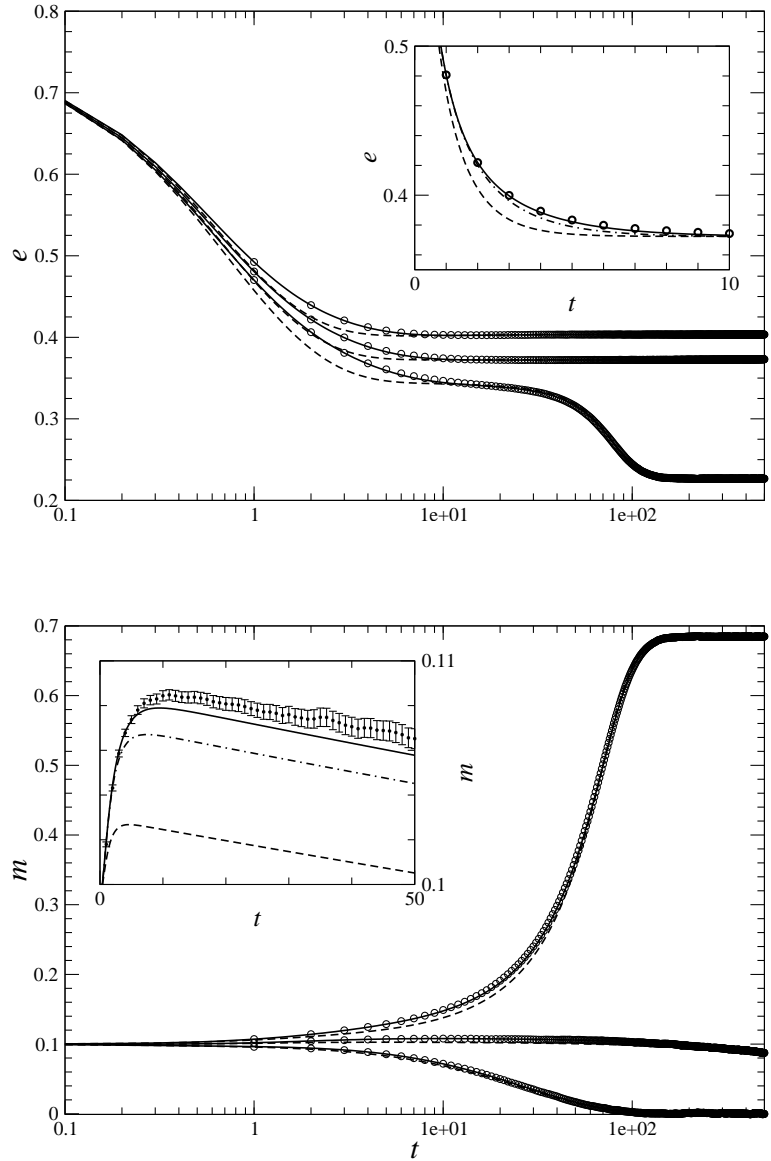

\begin{center}
\includegraphics[width=11cm]{bethe-ene.eps}
\vspace{1cm}
\includegraphics[width=11cm]{bethe-magne.eps}

\caption{
Les trois niveaux d'approximation sont repr\'esent\'es
par ordre croissant de pr\'ecision avec une ligne pointill\'ee, une
ligne mixte et une ligne pleine. Dans les figures centrales la ligne mixte
est indistinguable de la ligne pleine.
Les symboles sont les r\'esultats
de simulations num\'eriques sur des syst\`emes avec $L=3$, de taille
$N=3\cdot10^6$, des moyennes \'etant prises sur $200$ simulations 
ind\'ependantes.
Les barres d'erreur, quand elles ne sont pas pr\'ecis\'ees, sont
plus petites que les symboles. La condition initiale correspond \`a
des spins ind\'ependants de magn\'etisation moyenne 0.1.
Haut~: De bas en haut $\beta=1.2, \ln 3, 1.0$,
la temp\'erature interm\'ediaire \'etant la temp\'erature critique du
mod\`ele, trac\'e de la densit\'e d'\'energie en fonction du temps.
L'inset pr\'esente un grossissement de la zone de moins bon accord, 
\`a la temp\'erature critique. 
Bas~: idem pour la magn\'etisation en fonction du temps,
la courbe la plus haute \'etant celle \`a temp\'erature la plus basse.}
\label{fig:arbre-em}
\end{center}
\end{figure}

L'image g\'en\'erale qui d\'ecoule de la comparaison entre simulations et 
calculs est la suivante~: aux temps courts et aux temps
longs, les diff\'erents niveaux d'approximation sont en fait exacts. La raison
est qu'aux temps courts, pour ce type de conditions initiales,
les hypoth\`eses de fermeture de la hi\'erarchie d'\'equations, qui reposent
sur l'absence de corr\'elations, sont correctes. 
De m\^eme aux temps longs, comme la dynamique microscopique v\'erifie 
la condition de balance
d\'etaill\'ee, le syst\`eme \'evolue vers l'\'equilibre thermodynamique. On
a vu qu'alors la forme de $p_\sigma(u;t)$ suppos\'ee dans l'approximation
binomiale devenait exacte. Il en va de m\^eme pour les hypoth\`eses de
fermeture aux niveaux suivants d'approximation.
Par contre aux temps interm\'ediaires,
il y a comme pr\'evu des d\'eviations syst\'ematiques entre les pr\'edictions
des calculs approch\'es et le comportement observ\'e des simulations
num\'eriques. De mani\`ere satisfaisante, les niveaux d'approximation 
successifs, qui voient cro\^itre la finesse des observables,
fournissent des r\'esultats de plus en plus pr\'ecis.
La figure \ref{fig:arbre-em} pr\'esente un r\'esum\'e de cette \'etude.
L'accord est tr\`es satisfaisant, m\^eme pour les niveaux d'approximations
les plus simples.

Une r\'esolution analytique des \'equations diff\'erentielles coupl\'ees
qui apparaissent dans cette \'etude semble difficile. On peut n\'eanmoins
\'etudier assez simplement leurs solutions stationnaires. En particulier,
il est possible de v\'erifier explicitement que la condition de balance
d\'etaill\'ee implique que ces points fixes de l'\'evolution correspondent
bien aux \'etats d'\'equilibre thermodynamique. Ceci est possible ici
car les hypoth\`eses de fermeture deviennent exactes dans de telles conditions.
On v\'erifie aussi que dans le r\'egime de basse temp\'erature l'\'etat
paramagn\'etique est dynamiquement instable. Finalement, l'\'etude du
voisinage du point critique conduit \`a des r\'esultats tr\`es proches
de ceux du mod\`ele de Curie-Weiss~:

\vspace{2mm}

\begin{itemize}
\item Les temps de relaxation divergent \`a $\beta_c$ comme $A_{\pm} 
|\beta -\beta_c|^{-1}$. Les deux amplitudes $A_{\pm}$ pour la divergence
dans les phases ferromagn\'etique et paramagn\'etique d\'ependent des
d\'etails microscopiques de la dynamique, mais leur ration $A_+/A_-$ en est
ind\'ependant, et vaut \`a nouveau $1/2$.
\item Exactement \`a la temp\'erature critique, la 
magn\'etisation s'annule en $t^{-1/2}$ et l'\'energie tend vers sa
valeur d'\'equilibre comme $t^{-1}$.
\end{itemize}

\vspace{2mm}

Les d\'etails de ces calculs peuvent se trouver dans la publication. La
co\"incidence des exposants et du ratio d'amplitude avec ceux du mod\`ele de
Curie-Weiss \'etait attendue puisque ces deux mod\`eles sont de champ-moyen.
Toutefois il n'\'etait pas \'evident a priori que l'on puisse les expliciter
dans le cas du mod\`ele dilu\'e, pour lequel on n'a pas une description exacte
de la dynamique aux temps interm\'ediaires.

\subsection{Corr\'elations \`a deux temps}
On a donc vu comment caract\'eriser approximativement l'\'etat du syst\`eme
au cours de son \'evolution vers l'\'equilibre par un certain nombre 
d'observables macroscopiques. Cette description n'\'epuise pas
toutes les questions que l'on peut se poser sur la dynamique du syst\`eme.
D\'efinissons par exemple la fonction de corr\'elation \`a deux temps
\begin{equation}
C(t_2,t_1) = \frac{1}{N} \sum_i \sigma_i(t_2) \sigma_i(t_1) \ ,
\end{equation}
qui vaut $1$ si les configurations des spins aux deux temps $t_1$ et $t_2$ 
sont identiques (notamment pour $t_1=t_2$), $0$ si elles sont
compl\`etement d\'ecorrel\'ees. On prendra $t_2 \ge t_1$ dans la suite pour
simplifier les notations. Si on laisse le syst\`eme \'evoluer suffisamment
longtemps avant de commencer les mesures de corr\'elation ($t_1 \to \infty$),
on va obtenir une fonction de corr\'elation \`a l'\'equilibre qui ne d\'epend
que de la diff\'erence de temps entre les deux instants d'observation,
$C_{\rm eq}(\tau)=\underset{t_1 \to \infty}{\lim} C(t_1+\tau,t_1)$. M\^eme
si dans cette limite toutes les quantit\'es thermodynamiques ont atteint 
leurs valeurs d'\'equilibre, et que la fonction de corr\'elation y est 
stationnaire, celle-ci contient une information non triviale sur la vitesse 
\`a laquelle le syst\`eme \og oublie\fg\ le d\'etail microscopique de ses
configurations ant\'erieures.
Selon la temp\'erature, la fonction de corr\'elation 
d'\'equilibre d\'ecro\^it \`a $0$ (haute temp\'erature, phase paramagn\'etique)
ou vers le carr\'e de la magn\'etisation d'\'equilibre (basse temp\'erature,
phase ferromagn\'etique). Dans ce deuxi\`eme cas l'ergodicit\'e du syst\`eme
est bris\'ee, on reste dans un des deux \'etats purs avec un signe d\'efini
de la magn\'etisation. Cette brisure d'ergodicit\'e n'est compl\`ete que dans
la limite thermodynamique, dans un syst\`eme de taille finie l'ergodicit\'e est
restaur\'ee par les fluctuations de la magn\'etisation.

On peut en outre s'interroger sur le comportement des fonctions de 
corr\'elation \`a deux temps dans le r\'egime transitoire
qui pr\'ec\`ede l'\'equilibre. On a alors effectivement une d\'ependance 
dans les deux temps (et pas seulement dans leur diff\'erence), \`a cause de 
la pr\'eparation \`a l'instant initial qui brise l'invariance par
translation temporelle.

Une extension des m\'ethodes d'approximation pr\'esent\'ees
pr\'ec\'edemment permet de calculer, de mani\`ere approch\'ee, ces
fonctions de corr\'elation. L'id\'ee consiste toujours \`a projeter la 
dynamique sur un petit nombre d'observables macroscopiques, mais cette fois-ci
en incluant des quantit\'es qui gardent une trace de
la configuration microscopique \`a l'instant ant\'erieur $t_1$. Le plus
simple serait de suivre en fonction de $t$ l'ensemble de param\`etres
$\{ e(t), m(t) , C(t,t_1) \}$, et d'\'ecrire des \'equations 
diff\'erentielles sur ces quantit\'es. On trouve alors les \'equations 
(\ref{eq:bethe-evbino}) muni de l'approximation binomiale 
(\ref{eq:bethe-psigbin}) pour $p_\sigma(u,t)$, et une nouvelle
\'equation diff\'erentielle pour $C$,
\begin{equation}
\frac{d}{dt} C(t,t_1) = - 2 C(t,t_1) A_c(t) \ , \qquad 
A_c(t) = \sum_{u=0}^L W(u,\beta) [p_{-}(u;t) + p_{+}(u;t)] \ .
\end{equation}
L'interpr\'etation en est assez simple. $A_c(t)$ est la probabilit\'e
d'acceptation d'un flip au temps $t$. A chaque fois qu'une variable est
renvers\'ee, on suppose qu'elle a une probabilit\'e $(1+C)/2$ d'\^etre dans le
m\^eme \'etat aux instants $t$ et $t_1$, alors $C$ diminue de $2/N$, tandis
qu'avec probabilit\'e $(1-C)/2$ elle augmente de $2/N$.

\begin{figure}
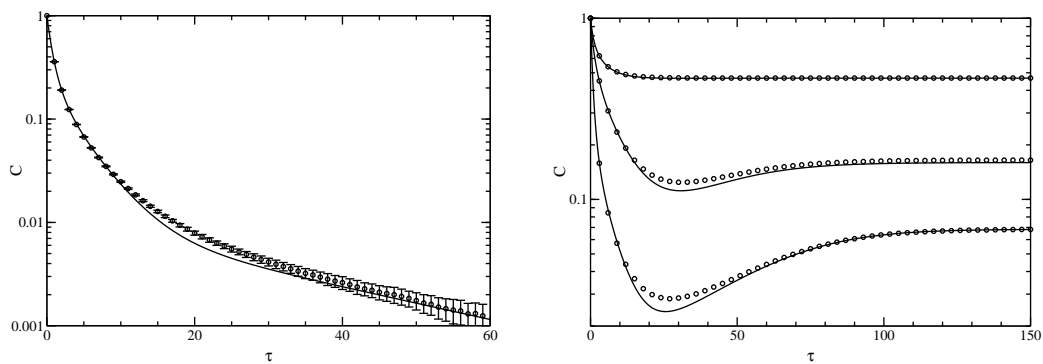

  \begin{center}
    \includegraphics[angle=0,width=6.5cm]{bethe-correl_b1.eps} 
\hspace{5mm}
    \includegraphics[angle=0,width=6.5cm]{bethe-correl_b12.eps}
\\[0.3cm]
  \end{center}
  \caption{Corr\'elations \`a deux temps $C(t_1+\tau ,t_1)$ 
    en fonction de $\tau$, pour $L=3$. Les traits pleins repr\'esentent
    les r\'esultats de l'\'etude analytique, les symboles sont des
    moyennes sur 200 simulations Monte-Carlo pour des syst\`emes de
    taille $N=10^7$. Gauche~: la d\'ecor\'elation avec la
    configuration initiale ($t_1=0$) dans la phase paramagn\'etique
    ($\beta=1$). Droite~: de bas en haut $t_1=0, 30, 150$ dans la phase
    ferromagn\'etique ($\beta=1.2$).}
  \label{fig:arbre-correl}
\end{figure}

Ce niveau d'approximation conduit \`a des r\'esultats m\'ediocres, mais on
peut l'am\'eliorer \`a l'image des approximations successives de la partie
pr\'ec\'edente. La quantit\'e qui s'est av\'er\'e fournir un bon compromis
entre qualit\'e des r\'esultats et simplicit\'e des calculs est 
$q_{\sigma_1 \sigma}(uu,us,su;t_1,t)$, fraction de sites qui ont
un spin $\sigma_1$ (resp. $\sigma) $\`a l'instant $t_1$ (resp. $t$), et dont,
parmi les $L$ liens qui l'entourent,  $uu$ sont frustr\'es
aux deux temps, $us$ sont frustr\'es \`a $t_1$ et satisfait \`a $t$,
$su$ dans la situation inverse, et le reste ($L-uu-us-su$) sont
satisfaits aux deux temps. De cette quantit\'e on peut tirer la fonction
de corr\'elation comme
\begin{eqnarray}
C(t,t_1) &=&\sum_{uu,us,su}(q_{++}(uu,us,su;t_1,t) + q_{--}(uu,us,su;t_1,t) 
\nonumber \\
&-& q_{+-}(uu,us,su;t_1,t) - q_{-+}(uu,us,su;t_1,t)) \ .
\end{eqnarray}
Il ne reste plus qu'\`a \'ecrire une \'equation d'\'evolution pour $q$, qui
s'obtient avec le m\^eme type de raisonnement que celui utilis\'e pour
le niveau d'approximation des voisins ind\'ependants.
On doit tenir compte de l'effet du renversement d'un spin sur
le type de la variable en question, $(\sigma_1,\sigma,uu,us,su) \to 
(\sigma_1,-\sigma,us,uu,ss=L-uu-us-su)$, et de l'effet sur ses voisines.
On obtient finalement~:
\begin{eqnarray}
&&\frac{d}{dt} q_{\sigma_1 \sigma}(uu,us,su;t_1,t)= \nonumber \\ 
&-& q_{\sigma_1 \sigma}(uu,us,su) W(uu+su)
+ q_{\sigma_1, -\sigma}(us,uu,ss) W(ss+us) \nonumber \\ 
&+&\frac{\langle uu W(uu+su) \rangle_{-\sigma_1-\sigma}}{\langle uu
  \rangle_{\sigma_1 \sigma}} [-uu \; q_{\sigma_1
    \sigma}(uu,us,su) + (uu+1) q_{\sigma_1 \sigma}(uu+1,us-1,su)
] \nonumber \\ 
&+&\frac{\langle us W(uu+su) \rangle_{-\sigma_1 \sigma}}{\langle us
  \rangle_{\sigma_1 \sigma}} [ -us \; q_{\sigma_1
    \sigma}(uu,us,su) + (us+1) q_{\sigma_1 \sigma}(uu-1,us+1,su)
] \nonumber \\ 
&+&\frac{\langle su W(uu+su) \rangle_{\sigma_1 -\sigma}}{\langle su
  \rangle_{\sigma_1 \sigma}} [ -su \; q_{\sigma_1
    \sigma}(uu,us,su) + (su+1) q_{\sigma_1 \sigma}(uu,us,su+1)
] \nonumber \\ 
&+&\frac{\langle ss W(uu+su) \rangle_{\sigma_1 \sigma}}{\langle ss
  \rangle_{\sigma_1 \sigma}} [ -ss \; q_{\sigma_1
    \sigma}(uu,us,su) + (ss+1) q_{\sigma_1 \sigma}(uu,us,su-1)
] \ ,
\end{eqnarray}
o\`u l'on a utilis\'e la notation
\renewcommand{\arraystretch}{.75}
\begin{equation}
\langle \bullet \rangle_{\sigma_1 \sigma} =
\underset{\begin{array}{c} uu,us,su \\
  uu+us+su \le L \end{array}}{\sum} \hspace{-1cm} \bullet \; \; q_{\sigma_1
\sigma}(uu,us,su;t_1,t) \ .
\end{equation}
Ce jeu d'\'equations diff\'erentielles coupl\'ees est compl\'et\'e par
des conditions aux bords quand $t=t_1$, et par des conditions de coh\'erence
pour que l'\'energie, soit au temps $t_1$ soit au temps $t$, soit
la m\^eme quand on la calcule comme le nombre de spins $-1$ autour des $+1$,
ou l'inverse. On peut alors int\'egrer num\'eriquement ces \'equations,
et les comparer avec des simulations Monte-Carlo. A nouveau la concordance
de ces deux approches est tr\`es satisfaisante, comme on peut le voir
sur la figure \ref{fig:arbre-correl}. 

Pour des diff\'erences de temps $\tau$ qui divergent, la
fonction de corr\'elation $C(t_1+\tau ,t_1)$ tend vers le produit
des magn\'etisations aux deux temps. Ceci explique pourquoi cette
valeur est correctement pr\'edite par le calcul approch\'e quand $t_1=0$
ou $t_1 \to \infty$, puisque dans ces deux cas les magn\'etisations en jeu
sont celles de la configuration initiale et/ou de l'\'equilibre, qui
sont correctement captur\'ees par l'approximation. Par contre pour une valeur 
interm\'ediaire $0 < t_1 < \infty$, la magn\'etisation au temps $t_1$ n'est
qu'approch\'ee, et la pr\'ediction pour $C(t_1+\infty,t_1)$ n'est pas
exacte. Ces arguments sont confirm\'ees par les r\'esultats de la figure 
\ref{fig:arbre-correl}.

\subsection{Perspectives}
Le mod\`ele que l'on vient d'\'etudier est certainement un des plus simples 
que l'on puisse imaginer dans la classe des probl\`emes de
dynamique sur des syst\`emes dilu\'es. En effet, il v\'erifiait la
condition de balance d\'etaill\'ee et l'\'etat d'\'equilibre thermodynamique
\'etait atteint sur des \'echelles de temps finies (dans la phase de
basse temp\'erature l'ergodicit\'e est bris\'ee de la m\^eme fa\c con que
dans le mod\`ele de Curie-Weiss, il n'y a pas de vieillissement par
croissance de domaine comme en dimension finie~\cite{Br-review}). 
L'aspect \og hors d'\'equilibre\fg\ de ce travail vient seulement de ce que 
l'on s'est attach\'e \`a d\'ecrire le r\'egime transitoire \`a partir de 
conditions initiales \'eloign\'ees de l'\'equilibre.
De plus, la pr\'esence de d\'esordre \'etait 
assez anecdotique, puisqu'il apparaissait seulement par l'interm\'ediaire 
des longues boucles.

On peut imaginer un grand nombre de situations plus compliqu\'ees. 
Dans la partie suivante on \'etudiera une dynamique d'origine algorithmique 
qui ne v\'erifie pas la condition de balance d\'etaill\'ee. En restant dans 
le cadre de dynamiques \og physiques\fg\ qui v\'erifient
cette condition, on peut tracer deux directions pour des travaux plus
pouss\'es~:

\vspace{2mm}

\begin{itemize}
\item
Une premi\`ere serait de conserver un mod\`ele avec des interactions 
ferromagn\'etique entre paires de spins, mais en introduisant du d\'esordre 
dans les couplages. On pourrait soit garder un mod\`ele \`a connectivit\'e 
fixe, et des couplages d'intensit\'e al\'eatoires, soit des couplages 
d'intensit\'e constante, mais en autorisant des fluctuations
dans la connectivit\'e des variables, en utilisant par exemple un graphe
al\'eatoire d'Erd\"os et R\'enyi. Ce dernier cas serait la version
ferromagn\'etique du mod\`ele de Viana-Bray. On s'attend alors \`a avoir un 
ph\'enom\`ene connu sous le nom de phase de 
Griffiths~\cite{Griffiths,BrHu,Br-Griffiths,RaSePa}: la relaxation dans la
phase paramagn\'etique est anormalement lente (plus lente qu'une
exponentielle), \`a cause de r\'egions rares dans le syst\`eme qui ont,
localement, une plus grande tendance \`a s'ordonner. Dans le cas du mod\`ele
de Viana-Bray, cela peut d\'ecouler de la pr\'esence de sites avec une grande
connectivit\'e (comme on l'a vu, de mani\`ere un peu caricaturale, dans 
la version sph\'erique du chapitre pr\'ec\'edent), pour un mod\`ele \`a
connectivit\'e fixe on peut avoir des grandes r\'egions o\`u l'intensit\'e
des couplages est plus grande que la moyenne.

\item
Une deuxi\`eme direction, plus ambitieuse encore, serait d'\'etudier
des mod\`eles dont la phase de basse temp\'erature est vitreuse. Pour cela
on peut mettre du d\'esordre dans le signe des couplages de mani\`ere \`a 
induire de
la frustration. En fait sur le graphe al\'eatoire r\'egulier, un mod\`ele 
purement antiferromagn\'etique devrait aussi \^etre vitreux, la frustration 
provenant des boucles de longueur impaires. Une variante qui pourrait sembler 
sans gravit\'e mais qui
fait tomber dans le cadre des syst\`emes vitreux
consiste \`a consid\'erer des interactions qui font intervenir 3, ou plus,
spins par interactions. Dans ce cas-l\`a, m\^eme des interactions 
ferromagn\'etiques conduisent \`a des ph\'enom\`enes vitreux par une
frustration dynamiquement induite~\cite{ferroglass}.

Dans tous ces cas l'\'equilibre ne sera pas atteint, la dynamique pr\'esentera
des caract\'eristiques de vieillissement dans ses fonctions de corr\'elation
et de r\'eponse. Une des difficult\'es \`a r\'esoudre pour
entreprendre l'\'etude de tels mod\`eles par l'approche pr\'esent\'ee ici
consiste \`a d\'eterminer une description macroscopique judicieuse du 
syst\`eme. Pour cela, la connection avec la th\'eorie dynamique 
des r\'epliques sera peut-\^etre utile, puisque ce formalisme permet de 
traiter en principe les cas de RSB.

Enfin, une \'etude r\'ecente~\cite{MoRi-cooling} de la phase de basse 
temp\'erature d'un verre de spin dilu\'e a mis en \'evidence
des effets tr\`es int\'eressants de d\'ependance de l'\'etat
stationnaire obtenu aux temps longs selon l'histoire du syst\`eme. Il
serait tr\`es appr\'eciable de pouvoir capturer de tels effets par une 
approche analytique, m\^eme aussi approch\'ee que celle pr\'esent\'ee ici.

\end{itemize}

\newpage

\section{Un algorithme de recherche locale}
\label{sec:walksat}

La fin de ce chapitre sera consacr\'ee \`a l'\'etude d'un algorithme de
recherche de solutions d'un probl\`eme d'optimisation combinatoire.
A l'inverse des dynamiques \'etudi\'ees jusqu'\`a maintenant, celle-ci
ne respecte pas de condition de balance d\'etaill\'ee, et le comportement
aux temps longs du syst\`eme ne correspond pas \`a une mesure
d'\'equilibre de Boltzmann. De plus on s'int\'eresse au comportement de
l'algorithme sur des probl\`emes tir\'es d'un ensemble al\'eatoire, ce
qui introduit un d\'esordre gel\'e dans la d\'efinition du processus
dynamique. Ces caract\'eristiques rendent ce probl\`eme plus difficile
que le mod\`ele ferromagn\'etique \`a connectivit\'e fixe, cependant il en
partage certaines propri\'et\'es~: on peut aussi le repr\'esenter comme
l'\'evolution d'un syst\`eme de spins d'Ising, o\`u \`a chaque pas de temps
un seul spin est renvers\'e. En outre les observables macroscopiques
int\'eressantes sont extensives et varient peu au cours d'un pas de temps
\'el\'ementaire, ce qui nous met en position d'utiliser une fois de plus
les m\'ethodes g\'en\'erales d\'evelopp\'ees en d\'ebut de chapitre.

Ce travail a d'abord \'et\'e publi\'e dans un journal de physique (publication 
\pubwsat),
puis r\'e\'ecrit pour le rendre plus accessible \`a la communaut\'e 
informaticienne
\`a laquelle il a \'et\'e pr\'esent\'e dans une conf\'erence, cf. 
la publication \pubsat . Il a aussi constitu\'e une partie de la revue
\pubptac . Une \'etude tr\`es similaire a \'et\'e conduite simultan\'ement par 
Barthel, Hartmann et Weigt~\cite{BaHaWe-wsat}.

\subsection{D\'efinitions}
\label{sec:ws-intro}

\noindent \underline{\emph{Le probl\`eme de la satisfiabilit\'e}}

\vspace{4mm}

Le probl\`eme d'optimisation auquel on s'attache ici est celui de la
$K$-satisfiabilit\'e.
On a $N$ variables bool\'eennes $x_i$ qui peuvent \^etre vraies ou fausses.
Une clause de longueur $K$ est le OU logique (disjonction, not\'ee $\vee$) 
de $K$ variables parmi les $x_i$,
certaines de ces variables pouvant \^etre ni\'ees (la n\'egation de vrai 
\'etant faux, et vice-versa). Une clause est donc vraie d\`es qu'une des 
variables est dans l'\'etat impos\'e par la clause. Par exemple, pour $K=3$,
$x_2 \vee \overline{x_5} \vee x_6$ est vraie d\`es que $x_2$ est vraie, ou
$x_5$ est fausse, ou $x_6$ est vraie. Une instance du probl\`eme est
une formule constitu\'ee par le ET logique 
(conjonction, not\'ee $\wedge$) d'un certain nombre $M$
de clauses. Une formule est donc vraie si et seulement si toutes ses clauses 
sont vraies. On dit qu'une formule est satisfiable s'il existe une valeur des
variables $x_i$ telle que la formule soit vraie. Une telle configuration,
est alors appel\'ee une solution de la formule. La plupart du temps il n'y a
pas unicit\'e de la solution.

Devant une formule donn\'ee, la premi\`ere question que l'on peut se poser est
de savoir si elle est satisfiable ou pas, et de prouver cette affirmation.
Si le nombre de variables et le nombre de clauses sont petits, on peut toujours
faire une recherche exhaustive de toutes les configurations pour v\'erifier
si l'une d'entre elle est une solution ou pas. Ceci est seulement possible pour
$N$ tr\`es faible, le nombre de configurations croissant comme $2^N$. Dans 
toutes les affirmations concernant la difficult\'e du probl\`eme, on 
sous-entendra que l'on s'int\'eresse \`a de grandes formules.

Prouver la satisfiabilit\'e ou l'insatisfiabilit\'e d'une formule sont deux
t\^aches tr\`es diff\'erentes.
Dans le premier cas il \og suffit\fg\ d'exhiber une solution, la v\'erification
que la formule est en effet satisfaite par cette configuration est possible en 
un nombre d'op\'erations qui cro\^it comme un polyn\^ome avec le nombre $N$
de variables. Il est bien s\^ur tr\`es difficile en g\'en\'eral de trouver 
effectivement une solution. En termes plus pr\'ecis, la $K$-satisfiabilit\'e
est un probl\`eme dit NP-complet pour $K \ge 3$, c'est-\`a-dire que si l'on
connaissait un algorithme capable de trouver une solution en un nombre 
polynomial d'op\'erations \'el\'ementaires pour \emph{n'importe quelle}
formule satisfiable, 
tous les probl\`emes d'optimisation de la famille dite NP seraient
aussi solubles en un temps polynomial. L'hypoth\`ese la plus probable \`a
l'heure actuelle est qu'un tel algorithme n'existe pas, c'est cependant
un probl\`eme ouvert des math\'ematiques. Pour plus de d\'etails sur
les d\'efinitions des diff\'erentes familles de probl\`emes d'optimisation
et sur la th\'eorie de la complexit\'e, on pourra consulter~\cite{PaSt}.

Montrer qu'une formule n'est pas satisfiable est conceptuellement 
plus difficile, il faut mettre en \'evidence une contradiction qui emp\^eche 
\emph{toutes} les configurations d'\^etre des solutions.

\vspace{8mm}

\noindent \underline{\emph{Diff\'erents types d'algorithmes}}

\vspace{4mm}

On dit qu'un algorithme est complet s'il est capable de donner le
statut (satisfiable ou pas) de toute formule qu'on lui pr\'esente, et de
justifier sa r\'eponse en exhibant soit une solution si la formule est
satisfiable, soit une contradiction dans le cas contraire. L'exemple le
plus connu est l'algorithme de Davis-Putnam-Loveland-Logeman (DPLL)~\cite{DPLL}
qui explore d'une mani\`ere syst\'ematique l'espace des configurations
des variables bool\'eennes, en \'eliminant le plus rapidement possible les
r\'egions o\`u il est s\^ur de ne pas trouver de solutions. On trouvera
plus de d\'etails et de r\'ef\'erences sur ce type d'algorithme dans la
revue \pubptac .

D'autres algorithmes, \og incomplets\fg , ne se prononcent avec
certitude que dans certains cas. Par exemple, un algorithme de recherche
locale comme celui que l'on va \'etudier dans la suite du chapitre, 
peut trouver une solution de la formule qu'on lui pr\'esente.
Dans ce cas-l\`a, celle-ci \'etait sans aucun doute satisfiable. Il se peut
aussi que l'algorithme, au bout d'un temps d\'efini \`a l'avance, n'ait pas
trouv\'e de solution. Alors on ne peut pas conclure~: soit la formule 
n'\'etait pas satisfiable, soit l'algorithme n'a pas \'et\'e assez astucieux
pour trouver une des solutions.

Signalons aussi que les m\'ethodes de la physique statistique des syst\`emes
d\'esordonn\'ees ont r\'ecemment conduit \`a un nouveau type d'algorithme 
incomplet, appel\'e \og survey propagation\fg\ ~\cite{MeZe-SP}. Celui-ci
exploite l'intuition sur la structure de l'espace des configurations
acquise gr\^ace \`a la m\'ethode des r\'epliques et de la cavit\'e pour
rep\'erer les variables cruciales du probl\`eme.

\vspace{8mm}

\noindent \underline{\emph{Un algorithme de recherche locale}}

\vspace{4mm}

On va consid\'erer dans la suite l'algorithme Pure Random WalkSAT (PRWSAT),
introduit par Papadimitriou~\cite{Pa-PRWSAT} pour $K=2$ en 1992. Il fonctionne
de la mani\`ere suivante~:

\begin{enumerate}
\item A l'instant initial, la valeur
des variables bool\'eennes $x_i$ est choisie al\'eatoirement, \'egale
\`a vrai ou faux avec probabilit\'e $1/2$. 

\item A chaque pas de temps ult\'erieur, si toutes les clauses sont 
satisfaites, on a trouv\'e une solution et l'algorithme se termine. Sinon,
on choisit al\'eatoirement et uniform\'ement une des clauses non satisfaites, 
puis (toujours al\'eatoirement et uniform\'ement ) une des variables de cette
clause, et on la renverse (elle passe de vraie \`a fausse, ou 
r\'eciproquement). 

\item On retourne au point pr\'ec\'edent, \`a moins qu'une limite sur le nombre
de pas de temps ait \'et\'e d\'epass\'ee. Dans ce cas, on quitte la boucle
sans pouvoir conclure sur l'existence ou pas d'une solution.
\end{enumerate}

La motivation du deuxi\`eme point est la suivante~: lorsqu'on renverse une 
variable d'une clause non satisfaite, elle devient forc\'ement
satisfaite. Bien s\^ur, il est possible que cette m\^eme variable appartienne
\`a d'autres clauses qui \'etaient auparavant satisfaites et qui ne le sont 
plus apr\`es le renversement. 

Il existe quelques r\'esultats rigoureux concernant cet algorithme,
valables quelque soit la formule \'etudi\'ee.
Le premier est d\^u \`a Papadimitriou~\cite{Pa-PRWSAT}~: si $K=2$,
c'est-\`a-dire si toutes les clauses comportent deux variables, et si
la formule admet au moins une solution, PRWSAT la trouve
en un temps d'ordre $N^2$, avec grande probabilit\'e (par rapport aux choix
al\'eatoires de la configuration initiale et des choix aux diff\'erents pas
de l'algorithme). Le probl\`eme de la 2-satisfiabilit\'e est en fait 
polynomial, et il existe des algorithmes d\'eterministes qui r\'esolvent toute
formule en un temps lin\'eaire, l'approche stochastique n'est donc pas
optimale ici.

Pour le cas plus int\'eressant de la 3-satisfiabilit\'e, qui est donc
NP-complet, Sch\"oning~\cite{Schoning} a montr\'e qu'une formule satisfiable 
\'etait
r\'esolue par PRWSAT en un temps born\'e par $(4/3)^N$.
La borne est exponentielle dans ce cas, ce qui est attendu \`a cause de la 
NP-compl\'etude du probl\`eme. Le $4/3$ a \'et\'e am\'elior\'e un petit peu 
depuis, gr\^ace \`a des choix plus \'elabor\'es de la condition 
initiale~\cite{Schoning2}. Il faut noter que cette borne est exponentiellement
meilleure que le temps $2^N$ n\'ecessaire pour une \'enum\'eration exhaustive
de toutes les configurations.

Les r\'esultats de Papadimitriou et Sch\"oning ont de profondes implications~:
m\^eme si cet algorithme n'est pas complet dans un sens strict puisqu'il
peut se tromper (ne pas trouver de solutions \`a une formule satisfiable), 
la probabilit\'e qu'il fasse une erreur peut-\^etre rendue
aussi faible que d\'esir\'ee. Il est donc \og probabilistiquement complet\fg\ .
On trouvera une discussion d\'etaill\'ee de ce type d'algorithmes 
dans~\cite{MoRa-random}.

\subsection{L'ensemble des formules al\'eatoires}

\noindent \underline{\emph{D\'efinition et propri\'et\'es statiques}}

\vspace{4mm}

La th\'eorie de la complexit\'e algorithmique bri\`evement \'evoqu\'ee 
ci-dessus s'int\'eresse \`a la difficult\'e d'un probl\`eme dans le pire 
des cas. S'il est vraisemblable que l'on ne puisse pas construire
d'algorithmes r\'esolvant n'importe quelle formule en un temps polynomial,
cela pourrait \^etre d\^u \`a quelques formules particuli\`erement
difficiles mais rares, alors que la majorit\'e des formules sont faciles.
Pour donner un sens plus pr\'ecis \`a ces id\'ees de complexit\'e typique,
un ensemble probabiliste de formules a \'et\'e d\'efini 
dans~\cite{MiSeLe-rsat}.

Une formule de cet ensemble est construite de la mani\`ere suivante.
Chacune des $M$ clauses
est g\'en\'er\'ee ind\'ependamment des autres, en choisissant un $K$-uplet
de variables uniform\'ement parmi les ${N \choose K}$ possibilit\'es, et
chacune des variables dans la clause est ni\'ee ou pas avec probabilit\'e 
$1/2$. Le r\'egime qui nous int\'eresse est celui de la limite thermodynamique
o\`u le nombre de variables $N$ et le nombre de clauses $M$ tendent
simultan\'ement vers l'infini avec un ratio $\alpha=M/N$ fix\'e. Les clauses
forment donc un hypergraphe poissonnien de connectivit\'e moyenne $\alpha K$.

La probabilit\'e qu'une formule ainsi g\'en\'er\'ee soit satisfiable
pr\'esente un comportement de seuil, ou transition de phase~: avec grande 
probabilit\'e (c'est-\`a-dire avec une probabilit\'e qui tend vers 1 dans la
limite thermodynamique), la formule est satisfiable si $\alpha<\alpha_c(K)$, 
non satisfiable si $\alpha>\alpha_c(K)$. Pour $K=3$, le seuil est \`a
$\alpha_c \sim 4.2$. Ce ph\'enom\`ene a d'abord \'et\'e constat\'e
num\'eriquement.

La preuve de l'existence du seuil de
satisfiabilit\'e n'est pas achev\'ee~\cite{Friedgut}, des bornes rigoureuses
ont cependant \'et\'e \'etablies~: s'il existe, le seuil est compris entre
3.145 ~\cite{Ac-bounds} et 4.506 ~\cite{Du-bounds}. D'autres 
travaux~\cite{AcMo,AcPe} ont
permis de resserrer l'\'ecart entre bornes inf\'erieures et sup\'erieures
dans la limite d'un grand nombre $K$ de variables par clause.

Ce probl\`eme a \'et\'e \'etudi\'e par des m\'ethodes de physique statistique,
en utilisant l'analogie d\'ecrite dans l'introduction avec les probl\`emes de
verres de spin. Le travail originel de Monasson et Zecchina~\cite{MoZe-KSAT}
reposait sur l'utilisation de la m\'ethode des r\'epliques avec l'hypoth\`ese
de sym\'etrie des r\'epliques, et montrait entre autres que le nombre de
solutions dans la phase satisfiable \'etait exponentiellement grand dans
la taille du syst\`eme. Les difficult\'es techniques de
la brisure de sym\'etrie des r\'epliques dans les syst\`emes
\`a la connectivit\'e finie~\cite{Mo-c-sigma}
ont retard\'es l'apparition de la solution \`a un pas de brisure.
Une \'etape interm\'ediaire a \'et\'e entreprise par Biroli, Monasson et 
Weigt~\cite{BiMoWe} qui ont trouv\'e une forme variationnelle de l'ansatz
1RSB, et montr\'e l'existence d'une deuxi\`eme transition dans la zone 
satisfiable $\alpha<\alpha_c$~: pour $\alpha$ tr\`es faible, l'ensemble des
solutions est r\'eparti uniform\'ement dans l'espace des configurations des
variables. Quand on augmente le ratio $\alpha$ de  contraintes par variables,
il appara\^it une structure dans l'ensemble des solutions, qui se regroupent
par amas de solutions proches, s\'epar\'ees par des zones vides de solution.
Cette transition est dite de \og clustering\fg\ .
Plus r\'ecemment, la reformulation des \'equations 1RSB par la m\'ethode
de la cavit\'e~\cite{MePa-bethe} a permis de les r\'esoudre num\'eriquement
par une m\'ethode de dynamique de populations. Cette m\'ethode a \'et\'e
appliqu\'ee au probl\`eme de la satisfiabilit\'e par M\'ezard et 
Zecchina~\cite{MeZe-SP}, et pr\'edit la valeur du seuil de satisfiabilit\'e
pour $K=3$ \`a $\alpha_c=4.267$. Par ailleurs Franz et Leone~\cite{FrLe} ont 
montr\'e, en utilisant la m\'ethode d'interpolation de 
Guerra~\cite{Gu-interpolation}, que les seuils de satisfiabilit\'e 
calcul\'es au niveau 1RSB \'etait des bornes sup\'erieures rigoureuses.

Il est plus difficile de donner une valeur pr\'ecise du seuil de clustering.
La solution 1RSB appara\^it \`a $\alpha \approx 3.86$, mais jusqu'\`a
$\alpha \approx 4.15$ elle est instable vis-\`a-vis d'une brisure compl\`ete 
de la sym\'etrie des r\'epliques~\cite{MoRi-stab,MoPaRi-stab,MeMeZe-threshold}.
Entre ces deux valeurs l'ensemble des solutions acquiert donc une structure,
mais elle est plus compliqu\'ee que l'image 1RSB de clusters de solutions.

\vspace{8mm}

\noindent \underline{\emph{Le comportement de PRWSAT sur des formules al\'eatoires}}

\vspace{4mm}

On va \'etudier dans la suite le comportement de l'algorithme PRWSAT dans
le cas o\`u la formule qu'on lui propose de r\'esoudre est tir\'ee
de l'ensemble al\'eatoire dont on vient de d\'ecrire les propri\'et\'es
statiques.

\begin{center}
\begin{figure}
\hskip -2cm
\includegraphics[width=8cm]{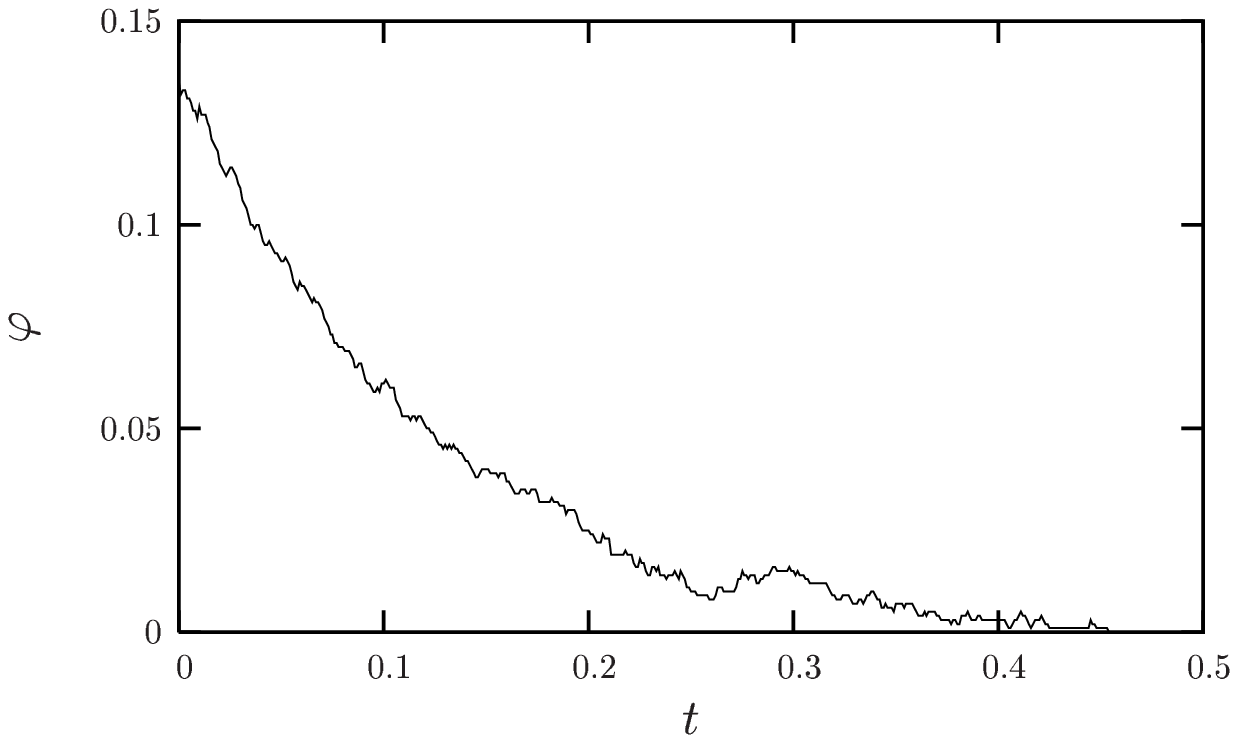}
\hskip 1cm
\includegraphics[width=8cm]{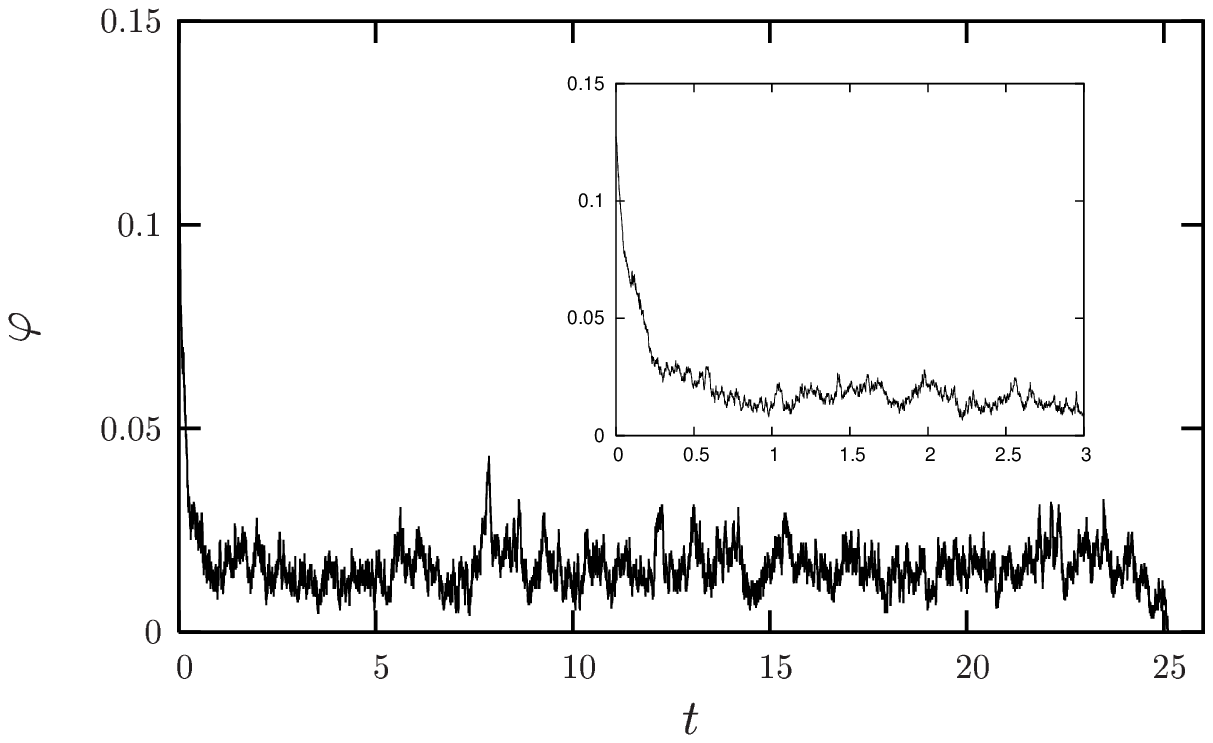}
\caption{D\'ecroissance de la fraction de clauses non satisfaites en fonction
du temps, pour deux formules de $N=500$ variables avec $K=3$. A gauche, 
$\alpha=2$, une
solution est trouv\'ee rapidement. A droite, $\alpha=3$, la d\'ecroissance
initiale (voir l'inset pour un grossissement) conduit \`a un plateau. Une 
fluctuation suffisamment grande finit par conduire \`a une solution.}
\label{fig:pheno-PRWSAT-1}
\end{figure}
\end{center}

Il est tr\`es facile de g\'en\'erer de telles formules et de simuler 
num\'eriquement le comportement de PRWSAT. Les deux 
trac\'es de la figure \ref{fig:pheno-PRWSAT-1} repr\'esentent la fraction
de clauses non satisfaites au cours de l'\'evolution de l'algorithme.
De mani\`ere \`a avoir une limite thermodynamique bien d\'efinie, on a plac\'e 
en ordonn\'ees la fraction de clauses
$\varphi$ et non le nombre total de clauses non satisfaites (\'energie
$E$), qui sont donc reli\'es par $\varphi=E/M$. Le temps est, pour la m\^eme
raison, d\'efini comme $t=T/M$, avec $T$ le nombre de pas discrets de 
l'algorithme. La valeur initiale de $\varphi$ s'interpr\`ete ais\'ement~:
la configuration initiale \'etant choisi al\'eatoirement, chaque clause
a une probabilit\'e $2^{-K}$ d'\^etre viol\'ee, puisqu'une seule
parmi les $2^K$ configurations des variables ne la satisfait pas.
On a donc $\varphi(t=0)=2^{-K}$, aux fluctuations de taille finie pr\`es.
Chacune des deux courbes de la figure \ref{fig:pheno-PRWSAT-1} a \'et\'e 
obtenue \`a partir d'une seule simulation, pour des des formules assez
petites ($N=500$).
Pour une premi\`ere valeur de $\alpha$, ici $\alpha=2$, la courbe
de gauche montre une d\'ecroissance relativement r\'eguli\`ere 
(aux fluctuations pr\`es) et rapide de l'\'energie. Lorsqu'elle
s'annule, l'algorithme a trouv\'e une solution de la formule et s'arr\^ete. 
La figure de droite, trac\'ee pour $\alpha=3$, a une allure bien diff\'erente.
Aux temps courts le comportement est similaire (voir l'inset), mais $\varphi$
ne d\'ecro\^it pas jusqu'\`a 0, et tend (en moyenne) vers une valeur positive. 
Comme le syst\`eme est de taille finie, il y a des fluctuations autour de ce
plateau. Au bout d'un certain temps, une de ces fluctuations est suffisamment
grande pour atteindre l'\'energie nulle, une solution est alors trouv\'ee.

\begin{center}
\begin{figure}
\includegraphics[width=9cm]{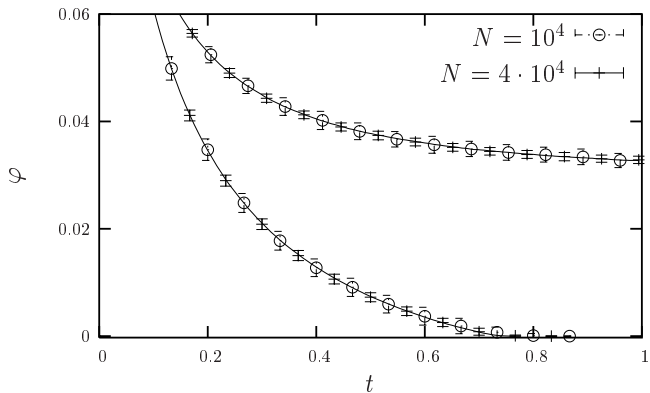}
\caption{D\'ecroissance de la fraction de clauses non satisfaites en fonction
du temps, moyenn\'ee sur 100 \'echantillons ind\'ependants pour $\alpha=2.4$ et
$\alpha=3.5$. Les barres d'erreur correspondent \`a $N=10^4$ et 
$N=4 \cdot 10^4$.}
\label{fig:pheno-PRWSAT-2}
\end{figure}
\end{center}

On s'attend \`a ce que dans la limite thermodynamique, l'\'evolution de 
$\varphi(t)$ pour une seule simulation soit
concentr\'ee autour de sa valeur moyenne (par rapport au choix de la formule
et aux choix stochastiques de l'algorithme), avec des fluctuations 
d'ordre $N^{-1/2}$. La figure \ref{fig:pheno-PRWSAT-2} permet de confirmer 
cette intuition. On a r\'ep\'et\'e cent simulations ind\'ependantes, en
tirant \`a chaque fois une nouvelle formule et une nouvelle histoire de 
l'algorithme, pour deux valeurs de $\alpha$ ($2.4$ et $3.5$) et deux valeurs
de $N$ ($10^4$ et $4 \cdot 10^4$). Pour chacun de ces groupes de cent 
simulations on a mesur\'e la moyenne et l'\'ecart quadratique moyen de 
$\varphi(t)$. Comme attendu, les valeurs moyennes sont quasiment 
ind\'ependantes de la taille du syst\`eme,
et les \'ecarts quadratiques sont approximativement deux fois plus faibles
pour la taille quatre fois plus grande.

La courbe autour de laquelle $\varphi(t)$ se concentre dans la limite
thermodynamique pr\'esente deux comportements tr\`es diff\'erents selon
la valeur de $\alpha$~: quand ce param\`etre est suffisamment faible, 
elle s'annule
en un temps fini que l'on notera $t_{\rm sol}(\alpha,K)$. A $\alpha$ plus
grand, elle reste toujours positive et tend vers une valeur de
plateau $\varphi_{\rm as}(\alpha,K)$. La valeur du param\`etre $\alpha$ qui
s\'epare ces deux r\'egimes sera not\'ee $\alpha_d(K)$ . 
Des simulations plus compl\`etes 
pr\'esent\'ees dans la suite montrent que $\alpha_d(K=3) \sim 2.7$.

Le ph\'enom\`ene de concentration implique par cons\'equent que pour 
$\alpha<\alpha_d$ une solution de la formule est typiquement trouv\'ee en
$M t_{\rm sol}(\alpha,K)$ pas de l'algorithme, qui a donc une complexit\'e
typique lin\'eaire pour ces formules. Par contre quand $\alpha> \alpha_d$,
une solution n'est trouv\'ee que par l'interm\'ediaire d'une grande fluctuation
de la densit\'e d'\'energie autour de sa valeur moyenne. Ces grandes
d\'eviations ayant des probabilit\'es exponentiellement faibles pour des
grands syst\`emes, le temps de r\'esolution va cro\^itre avec $N$ comme 
$\exp[\zeta(\alpha) N]$.

On peut parler de m\'etastabilit\'e pour le comportement \`a 
$\alpha_d < \alpha < \alpha_c$~: il existe des solutions avec grande 
probabilit\'e, qui sont les \'etats absorbants de la dynamique, mais le
temps pour les atteindre diverge dans la limite thermodynamique. Le 
syst\`eme reste donc pendant tr\`es longtemps dans un \'etat m\'etastable
d'\'energie $\varphi_{\rm as}>0$.
Cette m\'etastabilit\'e est similaire \`a celle d'autres syst\`emes physiques. 
En particulier, le processus de
contact~\cite{Liggett,DeMo-CP}
pr\'esente une ph\'enom\'enologie tr\`es proche. Dans ce mod\`ele,
on a des particules sur les sites d'un r\'eseau, avec au maximum
une particule par sommet. Chaque particule dispara\^it avec un taux
constant, et les sites vides deviennent occup\'es avec un taux proportionnel
au nombre de sommets voisins d\'ej\`a occup\'es par une particule. Il y a un
\'etat absorbant dans le syst\`eme, qui correspond \`a un r\'eseau 
compl\`etement vide. Selon la densit\'e de particules dans l'\'etat initial,
cet \'etat absorbant est atteint en un temps logarithmique dans la taille
du syst\`eme, ou bien sur des \'echelles de temps exponentiellement grandes
par l'interm\'ediaire de fluctuations. Il y a deux diff\'erences dans le cas
de PRWSAT~: l'\'etat absorbant a une grande d\'eg\'en\'erescence, puisque
toutes les solutions (en nombre typiquement exponentiel pour une formule
al\'eatoire) bloquent l'\'evolution de l'algorithme, et de plus il y a un
d\'esordre gel\'e dans la d\'efinition des r\`egles dynamiques, 
\`a cause du choix al\'eatoire de la formule. A ces diff\'erences pr\`es,
le comportement des deux syst\`emes est tr\`es similaire. 

Cette parenth\`ese referm\'ee, on va pr\'esenter dans les sections suivantes
les r\'esultats analytiques
obtenus dans le but d'expliquer les constatations num\'eriques. Le probl\`eme
serait compl\`etement r\'esolu si l'on pouvait calculer 
exactement la limite thermodynamique de la fonction $\varphi(t)$ pour toutes
les valeurs de $\alpha$, (on obtiendrait ainsi $\alpha_d$, $t_{\rm sol}$ et
$\varphi_{\rm as}$), ainsi que la loi de probabilit\'e des grandes d\'eviations
de $\varphi$ pour $\alpha > \alpha_d$ (ce qui permettrait de calculer 
$\zeta(\alpha)$, le taux de croissance exponentiel des temps de r\'esolution).
Ce programme est bien s\^ur trop ambitieux, les r\'esultats suivants sont
soit des d\'eveloppements soit des approximations, en assez bon accord avec
les simulations num\'eriques.

A ma connaissance il n'existe qu'un seul r\'esultat rigoureux concernant le
comportement de PRWSAT sur des formules al\'eatoires~: Alekhnovich et
Ben-Sasson~\cite{BenSasson} ont montr\'e que pour $\alpha<1.63$, une formule
de l'ensemble 3-SAT al\'eatoire \'etait r\'esolu presque toujours en un nombre
de pas qui cro\^it lin\'eairement avec le nombre de variables. Ce r\'esultat 
est bien en accord avec les simulations num\'eriques que l'on vient de 
pr\'esenter puisqu'on a trouv\'e un seuil dynamique $\alpha_d \sim 2.7 > 1.63$.
Signalons aussi un travail num\'erique sur la phase \`a petit 
$\alpha$~\cite{Parkes}.

\subsection{D\'eveloppements en clusters dans la phase lin\'eaire}
\label{sec:ws-devclu}
Un premier angle d'attaque repose sur l'utilisation de la m\'ethode
du d\'eveloppement en clusters pr\'esent\'ee dans la partie 
\ref{sec:devclusters}. Cette m\'ethode consistant \`a calculer une s\'erie
de Taylor autour de $\alpha=0$, on s'int\'eresse \`a la phase 
$\alpha < \alpha_d(K)$, dans laquelle l'algorithme trouve une solution
apr\`es un nombre de pas proche de la valeur moyenne 
$M t_{\rm sol}(\alpha,K)$. On va donc chercher un d\'eveloppement de la
fonction $t_{\rm sol}$ en puissances de $\alpha$.

Notons $T_{\rm sol}$ le nombre de pas effectu\'e par l'algorithme avant
de trouver une solution. Cette variable est al\'eatoire, et ce pour plusieurs
raisons~:

\vspace{2mm}

\begin{itemize}
\item Le choix de la formule $F$ dans l'ensemble $K$-sat al\'eatoire. Comme
on l'a vu la g\'en\'eration d'une formule consiste \`a choisir un hypergraphe 
poissonnien $G$ (par le choix des variables dans chaque clause), puis \`a 
choisir les n\'egations des variables dans les clauses.
\item Le choix de la configuration initiale des variables.
\item Les choix al\'eatoires que l'algorithme effectue \`a chaque pas de temps.
\end{itemize}

\vspace{2mm}

La d\'ecomposition de l'hypergraphe $G$ en ses composantes connexes $G_i$
se traduit
naturellement par une d\'ecomposition de la formule $F$ en sous-formules 
ind\'ependantes $F_i$. 
On s'aper\c coit alors que $T_{\rm sol}$ est la somme de 
variables al\'eatoires, une pour chaque composante connexe~: par d\'efinition, 
une solution de $F$ est trouv\'ee quand toutes les $F_i$ sont r\'esolues,
et le nombre total de pas de l'algorithme
est la somme des pas effectu\'es dans chacune des sous-formules.
Notons $\overline{T}_{\rm sol}(G)$
la valeur moyenne de $T_{\rm sol}$ par rapport aux choix des n\'egations des
variables, de la configuration initiale et des pas de temps de l'algorithme.
Cette quantit\'e est additive par rapport \`a la d\'ecomposition en 
clusters, et on peut donc appliquer le formalisme g\'en\'erique de la partie 
\ref{sec:devclusters}. La
seule t\^ache restant \`a effectuer est le d\'enombrement du nombre moyen
de pas de temps pour r\'esoudre chaque type de cluster, dont on va donner
quelques exemples maintenant.

Pour un cluster constitu\'e
d'une seule clause, avec probabilit\'e $1- 2^{-K}$ la configuration initiale
est d\'ej\`a une des solutions. Sinon, un seul renversement de 
variable sera suffisant pour satisfaire la clause. Le temps moyen pour
r\'esoudre un tel cluster est donc $2^{-K}$.

Consid\'erons maintenant un cluster fait de deux clauses. Deux cas sont \`a
consid\'erer~:

\vspace{2mm}

\begin{itemize}
\item
les deux clauses portent le m\^eme signe sur la variable commune, un
exemple pour $K=2$ est $(x_1 \vee \overline{x}_2)
\wedge(\overline{x}_2 \vee \overline{x}_3)$.
La configuration initiale viole les deux clauses avec probabilit\'e 
$2^{1-2K}$, une seule clause avec $2 \cdot 2^{-K} \cdot (1 - 2^{1-K})$, 
et aucune sinon. Si la configuration initiale viole une seule clause, une 
solution est forc\'ement trouv\'ee apr\`es le renversement d'une variable.
Si elle viole les deux clauses, une solution peut \^etre 
trouv\'ee en un seul pas de temps si c'est la variable commune qui est 
renvers\'ee la premi\`ere (avec donc probabilit\'e $1/K$), il faut sinon deux 
pas de temps. On a donc dans ce cas un temps moyen de~:
\begin{equation}
\frac{1}{2^{2K-1}}\left( 2^K - \frac{1}{K} \right) \ .
\end{equation}
\item
les choses se compliquent un peu si les
deux clauses ont des exigences contradictoires pour la variable commune,
par exemple $(\overline{x}_1 \vee \overline{x}_2) \wedge( x_2 \vee 
\overline{x}_3)$. En
effet, imaginons que l'on soit dans la configuration des variables telle
que la premi\`ere clause soit viol\'ee, et que la deuxi\`eme ne soit satisfaite
que par la variable commune (dans l'exemple ci-dessus, ce serait le cas si les
trois variables \'etaient vraies). L'algorithme choisit de renverser une des 
variables de la premi\`ere clause~; si par malheur il prend la variable commune
aux deux clauses, on se retrouve dans une situation sym\'etrique, o\`u c'est
la deuxi\`eme clause qui est viol\'ee, et la premi\`ere qui n'est satisfaite
que par la variable commune. L'algorithme peut donc \og h\'esiter\fg\ plusieurs
fois entre ces deux configurations avant de trouver une solution. On trouve
apr\`es un petit calcul que le nombre de pas moyen pour trouver une solution 
est ici~:
\begin{equation}
\frac{1}{2^{2K-1}}\left(2^K+\frac{2}{K-1} \right) \ .
\end{equation}
En prenant la moyenne de ces deux r\'esultats on obtient la ligne $c$ de la
table \ref{tab:PRWSAT-clusters}.
\end{itemize}

\vspace{2mm}

Ce dernier exemple met en \'evidence d'une part un des 
d\'efauts de cet algorithme tr\`es simplifi\'e,  d'autre part les 
complications qui apparaissent quand on fait ce type de d\'enombrement avec 
des clusters de plus en plus grands. Les r\'esultats de l'\'enum\'eration
pour des formules avec trois clauses au maximum sont r\'esum\'es dans la 
table \ref{tab:PRWSAT-clusters}, ce qui donne en utilisant 
(\ref{eq:gr-devalpha3}) la formule suivante~:
\begin{eqnarray}
t_{\rm sol} (\alpha , K)&=& \frac{1}{2^K} + \frac{K(K+1)}{K-1}
\frac{1}{2^{2K+1}} \, \alpha \label{eq:PRWSAT-dev_cluster_tsolK} \\
&+& \frac{4K^6 + K^5 +6 K^3 -10 K^2 + 2
K}{3(K-1)(2K-1)(K^2-2)} \frac{1}{2^{3K+1}} \, \alpha^2 
+ {\cal O}(\alpha^3) \, .
\nonumber
\end{eqnarray}
Notons qu'il y a un facteur $\alpha^{-1}$ par rapport \`a 
(\ref{eq:gr-devalpha3}) car on divise par $M$ au lieu de $N$ dans la 
d\'efinition de $t_{\rm sol}$.

\begin{center}
\begin{figure}
\hskip -2cm
\includegraphics[width=8cm]{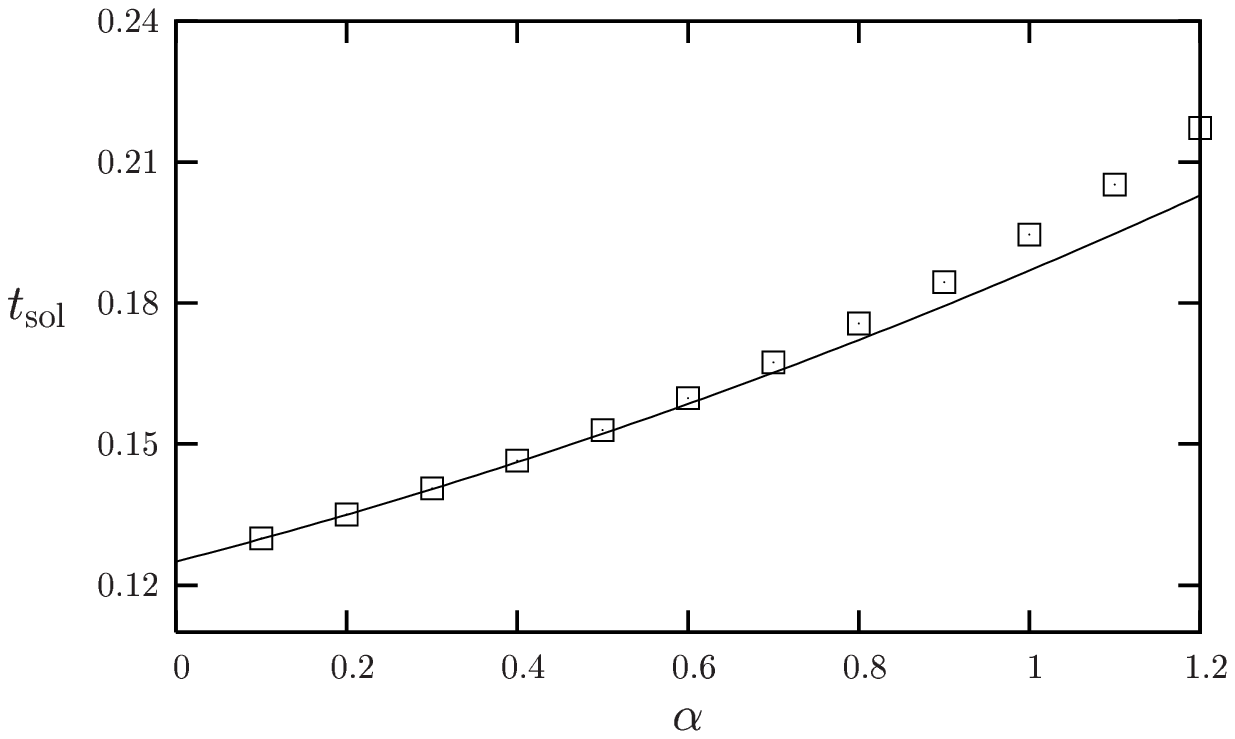}
\hskip 1cm
\includegraphics[width=8cm]{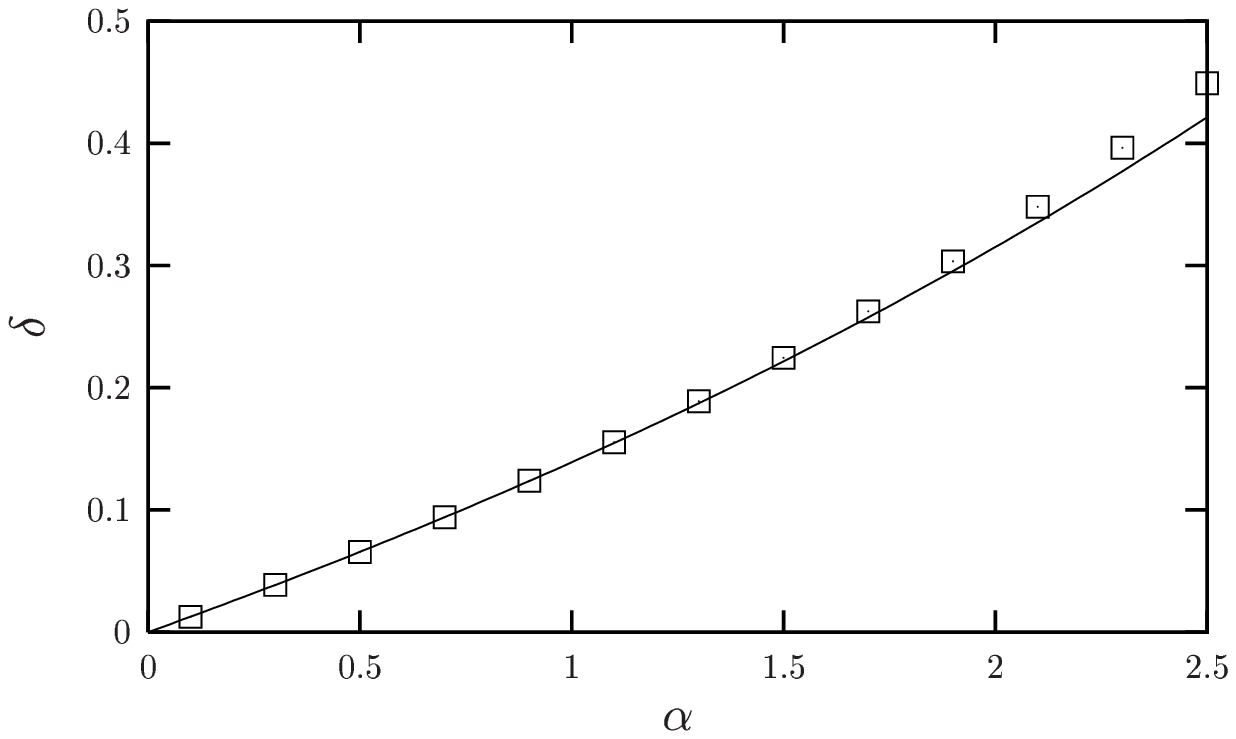}
\caption{A gauche, le temps moyen de d\'ecouverte d'une solution 
$t_{\rm sol}(\alpha , 3)$. Les symboles sont les r\'esultats de simulations
num\'eriques pour 100 \'echantillons de $10^6$ variables, les barres d'erreurs
sont plus petites que les symboles. La courbe
est la pr\'ediction du d\'eveloppement en clusters 
(\ref{eq:PRWSAT-dev_cluster_tsolK}). A droite, la distance de Hamming
moyenne entre la configuration initiale et la solution trouv\'ee, compar\'ee
\`a l'\'equation (\ref{eq:ws-dev_dist}).}
\label{fig:PRWSAT-tsol}
\end{figure}
\end{center}

Cette expression est compar\'ee aux r\'esultats de simulations num\'eriques
dans la figure \ref{fig:PRWSAT-tsol}. Comme attendu pour un d\'eveloppement
de Taylor au voisinage de 0, l'accord entre les deux se d\'egrade quand
$\alpha$ augmente. En particulier la divergence \`a $\alpha_d$ (non montr\'ee
sur la figure) n'est \'evidemment pas reproduite par un d\'eveloppement
polynomial.

Les simulations num\'eriques pour d\'eterminer $t_{\rm sol}$ 
peuvent \^etre faites
avec des syst\`emes de tr\`es grande taille, car dans ce r\'egime les
temps de calcul ne croissent que lin\'eairement avec le nombre de variables. 
Ceci permet de s'affranchir des effets de taille finie et de travailler dans
la \og limite thermodynamique num\'erique\fg\ . En particulier on peut 
se convaincre
raisonnablement que la fonction $t_{\rm sol} (\alpha , K)$ ne pr\'esente pas
de singularit\'e \`a la transition de percolation de l'hypergraphe
$\alpha_p=1/(K(K-1))$, et a priori elle est r\'eguli\`ere jusqu'\`a sa 
divergence \`a $\alpha_d$. Sur cet exemple, l'int\'er\^et de la m\'ethode
de d\'eveloppements en clusters
est assez clair, puisqu'on a pu faire des pr\'edictions de nature non triviale
(temps d'arr\^et d'une dynamique assez \'elabor\'ee) valables 
(perturbativement) dans toute la r\'egion $\alpha<\alpha_d$, \`a partir
de d\'enombrements sur des objets finis qui, s'ils sont p\'enibles \`a 
effectuer en pratique, ne pr\'esentent pas de difficult\'es de principe.

\begin{table}
\renewcommand{\arraystretch}{1.5}
$$\begin{array}{| c | c |}
\hline
\phantom{a} {\mbox{Type}} \phantom{a} & 
\phantom{a} (\overline{T}_{\rm sol})_t  \phantom{a} \\
\hline 
a & 0 \\ 
b & \frac{1}{2^K} \\ 
c & \frac{1}{2^{2K}}\left[2^{K+1} + \frac{K+1}{K(K-1)}\right] \\ 
d & \frac{1}{2^{3K}}\left[3 \cdot 2^{2K} + 2^{K+1} \frac{K+1}{K(K-1)} +
\frac{4 K^4 + 9 K^3 +9 K^2 + 6K -4}{3 K^2 (K-1) (2K -1)(K^2 -2)}
\right] \\
e & \frac{1}{2^{3K}}\left[3 \cdot 2^{2K} + 2^K \frac{3(K+1)}{K(K-1)} -
\frac{2K+1}{K^2(K-1)}\right] \\
\hline
\end{array}$$
\renewcommand{\arraystretch}{.66}
\caption{Contributions au d\'eveloppement en clusters pour le temps de
d\'ecouverte d'une solution (\ref{eq:PRWSAT-dev_cluster_tsolK}). 
La nomenclature des types est celle de la figure \ref{fig:gr_clusters}.}
\label{tab:PRWSAT-clusters}
\end{table}

Au del\`a du temps mis pour r\'esoudre une formule, on peut aussi s'interroger
sur la ressemblance entre la configuration initiale et la solution trouv\'ee.
On d\'efinit plus pr\'ecis\'ement la distance de Hamming $d$ entre ces deux 
configurations comme le nombre de variables qui sont diff\'erentes de l'une \`a
l'autre. On peut en fait calculer la moyenne de cette distance en utilisant
la m\^eme propri\'et\'e d'additivit\'e sur les clusters et en faisant des
\'enum\'erations similaires. On obtient pour le d\'eveloppement de 
$\delta(\alpha,K)$, la valeur typique de $d/N$~:
\begin{eqnarray}
& &\delta(\alpha , K) = 
\frac{1}{2^K} \, \alpha + \frac{1}{2^{2K+1}} \frac{K(K-1)}{K+1} \,  \alpha^2 
+ \label{eq:ws-dev_dist} \\
&&\frac{1}{2^{3K+1}} \frac{8 K^8 - 6 K^7 - 33 K^6 + 35 K^5 + 58 K^4 - 24 K^3 - 48 K^2 - 2 K}{3 (K+1)^2 (4K^2 -1)(K^2 - 2)} \, \alpha^3 + {\cal O}(\alpha^4) \ ,
\nonumber
\end{eqnarray}
qui est aussi en bon accord avec les simulations num\'eriques comme le montre
la figure \ref{fig:PRWSAT-tsol}.

Ce calcul \'etait motiv\'e par la constatation suivante. Il serait 
int\'eressant de localiser $\alpha_d(K)$ comme le point o\`u $t_{\rm sol}$ 
diverge. Comme on n'a qu'un petit nombre de termes du d\'eveloppement, 
il est difficile d'utiliser les techniques habituelles pour estimer le rayon 
de convergence de la s\'erie (\ref{eq:PRWSAT-dev_cluster_tsolK}). 
Au contraire, $\delta(\alpha,K)$ est par d\'efinition
toujours born\'e, donc une approximation polynomiale va \^etre de meilleure
qualit\'e. A premi\`ere vue, comme en se rapprochant de $\alpha_d$ le
temps n\'ecessaire pour trouver une solution, donc le nombre de fois o\`u 
les variables sont renvers\'ees, diverge, on peut penser que la solution aura 
perdu
toute corr\'elation avec la configuration initiale, et donc que 
$\delta(\alpha_d(K),K)=1/2$. Ceci fournirait un bon crit\`ere pour d\'eterminer
le seuil $\alpha_d$ \`a partir du d\'eveloppement (\ref{eq:ws-dev_dist}).
En fait cette condition n'est pas tout \`a fait vraie. Par exemple,
les variables qui n'appartiennent \`a aucune clause ne sont jamais renvers\'ees
au cours de l'algorithme, et il y en a une fraction finie $e^{-\alpha_d(K)K}$.
Une question ouverte serait donc de d\'eterminer 
$\delta(\alpha_d(K),K)$, autrement dit de savoir quelle partie d'une formule
au seuil dynamique se d\'ecorrelle au cours de l'\'evolution. On pourrait
par ailleurs imaginer que cette approche fournisse une borne rigoureuse sur
$\alpha_d$.

\subsection{Une caract\'erisation approch\'ee du comportement typique}
\label{sec:ws-approxtyp}

La m\'ethode de d\'eveloppement pr\'esent\'ee dans la section pr\'ec\'edente
n'est pas suffisante pour expliquer toutes les caract\'eristiques du
probl\`eme~: elle n'est pas capable de pr\'edire simplement la valeur
du seuil dynamique $\alpha_d(K)$, ni, pour $\alpha>\alpha_d(K)$, la valeur 
du plateau atteint aux temps longs et la loi des fluctuations autour de
celui-ci. On va donc utiliser maintenant une description approch\'ee de la 
dynamique qui capture qualitativement toutes les propri\'et\'es du probl\`eme, 
et qui est en assez bon accord quantitatif avec les donn\'ees num\'eriques.

L'id\'ee a d\'ej\`a \'et\'e exploit\'ee \`a plusieurs reprises dans ce 
chapitre~: on veut se d\'ebarrasser des d\'etails microscopiques du syst\`eme,
ici la configuration des variables, et se contenter d'une description en
termes d'observables macroscopiques. Dans le cas pr\'esent, la plus importante 
est le nombre de clauses non satisfaites dans la formule, dont on notera $E(T)$
la valeur apr\`es $T$ pas d'\'evolution de l'algorithme. Le processus
stochastique qui r\'egit l'\'evolution des variables microscopiques est
markovien d'apr\`es la d\'efinition de PRWSAT 
donn\'ee en \ref{sec:ws-intro}. La 
projection de la dynamique sur une variable macroscopique fait perdre ce
caract\`ere markovien, comme on l'a vu dans la premi\`ere partie de ce 
chapitre. On va cependant faire l'approximation, a priori assez brutale, que
l'\'evolution de $E(T)$ est markovienne.

On a donc besoin de la probabilit\'e de passer de $E$ \`a $E + \Delta$ clauses
non satisfaites au cours d'un pas de temps, probabilit\'e not\'ee 
$W(E+\Delta ,E)$.
A chaque pas de temps, une clause non satisfaite est choisie au hasard, ainsi
qu'une des variables de cette clause. Cette variable appartient 
\`a $n$ clauses en plus de la clause s\'electionn\'ee. En l'absence 
d'informations microscopiques plus pr\'ecises, on ne peut que supposer que 
la loi de probabilit\'e de $n$ est une loi de Poisson avec
param\`etre $\alpha K$, ce qui serait le cas si la clause \'etait 
s\'electionn\'ee de mani\`ere purement al\'eatoire (cf. 
Sec.~\ref{sec:gr-hyper}).
Parmi ces $n$ clauses, $u$ sont non satisfaites avant le
renversement de la variable. A nouveau, faute de plus d'informations,
on suppose que chacune des $n$ clauses a la probabilit\'e
$E(T)/M$ d'\^etre non satisfaite. $u$ a donc une distribution
binomiale de param\`etre
$E(T)/M$ parmi $n$ tentatives. Le renversement de la variable va permettre
de satisfaire les $u+1$ clauses non satisfaites autour de la variable 
consid\'er\'ee. Par ailleurs, certaines des $n-u$ clauses
satisfaites vont \^etre viol\'ees apr\`es le renversement. 
Cela concerne les clauses qui 
n'\'etaient satisfaites que par la variable flipp\'ee. Puisque l'on sait
seulement que ces clauses sont satisfaites, la probabilit\'e qu'elles le
soient par la variable renvers\'ee est $f = 1/(2^K-1)$ (cette notation
est introduite pour simplifier la discussion de la section \ref{sec:ws-XOR}).
En notant $\varphi=E/M$, on obtient donc~:
\begin{eqnarray}
W(E+\Delta,E)& = & \sum_{n=0}^\infty e^{-\alpha K}\frac{(\alpha K)^n}{n!} 
\sum_{u=0}^n
{n \choose u} \varphi^u (1-\varphi)^{n-u} \label{eq:ws-Wmarkov}\\
& &\hspace{1.5cm} \sum_{s'=0}^{n-u} {n-u \choose s'}
f^{s'} (1-f)^{n-u-s'}
\delta_{\Delta ,s'-u-1} \nonumber \ ,
\end{eqnarray}
qui a la forme $w(\Delta,\varphi)$ avec $\Delta$ d'ordre 1, et
une d\'ependance r\'eguli\`ere de $w$ en $\varphi$.

Une fois \'etablie cette approximation markovienne pour l'\'evolution
de $E(T)$, on peut utiliser le formalisme g\'en\'eral 
du chapitre. Commen\c cons par calculer le comportement typique de ce
processus stochastique.

\begin{figure}
\includegraphics[width=9cm]{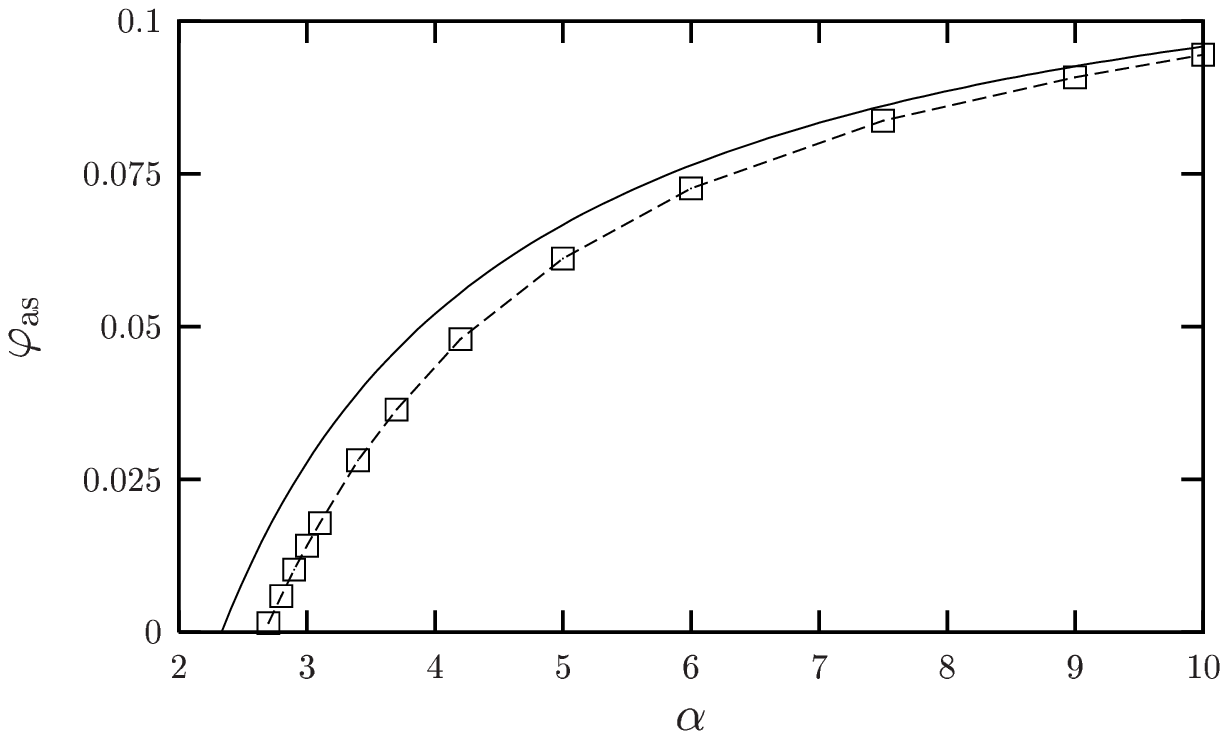}
\caption{La hauteur du plateau dans l'approximation markovienne (cf. eq. 
(\ref{eq:ws-markov-plateau})) pour $K=3$ (ligne pleine), compar\'ee \`a 
des simulations num\'eriques sur des formules de taille $N=10^5$ (symboles
reli\'es par la ligne pointill\'ee, les barres d'erreur sont plus petites que
les symboles).}
\label{fig:ws-plateau}
\end{figure}

La variation moyenne de $E$ au cours d'un pas de temps d\'emarrant \`a 
$E \sim M \varphi$ est
\begin{equation}
[\Delta]_{\varphi}= \sum_{\Delta=-\infty}^{\infty} w(\Delta,\varphi) \ \Delta =
-1 - \alpha K \varphi + f \alpha K (1-\varphi) \ .
\end{equation}
L'\'evolution moyenne $\varphi(t)$ est donc solution de l'\'equation 
diff\'erentielle ordinaire
\begin{equation}
\frac{d}{dt} \varphi (t) = [\Delta]_{\varphi(t)} = 
(- 1 + \alpha K f) - \alpha K (1+f) \varphi (t) \ . 
\label{eq:ws-eqsurphi}
\end{equation}
La condition initiale $\varphi(0)=2^{-K}$ (chaque clause a probabilit\'e 
$2^{-K}$ d'\^etre viol\'ee par une configuration al\'eatoire des variables)
peut se r\'ecrire $\varphi(0)=f/(1+f)$. On a donc
\begin{equation}
\varphi(t)=\frac{f}{1+f} \left( 1 + \frac{\alpha_d(K)}{\alpha} 
\left( e^{-\alpha K (1+f)t} - 1 \right) \right) \ , \qquad 
\alpha_d(K) = \frac{1}{f K} \ .
\end{equation}
Cette expression est en accord qualitatif avec les observations num\'eriques~:
pour $\alpha<\alpha_d(K)$, l'\'energie s'annule au bout d'un temps fini
\begin{equation}
t_{\rm sol}(\alpha,K) = \frac{-1}{\alpha K (1+f)} 
\ln \left( 1 -\frac{\alpha}{\alpha_d(K)} \right) \ ,
\label{eq:ws-markovtsol}
\end{equation}
alors que pour $\alpha > \alpha_d(K)$ on a une valeur asymptotique positive,
\begin{equation}
\varphi_{\rm as} = \frac{f}{1+f} \left(1 - 
\frac{ \alpha_d(K)}{\alpha} \right) \ .
\label{eq:ws-markov-plateau}
\end{equation}
Comme ici $f= 1/(2^K-1)$, la valeur du seuil pr\'evu dans le cadre de
cette approximation est
\begin{equation}
\alpha_d(K)= \frac{2^K - 1}{K} \ , 
\end{equation}
soit $7/3$ pour $K=3$ au lieu de la valeur observ\'ee num\'eriquement d'environ
$2.7$. La figure \ref{fig:ws-plateau} pr\'esente le r\'esultat de ce calcul
pour la valeur du plateau, compar\'e aux r\'esultats de simulations 
num\'eriques. L'accord n'est certes pas parfait, ce qui ne saurait \^etre
surprenant vu l'approximation que l'on a faite ici, mais n'est pas non plus
compl\`etement d\'eraisonnable.

\subsection{Calcul approch\'e des grandes d\'eviations}
\label{sec:ws-approxlarge}

Il reste maintenant \`a \'etudier les fluctuations autour du plateau dans 
le r\'egime $\alpha > \alpha_d(K)$. On s'int\'eressera en particulier
aux grandes d\'eviations qui conduisent \`a la d\'ecouverte d'une solution.

Restant dans le cadre de l'approximation markovienne introduite ci-dessus,
on remplace l'\'evolution microscopique du syst\`eme par les probabilit\'es 
de transition de l'\'equation (\ref{eq:ws-Wmarkov}).
L'\'etude des propri\'et\'es g\'en\'erales des processus markoviens locaux 
de la section \ref{sec:pml} nous a montr\'e comment calculer les grandes 
d\'eviations d'un tel processus. Introduisons donc la fonction de
grande d\'eviation $\pi$ comme
\begin{equation}
\mbox{Prob}(E,T) \sim \exp[-M \pi(\varphi,t)] \ , \quad \varphi=\frac{E}{M} \ ,
\quad t=\frac{T}{M} \ .
\end{equation}
Pour suivre les conventions utilis\'es dans cette partie on prend $M$ au lieu
de $N$ comme grand param\`etre, ce qui ne change pas la discussion puisque $M$
et $N$ sont du m\^eme ordre. La transform\'ee de
Legendre de $\pi$, not\'ee $g(\lambda,t)$, est solution de l'\'equation
aux d\'eriv\'ees partielles (\ref{eq:pml-eqsurg}) o\`u
\begin{equation}
e^{\ell(\lambda,\varphi)} = \sum_{\Delta=-\infty}^{\infty} w(\Delta,\varphi)
\ e^{\lambda \Delta} \ .
\end{equation}
On calcule ais\'ement la fonction $\ell$ \`a partir de la forme 
(\ref{eq:ws-Wmarkov}) pour $w$, ce qui conduit \`a l'\'equation suivante 
sur $g$~:
\begin{equation}
\frac{\partial}{\partial t} g(\lambda,t) = - \lambda 
- \alpha K f \left( 1-e^\lambda \right)
- \alpha K \left(1-f+f e^\lambda - e^{-\lambda}\right)  
\frac{\partial}{\partial \lambda} g(\lambda,t) \ .
\label{eq:ws-eqsurg}
\end{equation}
A titre de v\'erification, on peut r\'eobtenir l'\'equation 
(\ref{eq:ws-eqsurphi}) sur la valeur moyenne avec $\varphi = \partial g/ 
\partial \lambda|_{\lambda=0}$. L'\'equation aux d\'eriv\'ees partielles sur
$g$ doit \^etre compl\'et\'ee par une condition initiale. Les clauses
\'etant initialement non satisfaites avec probabilit\'e $f/(1+f)$, on a
\begin{equation}
g(\lambda,t=0) = \ln \left( 1+ \frac{f}{1+f} \left( e^\lambda - 1 \right) 
\right) \ .
\end{equation}
Notons $g_{\rm as}$ la solution stationnaire de (\ref{eq:ws-eqsurg}), atteinte
dans la limite $t \to \infty$. En utilisant le fait que $g$ est nulle pour
$\lambda =0$, on peut exprimer cette solution stationnaire comme
\begin{equation}
g_{\rm as}(\lambda) = - \int_0^\lambda dx \ \frac{x + \alpha K f (1-e^x)}
{\alpha K\left( 1 -f+ f e^x - e^{- x} \right) } \ .
\label{eq:ws-gas}
\end{equation}
Par transform\'ee de Legendre inverse de cette fonction on  
reconstruit alors la fonction de grande d\'eviation pour des temps longs, 
$\pi_{\rm as}(\varphi)$. En particulier, la probabilit\'e d'une fluctuation
vers une configuration d'\'energie nulle, 
$\epsilon_N=\exp[-M \pi_{\rm as}(0)]$,
s'obtient \`a partir de
\begin{equation}
\pi_{\rm as}(0) = - \min_{\lambda} g_{\rm as}(\lambda) = -g(\lambda^*) \ ,
\end{equation}
o\`u $\lambda^*$ est la valeur de $x$ qui annule l'int\'egrand de
l'\'equation (\ref{eq:ws-gas}). On trouve ainsi que $\pi_{\rm as}(0)$,
qui est nul pour $\alpha=\alpha_d$, cro\^it continument avec $\alpha$~:
plus l'\'energie du plateau est \'elev\'ee, moins probable est une fluctuation
jusqu'\`a 0.

Comme dans le calcul du temps d'ergodicit\'e du mod\`ele de Curie-Weiss,
les temps de r\'esolution par l'interm\'ediaire de ces fluctuations vont
\^etre inversement proportionnel \`a la probabilit\'e de ces grandes
fluctuations, et doivent donc diverger comme $\exp[M \pi_{\rm as}(0)]$.
On s'attend plus pr\'ecis\'ement \`a observer une distribution exponentielle
des temps de r\'esolution, de moyenne $\exp[M \pi_{\rm as}(0)]$.
Le raisonnement est le suivant. La probabilit\'e $\epsilon_N$ que l'on vient
de calculer est celle de d\'ecouverte d'une solution sur un intervalle
de temps $t$ grand mais fini dans la limite thermodynamique, autrement
dit en un nombre de pas lin\'eaire dans la taille du syst\`eme. D\'ecoupons
donc l'axe des temps $t$ en segments de longueur $C$, avec $C \gg 1$ mais 
ind\'ependant de
$N$, et supposons que sur chacun de ces intervalles une solution est trouv\'ee
avec probabilit\'e $\epsilon_N$, si elle n'a pas \'et\'e trouv\'e avant,
auquel cas l'algorithme se serait arr\^et\'e. 
La probabilit\'e que l'on obtienne une solution dans le $k+1$-\`eme 
intervalle est donc $\epsilon_N (1-\epsilon_N)^k$. 
Comme $\epsilon_N$ est 
exponentiellement petit, on doit avoir $k$ exponentiellement grand pour que
la limite thermodynamique de cette probabilit\'e ne soit pas trivialement
nulle. On a alors une distribution des temps de r\'esolution qui suit une loi
exponentielle ayant pour moyenne $\epsilon_N^{-1}$. Dans tout ce qui 
pr\'ec\`ede on a n\'eglig\'e tous les pr\'efacteurs, ce r\'esultat ne doit 
donc \^etre que l'ordre dominant dans la limite thermodynamique.

\begin{figure}
\includegraphics[width=9cm]{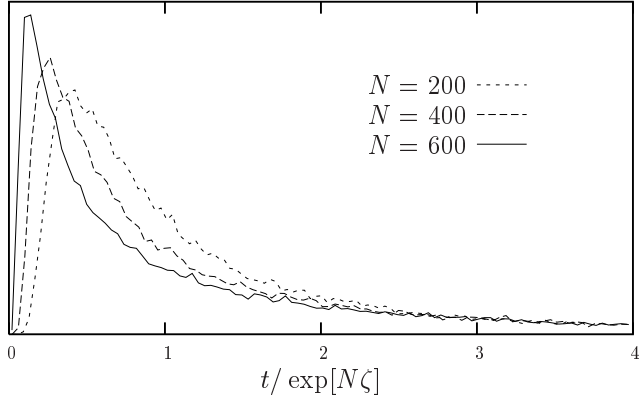}
\caption{Histogramme des temps de r\'esolution pour $\alpha=3>\alpha_d$,
sur l'\'echelle exponentiellement divergente $\exp[N\zeta]$. La
valeur de $\zeta$ a \'et\'e ajust\'ee pour faire se superposer les queues aux
diff\'erentes tailles.}
\label{fig:ws-Exp}
\end{figure}

La figure \ref{fig:ws-Exp} pr\'esente un histogramme des temps de r\'esolution
dans la phase $\alpha>\alpha_d$, pour diff\'erentes tailles des formules.
Le comportement est en accord qualitatif avec les pr\'evisions analytiques,
les temps typiques divergent exponentiellement avec $N$. Cependant le
taux $\zeta$ de divergence, qui devrait valoir $\alpha \pi_{\rm as}(0)$,
n'est pas quantitativement pr\'edit par l'approximation markovienne. 

\subsection{XORSAT}
\label{sec:ws-XOR}

XORSAT~\cite{CrDa-XORSAT} est une variante du probl\`eme de la
satisfiabilit\'e dans laquelle les variables au sein d'une clause sont
reli\'es par des OU EXCLUSIF logiques \`a la place du OU standard. En termes
de variables d'Ising ceci correspond \`a une interaction habituelle o\`u
l'\'energie est le produit de $K$ spins. Une clause est satisfaite si et 
seulement si le produit des spins de la clause est \'egal \`a une certaine 
variable gel\'ee \`a $\pm 1$. Parmi les $2^K$ configurations
des variables d'une clause, la moiti\'e la satisfait, l'autre moiti\'e
viole cette contrainte. Du point de vue de la complexit\'e dans le pire
des cas, XORSAT est un probl\`eme facile. On peut en effet le reformuler
comme un probl\`eme d'alg\`ebre lin\'eaire modulo 2, soluble par
des algorithmes polynomiaux.

Un ensemble al\'eatoire de formules de XORSAT peut \^etre d\'efini exactement
de la m\^eme fa\c con que pour la $K$-satisfiabilit\'e, les $K$ variables
de chaque clause \'etant choisies uniform\'ement parmi les ${N \choose K}$
$K$-uplets possibles. Il y a dans ce cas aussi un ph\'enom\`ene de seuil~:
pour $\alpha<\alpha_c$ presque toutes les formules ont des solutions, pour
$\alpha>\alpha_c$ presqu'aucune formule n'en a. La valeur du seuil est
$\alpha_c=0.918$ quand $K=3$.
Le clustering des solutions dans la phase satisfiable est aussi pr\'esent,
et se produit \`a $\alpha=0.818$. La pr\'esence de ces deux ph\'enom\`enes,
ainsi que les valeurs des seuils, ont \'et\'e prouv\'ees
rigoureusement~\cite{CoDuMaMo-XOR,MeRiZe-XOR}.

On peut utiliser l'algorithme PRWSAT sur les formules de XORSAT. Les 
simulations num\'eriques sur des formules al\'eatoires pr\'esentent le m\^eme 
type de comportement, avec une phase de basse concentration o\`u une solution
est trouv\'ee en un nombre lin\'eaire de pas de temps, alors qu'un plateau
dans l'\'energie se d\'eveloppe au dessus d'un seuil dynamique. L'\'etude
analytique dans le cadre de l'approximation markovienne est
tr\`es similaire \`a celle effectu\'ee pour la $K$-SAT. En fait la seule 
diff\'erence est que pour XORSAT, une clause satisfaite dont on renverse une
variable devient automatiquement non satisfaite. Autrement dit, toute
les formules des sections \ref{sec:ws-approxtyp} et \ref{sec:ws-approxlarge}
restent valable en prenant $f=1$. En particulier le seuil est dans ce cas
pr\'evu \`a $\alpha_d(K)=1/K$. L'accord est quantitativement meilleur pour
XORSAT, comme on pourra le constater sur les figures de la publication
\pubwsat .

\subsection{Limite de grand $K$}
\label{sec:ws-largeK}
L'approximation markovienne a permis de reproduire qualitativement les
r\'esultats des simulations num\'eriques, mais n'est \'evidemment pas
quantitativement exacte. De plus, cette approximation ne fait pas appara\^itre
de mani\`ere \'evidente un petit param\`etre qui contr\^olerait l'importance
des termes n\'eglig\'es. On peut cependant faire deux conjectures sur des
limites dans lesquelles l'approximation markovienne serait exacte.

Une premi\`ere situation est celle o\`u le ratio $\alpha =M/N$ devient tr\`es 
grand. Dans ce cas le calcul approch\'e pr\'edit que le plateau tend
vers $f/(1+f)$, qui n'est autre que la valeur de $\varphi$ pour une 
configuration
al\'eatoire des variables. Autrement dit, la formule est tellement 
surcontrainte que l'algorithme se contente de renverser des variables sans
faire d\'ecro\^itre l'\'energie. Ceci sugg\`ere donc que $1/\alpha$ pourrait
\^etre un petit param\`etre dans une am\'elioration syst\'ematique de
l'approximation. La courbe de la figure \ref{fig:ws-plateau} semble
indiquer que cette hypoth\`ese est correcte.

Une autre limite, plus riche, est celle d'un grand nombre $K$ de variables par
clauses. Le seuil dynamique est alors \'equivalent \`a $2^K/K$,
et l'on pose $\alpha = \alpha^* (2^K -1)/K $ pour avoir une limite non 
triviale.
Consid\'erons par exemple le temps de r\'esolution dans l'approximation 
markovienne donn\'e par l'\'equation (\ref{eq:ws-markovtsol}). Posons aussi
$t_{\rm sol}^*= 2^K t_{\rm sol}$. Il vient alors dans la limite 
$K \to \infty$~:
\begin{equation}
t_{\rm sol}^*(\alpha^*) = - \frac{1}{\alpha^*} \ln (1 - \alpha^*) = 1
+ \frac{1}{2} \alpha^* + \frac{1}{3} (\alpha^*)^2 + {\cal O}((\alpha^*)^3) \ . 
\end{equation}
Le point encourageant est que le d\'eveloppement en clusters
(\ref{eq:PRWSAT-dev_cluster_tsolK}), une fois exprim\'e en termes des 
quantit\'es r\'e\'echell\'ees $t_{\rm sol}^*$ et $\alpha^*$, conduit au m\^eme
r\'esultat dans la limite $K \to \infty$. Cette co\"incidence donne du
cr\'edit \`a l'hypoth\`ese d'exactitude de l'approximation markovienne dans
cette limite. 

\begin{figure}
\includegraphics[width=9cm]{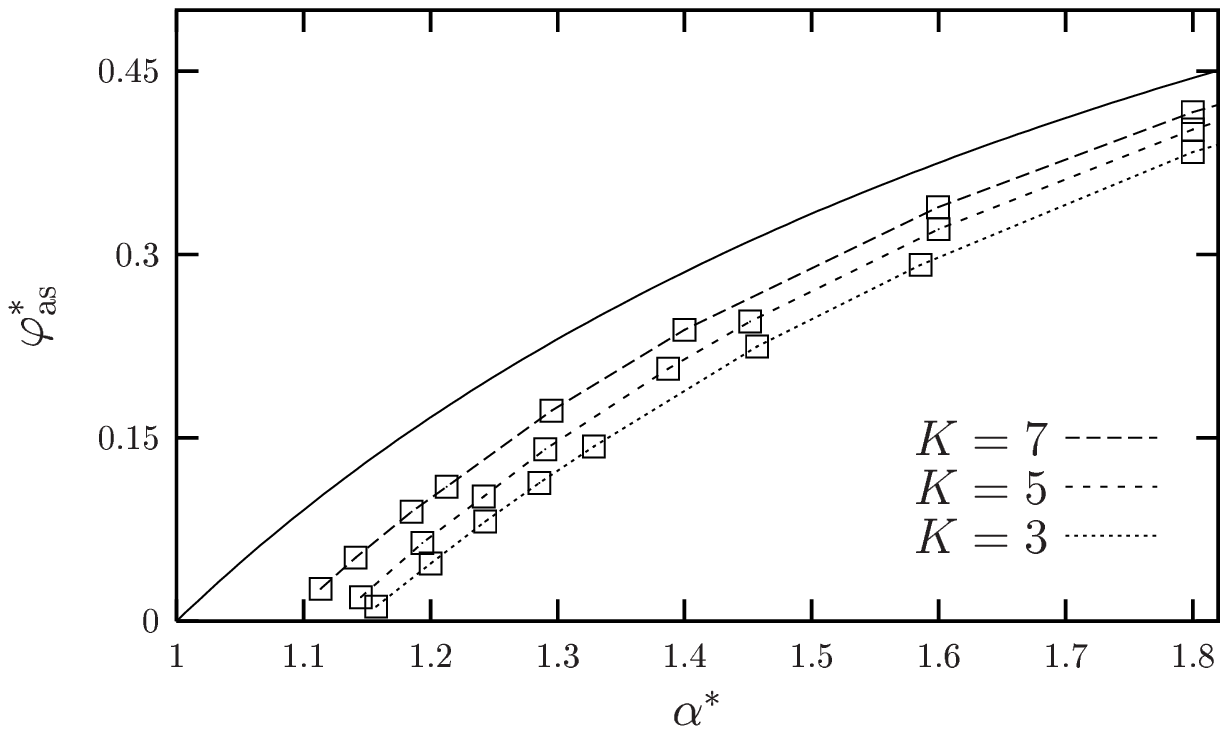}
\caption{La hauteur du plateau en unit\'es r\'e\'echell\'ees dans l'approximation
markovienne ($\varphi_{\rm as}^* = 1-(1/\alpha^*)$) en ligne pleine, 
compar\'ee \`a des simulations num\'eriques pour $K=3,5$ et 7.}
\label{fig:ws-plateau-largeK}
\end{figure}

La figure \ref{fig:ws-plateau-largeK} est un autre \'el\'ement en faveur de
cette hypoth\`ese. Elle pr\'esente des r\'esultats de simulations num\'eriques
pour la hauteur du plateau $\varphi_{\rm as}$ avec diff\'erentes valeurs de
$K$. On a pris les \'echelles $\varphi_{\rm as}^* = 2^K \varphi_{\rm as}$
et $\alpha^*$ pour pouvoir comparer les diff\'erentes valeurs de $K$ sur la
m\^eme courbe. On constate que quand $K$ augmente l'accord avec la pr\'ediction
de l'approximation markovienne s'am\'eliore.

\subsection{Perspectives}
\label{sec:ws-perspe}

Esquissons quelques directions dans lesquelles cette \'etude 
pourrait \^etre approfondie.

D'une part, une description 
quantitativement meilleure de la dynamique de cet algorithme serait 
souhaitable. Une possibilit\'e consiste \`a projeter sur une observable
macroscopique plus fine que celle utilis\'ee ici.
Dans la publication \pubwsat\ nous avons suivi cette id\'ee en distinguant
plusieurs types de clauses. La pr\'ecision de l'approximation en est
l\'eg\`erement am\'elior\'ee, mais la valeur pr\'edite pour le seuil
n'est pas modifi\'ee. Les auteurs de \cite{BaHaWe-wsat} ont utilis\'e
comme observable une fonction ressemblant au $p_\sigma(u;t)$ de la partie 
\ref{sec:ferro-arbre}. Cela conduisait \`a des r\'esultats meilleurs, au prix
de l'int\'egration num\'erique d'un jeu d'\'equations diff\'erentielles 
coupl\'ees. Il y a ici une difficult\'e suppl\'ementaire par rapport au
cas du ferromagn\'etique sur l'arbre de Bethe~: la connectivit\'e des
variables n'\'etant pas born\'e dans l'ensemble des formules al\'eatoires,
il appara\^it un nombre infini d'\'equations coupl\'ees. En pratique il faut
donc mettre une coupure pour les r\'esoudre num\'eriquement.
Une autre possibilit\'e consisterait \`a garder l'observable la plus simple
possible, mais \`a tenir compte des effets non markoviens dans son
\'evolution. On peut s'attendre alors \`a ce que l'\'equation donnant 
l'\'energie moyenne soit une g\'en\'eralisation de (\ref{eq:ws-eqsurphi}) 
du type
\begin{equation}
\frac{d}{dt} \varphi (t) = 
(- 1 + \alpha K f) - \alpha K (1+f) \varphi (t) + \int_0^t dt' \ F(t,t')
\varphi(t') \ ,
\end{equation}
o\`u $F(t,t')$ devrait s'exprimer de mani\`ere autocoh\'erente en termes
de $\varphi$. C'est ce type d'\'equation qui a \'et\'e obtenue
par Deroulers et Monasson dans leur \'etude du processus de 
contact~\cite{DeMo-CP}. Dans leur cas l'approximation markovienne 
correspondait \`a une limite de dimension $D$ infinie, et $F$ est 
d'ordre $1/D$. D'apr\`es la discussion de la partie \ref{sec:ws-largeK}
il est raisonnable de s'attendre ici \`a un noyau $F$ d'ordre $1/K$.
On peut esp\'erer obtenir un tel 
d\'eveloppement soit en adaptant la m\'ethode de \cite{DeMo-CP} 
\`a ce probl\`eme, plus difficile \`a cause de la
pr\'esence de d\'esordre gel\'e dans les r\`egles dynamiques,
soit en utilisant les op\'erateurs de projection de mani\`ere plus astucieuse.

Les appendices \ref{sec:ap-ws-ferrowsat} et \ref{sec:ap-ws-prwXfixed}
pr\'esentent deux petits calculs non publi\'es pour des variantes
simplifi\'ees du probl\`eme, qui pourraient constituer d'autres points
de d\'epart pour une am\'elioration syst\'ematique de l'approximation.

\vspace{4mm}

D'autre part, on peut aussi noter que le processus stochastique \'etudi\'e ici 
n'est que la version la plus simple d'une famille d'algorithmes, connue sous 
le nom g\'en\'erique de WalkSAT~\cite{SeKaCo}. Il est raisonnable de choisir 
une clause non satisfaite \`a chaque pas de temps d'un algorithme de
recherche locale, puisque n\'ecessairement une de ses variables
doit \^etre renvers\'ee avant qu'on trouve une solution. Mais on a une
tr\`es grande libert\'e pour choisir, au sein de cette clause, la variable \`a
renverser. Comme on l'a vu dans les \'enum\'erations de la partie 
\ref{sec:ws-devclu}, le choix purement al\'eatoire peut \^etre dangereux quand
il conduit \`a violer des clauses qui n'\'etaient jusque l\`a satisfaites
que par la variable que l'on va flipper. 
Des dizaines d'heuristiques diff\'erentes
ont \'et\'e invent\'ees pour am\'eliorer ce choix~\cite{AlSeKa,HoSt}.
Certaines d'entre elles sont \og markoviennes\fg , au sens o\`u elles
n'utilisent que l'information sur la configuration pr\'esente pour faire
le choix de la variable \`a flipper, d'autres gardent au contraire une
m\'emoire de l'\'evolution pass\'ee. On pr\'esente dans l'appendice 
\ref{sec:ap-ws-heur} deux de ces heuristiques plus \'elabor\'ees.

Au moins pour les heuristiques markoviennes, on peut s'attendre, et on a
en partie v\'erifi\'e num\'eriquement, \`a ce que l'image d\'ecrite ici 
d'un r\'egime \`a faible $\alpha$ o\`u les formules sont r\'esolues apr\`es 
un nombre de pas proportionnel \`a leur taille reste valable. Un argument en
faveur de cette hypoth\`ese est que le d\'eveloppement en cluster du temps de
r\'esolution reste faisable pour de telles heuristiques. La question se
pose alors naturellement de conna\^itre les seuils dynamiques des
diff\'erentes heuristiques, et de savoir s'il existe une barri\`ere
intrins\`eque, strictement inf\'erieure au seuil de satisfiabilit\'e, au del\`a
de laquelle aucun algorithme de recherche locale ne serait capable de trouver
une solution. Cette question, assez ouverte, est discut\'ee dans la revue
\pubptac . L'id\'ee selon laquelle le seuil de clustering doit jouer ce
r\^ole est partiellement bas\'e sur l'\'etude de la dynamique de Langevin
du mod\`ele $p$-spin sph\'erique~\cite{CuKu-sphe}. Toutefois, les algorithmes
de recherche locale ne v\'erifient pas de condition de balance d\'etaill\'ee,
il n'est donc pas \'evident que les caract\'eristiques du \og paysage 
d'\'energie libre\fg\ soient pertinents pour eux. De plus, une \'etude
r\'ecente de Montanari et Ricci-Tersenghi~\cite{MoRi-cooling} incite 
\`a reconsid\'erer les intuitions bas\'ees sur les r\'esultats du $p$-spin 
sph\'erique, qui semble \^etre un cas tr\`es particulier dans la famille
des mod\`eles d\'esordonn\'es en champ moyen.

\vspace{4mm}

Finalement, on peut faire deux remarques au vu de l'activit\'e dans la
communaut\'e informaticienne. Les formules de l'ensemble al\'eatoire
\'etudi\'e th\'eoriquement sont tr\`es diff\'erentes de celles rencontr\'ees
dans les applications pratiques de la satisfiabilit\'e. Ces derni\`eres
sont souvent produites par des logiciels qui convertissent un premier
probl\`eme en une formule de satisfiabilit\'e, celle-ci \'etant
ensuite soumise \`a l'algorithme proprement dit. Les formules vont donc porter 
la marque de cette traduction,
et une structure particuli\`ere, s\^urement tr\`es diff\'erente d'un
hypergraphe poissonien, doit appara\^itre. De plus, les
probl\`emes de d\'epart sont souvent d\'efinis dans un espace euclidien, par
exemple les probl\`emes de routage de circuit imprim\'e ont une structure
naturelle planaire. Il serait donc int\'eressant de d\'efinir un ensemble
al\'eatoire (car cela permet d'utiliser des m\'ethodes probabilistes) dont les
formules typiques ressembleraient un peu plus \`a celles int\'eressantes pour
les applications. Une \'etape interm\'ediaire consisterait peut-\^etre \`a
s'inspirer des graphes dits \og small-world\fg\ qui interpolent entre une
structure euclidienne et un voisinage al\'eatoire de type 
champ moyen~\cite{WaSt}. Une
solution encore plus satisfaisante de ce dilemme serait de pouvoir donner
des pr\'edictions formule par formule et non pas typiquement sur un
ensemble. L'utilisation de la m\'ethode de la cavit\'e dans l'algorithme
de survey propagation~\cite{MeZe-SP} semble ouvrir une porte dans cette
direction~; il n'est cependant pas encore \'evident que cette m\'ethode
soit g\'en\'eralisable
\`a des formules tr\`es diff\'erentes de l'ensemble
$K$-SAT al\'eatoire. La derni\`ere remarque concerne le probl\`eme dit
de MAX-K-SAT, qui consiste \`a trouver une configuration optimale,
c'est-\`a-dire minimisant le nombre de clauses viol\'ees, d'une formule non
satisfiable. C'est un probl\`eme tr\`es difficile; des versions moins
exigeantes, dites d'approximabilit\'e, se contentent de trouver une 
configuration satisfaisant plus qu'un certain pourcentage du nombre optimal
de contraintes satisfiables simultan\'ement. Les \'etudes num\'eriques de
PRWSAT pr\'esent\'ees ici montrent que cet algorithme trouve, en temps 
lin\'eaire et avec grande probabilit\'e pour des formules de l'ensemble
$K$-SAT al\'eatoire, une configuration violant moins que
$M(\varphi_{\rm as} + \epsilon )$ clauses, o\`u $\varphi_{\rm as}$ est la
hauteur du plateau et $\epsilon >0$ est arbitraire. Ce r\'esultat
serait peut-\^etre utile dans le contexte des algorithmes d'approximation.

\newpage 

\section{Appendice~: Une dynamique algorithmique exactement soluble}
\label{sec:ap-ws-ferrowsat}

Consid\'erons la variante suivante du mod\`ele de $K$-SAT al\'eatoire. 
Une formule est toujours constitu\'ee de $M$ clauses de longueur $K$, tir\'ees
uniform\'ement parmi les ${N \choose K}$ $K$-uplets possibles sur $N$ 
variables. La diff\'erence est que les lit\'eraux ne sont jamais ni\'es,
autrement dit la seule configuration qu'une clause interdit est celle o\`u les
$K$ variables qu'elle contient sont toutes fausses. Il est
clair que pour toute valeur de $\alpha=M/N$, la formule est satisfiable, 
puisqu'il suffit de prendre toutes les variables vraies (i.e. tous les spins 
$\sigma_i=+1$) pour obtenir une solution.

On peut \'etudier le comportement de l'algorithme PRWSAT sur ce mod\`ele-l\`a,
m\^eme si \'evidemment l'int\'er\^et en est assez acad\'emique. Supposons donc
qu'une configuration initiale al\'eatoire des spins soit choisie, et qu'\`a
chaque pas de temps une des clauses non satisfaites soit s\'electionn\'ee, puis
une de ses variables renvers\'ees.

D\'ecrivons l'\'etat du syst\`eme apr\`es $T$ pas de l'algorithme par $N(u,T)$,
nombre de variables entour\'ees par $u$ clauses non satisfaites. On posera
$t=T/N$ le temps continu dans la limite thermodynamique, et 
$p(u,t)=N(u,T=Nt)/N$ la fraction de variables de \og type\fg\ $u$. 
Les moyennes avec cette loi $p$ seront not\'ees~:
\begin{equation}
\langle \bullet \rangle_t = \sum_{u=0}^\infty \bullet \ p(u,t) \ .
\end{equation}
Comme c'est une clause non satisfaite qui est choisie \`a chaque pas de temps,
la variable renvers\'ee est de type $u$ avec une probabilit\'e proportionnelle
\`a $u p(u,t)$, donc par normalisation cette probabilit\'e est \'egale \`a
$u p(u,t)/\langle u \rangle_t$. La variable devient de type $0$ apr\`es son
renversement. Les $(K-1) u$ voisines sont
de type $u'$ avec probabilit\'e $u' p(u',t)/\langle u \rangle_t$ si l'on 
n\'eglige les corr\'elations entre voisins (cf. l'approximation des voisins
ind\'ependants de la partie \ref{sec:bethe-approx}) et voient leur type 
diminuer de $1$. Regroupant ces diff\'erentes contributions, on obtient 
l'\'equation d'\'evolution~:
\begin{equation}
\frac{d}{dt}p(u,t) = \delta_{u,0} - \frac{u p(u,t)}{\langle u \rangle_t} +
(K-1) \frac{\langle u^2 \rangle_t}{\langle u \rangle_t^2} [ (u+1) p(u+1,t)
- u p(u,t)] \ .
\label{eq:ferrowsat-evol}
\end{equation}
Calculons la condition initiale $p(u,0)$.
Avec probabilit\'e $1/2$ la variable est vraie ($\sigma_i=+1$), et donc elle
est de type $0$. Si la variable est fausse, elle appartient \`a $n$ clauses 
avec une loi de Poisson de param\`etre $\alpha K$, chacune de ces $n$ clauses
est non satisfaite avec probabilit\'e $2^{1-K}$ puisqu'il faut pour cela 
que les autres
variables soient elles aussi fausses. On a donc
\begin{eqnarray}
p(u,0) &=& \frac{1}{2} \delta_{u,0} + \frac{1}{2} \sum_{n=u}^\infty
e^{-\alpha K} \frac{(\alpha K)^n}{n!} {n \choose u} 
\left(2^{1-K}\right)^u \left(1-2^{1-K}\right)^{n-u} \\
&=& a(0) \delta_{u,0} + (1-a(0)) e^{-b(0)} \frac{b(0)^u}{u!} \ ,
\end{eqnarray}
avec $a(0)=1/2$ et $b(0)=\alpha K 2^{1-K}$. 

On trouve en fait une solution de l'\'equation d'\'evolution 
(\ref{eq:ferrowsat-evol}) avec
\begin{equation}
p(u,t)= a(t) \delta_{u,0} + (1-a(t)) e^{-b(t)} \frac{b(t)^u}{u!} \ ,
\end{equation}
o\`u les param\`etres $a$ et $b$ d\'ependent du temps. En ins\'erant cette
forme dans (\ref{eq:ferrowsat-evol}) on obtient des \'equations
diff\'erentielles pour $a$ et $b$, qui conduisent \`a
\begin{equation}
a(t) = \frac{1}{2} + t \quad , \quad
b(t) = -\frac{K}{K-1} + \left( \alpha K + 2^{K-1} \frac{K}{K-1} \right)
\left( \frac{1}{2} - t\right)^{K-1} \ ,
\end{equation}
et donc finalement la fraction de clauses non satisfaites 
$\varphi(t) = \langle u \rangle_t / (\alpha K)$ vaut
\begin{equation}
\varphi(t) = -\frac{1}{\alpha (K-1)} \left( \frac{1}{2} -t \right)
+\left( 1 + \frac{2^{K-1}}{\alpha (K-1)} \right) 
\left( \frac{1}{2} -t \right)^K \ .
\label{eq:ferrowsat-phi}
\end{equation}
Cette fonction d\'ecro\^it de $2^{-K}$ \`a $t=0$ jusqu'\`a $0$ pour 
$t=t_{\rm sol}$, avec
\begin{equation}
t_{\rm sol} = \frac{1}{2} - 
\frac{1}{\left(2^{K-1}+ \alpha(K-1)\right)^{1/(K-1)}} \ .
\end{equation} 

Les simulations num\'eriques sont en parfait accord avec le calcul. 
Il est sans doute exact~: les variables \'etant flipp\'ees au maximum une 
fois, 
il n'y a pas de corr\'elation entre clauses voisines et l'approximation
des voisins ind\'ependants doit \^etre correcte. Une autre mani\`ere de
rendre rigoureux ce calcul consiste \`a voir l'\'evolution de l'algorithme
comme un processus de d\'ecimation de graphe. En effet, au bout de $T$ pas
de temps le nombre de variables fausses (de spins $-1$) est
$N(T)=N\times(\frac{1}{2}-t)$, car les variables flipp\'ees le sont toujours 
de fausses vers vraies. De plus, l'ensemble des clauses fausses forme un 
hypergraphe poissonnien de $M(T)$ clauses sur $N(T)$ variables. A chaque
pas de temps on enl\`eve une variable, et $n+1$ clauses, o\`u $n$ est
tir\'ee avec une loi de Poisson de param\`etre $K \alpha(t)$~: 
$\alpha(t)=M(T)/N(T)$ est la densit\'e de clauses dans le sous-graphe
compos\'e seulement des clauses viol\'ees. En calculant le nombre moyen
de clauses qui survivent \`a la d\'ecimation jusqu'\`a l'instant $t$, on
retrouve le r\'esultat (\ref{eq:ferrowsat-phi}).

L'int\'er\^et de ce calcul ne r\'eside pas dans le r\'esultat, mais dans
la possibilit\'e que l'on puisse s'en servir comme point de d\'epart pour un
d\'eveloppement perturbatif~: si dans la g\'en\'eration d'une formule de
$K$-SAT on choisit de nier un lit\'eral avec probabilit\'e 
$\epsilon/2$, l'ensemble al\'eatoire habituel correspond \`a $\epsilon=1$, 
le calcul exact que l'on vient de faire \`a $\epsilon=0$. Peut-\^etre 
pourrait-on syst\'ematiser un d\'eveloppement en puissances de $\epsilon$. 

\newpage

\section{Appendice~: Une deuxi\`eme variante}
\label{sec:ap-ws-prwXfixed}

Je voudrais pr\'esenter ici des r\'esultats num\'eriques et analytiques
concernant la dynamique de PRWSAT sur une autre variante du probl\`eme de
satisfiabilit\'e.

On va s'int\'eresser \`a des formules de $K$-XORSAT, en d'autres termes un
mod\`ele du type $K$-spin dilu\'e, mais au lieu de prendre un hypergraphe
poissonien on va utiliser un hypergraphe \`a connectivit\'e fixe~: chaque
variable appartient \`a $L=l+1$ clauses. Pour $L \ge K$ les formules ne sont
jamais satisfiables, on peut cependant s'int\'eresser \`a la dynamique de
PRWSAT sur ce probl\`eme, et en particulier \`a l'\'energie stationnaire
atteinte aux temps longs.

Je ne d\'etaillerais pas les calculs qui sont tr\`es proches de ceux
pr\'esent\'es dans le corps du chapitre. En d\'efinissant $\varphi(t)$ la
fraction de clauses non satisfaites apr\`es $Mt$ pas de temps, on obtient
dans l'approximation binomiale (c'est-\`a-dire l'approximation markovienne
pour une projection sur le nombre de clauses non satisfaites)~:
\begin{equation}
\varphi(t) = \frac{1}{2} + \frac{1}{2l} \left( e^{-2 l t}-1 \right) \ .
\end{equation}
Dans l'approximation des voisins ind\'ependants on suit $p(u)$, la fraction
de sites entour\'ees de $u$ clauses frustr\'ees, dont on sous-entend la
d\'ependance temporelle pour all\'eger les notations. On d\'efinit aussi
\begin{equation}
\langle \bullet \rangle = \sum_{u=0}^L \bullet \ p(u) \ , \qquad \mbox{tel que}
\ \ \ \varphi = \frac{1}{L} \langle u \rangle \ .
\label{eq:ws-prwXfixed-1}
\end{equation}
On obtient \`a ce niveau d'approximation une \'equation d'\'evolution pour 
$p$~:
\begin{eqnarray}
\frac{d}{dt} p(u) &=& \frac{L}{K \langle u \rangle} [ - u p(u) + (L-u) p(L-u)] 
\nonumber \\
&+& \frac{L (K-1) \langle u^2 \rangle  }{K \langle u \rangle^2 } 
[ (u+1) p(u+1) - u p(u) ] \nonumber   \\
&+& \frac{L (K-1) \langle u (L-u) \rangle }{K \langle L-u \rangle 
\langle u \rangle} [ (L-u+1) p(u-1) - (L-u) p(u) ] \ ,
\label{eq:ws-prwXfixed-2}
\end{eqnarray}
que l'on peut int\'egrer num\'eriquement.
\begin{figure}[h]
\includegraphics[width=10cm]{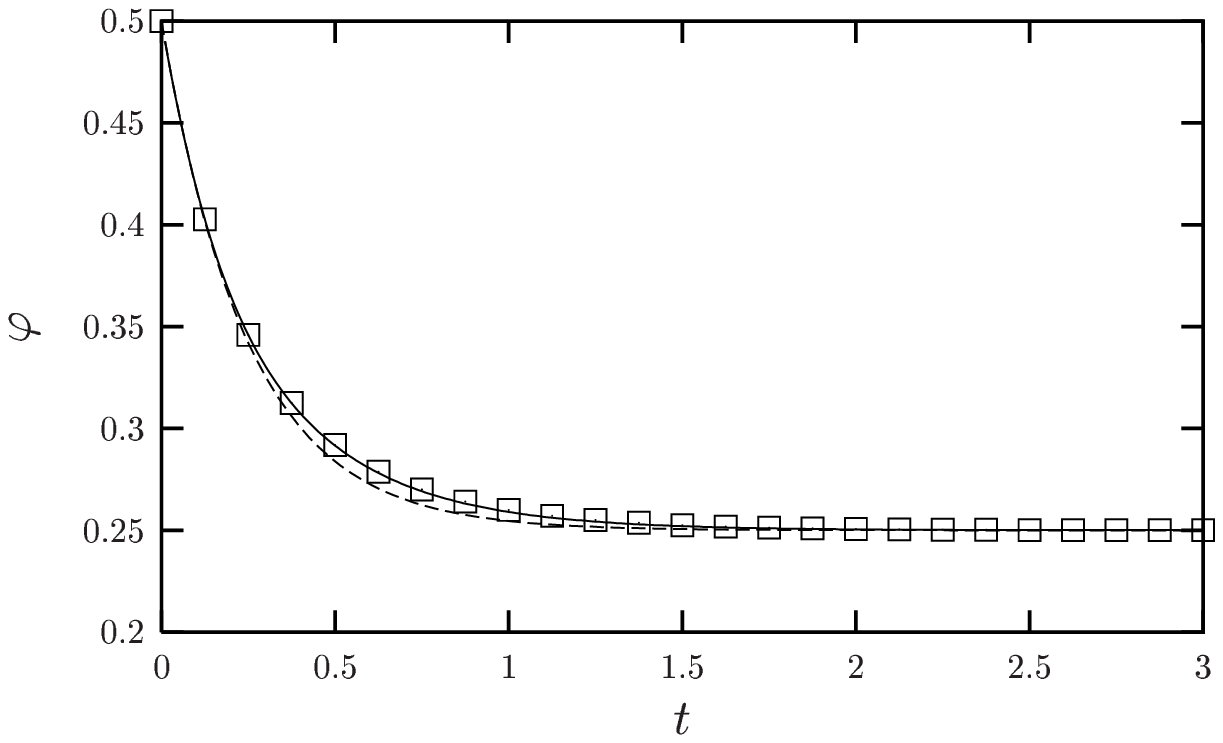}
\caption{Evolution de l'\'energie pour PRWSAT sur des formules de XORSAT \`a 
connectivit\'e fixe. Symboles~: moyennes sur 100 simulations avec $K=3$,
$L=3$, pour des formules de $N=10^7$ variables, les barres d'erreur sont
plus petites que les symboles. Ligne pointill\'ee~: approximation binomiale
(\ref{eq:ws-prwXfixed-1}). Ligne pleine~: approximation des voisins
ind\'ependants (\ref{eq:ws-prwXfixed-2}).}
\label{fig:ws-prwXfixed}
\end{figure}

La figure \ref{fig:ws-prwXfixed} pr\'esente les r\'esultats des simulations
num\'eriques compar\'ees \`a ces deux niveaux d'approximation. Comme l'on
pouvait s'y attendre, la deuxi\`eme approximation conduit \`a un meilleur
accord avec les simulations num\'eriques. Pour $t \approx 0.5$ il reste
cependant une erreur syst\'ematique, trop petite pour \^etre vue sur la
figure.

Le point qui me para\^it le plus int\'eressant est la limite des temps
longs. L'approximation binomiale pr\'edit
\begin{equation}
\varphi_{\rm as} = \lim_{t \to \infty} \varphi(t) = \frac{L-2}{2(L-1)} \ .
\label{eq:ws-prwXfixed-as}
\end{equation}
La solution stationnaire des \'equations (\ref{eq:ws-prwXfixed-2}) n'a
\emph{pas} une forme binomiale (sauf dans la limite $K \to \infty$), 
mais conduit cependant \`a la \emph{m\^eme}
fraction de clauses viol\'ees que l'approximation binomiale. De plus, ce
r\'esultat (\ref{eq:ws-prwXfixed-as}) est compatible, aux fluctuations de
taille finie pr\`es, avec les simulations num\'eriques. Ces derni\`eres
\'etant \'et\'e r\'ealis\'es sur des syst\`emes de tr\`es grande taille
($N=10^7$, on peut difficilement faire plus \`a cause de la taille de la
m\'emoire vive des ordinateurs actuels), il est raisonnable de faire la
conjecture que le r\'esultat (\ref{eq:ws-prwXfixed-as}) est exact.

On aurait d\'etermin\'e 
alors une des caract\'eristiques de l'\'etat stationnaire de ce
processus hors-d'\'equilibre (rappelons que la dynamique de PRWSAT ne
v\'erifie pas la condition de balance d\'etaill\'ee), bien que l'\'etat
stationnaire soit non trivial, la forme de $p$ n'\'etant pas
binomiale. Dans une comparaison un peu hasardeuse, on peut faire un 
rapprochement avec le processus d'exclusion compl\`etement asym\'etrique~:
le diagramme des phases de ce dernier probl\`eme est correctement pr\'edit par
une approximation de champ moyen. Pourtant son \'etat stationnaire est plus
riche que ne le sugg\`ere la solution de champ moyen, comme l'indique
sa repr\'esentation exacte en termes de produits de matrices v\'erifiant
une certaine alg\`ebre~\cite{TASEP}. On peut se demander si cette similitude
est purement fortuite ou si les m\'ethodes d\'evelopp\'ees dans le cadre
du processus d'exclusion pourrait \^etre adapt\'ees aux probl\`emes
algorithmiques.

\newpage

\section{Appendice~: Deux heuristiques plus performantes}
\label{sec:ap-ws-heur}

\vspace{8mm}

\noindent\underline{\emph{L'algorithme WalkSAT/SKC }}

\vspace{4mm}

Comme on l'a discut\'e dans la section \ref{sec:ws-perspe}, le choix de
la variable renvers\'ee dans une clause non satisfaite peut donner lieu \`a
des strat\'egies plus ou moins raffin\'ees. De mani\`ere g\'en\'erale, 
leur objectif est d'\'eviter de flipper une variable lorsque celle-ci
appartient \`a une clause satisfaite seulement par la variable en
question.

\begin{figure}[h]
\includegraphics[width=10cm]{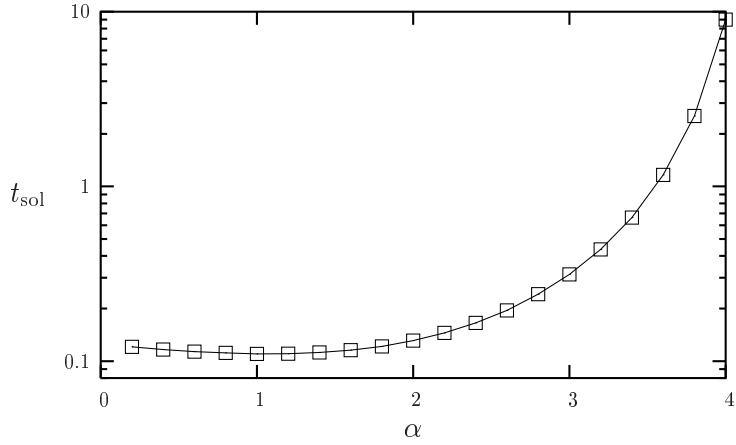}
\caption{Temps de r\'esolution dans la phase lin\'eaire pour l'heuristique 
WalkSAT/SKC. Simulations num\'eriques r\'ealis\'ees sur des formules de 
$N=10^6$ variables pour $K=3$, la param\`etre de bruit valant $p=0.5$. La 
courbe sert seulement de guide pour l'\oe il.}
\label{fig:ws-skc-time}
\end{figure}

Parmi les heuristiques que l'on a qualifi\'ees de markoviennes, consid\'erons
celle nomm\'ee WalkSAT/SKC~\cite{SeKaCo}~: 
\`a chaque pas de temps de l'algorithme une
des clauses non satisfaites est choisie au hasard. On examine ensuite ses
$K$ variables, et pour chacune on d\'efinit son \emph{breakcount} comme 
le nombre de clauses satisfaites qui deviendraient non satisfaites si elle
\'etait renvers\'ee. S'il y a des variables
qui ont un \emph{breakcount} nul, on flippe au hasard une de celles-ci. Si
au contraire elles ont toutes un \emph{breakcount} strictement positif~:
\begin{itemize}
\item Avec une certaine probabilit\'e $p$, on flippe une des $K$ variables
au hasard.
\item Avec probabilit\'e $1-p$, on flippe une des variables ayant un
\emph{breakcount} minimal.
\end{itemize}
Des exp\'eriences num\'eriques montrent qu'au moins pour $\alpha \le 4$, le
nombre de pas $T_{\rm sol}$ n\'ecessaires pour trouver une solution 
\`a une formule al\'eatoire cro\^it lin\'eairement avec la taille des 
formules. On a repr\'esent\'e sur la figure \ref{fig:ws-skc-time} les
moyennes de ces temps d\'etermin\'es num\'eriquement, 
en utilisant \`a nouveau l'unit\'e $t_{\rm sol}=T_{\rm sol}/M$. 

Comme dans le cas de PRWSAT, la distribution de $t_{\rm sol}$ est fortement
piqu\'ee autour de sa valeur moyenne dans la limite thermodynamique.
On peut aussi calculer cette fonction ordre par ordre en $\alpha$ 
gr\^ace \`a la m\'ethode du d\'eveloppement en 
clusters, le temps de r\'esolution moyen \'etant ici encore une fonction 
additive gr\^ace au caract\`ere markovien de l'heuristique de choix de la
variable flipp\'ee.

Cette modification rend donc l'algorithme beaucoup plus performant, puisqu'on
est pass\'e de $\alpha_d \approx 2.7$ \`a $\alpha_d \ge 4$, donc un seuil
dynamique tr\`es proche du seuil de satisfiabilit\'e $\alpha_c \approx 4.26$.
Il est assez difficile d'estimer avec beaucoup de pr\'ecision le seuil 
$\alpha_d$. Les simulations num\'eriques sur des formules de taille $N=10^6$
sugg\`erent $\alpha_d \approx 4.19$ (cette valeur est aussi avanc\'ee 
dans~\cite{RRT}), mais paradoxalement ces tailles sont
peut-\^etre trop petites pour \^etre d\'ebarrass\'ees des effets de taille
finie (bien que l'on atteigne quasiment la limite de m\'emoire vive
des ordinateurs actuels). En effet, les calculs statiques~\cite{MoPaRi-stab} 
pr\'edisent pour ces valeurs de $\alpha$ des \og \'etats m\'etastables\fg\ 
dont l'\'energie pourrait \^etre de l'ordre de grandeur du bruit de 
taille finie. On ne peut donc pas trancher ici \`a propos de la pertinence 
ou pas des pr\'edictions statiques pour l'existence d'un seuil intrins\`eque
\`a tous les algorithmes de recherche locale.

\vspace{8mm}

\noindent\underline{\emph{L'algorithme RRT (Record to Record Travel)}}

\vspace{4mm}

Une autre variante, introduite par Seitz et Orponen~\cite{RRT}, 
tombe dans la classe des heuristiques non markoviennes. 
La modification de PRWSAT est la suivante~:
\begin{itemize}

\item On garde en m\'emoire au cours de l'\'evolution de l'algorithme la 
valeur $E_{\rm min}$, qui est la plus basse valeur de l'\'energie (nombre de
clauses viol\'ees) que l'on ait rencontr\'e depuis le d\'ebut de l'ex\'ecution.

\item A chaque pas de temps, on choisit une clause viol\'ee au hasard, ainsi
qu'une de ses $K$ variables uniform\'ement (comme dans PRWSAT). Cette variable
n'est renvers\'ee que si l'\'energie apr\`es le renversement est plus petite
que $E_{\rm min} + d$, o\`u $d$ est un param\`etre fix\'e pour toute 
l'\'evolution. Sinon on ne change pas la configuration des variables.
\end{itemize}

$d$ est une mesure du \og laxisme\fg\ de l'algorithme~: si ce param\`etre
est tr\`es grand la plupart des mouvements vont \^etre accept\'es.

\begin{figure}[h]

\includegraphics[width=12cm]{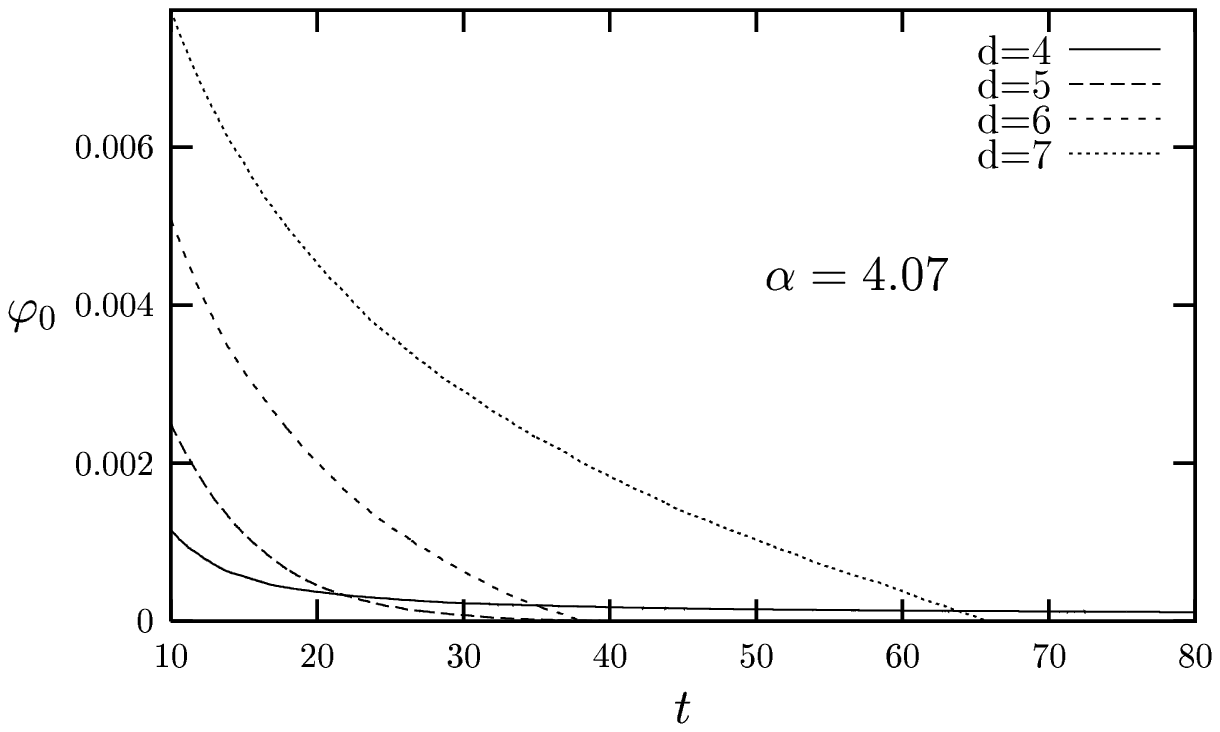}

\vspace{8mm}

\includegraphics[width=12cm]{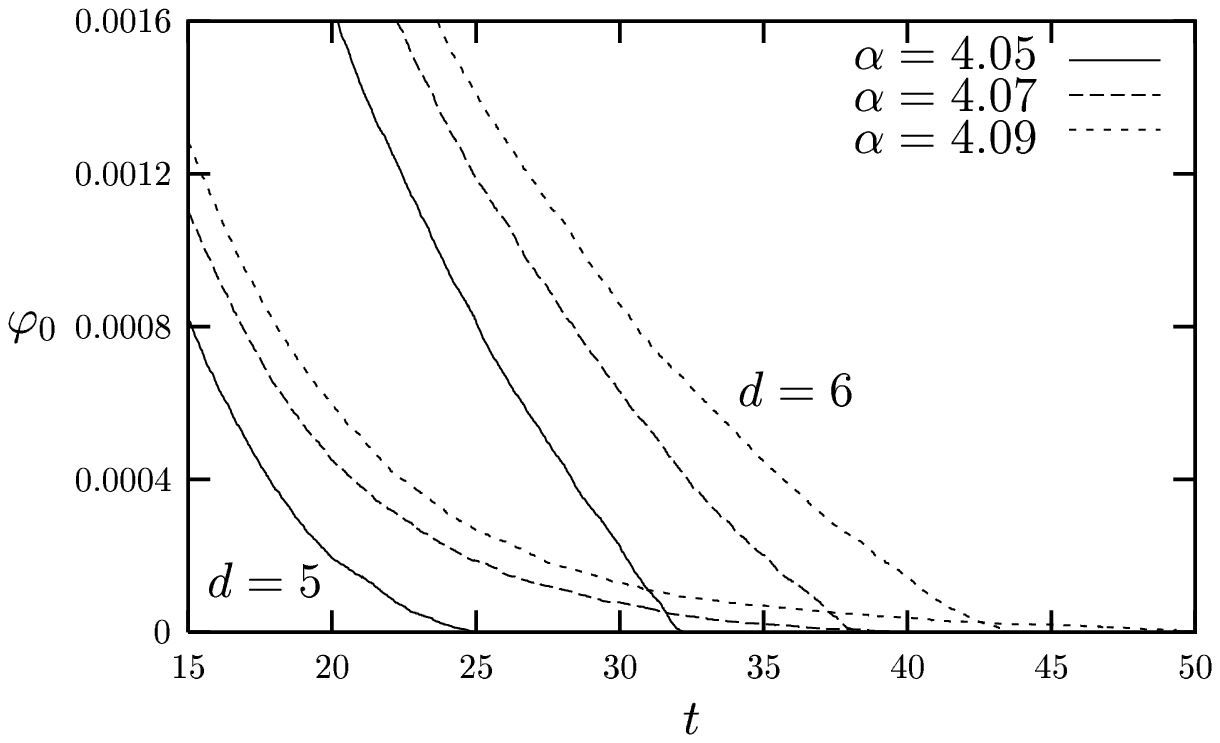}
\caption{D\'ecroissance de la fraction de clauses non satisfaites en fonction 
du temps pour l'algorithme RRT, \`a diff\'erentes valeurs de $\alpha$ et de 
$d$. Simulations num\'eriques r\'ealis\'ees sur des formules de 
$N=10^6$ variables pour $K=3$.}
\label{fig:ws-rrt}
\end{figure}

La figure \ref{fig:ws-rrt} montre l'\'evolution de l'\'energie pour 
diff\'erentes valeurs de $\alpha$ et de $d$. Examinons d'abord la figure
du haut, qui se concentre sur une seule valeur de $\alpha$. On a seulement
repr\'esent\'e la fin de l'\'evolution, pour que la figure soit plus claire.
Pour $t=10$, les \'energies croissent avec $d$~: \`a cet instant l'\'evolution
la moins permissive a conduit \`a la meilleure configuration. 
Cependant aux temps plus longs, cette courbe reste bloqu\'ee \`a une 
\'energie strictement positive, alors que les autres \'evolutions, 
plus laxistes, finissent par trouver une solution et s'av\`erent donc
plus efficaces. On constate cependant que la courbe avec le plus grand $d$ 
met le plus de temps pour trouver une solution. Il y a donc, pour une valeur 
de $\alpha$ donn\'ee, une tol\'erance $d$ optimale pour trouver une solution 
le plus rapidement possible. La partie du bas de la figure 
\ref{fig:ws-rrt} compl\`ete
cette description~: pour $\alpha=4.05$ c'est $d=5$ qui est optimal, pour 
$\alpha=4.07$ les choix $d=5$ et $d=6$ se valent, tandis que
pour $\alpha=4.09$ le param\`etre $d=6$ devient optimal.

\begin{figure}

\includegraphics[width=12cm]{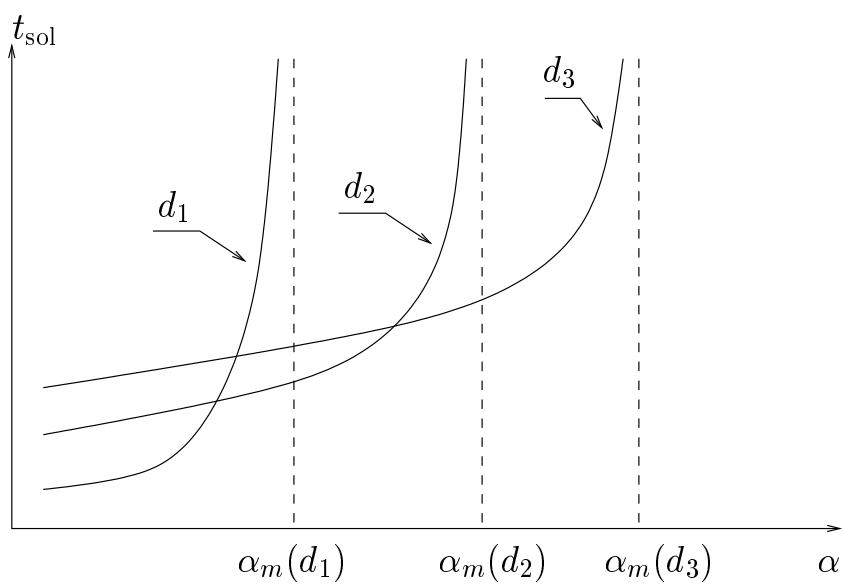}

\caption{Sch\'ematisation du comportement du temps de r\'esolution de RRT 
en fonction de $\alpha$ pour trois valeurs du param\`etre, $d_1<d_2<d_3$.}
\label{fig:ws-rrt-sketch}
\end{figure}

Le sch\'ema \ref{fig:ws-rrt-sketch} r\'esume le comportement de l'algorithme~:
pour une tol\'erance $d$ donn\'ee, le temps de r\'esolution (compt\'e dans
l'unit\'e r\'eduite nombre de pas divis\'e par nombre de clauses) cro\^it avec
$\alpha$ et diverge \`a une valeur $\alpha_m(d)$. Cette valeur seuil est
d'autant plus grande que l'\'evolution est permissive.

Dans leur article, Seitz et Orponen ont d\'etermin\'e $\alpha_m(d)$ pour $d$
entre 5 et 9 par un ajustement de donn\'ees exp\'erimentales sur la
divergence des temps de r\'esolution en fonction de $\alpha$, puis extrapol\'e 
le comportement de $\alpha_m(d)$ quand $d$ tend vers l'infini. Cela
conduit \`a $\alpha_m(\infty) \approx 4.26$, autrement dit cet algorithme
serait capable de r\'esoudre en un nombre de pas $Nf(\alpha)$  lin\'eaire 
dans le nombre de variables, jusqu'au seuil de satisfiabilit\'e $\alpha_c$ 
(avec cependant $f$ qui diverge \`a $\alpha_c$). Cette hypoth\`ese est assez
frappante, \`a nouveau de possibles effets de taille finie sont difficiles
\`a estimer ici.

Soulignons que $d$, qui est d'ordre 1, est une tol\'erance dans le nombre de 
clauses non satisfaites, lui m\^eme d'ordre $N$. Modifier d'une quantit\'e
finie la hauteur des \og barri\`eres\fg\ que l'on s'autorise \`a franchir 
modifie radicalement le comportement de cet algorithme, dont une description
analytique semble difficile \`a l'heure actuelle.

\chapter{Autour d'un th\'eor\`eme de fluctuation}
\label{sec:ch-ft}
\markboth{\hspace{3mm} \hrulefill \hspace{3mm} Ch. 5~: Autour d'un th\'eor\`eme de fluctuation}{Ch. 5~: Autour d'un th\'eor\`eme de fluctuation\hspace{3mm} \hrulefill \hspace{3mm} }

La dynamique des syst\`emes \`a l'\'equilibre thermique v\'erifie des
propri\'et\'es caract\'eristiques comme l'invariance par translation
dans le temps et le th\'eor\`eme de fluctuation-dissipation (FDT) qui
relie fonctions de r\'eponse et de corr\'elation. Parmi les diff\'erentes
familles de syst\`emes hors d'\'equilibre, les verres de spin pr\'esentent
des violations de ces deux propri\'et\'es d'un type particulier (vieillissement
et apparition de temp\'eratures effectives). Ce ph\'enom\`ene a \'et\'e
initialement \'etudi\'e dans le cadre des mod\`eles de champ moyen 
compl\`etement connect\'es. Pour cette famille de mod\`ele, toutes les 
fonctions de corr\'elation et de r\'eponse d\'ecoulent de celles \`a deux 
temps, comme on l'a vu dans la partie \ref{sec:msrgene}~\footnote{Ceci n'est 
en toute rigueur vrai que pour les mod\`eles
sph\'eriques, et pour les contributions dominantes dans la limite 
thermodynamique. Le calcul de la partie connexe des fonctions \`a quatre 
temps du mod\`ele SK sph\'erique peut se trouver dans~\cite{CaPaRa}. Dans ce 
mod\`ele compl\`etement connect\'e ces parties connexes sont des corrections 
d'ordre $1/N$.}.
En toute logique, celles-ci ont \'et\'e les objets d'\'etude principaux des
investigations th\'eoriques.

Dans le cas des mod\`eles dilu\'es, il faut en principe conna\^itre toute 
la hi\'erarchie des fonctions de corr\'elation et de r\'eponse pour 
caract\'eriser le syst\`eme. Cette constatation a motiv\'e l'\'etude 
pr\'esent\'ee dans ce chapitre, qui
a fait l'objet d'une partie de la publication \pubjsp . 

La premi\`ere partie rappelle des r\'esultats classiques
sur les propri\'et\'es d'\'equilibre des fonctions de corr\'elation et de
r\'eponse \`a deux temps. On \'etudie ensuite la g\'en\'eralisation
de ces propri\'et\'es aux fonctions \`a plus de deux temps.
La troisi\`eme partie est consacr\'ee \`a une version du th\'eor\`eme 
de fluctuation qui unifie toutes ces relations. Finalement
des conjectures sur la modification de ces r\'esultats pour des
syst\`emes hors d'\'equilibre du type verres de spin dilu\'es sont avanc\'ees.

\section{Propri\'et\'es d'\'equilibre des fonctions \`a deux temps}

\subsection{Enonc\'es}

Consid\'erons un syst\`eme physique en contact avec un thermostat \`a la
temp\'erature $T$. Pour deux observables g\'en\'eriques $A$ et $B$ du
syst\`eme, leur fonction de corr\'elation \`a deux temps est d\'efinie comme
\begin{equation}
C_{AB}(t,t') = \langle A(t) B(t') \rangle \ .
\end{equation}
La notation $\langle \bullet \rangle$ d\'esigne une moyenne d'ensemble,
c'est-\`a-dire sur la r\'ep\'etition d'un grand nombre de mesures.

Supposons que l'on rajoute un champ ext\'erieur $h(t)$, coupl\'e lin\'eairement
\`a l'observable $B$~: l'\'energie du syst\`eme est modifi\'ee par un
terme $-hB$. Dans le cas de syst\`eme de spins homog\`enes, il est naturel de 
consid\'erer un champ magn\'etique coupl\'e \`a la 
magn\'etisation totale. Pour des syst\`emes d\'esordonn\'es, il peut \^etre 
n\'ecessaire de consid\'erer des champs variables dans l'espace.

On notera $\langle \bullet \rangle_h$ les moyennes d'ensemble en pr\'esence
de la perturbation ext\'erieure. Si cette derni\`ere est suffisamment faible,
la th\'eorie de la r\'eponse lin\'eaire s'applique~:
\begin{equation}
\langle A(t) \rangle_h = \langle A(t) \rangle 
+ \int_{t_0}^{t} dt' \ R_{AB}(t;t') h(t')
+ {\cal O}(h^2) \ ,
\label{eq:fdt-integ}
\end{equation}
o\`u $t_0$ est le temps initial de pr\'eparation du syst\`eme. 
Cette relation d\'efinit la fonction de r\'eponse comme
\begin{equation}
R_{AB}(t;t') = \left. \frac{\delta}{\delta h(t')}  \langle A(t) \rangle_h
\right|_{h=0} \ .
\label{eq:fdt-defrab}
\end{equation}
La fonction de r\'eponse $R_{AB}(t;t')$ mesure donc la variation moyenne
de l'observable $A$ \`a l'instant $t$ pour une perturbation coupl\'ee \`a $B$ 
appliqu\'ee pendant un intervalle de temps infinit\'esimal autour de $t'$.

Enon\c cons maintenant les propri\'et\'es de ces fonctions de corr\'elation
et de r\'eponse.

\vspace{8mm}

\noindent\underline{\emph{Causalit\'e}\hspace{-1.5mm}\phantom{p}}

\vspace{4mm}

La borne sup\'erieure de l'int\'egration dans (\ref{eq:fdt-integ}) est
prise en $t$ par causalit\'e~: une perturbation ne peut modifier le syst\`eme
avant d'avoir \'et\'e appliqu\'ee. De mani\`ere \'equivalente, 
$R_{AB}(t;t')=0$ si $t' > t$. Ces exigences de causalit\'e restent valables
hors de l'\'equilibre.

\vspace{8mm}

\noindent\underline{\emph{Invariance par translation temporelle}}

\vspace{4mm}

Pour un syst\`eme \`a l'\'equilibre, toutes les fonctions de corr\'elation
et de r\'eponse sont invariantes par translation temporelle, en particulier
celles \`a deux temps ne sont fonction que de la diff\'erence entre les
deux temps, $C_{AB}(t,t')=C_{AB}(t-t')$ et 
$R_{AB}(t;t')=R_{AB}(t-t')$. Cette situation est v\'erifi\'ee si
le syst\`eme est pr\'epar\'e \`a l'instant initial dans une configuration
typique de la distribution de Gibbs-Boltzmann, ou bien si l'on laisse
le syst\`eme relaxer suffisamment longtemps apr\`es sa mise en contact
avec le thermostat.

\vspace{8mm}

\noindent\underline{\emph{Th\'eor\`eme de fluctuation-dissipation}}

\vspace{4mm}

Les fonctions de corr\'elation et de r\'eponse d'\'equilibre ne sont pas 
ind\'ependantes. Le th\'eor\`eme de fluctuation-dissipation impose
en effet la relation suivante~:
\begin{equation}
R_{AB}(\tau) = - \frac{1}{T} \frac{dC_{AB}(\tau)}{d\tau} 
\qquad \mbox{pour} \ \ \tau > 0 \ .
\end{equation}
Cette relation est assez remarquable car elle relie des propri\'et\'es de
natures diff\'erentes (\og fluctuation\fg\ : d\'ecroissance de la fonction
de corr\'elation au cours du temps en l'absence de perturbations ext\'erieures
{\em vs} \og dissipation\fg\ : r\'eponse du syst\`eme \`a une perturbation, 
dont le travail fourni doit \^etre dissip\'e).
Elle est par ailleurs ind\'ependante du syst\`eme consid\'er\'e, et ne fait 
intervenir que la temp\'erature du bain thermique.

\vspace{8mm}

\noindent\underline{\emph{Relation d'Onsager}}

\vspace{4mm}

Enfin, les relations de r\'eciprocit\'e d'Onsager impliquent
\begin{equation}
C_{AB}(t,t')=C_{BA}(t,t') \ ,
\end{equation}
ce qui n'est pas compl\`etement trivial si les observables $A$ et $B$ sont
distinctes.

\subsection{Une preuve}

Dans la publication \pubjsp\ ces propri\'et\'es sont d\'emontr\'ees pour une 
dynamique microscopique de variables continues qui \'evoluent selon 
l'\'equation de Langevin. Pour compl\'eter cette approche et insister sur la 
g\'en\'eralit\'e des r\'esultats on va utiliser une mod\'elisation
l\'eg\`erement diff\'erentes.

\vspace{8mm}

\noindent\underline{\emph{D\'efinitions}}

\vspace{4mm}

On suppose que le syst\`eme a un espace de configurations $\vec{\sigma}$ 
discr\`etes, et que l'influence du thermostat se traduit par une
\'evolution microscopique stochastique, selon l'\'equation ma\^itresse
en temps continu~:
\begin{equation}
\frac{d}{dt} P(\vec{\sigma},t) = \sum_{\vec{\sigma}'} 
W(\vec{\sigma},\vec{\sigma}') P(\vec{\sigma}',t) \ .
\end{equation}
$W(\vec{\sigma},\vec{\sigma}')$ repr\'esente le taux de transition entre les 
configurations $\vec{\sigma}'$ et $\vec{\sigma}$.
$P(\vec{\sigma},t)$ est la probabilit\'e que le syst\`eme soit dans la 
configuration $\vec{\sigma}$ \`a l'instant $t$, les moyennes d'ensemble seront
donc effectu\'es selon cette probabilit\'e. De plus les observables sont
de simples fonctions de la configuration, $A(t)=A(\vec{\sigma}(t))$.

La conservation des probabilit\'es implique une condition sur les
taux de transition,
\begin{equation}
\sum_{\vec{\sigma}} W(\vec{\sigma},\vec{\sigma}') = 0 \ .
\end{equation}
On prend donc pour les \'elements diagonaux de $W$~:
\begin{equation}
W(\vec{\sigma},\vec{\sigma}) = - \sum_{\vec{\sigma}' \neq \vec{\sigma}}
W(\vec{\sigma}',\vec{\sigma}) \ .
\end{equation}
Le bain thermique ext\'erieur au syst\`eme l'entra\^ine vers la
distribution d'\'equilibre de Gibbs-Boltzmann
$P_{\rm eq}(\vec{\sigma}) =(1/Z)\exp[-\beta H(\vec{\sigma})]$.
Pour que cette loi de probabilit\'e soit une solution stationnaire de 
l'\'equation ma\^itresse, on suppose que les taux de transition v\'erifient 
la condition de balance d\'etaill\'ee
\begin{equation}
W(\vec{\sigma},\vec{\sigma}') P_{\rm eq}(\vec{\sigma}') = 
W(\vec{\sigma}',\vec{\sigma}) P_{\rm eq}(\vec{\sigma}) \qquad 
\forall (\vec{\sigma},\vec{\sigma}') \ .
\end{equation}
Cette condition est suffisante (mais pas n\'ecessaire) pour que l'\'equilibre
thermique soit un point fixe de l'\'evolution. On reviendra plus tard sur
sa signification microscopique.

Introduisons aussi la probabilit\'e conditionnelle 
$P(\vec{\sigma},t | \vec{\sigma}',t')$ d'observer la configuration
$\vec{\sigma}$ \`a l'instant $t$ sachant que le syst\`eme \'etait en 
$\vec{\sigma}'$ \`a $t'$. Si les
taux de transition $W$ sont ind\'ependants du temps, cette probabilit\'e
conditionnelle n'est fonction que de la diff\'erence $t-t'$, on la notera
alors $T(\vec{\sigma},\vec{\sigma}',\tau=t-t')$. Elle v\'erifie
l'\'equation ma\^itresse sous la forme
\begin{equation}
\frac{d}{d\tau} T(\vec{\sigma},\vec{\sigma}',\tau) = \sum_{\vec{\sigma}''}
W(\vec{\sigma},\vec{\sigma}'') T(\vec{\sigma}'',\vec{\sigma}',\tau) \ .
\end{equation}

Explicitons avec ces notations une fonction de corr\'elation \`a deux temps,
pour $t>t'>t_0$,
\begin{equation}
\langle A(t) B(t') \rangle = 
\sum_{\vec{\sigma}_2,\vec{\sigma}_1,\vec{\sigma}_0}
A(\vec{\sigma}_2)  T(\vec{\sigma}_2,\vec{\sigma}_1,t-t') B(\vec{\sigma}_1)
T(\vec{\sigma}_1,\vec{\sigma}_0,t'-t_0) P_0(\vec{\sigma}_0) \ ,
\label{eq:fdt-excorrel}
\end{equation}
o\`u $P_0$ est la distribution de probabilit\'e au temps initial $t_0$.

On va introduire une notation matricielle plus compacte qui simplifiera les
g\'en\'eralisations de la partie suivante, et qui ressemble \`a celle 
utilis\'ee dans la publication \pubjsp . Consid\'erons des matrices 
indic\'ees par les configurations $\vec{\sigma}$, et en particulier 
$\hat{T}$ et $\hat{W}$ qui correspondent \`a la probabilit\'e conditionnelle 
et aux taux de transition~:
\begin{equation}
(\hat{T}(\tau))_{\vec{\sigma}_1 \vec{\sigma}_2} = 
T(\vec{\sigma}_1,\vec{\sigma}_2,\tau) \ , \qquad
\hat{W}_{\vec{\sigma}_1 \vec{\sigma}_2} = 
W(\vec{\sigma}_1,\vec{\sigma}_2) \ .
\end{equation}
Ces matrices sont reli\'ees par l'\'equation ma\^itresse
$d\hat{T}/d\tau=\hat{W}\hat{T}$, avec $\hat{T}(\tau=0)=\hat{1}$ la matrice
identit\'e. La solution de cette \'equation s'\'ecrit donc formellement
$\hat{T}(\tau)=\exp[\hat{W} \tau]$.

On notera aussi sous forme de matrices diagonales les observables et les
distributions de probabilit\'e,
\begin{equation}
\hat{A}_{\vec{\sigma}_1 \vec{\sigma}_2} = 
\delta_{\vec{\sigma}_1 \vec{\sigma}_2} A(\vec{\sigma}_1) \ , \qquad
(\hat{p}_0)_{\vec{\sigma}_1 \vec{\sigma}_2} = 
\delta_{\vec{\sigma}_1 \vec{\sigma}_2} P_0(\vec{\sigma}_1) \ , \qquad  
(\hat{p}_{\rm eq})_{\vec{\sigma}_1 \vec{\sigma}_2} = 
\delta_{\vec{\sigma}_1 \vec{\sigma}_2} P_{\rm eq}(\vec{\sigma}_1) \ .
\end{equation}
La notation de \og bra-kets\fg\ \`a la Dirac sera utile dans la suite. On
d\'efinit en particulier $\langle - |$ le vecteur ligne dont tous les 
\'el\'ements valent $1$, et $| - \rangle$ son transpos\'e. On a donc
\begin{equation}
\langle - | \hat{M} | - \rangle = \sum_{\vec{\sigma}_1 \vec{\sigma}_2}
\hat{M}_{\vec{\sigma}_1 \vec{\sigma}_2} \ .
\end{equation}
Ces d\'efinitions permettent de r\'e\'ecrire la fonction de 
corr\'elation (\ref{eq:fdt-excorrel}) comme
\begin{equation}
\langle A(t) B(t') \rangle = \langle - | \hat{A} \hat{T}(t-t') \hat{B} 
\hat{T}(t'-t_0) \hat{p}_0 |-\rangle \ .
\label{eq:fdt-excorrel2}
\end{equation}

Dans cette notation matricielle~:
\begin{itemize}
\item
la conservation des probabilit\'es s'exprime par~: $\langle - | \hat{W} =0$, et
de mani\`ere \'equivalente $^t \hat{W} |-\rangle =0$. 
$^t \bullet$ d\'esigne l'op\'eration de transposition matricielle.  
\item
la condition de balance d\'etaill\'ee devient
$\hat{p}_{\rm eq} {^t \hat{W}} = \hat{W} \hat{p}_{\rm eq}$. Comme 
$\hat{T}(\tau)= \exp[ \tau \hat{W} ]$, en d\'eveloppant l'exponentielle 
en s\'erie on a aussi
\begin{equation}
\hat{p}_{\rm eq} {^t \hat{T}(\tau)} = \hat{T}(\tau) \hat{p}_{\rm eq} \ .
\label{eq:fdt-bdsurT}
\end{equation}
\item 
la stationnarit\'e de la distribution de Gibbs-Boltzmann s'\'ecrit ici 
$\hat{W} \hat{p}_{\rm eq} |-\rangle = 0$,
et donc $\hat{T}(\tau) \hat{p}_{\rm eq} |-\rangle =\hat{p}_{\rm eq} |-\rangle$.
\end{itemize}

Il reste maintenant \`a consid\'erer l'effet d'un champ ext\'erieur $h$ 
coupl\'e lin\'eairement \`a une observable $B$ du syst\`eme.
Faisons l'hypoth\`ese que les taux de transition $\hat{W}_h$ en 
pr\'esence du champ v\'erifient la condition de balance d\'etaill\'ee par
rapport \`a ce nouvel hamiltonien,
\begin{equation}
\hat{W}_h \hat{p}_{{\rm eq},h} = \hat{p}_{{\rm eq},h} {^t \hat{W}_h} \ .
\end{equation}
On v\'erifie alors ais\'ement l'\'equation suivante sur la d\'eriv\'ee \`a
champ nul de $\hat{W}_h$,
\begin{equation}
\hat{W}'=\left. \frac{\partial \hat{W}_h}{\partial h}\right|_{h=0} 
\ , \qquad
\hat{W}' \hat{p}_{\rm eq} + \beta \hat{W} \hat{B} \hat{p}_{\rm eq} = 
\hat{p}_{\rm eq} {^t \hat{W}'} + \beta \hat{B} \hat{p}_{\rm eq} {^t \hat{W}}\ .
\label{eq:fdt-propwp}
\end{equation}
On a, comme pour la matrice $\hat{W}$, des propri\'et\'es dues \`a la 
conservation des probabilit\'es pour tout champ ext\'erieur qui impliquent
$\langle - | \hat{W}' =0$ et $^t \hat{W}' |-\rangle =0$.

La fonction de r\'eponse (\ref{eq:fdt-defrab}) est d\'efinie par
une d\'erivation fonctionnelle par rapport au champ ext\'erieur. Il faut
donc prendre un champ d'intensit\'e $h/\Delta t$ pendant un intervalle 
$\Delta t$ autour du temps de d\'erivation,
puis d\'eriver par rapport \`a $h$ et prendre la limite $\Delta t \to 0$. On
peut se convaincre que dans le formalisme utilis\'e ici, cela revient \`a 
ins\'erer la matrice $\hat{W}'$ dans le bra-ket \`a l'instant correspondant.
Par exemple, la fonction (\ref{eq:fdt-defrab}) s'exprime pour $t>t'$ comme
\begin{equation}
\left. \frac{\delta}{\delta h(t')}  \langle A(t) \rangle_h
\right|_{h=0} = \langle - | \hat{A} \hat{T}(t - t') \hat{W}' 
\hat{T}(t' - t_0) \hat{p}_0 | - \rangle \ .
\label{eq:fdt-exrep}
\end{equation}

Ces longues d\'efinitions pr\'eliminaires \'etant pos\'es, les
d\'emonstrations sont quasi-imm\'ediates.

\vspace{8mm}

\noindent\underline{\emph{Causalit\'e}\hspace{-1.5mm}\phantom{p}}

\vspace{4mm}

Si $t'>t$, la fonction de r\'eponse (\ref{eq:fdt-exrep}) devient
$\langle - | \hat{W}' \hat{T}(t' - t) \hat{A} \hat{T}(t - t_0) \hat{p}_0 
| - \rangle $ . Or $\langle - | \hat{W}'=0$, on a donc bien annulation de la 
r\'eponse \`a une excitation post\'erieure \`a l'observation, que le 
syst\`eme soit \'equilibr\'e ou pas.

\vspace{3mm}

Pour la preuve des propri\'et\'es d'\'equilibre, on suppose qu'au temps
$t_0$ le syst\`eme est \'equilibr\'e, $\hat{p}_0=\hat{p}_{\rm eq}$. 

\vspace{8mm}

\noindent\underline{\emph{Invariance par translation temporelle}}

\vspace{4mm}

On peut lire la propri\'et\'e d'invariance par translation temporelle sur
les \'equations (\ref{eq:fdt-excorrel2}) et (\ref{eq:fdt-exrep})~: si
$\hat{p}_0=\hat{p}_{\rm eq}$, $\hat{T}(t' - t_0) \hat{p}_0 | - \rangle = 
\hat{p}_{\rm eq} | - \rangle$ et les fonctions ne d\'ependent que de la 
diff\'erence des temps $\tau=t-t'$~:

\begin{eqnarray}
C_{AB}(\tau) &=& \langle - | \hat{A} \hat{T}(\tau) \hat{B} 
\hat{p}_{\rm eq} |-\rangle \ , \label{eq:fdt-correlstat} \\
R_{AB}(\tau) &=& \langle - | \hat{A} \hat{T}(\tau) \hat{W}' 
\hat{p}_{\rm eq} | - \rangle \ .
\end{eqnarray}

\vspace{8mm}

\noindent\underline{\emph{Th\'eor\`eme de fluctuation-dissipation}}

\vspace{4mm}

Le th\'eor\`eme de fluctuation-dissipation s'obtient en ins\'erant la 
propri\'et\'e (\ref{eq:fdt-propwp}) dans l'expression de la r\'eponse
que l'on vient d'\'etablir, et en utilisant $^t \hat{W}'|-\rangle = 
{^t \hat{W} |-\rangle} =0$ pour simplifier le r\'esultat~:
\begin{equation}
\langle - | \hat{A} \hat{T}(\tau) \hat{W}' \hat{p}_{\rm eq} | - \rangle 
= - \beta \langle - | \hat{A} \hat{T}(\tau) \hat{W} B \hat{p}_{\rm eq} | - 
\rangle \ .
\label{eq:fdt-rep2temps}
\end{equation}
Comme $d\hat{T}(\tau)/d\tau = \hat{T}(\tau)\hat{W}$, on reconnait dans 
(\ref{eq:fdt-rep2temps}) la d\'eriv\'ee temporelle de la fonction de 
corr\'elation, ce qui prouve
\begin{equation}
R_{AB}(\tau)= -\frac{1}{T} \frac{d C_{AB}(\tau)}{d\tau} \ .
\end{equation}

\vspace{8mm}

\noindent\underline{\emph{Relation d'Onsager}}

\vspace{4mm}

Finalement, la relation d'Onsager sur la fonction de corr\'elation s'obtient
en prenant la transpos\'ee de l'expression matricielle 
(\ref{eq:fdt-correlstat}),
\begin{equation}
\langle - | \hat{A} \hat{T}(\tau) \hat{B} \hat{p}_{\rm eq} |-\rangle
= \langle - | \hat{p}_{\rm eq} \hat{B} {^t \hat{T}(\tau)} \hat{A} |-\rangle
= \langle - | \hat{B} \hat{T}(\tau) \hat{A} \hat{p}_{\rm eq} |-\rangle \ ,
\end{equation}
ce qui prouve $C_{AB}(t,t')=C_{BA}(t,t')$. On a utilis\'e ici la 
condition de balance d\'etaill\'ee sous la forme (\ref{eq:fdt-bdsurT}).

\newpage

\section{G\'en\'eralisations pour des fonctions \`a $n$ temps}

Le formalisme d\'evelopp\'e dans la partie pr\'ec\'edente a le m\'erite de
faciliter la g\'en\'eralisation de ces r\'esultats a des fonctions de 
corr\'elation et de r\'eponse \`a un nombre arbitraire de temps. 
D\'efinissons par exemple une fonction de corr\'elation \`a $n$ points,
\begin{equation}
C(t_n,\dots,t_1) = \langle A_n(t_n) A_{n-1}(t_{n-1}) \dots A_1(t_1) \rangle \ ,
\end{equation}
o\`u les $A_i$ sont des observables quelconques du syst\`eme. On va supposer
sans perdre de g\'en\'eralit\'e que les temps sont class\'es selon 
$t_n \ge t_{n-1} \ge \dots \ge t_1$. On a alors, en supposant que le syst\`eme
est \'equilibr\'e \`a un temps $t_0<t_1$, 
\begin{equation}
C(t_n,\dots,t_1) = \langle - |
\hat{A}_n \hat{T}(t_n-t_{n-1}) \hat{A}_{n-1} \dots \hat{A}_2 \hat{T}(t_2-t_1)
\hat{A}_1 \hat{p}_{\rm eq} |- \rangle \ .
\end{equation}
La propri\'et\'e d'invariance par translation temporelle est claire~: si l'on 
ajoute la m\^eme quantit\'e \`a tous les temps $t_i$, la corr\'elation n'est
pas modifi\'ee.

\vspace{8mm}

\noindent\underline{\emph{Relation d'Onsager}}

\vspace{4mm}

\begin{figure}[hb]
\includegraphics[width=10cm]{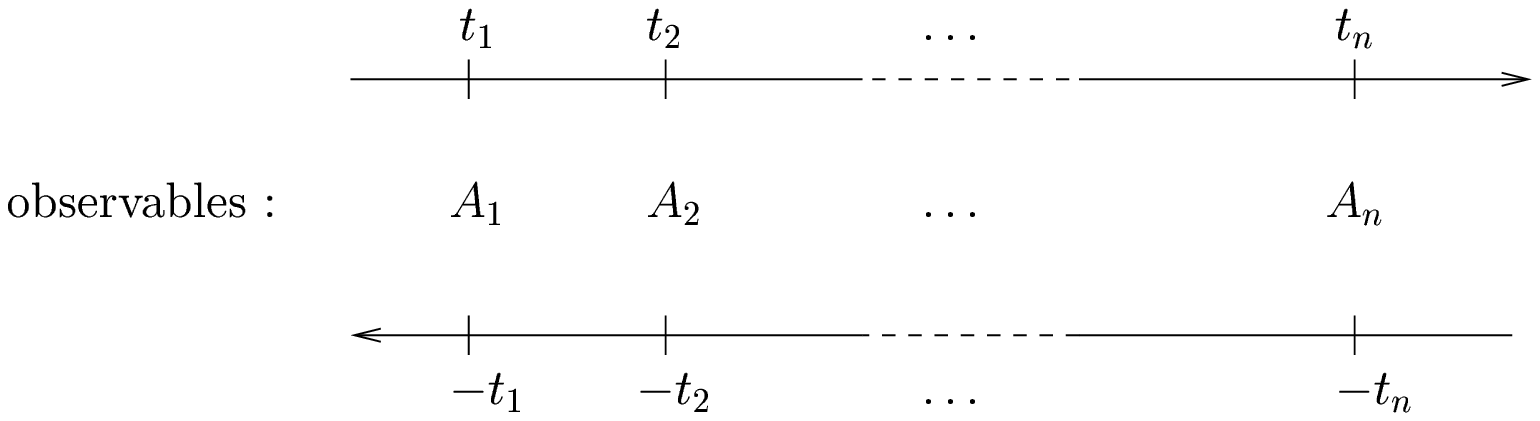}
\caption{Sch\'ematisation de la relation d'Onsager entre $n$ observables, cf.
eq. (\ref{eq:fdt-onsager-ntemps}). On a pris $t_*=0$ et $t_n=-t_1$ pour
simplifier le dessin.}
\label{fig:fdt-onsager-ntemps}
\end{figure}

On peut aussi construire une relation d'Onsager sur cette fonction de 
corr\'elation. Prenons, \`a l'image de la d\'emonstration faite pour la 
fonction \`a deux temps, la transpos\'ee de cette expression,
\begin{eqnarray}
&&\langle - |
\hat{A}_n \hat{T}(t_n-t_{n-1}) \hat{A}_{n-1} \dots \hat{A}_2 \hat{T}(t_2-t_1)
\hat{A}_1 \hat{p}_{\rm eq} |- \rangle \\ &=& 
\langle - | \hat{p}_{\rm eq} \hat{A}_1 {^t \hat{T}(t_2-t_1)} \hat{A}_2 \dots
\hat{A}_{n-1} {^t \hat{T}(t_n-t_n-1)} \hat{A}_n |-\rangle \\
&=& \langle - | \hat{A}_1 \hat{T}(t_2-t_1) \hat{A}_2 \dots
\hat{A}_{n-1} \hat{T}(t_n-t_n-1) \hat{A}_n \hat{p}_{\rm eq} |-\rangle 
\end{eqnarray}
On reconnait une fonction de corr\'elation o\`u l'ordre temporel des 
observables a \'et\'e renvers\'e. D\'efinissons pour \^etre plus pr\'ecis
une op\'eration de renversement temporel autour d'un point $t_*$,
\begin{equation}
t^R = 2 t_*-t \ .
\end{equation}
Gr\^ace \`a la propri\'et\'e d'invariance par translation, le choix de
$t_*$ est compl\`etement arbitraire. On peut alors \'ecrire la relation
pr\'ec\'edente sous la forme
\begin{equation}
\langle A_n(t_n) A_{n-1}(t_{n-1}) \dots A_2(t_2) A_1(t_1) \rangle
=\langle A_1(t_1^R) A_2(t_2^R) \dots A_{n-1}(t_{n-1}^R) A_n(t_n^R) \rangle \ ,
\label{eq:fdt-onsager-ntemps}
\end{equation}
ce que l'on visualise facilement sur le sch\'ema de la 
figure~\ref{fig:fdt-onsager-ntemps}. 

\newpage

\noindent\underline{\emph{Th\'eor\`eme de fluctuation-dissipation}}

\vspace{4mm}

Comme d\'ecrit dans le chapitre \ref{sec:continu}, on peut d\'efinir plusieurs 
fonctions de r\'eponse \`a $n$ temps, selon le nombre de perturbations $k$ et 
le nombre d'observables $n-k$ avec $k\in [1,n-1]$,
\begin{equation}
\left. \frac{\delta^ k}{\delta h_1(t_1) \dots \delta h_k(t_k)}
\langle A_{k+1}(t_{k+1}) \dots A_n(t_n) \rangle \right|_{h=0} \ .
\end{equation}
Ici chaque champ $h_i$ peut \^etre coupl\'e \`a une observable $B_i$ 
diff\'erente. On peut facilement montrer que ces fonctions sont causales,
c'est-\`a-dire qu'elles s'annulent d\`es qu'un des temps de perturbation
$\{ t_1,\dots,t_k \}$ est plus grand que le maximum des temps d'observation
$\{ t_{k+1},\dots,t_n \}$. En effet on a alors l'insertion d'un $\hat{W}'$
dans un bra-ket \`a gauche des $\hat{A}_i$, ce qui annule la fonction 
correspondante \`a cause de $\langle - |\hat{W}'=0$.

Commen\c cons par discuter le cas des fonctions \`a trois temps, et plus
particuli\`erement la r\'eponse d'une perturbation sur la moyenne de deux
observables, i.e. $n=3 \ , \ k=1$ dans la d\'efinition ci-dessus. Posons
\begin{equation}
C(t_1,t_2,t_p) = \langle A_1(t_1) A_2(t_2) B(t_p) \rangle \ , \qquad
R(t_1,t_2;t_p) = \left. \frac{\delta}{\delta h(t_p)} \langle A_1(t_1) A_2(t_2)
\rangle \right|_{h=0} \ ,
\end{equation}
avec $h$ un champ coupl\'e \`a $B$. On supposera sans perte de g\'en\'eralit\'e
que $t_2>t_1$. Par contre
$t_p$ peut se situer dans diff\'erents secteurs~:
\begin{itemize}
\item Si $t_p>t_2$, $R=0$ par causalit\'e.

\item Si $t_p<t_1$, autrement dit si la perturbation est ant\'erieure aux
deux temps d'observation,
\begin{equation}
R(t_1,t_2;t_p) = \langle - | \hat{A}_2 \hat{T}(t_2-t_1) \hat{A}_1
\hat{T}(t_1-t_p) \hat{W}' \hat{p}_{\rm eq} |-\rangle \ .
\end{equation}
On peut alors suivre les m\^emes \'etapes que celles conduisant \`a la preuve
du FDT \`a deux points en transformant $\hat{W}' \hat{p}_{\rm eq} |-\rangle$,
et obtenir
\begin{equation}
R(t_1,t_2;t_p) = \frac{1}{T} \frac{\partial}{\partial t_p} C(t_1,t_2,t_p) 
\qquad
\mbox{pour} \ \ t_2>t_1>t_p \ . 
\label{eq:fdt-3pt1}
\end{equation}
Cette relation est une g\'en\'eralisation naturelle du FDT habituel.

\item Si le syst\`eme est perturb\'e entre les deux temps d'observation, 
$t_2>t_p>t_1$, on a 

\vspace{1mm}

\begin{eqnarray}
R(t_1,t_2;t_p) &=& \langle - | \hat{A}_2 \hat{T}(t_2-t_p) \hat{W}'
\hat{T}(t_p-t_1) \hat{A}_1 \hat{p}_{\rm eq} |-\rangle \\
&=& \langle - | \hat{A}_1 \hat{T}(t_p-t_1) \hat{p}_{\rm eq} {^t \hat{W}'}
\hat{p}_{\rm eq}^{-1}
\hat{T}(t_2-t_p) \hat{A}_2 \hat{p}_{\rm eq} |-\rangle
\ .
\end{eqnarray}

\vspace{1mm}

La deuxi\`eme ligne a \'et\'e obtenue en prenant la transpos\'ee de la
premi\`ere. On peut maintenant utiliser (\ref{eq:fdt-propwp}) pour transformer
ceci en

\vspace{1mm}

\begin{eqnarray}
R(t_1,t_2;t_p) &=& \langle - | \hat{A}_1 \hat{T}(t_p-t_1) \hat{W}'
\hat{T}(t_2-t_p) \hat{A}_2 \hat{p}_{\rm eq} |-\rangle \\
&+& \beta \langle - | \hat{A}_1 \hat{T}(t_p-t_1) (\hat{W} \hat{B} -
\hat{B} \hat{W})
\hat{T}(t_2-t_p) \hat{A}_2 \hat{p}_{\rm eq} |-\rangle
\end{eqnarray}

\vspace{1mm}

La premi\`ere ligne de cette \'equation correspond \`a une fonction de 
r\'eponse avec des arguments temporels renvers\'es dans le temps, tandis
que la deuxi\`eme peut se r\'eecrire comme une d\'eriv\'ee temporelle de la
fonction de corr\'elation. On obtient finalement
\begin{equation}
R(t_1,t_2;t_p) - R(t_1^R,t_2^R;t_p^R) = \frac{1}{T} 
\frac{\partial}{\partial t_p} C(t_1,t_2,t_p) \ ,
\label{eq:fdt-3pt2}
\end{equation}
o\`u l'on a r\'eutilis\'e l'op\'eration de renversement temporel. 
Il n'\'etait pas compl\`etement \'evident de pr\'evoir cette relation qui
fait intervenir deux fonctions de r\'eponse, au vu du FDT sur les fonctions
\`a deux temps.

Cette forme contient en fait (\ref{eq:fdt-3pt1}) comme cas particulier~:
si $t_p<t_1$, $t_p^R$ est plus grand que $t_1^R$ et $t_2^R$. La deuxi\`eme 
fonction de r\'eponse de (\ref{eq:fdt-3pt2}) s'annule alors.
\end{itemize}

On peut sans difficult\'e g\'en\'eraliser cette relation au cas d'un produit
de $m$ observables avec une perturbation appliqu\'ee au temps $t_p$,
\begin{equation}
R(t_1,\dots,t_m;t_p) - R(t_1^R,\dots,t_m^R;t_p^R) = \frac{1}{T} 
\frac{\partial}{\partial t_p} C(t_1,\dots,t_m,t_p) \ .
\end{equation}

Il resterait \`a \'etudier les relations sur les fonctions de r\'eponse
\`a plusieurs perturbations. Dans la publication on pourra trouver le cas
de la r\'eponse d'une observable \`a deux perturbations.

\newpage

\section{Un th\'eor\`eme de fluctuation}
\label{sec-ftt}

Dans les d\'emonstrations de la partie pr\'ec\'edente on a souvent utilis\'e
l'op\'eration de transposition matricielle, qui correspond physiquement
\`a un renversement temporel. La condition de balance d\'etaill\'ee est
profond\'ement reli\'ee \`a cette notion~: elle
exprime la r\'eversibilit\'e microscopique des transitions entre deux
configurations. En cons\'equence le flux de probabilit\'e entre deux 
configurations microscopiques s'annule pour un ensemble de syst\`emes \`a 
l'\'equilibre, et l'on ne peut plus distinguer le sens temporel de
l'\'evolution.

Les relations d'Onsager sont une reformulation de cette
invariance, et les fonctions de r\'eponse sont dans un certain sens une
mesure de l'irr\'eversibilit\'e de l'\'evolution induite par les perturbations
ext\'erieures.

A la lumi\`ere de ces remarques, il n'est pas \'etonnant que l'on puisse 
reformuler toutes les propri\'et\'es d\'emontr\'ees jusqu'ici dans une 
expression qui fait jouer un r\^ole central au renversement temporel de
l'\'evolution. Ce principe, que l'on va pr\'esenter dans cette partie,
ressemble beaucoup \`a certaines formes du th\'eor\`eme de 
fluctuation~\cite{EvCoMo-FT,GaCo-FT} pour des dynamiques microscopiques
stochastiques~\cite{Ku-FT,LeSp-FT}. On pourra consulter~\cite{Ri-FT} pour une
revue de ces travaux, et \cite{Ga-FT} pour une mise au point sur le bon
usage du nom de th\'eor\`eme de fluctuation.

Nous nous sommes aper\c cus, malheureusement apr\`es la publication de
l'article, que ce r\'esultat avait \'et\'e obtenu ant\'erieurement par 
Crooks~\cite{Cr-FT}.

On va \`a nouveau changer un petit peu la mod\'elisation et les notations par
rapport \`a la publication et \`a la partie pr\'ec\'edente. On consid\`ere
maintenant un processus stochastique o\`u l'espace des configurations et le
temps sont discrets. L'\'evolution du syst\`eme est r\'egie par 
l'\'equation-ma\^itresse
\begin{equation}
P(\vec{\sigma},T+1) = \sum_{\vec{\sigma}'} W(\vec{\sigma},\vec{\sigma}',T) 
P(\vec{\sigma}',T) \ .
\end{equation}
Les $W$ sont ici des probabilit\'es de transition et non plus des taux de
transition par unit\'e de temps.
On suppose qu'un champ ext\'erieur $h(T)$ est coupl\'e
\`a une observable $B$, et que la d\'ependance temporelle explicite des
probabilit\'es de transition ne se fait que par l'interm\'ediaire de ce champ~:
\begin{equation}
W(\vec{\sigma},\vec{\sigma}',T) = \tilde{W}(\vec{\sigma},\vec{\sigma}';h(T))
\ .
\end{equation}
Fixons-nous un intervalle de temps $[-M,M]$ pendant lequel on observe le
syst\`eme. On appelera trajectoire l'ensemble des configurations occup\'ees
successivement par le syst\`eme au cours de cet intervalle de temps, 
$\underline{\sigma}=\{\vec{\sigma}_{-M},\vec{\sigma}_{-M+1},\dots,
\vec{\sigma}_M \}$. D\'efinissons aussi la trajectoire du champ ext\'erieur 
$\underline{h}=\{ h_{-M},\dots,h_{M-1} \}$. La probabilit\'e d'observation 
d'une trajectoire des configurations \'etant donn\'e une trajectoire du champ 
ext\'erieur s'exprime comme le produit des probabilit\'es de transition,
\begin{eqnarray}
P(\underline{\sigma}|\underline{h}) &=& 
\tilde{W}(\vec{\sigma}_M,\vec{\sigma}_{M-1};h_{M-1})
\tilde{W}(\vec{\sigma}_{M-1},\vec{\sigma}_{M-2};h_{M-2})
\dots \\
& & \hspace{1.5cm} \dots 
\tilde{W}(\vec{\sigma}_{-M+1},\vec{\sigma}_{-M};h_{-M}) 
P_{\rm in}(\vec{\sigma}_{-M}) \ ,
\end{eqnarray}
$P_{\rm in}$ d\'esignant la loi de probabilit\'e des configurations au temps 
$-M$. On note maintenant
$\underline{\sigma}^R=\{ \vec{\sigma}_M , \dots , \vec{\sigma}_{-M} \}$
et $\underline{h}^R = \{ h_{M-1},\dots,h_{-M} \}$ les trajectoires renvers\'ees
dans le temps autour de $T_*=0$.
Le quotient de la probabilit\'e d'observation d'une trajectoire 
$\underline\sigma$ dans un champ $\underline h$ par la quantit\'e 
correspondante apr\`es renversement temporel des trajectoires s'\'ecrit~:
\begin{equation}
\frac{P(\underline{\sigma}|\underline{h})}
{P(\underline{\sigma}^R|\underline{h}^R)} = \left( \prod_{i=-M}^{M-1}
\frac{\tilde{W}(\vec{\sigma}_{i+1},\vec{\sigma}_i;h_i)}
{\tilde{W}(\vec{\sigma}_{i},\vec{\sigma}_{i+1};h_i)} \right)
\frac{P_{\rm in}(\vec{\sigma}_{-M})}{P_{\rm in}(\vec{\sigma}_{M})} \ .
\label{eq:fdt-ft1}
\end{equation}
Faisons deux hypoth\`eses suppl\'ementaires~:
\begin{itemize}
\item Au temps $-M$ le syst\`eme est \'equilibr\'e par rapport \`a la mesure
de Gibbs en champ nul, i.e. 
$P_{\rm in}(\vec{\sigma})=\exp[-\beta H(\vec{\sigma})]/Z$.
On peut imaginer que le syst\`eme ait \'et\'e pr\'epar\'e bien avant et qu'on
l'ait laiss\'e \'evoluer sans champ jusqu'au temps $-M$.

\item Les probabilit\'es de transition $\tilde{W}$ v\'erifient la condition
de balance d\'etaill\'ee par rapport \`a l'hamiltonien total $H - h B$~:
\begin{equation}
\tilde{W}(\vec{\sigma},\vec{\sigma}',h) e^{-\beta H(\vec{\sigma}')+ \beta h
B(\vec{\sigma}')} = 
\tilde{W}(\vec{\sigma}',\vec{\sigma},h) e^{-\beta H(\vec{\sigma})+ \beta h
B(\vec{\sigma})} \ .
\end{equation}
\end{itemize}

Avec ces deux nouvelles hypoth\`eses, (\ref{eq:fdt-ft1}) se simplifie en
\begin{equation}
\frac{P(\underline{\sigma}|\underline{h})}
{P(\underline{\sigma}^R|\underline{h}^R)} = 
\exp \left[ \beta \sum_{i=-M}^{M-1} (B(\vec{\sigma}_{i+1}) - 
B(\vec{\sigma}_i)) h_i \right] \ .
\label{eq:fdt-ft-discret}
\end{equation}
On aurait obtenu une forme analogue en partant d'une dynamique en temps 
continu~:
\begin{equation}
\frac{P(\underline{\sigma}|\underline{h})}
{P(\underline{\sigma}^R|\underline{h}^R)} = 
\exp \left[ \beta \int_{t=-t_M}^{t_M} dt \ h(t) \dot{B}(t) \right] \ ,
\label{eq:fdt-ft}
\end{equation}
o\`u $\dot{B}(t)$ d\'esigne la d\'eriv\'ee de $B(\vec{\sigma}(t))$ le
long de la trajectoire $\underline{\sigma}$.

Cette relation exprime en champ nul la micro-r\'eversibilit\'e de
l'\'evolution, et donc l'\'equiprobabilit\'e d'une trajectoire et
de sa renvers\'ee temporelle. Un champ ext\'erieur non nul brise cette
invariance~; on peut interpr\'eter~\cite{Cr-FT} l'exposant de 
(\ref{eq:fdt-ft-discret}) comme la partie du travail du champ ext\'erieur 
que le bain thermique doit dissiper, ce qui entra\^ine l'irr\'eversibilit\'e
du processus.

Ce th\'eor\`eme de fluctuation contient toutes les relations de la partie 
pr\'ec\'edente. Consid\'erons pour commencer le cas o\`u le champ $h$ est
nul. On peut facilement en d\'eduire les relations 
d'Onsager en multipliant les deux membres de l'\'equation par le produit
des observables concern\'ees~:
\begin{equation}
P(\underline{\sigma}|\underline{h}=0) 
A_n(\vec{\sigma}(t_n)) \dots A_1(\vec{\sigma}(t_1)) = 
P(\underline{\sigma}^R|\underline{h}=0) 
A_n(\vec{\sigma}(t_n)) \dots A_1(\vec{\sigma}(t_1)) \ .
\end{equation}
Sommant ensuite sur les trajectoires $\underline{\sigma}$ il vient
\begin{equation}
\langle A_n(t_n) \dots A_1(t_1) \rangle = 
\langle A_n(t_n) \dots A_1(t_1) \rangle_R = 
\langle A_1(t_1^R) \dots A_n(t_n^R) \rangle \ . 
\end{equation}
La notation $\langle \bullet \rangle_R$ d\'esigne ici la somme sur les 
trajectoires $\underline{\sigma}^R$, et la deuxi\`eme \'egalit\'e vient du
changement de variables d'int\'egration $\underline{\sigma}^R 
\to \underline{\sigma}$. On a ainsi r\'eobtenu la relation d'Onsager 
(\ref{eq:fdt-onsager-ntemps}).

On peut aussi d\'eduire toutes les relations de fluctation-dissipation 
de l'\'equation (\ref{eq:fdt-ft}). La m\'ethode, que l'on n'explicitera ici
que pour la fonction \`a deux points, consiste \`a nouveau \`a multiplier
les deux membres de l'\'equation par les observables et \`a sommer sur
les trajectoires~:
\begin{equation}
\left< A(t_1) \exp \left[- \beta \int dt \ h(t) \dot{B}(t) \right] \right>
= \langle A(t_1) \rangle_R \ .
\end{equation}
Prenant la d\'eriv\'ee fonctionnelle \`a champ nul de cette \'equation, il
vient en prenant soin de la d\'ependance temporelle renvers\'ee du deuxi\`eme
membre
\begin{equation}
- \beta \langle A(t_1) \dot{B}(t_2) \rangle + \frac{\delta}{\delta h(t_2)}
\langle A(t_1) \rangle = \frac{\delta}{\delta h(-t_2)} \langle A(-t_1) \rangle
\ ,
\end{equation}
qui est bien la forme habituelle du FDT, puisqu'une des deux r\'eponses 
s'annule par causalit\'e selon que $t_1>t_2$ ou $t_2>t_1$. On peut suivre la
m\^eme proc\'edure pour retrouver les relations entre corr\'elations et 
r\'eponses \`a trois temps obtenues dans la partie pr\'ec\'edente, ainsi que
leurs g\'en\'eralisations \`a un nombre arbitraire de temps.

\newpage

\section{G\'en\'eralisations hors d'\'equilibre}
\label{sec:ft-outofeq}

La motivation de l'\'etude pr\'esent\'ee dans ce chapitre r\'eside dans la
constatation que la dynamique des mod\`eles de spin dilu\'es ne peut
pas s'exprimer en termes des fonctions \`a deux temps uniquement. On a
explor\'e pour l'instant les propri\'et\'es qu'auraient les fonctions de
corr\'elation et de r\'eponse \`a plus de deux temps si le syst\`eme atteignait
l'\'equilibre. On s'attend cependant \`a ce que la phase de basse temp\'erature
de ces mod\`eles aient des propri\'et\'es vitreuses, et donc que le syst\`eme
reste hors d'\'equilibre \`a tous les temps.
On s'int\'eressera donc dans la fin de ce chapitre \`a la forme que prennent 
les fonctions \`a plusieurs temps dans cette phase de basse temp\'erature. 

Je commence avant cela par faire un rappel
sommaire de l'image de la dynamique hors d'\'equilibre des verres de spin qui
a \'emerg\'e des travaux sur les mod\`eles compl\`etement connect\'es.

\subsection{Le sc\'enario hors d'\'equilibre \`a une \'echelle de 
corr\'elation}

Le cadre th\'eorique dans lequel on se place ici a vu le jour avec
l'article de Cugliandolo et Kurchan~\cite{CuKu-sphe} sur la dynamique de
basse temp\'erature du mod\`ele $p$-spin sph\'erique. En effet, jusque l\`a
les \'etudes dynamiques de ces mod\`eles s'\'etaient cantonn\'es \`a la phase
d'\'equilibre \`a haute temp\'erature. 

Les \'equations sur les fonctions de corr\'elation et de r\'eponse de ce
mod\`ele, pour une pr\'eparation initiale imitant un refroidissement 
instantan\'e depuis la temp\'erature infinie vers celle du bain ext\'erieur,
ont \'et\'e donn\'ees dans la partie \ref{sec:msrgene}. Leur 
r\'esolution, dont le principe est expliqu\'e en grand d\'etail dans 
\cite{Cu-Houches}, r\'ev\`ele l'existence d'une temp\'erature de transition 
$T_d$ entre deux r\'egimes.

\begin{figure}
\includegraphics[width=13cm]{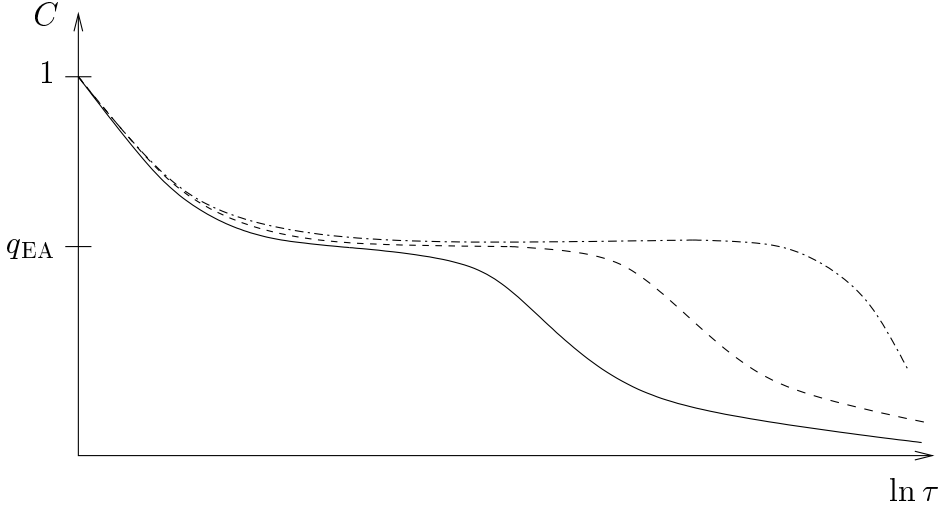}
\caption{Allure de la fonction de corr\'elation dans le sc\'enario \`a une
\'echelle de corr\'elation. Les diff\'erentes courbes correspondent soit
\`a diff\'erentes temp\'eratures pour $T>T_d$, soit \`a diff\'erents temps
d'attente pour une temp\'erature $T<T_d$ fix\'ee.}
\label{fig:ft-sketchC}
\end{figure}

A haute temp\'erature, la configuration initiale relaxe rapidement, et
apr\`es ce court r\'egime transitoire les fonctions de corr\'elation et
de r\'eponse sont stationnaires. Le th\'eor\`eme de fluctuation-dissipation
est aussi v\'erifi\'e, l'\'equilibre est donc atteint. Quand on se rapproche 
de $T_d$ par valeurs 
sup\'erieures, la fonction de corr\'elation 
prend une allure
particuli\`ere, sch\'ematis\'ee sur la figure \ref{fig:ft-sketchC}. La
d\'ecroissance se fait en deux temps, d'abord de la valeur initiale jusqu'\`a 
une valeur de plateau $q_{\rm EA}$, puis du plateau \`a 0. La dur\'ee de
ce plateau augmente quand on diminue la temp\'erature, et finit par diverger 
\`a $T_d$. Quelques remarques s'imposent ici. Tout d'abord, cette description
ne concerne que le cas $p \ge 3$~: on a vu dans la partie \ref{sec:stadynp2}
que pour $p=2$, c'est-\`a-dire la version sph\'erique du mod\`ele de
Sherrington-Kirpatrick, le param\`etre d'ordre $q_{\rm EA}$ croissait
continument \`a la temp\'erature de transition. Il n'y a donc pas l'apparition
d'un tel plateau dans la phase de haute temp\'erature. Signalons aussi que
la version sph\'erique du mod\`ele $p$-spin a \'et\'e introduite par
Crisanti et Sommers~\cite{CrSo-stat}, ces auteurs ayant aussi \'etudi\'e la
dynamique de haute temp\'erature dans \cite{CrSo-dyn}. Finalement, on peut
souligner la similitude de cette forme de la fonction de corr\'elation avec
celles pr\'edites pour les liquides structuraux par la th\'eorie du couplage
de modes (MCT)~\cite{Go-MCT}. Cette similitude n'est pas accidentelle~: comme 
l'ont remarqu\'e Kirkpatrick, Thirumalai et Wolynes~\cite{KiTh1,KiTh2,KiWo}, 
les \'equations dynamiques du mod\`ele $p$-spin sph\'erique ont la m\^eme forme
que les versions sch\'ematiques de la MCT. On trouvera une discussion plus
d\'etaill\'ee de cette relation, et de ses possibles implications \`a basse
temp\'erature, dans \cite{BoCuKuMe-MCA}.

Concentrons-nous maintenant sur la phase de basse temp\'erature. Dans ce cas
la dynamique est hors d'\'equilibre. L'invariance par translation
temporelle est bris\'ee par le temps initial de pr\'eparation du syst\`eme,
qui n'est jamais \og oubli\'e\fg .
Il est donc n\'ecessaire de conserver explicitement le temps d'attente
$t_w$ pass\'e dans la phase de basse temp\'erature avant le d\'ebut des
mesures d'auto-corr\'elation et de r\'eponse. Il se trouve que la fonction
de corr\'elation $C(t_w+\tau,t_w)$, pour des temps d'attente $t_w$ suffisamment
longs, a aussi l'allure donn\'ee sur la figure \ref{fig:ft-sketchC}.
Autrement dit il y a une premi\`ere d\'ecorr\'elation vers le plateau 
$q_{\rm EA}$~\footnote{Ce param\`etre d'ordre d\'epend de la temp\'erature et
cro\^it discontinument quand on passe en dessous de $T_d$, pour all\'eger les 
notations je laisse implicite cette d\'ependance en temp\'erature.} 
qui est ind\'ependante du temps d'attente, et la d\'ecroissance finale de la
corr\'elation se fait sur des \'echelles de temps d'autant plus grandes que
le syst\`eme est rest\'e longtemps dans la phase de basse temp\'erature.
Les syst\`emes les plus \og vieux\fg\ \'evoluent le plus lentement dans ce
r\'egime. La valeur de la corr\'elation $q_{\rm EA}$ permet donc de
distinguer entre un r\'egime rapide, stationnaire, et un r\'egime lent
qui d\'epend de l'\^age du syst\`eme. Plus pr\'ecis\'ement, on peut
utiliser la param\'etrisation suivante de la fonction de corr\'elation~:
\begin{equation}
C(t_w+\tau,t_w) \approx C_{\rm fast}(\tau) 
+ C_{\rm slow} \left( \frac{l(t_w + \tau )}{l(t_w)} \right) \ ,
\end{equation}
o\`u $l(t)$ est une fonction croissante.
Cette d\'ecomposition n'est valable que dans la limite $t_w \to \infty$. Je
me permettrai cependant d'utiliser un signe d'\'egalit\'e dans la suite pour
simplifier les \'ecritures. Pour un temps d'attente $t_w$ donn\'e, les
temps ult\'erieurs se divisent en deux \'epoques~: si $\tau$ n'est pas trop
grand, la partie lente de la corr\'elation est quasiment constante, \'egale
\`a $q_{\rm EA}$, tandis que la partie rapide $C_{\rm fast}(\tau)$ d\'ecro\^it
de $1-q_{\rm EA}$ \`a 0. Pour des temps beaucoup plus grand, la partie rapide
s'est annul\'e, toute la d\'ependance temporelle vient alors de la partie 
lente de la corr\'elation. On dira dans la suite que deux temps $t_1$ et $t_2$
sont proches (resp. \'eloign\'es) si $C(t_1,t_2)>q_{\rm EA}$ (resp. 
$C(t_1,t_2)<q_{\rm EA}$).

La fonction $l(t)$ encode la d\'ependance de la dynamique vis \`a vis de
l'\^age du syst\`eme.
Dans la plupart des cas elle n'a pas \'et\'e 
d\'etermin\'ee analytiquement~; le traitement des \'equations dans la limite
des longs temps d'attente repose sur l'abandon de certains termes, ce qui
fait appara\^itre une invariance par reparam\'etrisation de la fonction $l$.

Le sc\'enario pr\'esent\'e ici pour la d\'ependance temporelle de la 
fonction de corr\'elation est dit \`a brisure faible d'ergodicit\'e 
(WEB)~\cite{Bo-WEB,BoDe-WEB}. En effet, la d\'ecroissance de la partie rapide 
de la corr\'elation ne se fait
que jusqu'\`a une valeur positive de la corr\'elation, comme si
l'ergodicit\'e \'etait bris\'e par un confinement du syst\`eme dans une partie
de l'espace des phases de taille $q_{\rm EA}$. Cette brisure n'est pas 
compl\`ete
car sur des \'echelles de temps beaucoup plus longues le syst\`eme arrive \`a
\'echapper \`a ce confinement, la corr\'elation finissant par d\'ecro\^itre
jusqu'\`a 0.

La fonction de r\'eponse pr\'esente aussi une d\'ecomposition similaire
en deux contributions~:
\begin{equation}
R(t_w+\tau;t_w) = R_{\rm fast}(\tau) + \frac{l'(t_w)}{l(t_w)} 
R_{\rm slow} \left( \frac{l(t_w + \tau)}{l(t_w)} \right) \ .
\end{equation}
Le quotient $l'/l$ s'annule pour des longs temps d'attente, ceci fait partie
du sc\'enario \og \`a faible m\'emoire \`a long terme\fg\ (WLTM).

Les fonctions de corr\'elation et de r\'eponse pour des faibles diff\'erences 
de temps v\'erifient le th\'eor\`eme de fluctuation-dissipation avec
la temp\'erature $T$ du bain thermique ext\'erieur~:
\begin{equation}
R_{\rm fast}(\tau) = - \beta C'_{\rm fast}(\tau) \ .
\end{equation}
Ce th\'eor\`eme est viol\'e dans le r\'egime vieillissant~: la temp\'erature 
du thermostat est remplac\'ee par une temp\'erature effective g\'en\'er\'ee
spontan\'ement par le syst\`eme. Pour deux temps $t>t_w$ \'eloign\'es, on a
\begin{equation}
R(t;t_w) =  \beta_{\rm eff} \frac{\partial}{\partial t_w} C(t,t_w) \ .
\end{equation}
Cette temp\'erature effective, qui peut s'exprimer en fonction des 
caract\'eristiques du mod\`ele, poss\`ede certaines propri\'et\'es attendues
pour une \og temp\'erature\fg , comme le contr\^ole du sens des \'echanges 
d'\'energie entre deux syst\`emes.
Ces propri\'et\'es sont l'objet notamment de~\cite{CuKuPe-Teff}.

Le sc\'enario pr\'esent\'e ici est le plus simple dans la famille des 
dynamiques
hors d'\'equilibre de verres de spins champ moyen~: il n'y a qu'une valeur
$q_{\rm EA}$, donc seulement deux r\'egimes temporels (un d'\'equilibre, un
vieillissant) et deux temp\'eratures (celle du bain ext\'erieur, et celle
auto-induite sur les degr\'es de libert\'e lents).
On peut rencontrer d'autres situations plus compliqu\'ees, notamment dans la
version soft-spin du mod\`ele de Sherrington-Kirkpatrick~\cite{CuKu-SK}.
Il appara\^it alors une infinit\'e d'\'echel\-les de corr\'elation.
Cette diff\'erence de comportement est reli\'ee aux deux types de brisure
de la sym\'etrie des r\'epliques rencontr\'es dans les \'etudes statiques.
Pour le $p$-spin sph\'erique un seul pas de brisure est 
suffisant, alors que le SK pr\'esente une brisure compl\`ete de la sym\'etrie.

On utilisera dans la suite le sc\'enario \`a une \'echelle de corr\'elation, 
l'adaptation des r\'esultats au sc\'enario du SK \'etant par ailleurs possible.

\subsection{Cons\'equences sur les fonctions \`a trois temps}

On a donc une forme tr\`es particuli\`ere pour la violation des th\'eor\`emes
d'\'equilibre dans la dynamique de basse temp\'erature du mod\`ele $p$-spin
sph\'erique. Il serait int\'eressant d'avoir un mod\`ele dans lequel 
les fonctions \`a plus de deux temps ne d\'ecoulent pas directement de
celles \`a deux temps, et pour lequel on puisse expliciter les \'equations 
r\'egissant les fonctions de corr\'elation et de r\'eponse \`a plus de deux 
temps. Il conviendrait alors de
chercher l'analogue des formes asymptotiques trouv\'ees sur les fonctions
\`a deux temps dans le mod\`ele compl\`etement connect\'e. On va se contenter
d'une approche un peu d\'etourn\'ee~: supposons que l'on ait une variable
scalaire qui \'evolue selon un processus stochastique gaussien (par exemple
une \'equation single-spin pour une variable effective). On peut alors calculer
tous les cumulants de ce processus en fonction des premiers cumulants.
Si l'on prend pour ces derniers le sc\'enario hors d'\'equilibre \`a une 
\'echelle de temps, on peut en explorer les cons\'equences sur les 
fonctions \`a $n$ points. On supposera finalement que ces relations resteront
vraies dans un mod\`ele non gaussien.

D\'efinissons donc un processus gaussien $\phi(t)$ coupl\'e lin\'eairement
\`a un champ ext\'erieur $h(t)$. On suppose sa moyenne non nulle pour avoir
des fonctions \`a trois points non triviales, dans le cas contraire il
faudrait \'ecrire les relations \`a quatre points ce qui alourdirait la
discussion. Le processus est compl\`etement caract\'eris\'e par la donn\'ee
de sa moyenne, de la corr\'elation \`a champ nul et de la fonction de 
r\'eponse~:
\begin{equation}
\langle \phi(t) \rangle = M(t) \ , \qquad
\langle \phi(t) \phi(t') \rangle = C(t,t') \ , \qquad
\left. \frac{\delta}{\delta h(t')} \langle \phi(t) \rangle \right|_{h=0}= 
R(t;t') \ .
\end{equation}
Un calcul imm\'ediat d'int\'egrales gaussiennes permet d'exprimer les 
fonctions \`a trois temps~:
\begin{eqnarray}
C(t_1,t_2,t_3) &=& M(t_1) C(t_2,t_3) 
+ M(t_2) C(t_1,t_3) + M(t_3) C(t_1,t_2) \ , \nonumber \\
R(t_1,t_2;t_3) &=& \left. \frac{\delta}{\delta h(t_3)} \langle \phi(t_1) 
\phi(t_2) \rangle \right|_{h=0} = 
M(t_1) R(t_2;t_3) + M(t_2) R(t_1;t_3) \ .
\label{eq:ft-gaussien}
\end{eqnarray}
Supposons que tous les temps mis en jeu soient grands devant un temps 
microscopique, de sorte que la moyenne $M(t)$ ait atteint une valeur
asymptotique $M_{\rm as}$, et que la d\'ecomposition des fonctions \`a
deux temps en une partie rapide et une partie lente soit valable.
On cherche donc les relations entre les fonctions \`a trois points que
cette d\'ecomposition induit. Rappelons que l'on a d\'emontr\'e \`a
l'\'equilibre~:
\begin{equation}
\beta \frac{\partial}{\partial t_p} C(t_1,t_2,t_p) = R(t_1,t_2;t_p) -
R(t_1^R,t_2^R;t_p^R) \ ,
\label{fdt-3pteq}
\end{equation}
o\`u la notation $t^R$ d\'esigne le renversement temporel par rapport
\`a un temps $t_*$ arbitraire, $t^R=2 t_*-t$. Pour le processus gaussien 
\'etudi\'e ici,
il suffit d'ins\'erer les formes asymptotiques des fonctions \`a deux points
dans les relations (\ref{eq:ft-gaussien}) pour \'etablir les
g\'en\'eralisations de la relation d'\'equilibre (\ref{fdt-3pteq}). 
Les diff\'erents cas \`a consid\'erer sont d\'etaill\'es dans la publication
\pubjsp , je donne ici un r\'esum\'e des r\'esultats obtenus~:

\vspace{2mm}

\begin{itemize}
\item Supposons d'abord que le temps de perturbation $t_p$ soit ant\'erieur
aux deux temps d'observation, $t_p<t_1<t_2$. Il faut
alors distinguer deux possibilit\'es~:

\vspace{2mm}

\begin{itemize}
\item si $t_p$ et $t_1$ sont proches, c'est \`a dire $C(t_p,t_1)>q_{\rm EA}$,
le FDT est v\'erifi\'e avec la temp\'erature du bain ext\'erieur,
\begin{equation}
\beta \frac{\partial}{\partial t_p} C(t_1,t_2,t_p) = R(t_1,t_2;t_p) \ .
\end{equation}
\item si $t_p$ et $t_1$ sont \'eloign\'es, i.e. $C(t_p,t_1)<q_{\rm EA}$, 
c'est la temp\'erature effective qui contr\^ole
la relation de fluctuation-dissipation~:
\begin{equation}
\beta_{\rm eff} 
\frac{\partial}{\partial t_p} C(t_1,t_2,t_p) = R(t_1,t_2;t_p) \ .
\label{fdt-3pt-noneq1}
\end{equation}
\end{itemize}
Notons que le temps $t_2$ est \og spectateur\fg\ ici~: qu'il soit proche ou
loin de $t_1$ ne change pas la forme de la relation.

\vspace{2mm}

\item Int\'eressons-nous maintenant au cas $t_1<t_p<t_2$ d'une perturbation 
faite \`a un temps interm\'ediaire. On trouve que la
relation d'\'equilibre (\ref{fdt-3pteq}) est v\'erifi\'e d\`es qu'au moins un
des deux temps $t_1$ ou $t_2$ est proche de $t_p$. Par contre si les deux
en sont \'eloign\'es, c'est-\`a-dire si $C(t_p,t_1)<q_{\rm EA}$ et 
$C(t_p,t_2)<q_{\rm EA}$, la g\'en\'eralisation hors-d'\'equilibre prend la 
forme~:
\begin{equation}
\beta_{\rm eff} \frac{\partial}{\partial t_p} C(t_1,t_2,t_p) = R(t_1,t_2;t_p) -
R(t_1^{Rl},t_2^{Rl};t_p^{Rl}) \ .
\label{fdt-3pt-noneq2}
\end{equation}
Il y a deux modifications par rapport \`a (\ref{fdt-3pteq})~: comme on pouvait
s'y attendre, la temp\'erature effective remplace celle du bain thermique. 
Le point
le plus original ici est que l'on doit d\'efinir une nouvelle op\'eration
de \og renversement temporel dans l'\'echelle vieillissante\fg , 
\begin{equation}
t^{Rl} = l^{-1} \left( \frac{l(t_*)^2}{l(t)} \right) \ ,
\label{eq:tr-outofeq}
\end{equation}
o\`u $t_*$ est un temps arbitraire, \'eloign\'e de $t$, laiss\'e invariant par 
cette op\'eration de renversement. $l^{-1}$ d\'esigne ici la r\'eciproque 
de $l$, autrement dit le temps $t$ et son renvers\'e $t^{Rl}$ sont reli\'es par
$l(t)l(t^{Rl})=l(t_*)^2$. On v\'erifie ais\'ement que pour la forme
d'\'equilibre $l(t)=e^t$ cette d\'efinition co\"incide avec celle
du renversement temporel standard.
\end{itemize}

On pourrait de la m\^eme fa\c con \'etudier les cons\'equences du sc\'enario
hors d'\'equilibre sur des fonctions avec un plus grand nombre de temps.

\subsection{G\'en\'eralisation du th\'eor\`eme de fluctuation}

La section \ref{sec-ftt} pr\'esentait une formulation compacte des 
propri\'et\'es de la dynamique d'\'equilibre \`a l'aide d'un th\'eor\`eme de 
fluctuation. Celui-ci \'etait essentiellement une quantification de la brisure 
d'invariance par renversement temporel que le travail d'un champ ext\'erieur 
induit. On va voir maintenant que les relations hors d'\'equilibre 
postul\'ees dans le paragraphe pr\'ec\'edent d\'ecoulent, elles aussi, 
d'une modification du th\'eor\`eme de fluctuation.

D\'ecomposons pour cela l'\'evolution microscopique du syst\`eme en une partie
rapide, d'\'equilibre, et une partie lente vieillissante. Une telle 
d\'ecomposition a \'et\'e utilis\'e par Franz et Virasoro 
dans~\cite{FrVi-quasi}. L'id\'ee consiste \`a d\'efinir la partie lente
de l'\'evolution comme une moyenne sur un intervalle de temps suffisamment 
grand~:
\begin{equation}
\phi(t) = \phi_{\rm f}(t) + \phi_{\rm s}(t) \ , \qquad
\phi_{\rm s}(t) = \frac{1}{\Delta(t)} \int_t^{t+\Delta(t)} dt' \, \phi(t') 
\ ,
\end{equation}
o\`u $\Delta(t)$ est d\'efini par $C(t,\Delta(t))=q_{\rm EA}$. Le degr\'e de
libert\'e lent $\phi_{\rm s}(t)$ est donc la moyenne sur tous les temps proches
de $t$. Le champ ext\'erieur $h(t)$ peut \^etre de la m\^eme mani\`ere 
d\'ecompos\'e en une partie rapide et une partie lente. 

On peut v\'erifier alors que les relations hors d'\'equilibre s'obtiennent 
en supposant que~:

\vspace{2mm}

\begin{itemize}
\item La partie rapide $\phi_{\rm f}(t)$ v\'erifie le th\'eor\`eme de
fluctuation habituel (cf. eq. (\ref{eq:fdt-ft})) avec la temp\'erature $T$ du
thermostat, seule la partie rapide $h_{\rm f}$ du champ ext\'erieur agissant 
sur $\phi_{\rm f}(t)$.
\item Les degr\'es de libert\'e lents du syst\`eme sont soumis \`a un
principe g\'en\'eralis\'e,
\begin{equation}
\frac{P(\underline{\phi_{\rm s}}|\underline{h_{\rm s}})}
{P(\underline{\phi_{\rm s}}^{Rl}|\underline{h_{\rm s}}^{Rl})} = 
\exp \left[ \beta_{\rm eff} \int dt \, h_{\rm s}(t) 
\dot{\phi}_{\rm s}(t) \right] \ .
\label{eq:ft-outofeq}
\end{equation}
La temp\'erature du bain a \'et\'e remplac\'ee par la temp\'erature effective, 
et l'op\'eration de renversement temporel par sa contrepartie dans le domaine
vieillissant d\'efinie par l'\'equation (\ref{eq:tr-outofeq}).
\end{itemize}

\vspace{2mm}

A nouveau ce r\'esultat n'est qu'une conjecture, et il serait int\'eressant 
de tester sa v\'eracit\'e sur des mod\`eles suffisamment simples pour \^etre
solubles, mais non trivialement gaussiens.

\subsection{Perspectives}

La d\'emarche suivie pour \'etablir ces relations hors d'\'equilibre est
criticable~: les relations hors d'\'equilibre (\ref{fdt-3pt-noneq1}) et 
(\ref{fdt-3pt-noneq2}) sont ici des cons\'equences directes de l'hypoth\`ese
faite sur les fonctions \`a deux temps. La conjecture
avanc\'ee dans la publication \pubjsp\ est qu'il existe des mod\`eles non
triviaux pour lesquels ces relations restent vraies. Le travail expos\'e dans 
la partie \ref{sec:msrgene} montre que cette hypoth\`ese est au moins 
coh\'erente pour les mod\`eles $p$-spin dilu\'es. En effet, m\^eme si l'on ne 
sait pas r\'esoudre l'\'equation de point-col sur le param\`etre d'ordre dans 
ce cas-l\`a, les fonctions \`a $n$ points peuvent formellement s'exprimer 
par un d\'eveloppement diagrammatique perturbatif autour d'une th\'eorie
gaussienne. Ces diagrammes sont construits \`a partir des op\'erations de
convolution et de produit direct de noyaux supersym\'etriques. Or ces
deux op\'erations conservent les d\'ecompositions en diff\'erentes \'epoques 
temporelles~: si les fonctions de r\'eponse et de corr\'elation de deux
noyaux $F_1(a,b)$ et $F_2(a,b)$ ont une partie rapide et une partie lente
s\'epar\'ees par une certaine \'echelle de corr\'elation, les produits
$F_1\otimes F_2$ et $F_1 \bullet F_2$ vont aussi pr\'esenter le m\^eme type
de s\'eparation d'\'echelles.

Une direction alternative serait de tester ces pr\'edictions \`a l'aide de
simulations num\'eriques. Parmi les plus simples de ces v\'erifications,
on peut remarquer que le th\'eor\`eme de fluctuation g\'en\'eralis\'e,
pris sans champ ext\'erieur, donne lieu \`a des relations d'Onsager dans le 
r\'egime vieillissant. On a notamment pour trois temps $t_1<t_2<t_3$ tous 
\'eloign\'es les uns des autres~:
\begin{equation}
C(t_1,t_2,t_3) = C(t_1,t_2^{Rl},t_3) \ , 
\end{equation}
o\`u $t_2^{Rl}$ est d\'efini par $C(t_1,t_2)=C(t_2^{Rl},t_3)$. Ce type de 
relation devrait \^etre assez facile \`a tester dans des simulations 
num\'eriques de verres de spin dilu\'es, et donnerait plus de cr\'edit \`a
cette assez surprenante id\'ee de \og renversement temporel dans l'\'echelle
vieillissante\fg .

Signalons enfin deux \'etudes r\'ecentes concernant, dans des contextes un peu
diff\'erents, des g\'en\'eralisations du th\'eor\`eme de fluctuation dans
des situations hors d'\'equilibre~\cite{Za-FT,Ri-FToutofeq}.

\newpage
\pagestyle{empty}
\mbox{}
\newpage
\pagestyle{plain}

\chapter{Conclusion}

Ici s'ach\`eve la pr\'esentation des r\'esultats obtenus au cours de cette
th\`ese consacr\'ee \`a la dynamique hors d'\'equilibre des syst\`emes 
dilu\'es. Comme on l'a vu au cours des diff\'erents chapitres, la 
connectivit\'e finie de ces syst\`emes est \`a l'origine d'un certain nombre
de difficult\'es techniques. Plusieurs sch\'emas d'approximation ont par
cons\'equent \'et\'e propos\'es et partiellement exploit\'es. On peut
esp\'erer que ces travaux pr\'eliminaires pourront \^etre approfondis et
conduire \`a des pr\'edictions physiques plus compl\`etes sur le comportement
de cette famille de syst\`eme, qui est loin d'avoir \'et\'e \'elucid\'e ici.

\vspace{2mm}

J'ai essay\'e de donner \`a la fin de chaque partie des directions
d'approfondissements possibles, je me contenterai d'en reprendre quelques unes
ici~:

\vspace{2mm}

\begin{itemize}
\item La formulation supersym\'etrique de l'int\'egrale de chemin de 
Martin-Siggia-Rose compl\'et\'ee par l'introduction d'un param\`etre d'ordre
inspir\'e de la th\'eorie statique (section \ref{sec:msrgene}) a permis
d'\'ecrire formellement l'\'equation r\'egissant la dynamique des versions
soft-spin des mod\`eles dilu\'es. Il serait certainement int\'eressant de
pousser plus avant cette approche~; l'approximation \`a un seul d\'efaut de 
Biroli et Monasson, pr\'esent\'ee dans le cadre des matrices al\'eatoires,
devrait pouvoir \^etre adapt\'ee aux calculs dynamiques. Les r\'esultats
obtenus dans le chapitre \ref{sec:ch-ft} sur la forme hors d'\'equilibre
des fonctions \`a $n$ points seront peut-\^etre utiles dans cette tentative.

\item Le travail sur le ferromagn\'etique \`a connectivit\'e fixe (section 
\ref{sec:ferro-arbre}) pourrait par ailleurs constituer un point de d\'epart 
pour une investigation de la phase de Griffiths dans les mod\`eles \`a 
connectivit\'e fluctuante.

\item L'\'etude de l'algorithme d'optimisation expos\'e dans la partie
\ref{sec:walksat} ouvre de nombreuses portes, tant vers des travaux analytiques
pour une meilleure description quantitative de ce processus 
\og non-physique\fg\ que vers des investigations num\'eriques d'algorithmes
plus performants. Une des questions importantes \`a clarifier dans cette
perspective concerne le lien entre la structure du paysage des configurations
microscopiques et les propri\'et\'es de tels algorithmes qui ne respectent pas
les conditions physiques de type balance d\'etaill\'ee.
Il serait souhaitable que des \'echanges fructueux s'\'etablissent 
entre la communaut\'e
de physique statistique et celle d'informatique, dont les objets d'\'etude
sont \'etroitement li\'es. La confrontation d'approches et de
m\'ethodes assez radicalement diff\'erentes peut s\^urement apporter
beaucoup aux deux domaines.

\end{itemize}

\vspace{2mm}

Je voudrais finalement souligner un point qui n'a \'et\'e que tr\`es peu 
abord\'e dans ce manuscrit. Les syst\`emes dilu\'es sont des mod\`eles
de champ moyen, et toutes les paires de sites ont la m\^eme probabilit\'e 
a priori d'\^etre en interaction. Si l'on consid\`ere un \'echantillon
donn\'e, on peut par contre d\'efinir une distance entre deux sites comme
la longueur minimum des chemins qui les relient. On a donc une notion
de \og g\'eom\'etrie\fg , qui permet de d\'efinir par exemple des
corr\'elations spatiales dans le syst\`eme. Cet aspect a \'et\'e perdu
dans les \'etudes pr\'esent\'ees ici, par exemple dans l'approche fonctionnelle
de la section \ref{sec:msrgene}, car on s'est int\'eress\'e aux propri\'et\'es 
moyenn\'ees sur l'ensemble des \'echantillons. La n\'ecessit\'e et/ou la 
possibilit\'e de travailler sur un \'echantillon donn\'e dans ces syst\`emes 
dilu\'es sugg\`ere des perspectives assez fascinantes. Les applications de la 
m\'ethode de cavit\'e l'ont montr\'e pour les propri\'et\'es statiques, il 
reste des possibilit\'es pour \'etendre cette d\'emarche \`a la dynamique.

\newpage

\addcontentsline{toc}{chapter}{Bibliographie}

\bibliography{these}

\newpage
\pagestyle{empty}
\mbox{}
\newpage
\pagestyle{empty}

\chapter*{Table des publications}
\addcontentsline{toc}{chapter}{Table des publications}

\noindent{\Large \bf Publications dans des journaux de physique}

\vskip 8mm

\noindent [\pubclusters] G. Semerjian and L.F. Cugliandolo, Cluster expansions 
in dilute systems: Applications to satisfiability problems and spin glasses,
\emph{Phys. Rev. E} {\bf 64 }, 036115 (2001).

\vskip 4mm

\noindent [\pubmatrix] G. Semerjian and L.F. Cugliandolo, Sparse random 
matrices: the eigenvalue spectrum revisited, \emph{J. Phys. A} {\bf 35}, 4837 
(2002).

\vskip 4mm

\noindent [\pubsphe] G. Semerjian and L.F. Cugliandolo, Dynamics of dilute 
disordered models: A solvable case, 
\emph{Europhys. Lett.} {\bf 61}, 247 (2003).

\vskip 4mm

\noindent [\pubwsat] G. Semerjian and R. Monasson, Relaxation and metastability
in a local search procedure for the random satisfiability problem, 
\emph{Phys. Rev. E} {\bf 67}, 066103 (2003).

\vskip 4mm

\noindent [\pubjsp] G. Semerjian, L.F. Cugliandolo and A. Montanari, On the 
stochastic dynamics of disordered spin models, \emph{J. Stat. Phys} {\bf 115}, 
493 (2004).

\vskip 4mm

\noindent [\pubbethe] G. Semerjian and M. Weigt,  Approximation schemes for the
dynamics of diluted spin models: the Ising ferromagnet on a Bethe lattice, 
\emph{J. Phys. A} {\bf 37}, 5525 (2004).

\vskip 12mm

\noindent{\Large \bf 
Publications dans des actes de conf\'erences d'informatique}

\vskip 8mm

\noindent [\pubsat] G. Semerjian and R. Monasson, A Study of Pure Random Walk 
on Random Satisfiability Problems with \og Physical\fg\ Methods, 
\emph{Proceedings of the SAT 2003 conference}, 
E. Giunchiglia and A. Tacchella eds.,
\emph{Lecture Notes in Computer Science} {\bf 2919}, 120 (2004).

\vskip 4mm

\noindent [\pubptac] S. Cocco, R. Monasson, A. Montanari and G. Semerjian,
Approximate analysis of search algorithms with \og physical\fg methods,
\emph{Computational Complexity and Statistical Physics},
A. Percus, G. Istrate and C. Moore eds., Oxford, in press.

\newpage

\mbox{}

\newpage

\mbox{}

\newpage

\pagestyle{empty}

\noindent {\Huge \bf R\'esum\'e}

\vspace{1cm}

Cette th\`ese est consacr\'ee \`a l'\'etude des propri\'et\'es dynamiques
des mod\`eles dilu\'es. Ces derniers sont des syst\`emes de physique
statistique de type champ moyen, mais dont la connectivit\'e locale est finie.
Leur \'etude est notamment motiv\'ee par l'\'etroite analogie qui 
les relient aux probl\`emes d'optimisation combinatoire, la 
$K$-satisfiabilit\'e al\'eatoire par exemple.

On expose plusieurs approches analytiques visant \`a d\'ecrire le
r\'egime hors d'\'equilibre de ces syst\`emes, qu'il soit d\^u \`a une
divergence des temps de relaxation dans une phase vitreuse, \`a l'absence de 
condition de balance d\'etaill\'ee pour un algorithme d'optimisation, 
ou \`a une pr\'eparation initiale dans une configuration loin de l'\'equilibre 
pour un ferromagn\'etique. Au cours de ces \'etudes on rencontrera \'egalement
un probl\`eme de matrices al\'eatoires, et une g\'en\'eralisation du 
th\'eor\`eme de fluctuation-dissipation aux fonctions \`a $n$ temps.

\vspace{4cm}

\noindent {\Huge \bf Abstract}

\vspace{1cm}
This thesis is devoted to the study of dynamical properties of diluted models.
These are mean field statistical mechanics systems, but with finite local
connectivity. Among other reasons, the interest for these models arises from
their deep relationship with combinatorial optimization problems, random 
$K$-satisfiability for instance.

Several analytical descriptions of their out of equilibrium regime are 
described. This regime can be due to long relaxation times in glassy phases,
lack of detailed balance condition for optimization algorithms, or transient
relaxation from an arbitrary initial condition for ferromagnets. In the course
of these studies some attention will also be given to random matrix theory,
and to a generalization of fluctuation-dissipation theorem for $n$-times
functions.

\end{document}